\documentclass[onecolumn,notitlepage,floats,floatfix,amssymb,aps,prd,showpacs,superscriptaddress,groupedaddress,nofootinbib]{revtex4-2}  

\usepackage{moresize}
\usepackage[T1]{fontenc}
\usepackage{lmodern} 
\usepackage{graphicx}  
\usepackage{dcolumn}   
\usepackage{bm}        
\usepackage{amssymb}   
\usepackage{amsmath}
\usepackage{bbold}
\usepackage{array}
\usepackage{makecell}
\usepackage{braket}
\usepackage{longtable}
\usepackage{supertabular,booktabs}

\usepackage{titlesec} 
\usepackage[colorlinks,citecolor=blue,urlcolor=blue,hypertexnames=true]{hyperref}
\usepackage{subfigure}

\usepackage[usenames,dvipsnames,svgnames,table]{xcolor} 

\renewcommand{\arraystretch}{2}

\newcommand{\be}{\begin{equation}}
\newcommand{\ee}{\end{equation}}
\newcommand{\bea}{\begin{eqnarray}}
\newcommand{\eea}{\end{eqnarray}}
\newcommand{\bml}{\begin{subequations}}
\newcommand{\eml}{\end{subequations}}
\newcommand{\bfig}{\begin{figure}}
\newcommand{\efig}{\end{figure}}

\newcommand{\del}{\delta}

\newcommand{\slashed}{\hspace{-1.1ex}/}

\newcommand{\bmat}{\begin{pmatrix}}
\newcommand{\emat}{\end{pmatrix}}
\usepackage{graphicx,slashed,booktabs,xcolor,multirow,float,
amsfonts,bbold,mathtools,sidecap,tikz,bm,enumitem}
\usepackage{multirow}
\usepackage{bbding}
\usepackage{titlesec}
\usepackage{hyperref}
\usepackage{wasysym}
\usepackage{amssymb}
\usepackage{pifont}
\newcommand{\grad}{\nabla}

\renewcommand{\leq}{\leqslant}
\renewcommand{\geq}{\geqslant}

\usepackage[dvipsnames, usenames]{xcolor}

\definecolor{linkcolor}{rgb}{0.55, 0.13, .32}

\definecolor{oucrimsonred}{rgb}{0.6, 0.0, 0.0}
\definecolor{persianblue}{rgb}{0.11, 0.22, 0.73}
\definecolor{forestgreen}{rgb}{0.13,0.35,0.13}
\definecolor{lightgray}{rgb}{0.83, 0.83, 0.83}
 \hypersetup{colorlinks, citecolor=oucrimsonred, linkcolor=persianblue, urlcolor=oucrimsonred}
\definecolor{cornellred}{rgb}{0.7, 0.11, 0.11}
\definecolor{navyblue}{rgb}{0.0, 0.0, 0.5}
\definecolor{amethyst}{rgb}{0.6, 0.4, 0.8}
\definecolor{yellow}{rgb}{1.0, 1.0, 0.0}
\definecolor{firebrick}{rgb}{0.7, 0.13, 0.13}
\definecolor{tangerineyellow}{rgb}{1.0, 0.8, 0.0}
\definecolor{deepfuchsia}{rgb}{0.76, 0.33, 0.76}
\definecolor{amber}{rgb}{1.0, 0.75, 0.0}
\definecolor{VioletRed4}{rgb}{0.55, 0.13, .32}
\definecolor{indiagreen}{rgb}{0.07, 0.53, 0.03}
\definecolor{VioletRed4}{rgb}{0.55, 0.13, .32}

\usepackage{hyperref}

\usepackage{graphics,appendix,afterpage,makecell}

\usepackage{bbold}
\usepackage{tikz}
\usepackage{adjustbox}
\usepackage{tikz-feynman}
\usepackage{tcolorbox}

\definecolor{oucrimsonred}{rgb}{0.6, 0.0, 0.0}
\definecolor{persianblue}{rgb}{0.11, 0.22, 0.73}
\definecolor{forestgreen}{rgb}{0.13,0.35,0.13}
\definecolor{lightgray}{rgb}{0.83, 0.83, 0.83}
 \hypersetup{colorlinks, citecolor=oucrimsonred, linkcolor=persianblue, urlcolor=oucrimsonred}
\definecolor{cornellred}{rgb}{0.7, 0.11, 0.11}
\definecolor{navyblue}{rgb}{0.0, 0.0, 0.5}
\definecolor{amethyst}{rgb}{0.6, 0.4, 0.8}
\definecolor{yellow}{rgb}{1.0, 1.0, 0.0}
\definecolor{firebrick}{rgb}{0.7, 0.13, 0.13}
\definecolor{tangerineyellow}{rgb}{1.0, 0.8, 0.0}
\definecolor{deepfuchsia}{rgb}{0.76, 0.33, 0.76}
\definecolor{amber}{rgb}{1.0, 0.75, 0.0}
\definecolor{VioletRed4}{rgb}{0.55, 0.13, .32}
\definecolor{indiagreen}{rgb}{0.07, 0.53, 0.03}
\definecolor{VioletRed4}{rgb}{0.55, 0.13, .32}

\definecolor{oucrimsonred}{rgb}{0.6, 0.0, 0.0}
\newcommand\vertarrowbox[3][6ex]{%
  \begin{array}[t]{@{}c@{}} #2 \\
  \left\uparrow\vcenter{\hrule height #1}\right.\kern-\nulldelimiterspace\\
  \makebox[0pt]{\scriptsize#3}
  \end{array}%
}

\definecolor{mtcolor}{rgb}{.8,.3,.1}

\definecolor{violachiaro}{rgb}{1,0.6,1}

\definecolor{gbcolor}{rgb}{.43,.22,.12}
 
\definecolor{gbcolor2}{rgb}{.9,.2,.6}
\definecolor{gbcolor3}{rgb}{.3,.2,.6}

\definecolor{verdechiaro}{rgb}{0.6,1,0.6}
\definecolor{giallochiaro}{rgb}{1,1,0.6}
\definecolor{bluscuro}{rgb}{0.15, 0.2, 0.9}
\definecolor{verdes}{rgb}{0.1, 0.5, 0.1}%
\definecolor{tangerineyellow}{rgb}{1.0, 0.8, 0.0}
\definecolor{smokyblack}{rgb}{0.06, 0.05, 0.03}

\definecolor{americanrose}{rgb}{1.0, 0.01, 0.24}
\definecolor{cobalt}{rgb}{0.0, 0.28, 0.67}
\definecolor{brandeisblue}{rgb}{0.0, 0.44, 1.0}
\definecolor{mycolor}{rgb}{0.0, 0.0, 0.5}
\definecolor{oxfordblue}{rgb}{0.0, 0.13, 0.28}
\definecolor{azure}{rgb}{0.0, 0.5, 1.0}
\definecolor{turquoiseblue}{rgb}{0.0, 1.0, 0.94}
\newtcolorbox{mynewbox}[1]{colback=white!5!white,colframe=azure!75!black,fonttitle=\bfseries,title=#1}
\newtcolorbox{mybox}{colback=mycolor!5!white,colframe=azure!75!black}
\newtcolorbox{mynamedbox}[1]{colback=mycolor!5!white,colframe=azure!75!black,title=#1}
\definecolor{venetianred}{rgb}{0.78, 0.03, 0.08}
\newtcolorbox{mynamedbox1}[1]{colback=venetianred!5!white,colframe=venetianred!80!black,title=#1}
\newtcolorbox{mynamedbox2}[1]{colback=azure!5!white,colframe=azure!80!black,title=#1}

\definecolor{rossocorsa}{rgb}{0.83, 0.0, 0.0}

\tikzset{->-/.style={decoration={
  markings,
  mark=at position #1 with {\arrow{>}}},postaction={decorate}}}
\tikzset{-<-/.style={decoration={
  markings,
  mark=at position #1 with {\arrow{<}}},postaction={decorate}}} 

\def\be{\begin{equation}}
\def\ee{\end{equation}}
\def\ba{\begin{eqnarray}}
\def\ea{\end{eqnarray}}

\def\L*{{\cal L}_*}
\def\L{\mathcal{L}}
\def\({\left(}
\def\){\right)}

\def\<{\langle}
\def\>{\rangle}

\def\cs2{c_{s}^{2}}

 \def\be   {\begin{equation}}   \def\ee   {\end{equation}}
 \def\ba   {\begin{array}}      \def\ea   {\end{array}}
 \def\bea  {\begin{eqnarray}}   \def\eea  {\end{eqnarray}}
 \def\bean {\begin{eqnarray*}}  \def\eean {\end{eqnarray*}}

\titleclass{\subsubsubsection}{straight}[\subsection]

\newcounter{subsubsubsection}[subsubsection]
\renewcommand\thesubsubsubsection{\thesubsubsection.\arabic{subsubsubsection}}

\titleformat{\subsubsubsection}
  {\normalfont\normalsize\bfseries}{\thesubsubsubsection}{1em}{}
\titlespacing*{\subsubsubsection}
{0pt}{3.25ex plus 1ex minus .2ex}{1.5ex plus .2ex}

\makeatletter
\renewcommand\paragraph{\@startsection{paragraph}{5}{\z@}%
  {3.25ex \@plus1ex \@minus.2ex}%
  {-1em}%
  {\normalfont\normalsize\bfseries}}
\renewcommand\subparagraph{\@startsection{subparagraph}{6}{\parindent}%
  {3.25ex \@plus1ex \@minus .2ex}%
  {-1em}%
  {\normalfont\normalsize\bfseries}}
\def\toclevel@subsubsubsection{4}
\def\toclevel@paragraph{5}
\def\toclevel@paragraph{6}
\def\l@subsubsubsection{\@dottedtocline{4}{7em}{4em}}
\def\l@paragraph{\@dottedtocline{5}{10em}{5em}}
\def\l@subparagraph{\@dottedtocline{6}{14em}{6em}}
\makeatother

\begin{document}


\definecolor{lime}{HTML}{A6CE39}
\DeclareRobustCommand{\orcidicon}{\hspace{-2.1mm}
\begin{tikzpicture}
\draw[lime,fill=lime] (0,0.0) circle [radius=0.13] node[white] {{\fontfamily{qag}\selectfont \tiny \,ID}}; \draw[white, fill=white] (-0.0525,0.095) circle [radius=0.007]; 
\end{tikzpicture} \hspace{-3.7mm} }
\foreach \x in {A, ..., Z}{\expandafter\xdef\csname orcid\x\endcsname{\noexpand\href{https://orcid.org/\csname orcidauthor\x\endcsname} {\noexpand\orcidicon}}}
\newcommand{\orcidauthorA}{0000-0002-0459-3873}
\newcommand{\orcidauthorC}{0009-0003-9227-8615}
\newcommand{\orcidauthorD}{0009-0008-4326-5988}
\newcommand{\orcidauthorE}{0000-0003-1081-0632}


\title{\textcolor{Sepia}{\textbf \huge\Large\LARGE  
Primordial Black Holes from Effective Field Theory of Stochastic Single Field  Inflation 
at NNNLO
} 
}


\author{{\large  Sayantan Choudhury\orcidA{}${}^{1}$}}
\email{sayantan\_ccsp@sgtuniversity.org,  sayanphysicsisi@gmail.com (Corresponding author)}
\author{{\large  Ahaskar Karde\orcidC{}${}^{1}$}}
\email{kardeahaskar@gmail.com}
\author{\large Pankaj Padiyar\orcidD{}${}^{1,2}$}
\email{padiyarpankaj1234@gmail.com, pankajp20@iiserbpr.ac.in }
\author{ \large M.~Sami\orcidE{}${}^{1,3,4}$}
\email{ sami\_ccsp@sgtuniversity.org,  samijamia@gmail.com}

\affiliation{ ${}^{1}$Centre For Cosmology and Science Popularization (CCSP),\\
        SGT University, Gurugram, Delhi- NCR, Haryana- 122505, India.}
\affiliation{${}^{2}$Department of Physical Sciences, Indian Institute of Science Education and Research
Berhampur,
Transit Campus, Government ITI, Berhampur- 760010, Odisha, India,}
\affiliation{${}^{3}$Center for Theoretical Physics, Eurasian National University, Astana 010008, Kazakhstan.}
	\affiliation{${}^{4}$Chinese Academy of Sciences,52 Sanlihe Rd, Xicheng District, Beijing.}

\begin{abstract}

We present a study of the Effective Field Theory (EFT) generalization of stochastic inflation in a model-independent single-field framework and its impact on primordial black hole (PBH) formation. We show how the Langevin equations for the ``soft'' modes in quasi de Sitter background is described by the Infra-Red (IR) contributions of scalar perturbations, and the subsequent Fokker-Planck equation driving the probability distribution for the stochastic duration ${\cal N}$, significantly modify in the present EFT picture. An explicit perturbative analysis of the distribution function by implementing the stochastic-$\delta N$ formalism is performed up to the next-to-next-to-next-to-leading order (NNNLO) for both the classical-drift and quantum-diffusion dominated regimes. In the drift-dominated limit, we perturbatively analyse the local non-Gaussianity parameters $(f_{\rm NL}, g_{\rm NL}, \tau_{\rm NL})$ with the EFT-induced modifications. In the diffusion-dominated limit, we numerically compute the probability distribution featuring exponential tails at each order of perturbative treatment. 


\end{abstract}

\maketitle
\tableofcontents
\newpage


\section{Introduction}
\label{s1}

Cosmological inflation is a leading paradigm for the very early universe that provides a seeding mechanism for generating present-day large-scale structures from primordial quantum fluctuations. These fluctuations, generally associated with a scalar field taking part during inflation, begin their journey initially from the small-scale quantum regime to later transition into the large-scale, classical regime. An interesting consequence of the primordial fluctuations in the early universe, mostly those generated near the end of inflation at small scales, is the formation of objects known as primordial black holes (PBHs). The formation of PBHs \cite{Zeldovich:1967lct,Hawking:1974rv,Carr:1974nx,Carr:1975qj,Chapline:1975ojl,Carr:1993aq,Choudhury:2011jt,Yokoyama:1998pt,Kawasaki:1998vx,Rubin:2001yw,Khlopov:2002yi,Khlopov:2004sc,Saito:2008em,Khlopov:2008qy,Carr:2009jm,Choudhury:2011jt,Lyth:2011kj,Drees:2011yz,Drees:2011hb,Ezquiaga:2017fvi,Kannike:2017bxn,Hertzberg:2017dkh,Pi:2017gih,Gao:2018pvq,Dalianis:2018frf,Cicoli:2018asa,Ozsoy:2018flq,Byrnes:2018txb,Ballesteros:2018wlw,Belotsky:2018wph,Martin:2019nuw,Ezquiaga:2019ftu,Motohashi:2019rhu,Fu:2019ttf,Ashoorioon:2019xqc,Auclair:2020csm,Vennin:2020kng,Nanopoulos:2020nnh,Inomata:2021uqj,Stamou:2021qdk,Ng:2021hll,Wang:2021kbh,Kawai:2021edk,Solbi:2021rse,Ballesteros:2021fsp,Rigopoulos:2021nhv,Animali:2022otk,Frolovsky:2022ewg,Escriva:2022duf,Ozsoy:2023ryl,Ivanov:1994pa,Afshordi:2003zb,Frampton:2010sw,Carr:2016drx,Kawasaki:2016pql,Inomata:2017okj,Espinosa:2017sgp,Ballesteros:2017fsr,Sasaki:2018dmp,Ballesteros:2019hus,Dalianis:2019asr,Cheong:2019vzl,Green:2020jor,Carr:2020xqk,Ballesteros:2020qam,Carr:2020gox,Ozsoy:2020kat,Baumann:2007zm,Saito:2008jc,Saito:2009jt,Choudhury:2013woa,Sasaki:2016jop,Raidal:2017mfl,Papanikolaou:2020qtd,Ali-Haimoud:2017rtz,Di:2017ndc,Raidal:2018bbj,Cheng:2018yyr,Vaskonen:2019jpv,Drees:2019xpp,Hall:2020daa,Ballesteros:2020qam,Carr:2020gox,Ozsoy:2020kat,Ashoorioon:2020hln,Papanikolaou:2020qtd,Wu:2021zta,Kimura:2021sqz,Solbi:2021wbo,Teimoori:2021pte,Cicoli:2022sih,Ashoorioon:2022raz,Papanikolaou:2022chm,Papanikolaou:2023crz,Wang:2022nml,ZhengRuiFeng:2021zoz,Cohen:2022clv,Cicoli:2022sih,Brown:2017osf,Palma:2020ejf,Geller:2022nkr,Braglia:2022phb,Frolovsky:2023xid,Aldabergenov:2023yrk,Aoki:2022bvj,Frolovsky:2022qpg,Aldabergenov:2022rfc,Ishikawa:2021xya,Gundhi:2020kzm,Aldabergenov:2020bpt,Cai:2018dig,Cheng:2021lif,Balaji:2022rsy,Qin:2023lgo,Riotto:2023hoz,Riotto:2023gpm,Papanikolaou:2022did,Choudhury:2011jt,Choudhury:2023vuj, Choudhury:2023jlt, Choudhury:2023rks,Choudhury:2023hvf,Choudhury:2023kdb,Choudhury:2023hfm,Bhattacharya:2023ysp,Choudhury:2023fwk,Choudhury:2023fjs,Choudhury:2024one,Harada:2013epa,Harada:2017fjm,Kokubu:2018fxy,Gu:2023mmd,Saburov:2023buy,Stamou:2023vxu,Libanore:2023ovr,Friedlander:2023qmc,Chen:2023lou,Cai:2023uhc,Karam:2023haj,Iacconi:2023slv,Gehrman:2023esa,Padilla:2023lbv,Xie:2023cwi,Meng:2022low,Qiu:2022klm,Mu:2022dku,Fu:2022ypp,Davies:2023hhn,Firouzjahi:2023ahg,Firouzjahi:2023aum, Iacconi:2023ggt,Davies:2023hhn,Jackson:2023obv,Riotto:2024ayo,Choudhury:2024dei,Choudhury:2024dzw,Choudhury:2024aji,Choudhury:2024kjj} is linked to the large fluctuations at smaller scales that, after re-entry into the horizon, generate regions of overdensities and underdensities in the content of the universe, which gravitationally collapse after crossing a certain threshold, to produce PBHs. The interest in PBHs has seen a rapid rise due to their strong candidacy as dark matter, and the fluctuations giving rise to them can also participate in the production of primordial gravitational waves whose signatures can be observed today, see the refs. \cite{NANOGrav:2023gor, NANOGrav:2023hde, NANOGrav:2023ctt, NANOGrav:2023hvm, NANOGrav:2023hfp, NANOGrav:2023tcn, NANOGrav:2023pdq, NANOGrav:2023icp,EPTA:2023fyk, EPTA:2023sfo, EPTA:2023akd, EPTA:2023gyr, EPTA:2023xxk, EPTA:2023xiy,Reardon:2023gzh, Reardon:2023zen, Zic:2023gta,Xu:2023wog,LISACosmologyWorkingGroup:2023njw, Inomata:2023zup,Choudhury:2023hfm,Bhattacharya:2023ysp,Choudhury:2023fwk,Choudhury:2023fjs,Franciolini:2023pbf,Inomata:2023zup,Wang:2023ost,Balaji:2023ehk,Gorji:2023sil,Choudhury:2023kam,Yi:2023mbm,Cai:2023dls,Cai:2023uhc,Huang:2023chx,Huang:2023mwy,Frosina:2023nxu,Zhu:2023faa,Cheung:2023ihl,Gouttenoire:2023bqy,Salvio:2023ynn,Yi:2023npi,Di:2017ndc,Ballesteros:2018wlw,Sasaki:2018dmp,Fu:2019ttf,Ballesteros:2020qam,Papanikolaou:2020qtd,Inomata:2021uqj,Wang:2022nml,Ashoorioon:2022raz,Frolovsky:2022ewg,Domenech:2021ztg,Yuan:2021qgz,Chen:2019xse,Cang:2022jyc,Heydari:2023rmq,Bhaumik:2023wmw,Chen:2024gqn} for recent work covering such interests. Out of the possible mechanisms for PBH formation there is one that involves a section of the inflationary potential at the small scales to develop an almost flat region that significantly enhances the scalar-field fluctuations taking part in forming PBHs. Such a regime during inflation is also recognized as a period of ultra-slow roll (USR), where the quantum diffusion effects begin to dominate and in turn contribute to the overall dynamics of the large-scale classical perturbations, after horizon exit. A wide variety models featuring such a region have been analyzed from the perspective of stochastic inflation and their implications on PBH production \cite{Vennin:2024yzl,Animali:2024jiz,LISACosmologyWorkingGroup:2023njw,Animali:2022otk,Ezquiaga:2022qpw,Jackson:2022unc,Tada:2021zzj,Pattison:2021oen,Ando:2020fjm,Vennin:2020kng,Ezquiaga:2019ftu,Pattison:2019hef,Noorbala:2018zlv,Pattison:2017mbe,Grain:2017dqa,Hardwick:2017fjo,Mishra:2023lhe}. 

\textcolor{black}{In the paper \cite{Starobinsky:1986fx}, Starobinsky first introduced the stochastic inflationary paradigm as a new way of understanding the dynamics of large-scale fluctuations that get affected by the presence of noise terms coming from the quantum-to-classical transition of the small wavelength modes of primordial fluctuations. See also refs.\cite{Gorbenko:2019rza,Cohen:2021fzf,Cohen:2022clv,Green:2022ovz,Cohen:2021jbo,Cohen:2020php} on ``Soft de Sitter Effective Theory'' (SdSET) which strengthens the building blocks of the stochastic inflationary paradigm. The word ``soft'' here signifies the low-energy component after splitting of the gauge-invariant variable, that is the comoving curvature perturbation $\zeta$ in the EFT of inflation, into its long and short wavelength parts. The short-wavelength parts would suffer coarse-graining from stochastic effects and later join the long-wavelength dynamics after horizon crossing. Recently, the stochastic inflation formalism has been applied to a plethora of settings and has also found significant applications in the study of PBH production \cite{Vennin:2024yzl,Animali:2024jiz,LISACosmologyWorkingGroup:2023njw,Animali:2022otk,Ezquiaga:2022qpw,Jackson:2022unc,Tada:2021zzj,Pattison:2021oen,Ando:2020fjm,Vennin:2020kng,Ezquiaga:2019ftu,Pattison:2019hef,Noorbala:2018zlv,Pattison:2017mbe,Grain:2017dqa,Hardwick:2017fjo}.} In this work, we aim to generalize this picture of PBH formation from a USR phase in a model-independent manner by building a soft de Sitter Effective Field Theory (EFT) formulation of stochastic single-field inflation without explicitly inserting any scalar field in the framework. 

The EFT of inflation framework \cite{Weinberg:2008hq,Cheung:2007st,Choudhury:2017glj,Delacretaz:2016nhw,Naskar:2017ekm} provides an opportunity to construct a detailed technical description of the UV complete theory in a model-independent manner. Here we can study the dynamics of the metric perturbations around a quasi de Sitter spacetime without worrying about a specific model potential driving the scalar field. The EFT action of interest is built around the symmetries manifesting themselves into the structure of the higher-dimensional operators present within the action. We adhere to the well-known St$\ddot{u}$ckelberg approach, while working with the unitary gauge, from which restoring the gauge invariance in the EFT action gives rise to a new scalar degree of freedom, dubbed the Goldstone mode, that non-linearly transforms under the broken-time diffeomorphism symmetry. 
\textcolor{black}{In the stochastic inflation picture, we deal with the dynamics of large-scale metric perturbations that encounter the classical noises resulting from stochastic effects near horizon crossing. Here we present} how to embed the underlying EFT principles into deriving the evolution equation for the same large-scale comoving curvature perturbations, labeled as $\zeta$, which results from converting the initial quantum fluctuations at the end of inflation. \textcolor{black}{The short-scale or Ultra-Violet (UV) fluctuation component of $\zeta$} undergoes a process referred to as coarse-graining in the presence of the stochastic effects dominant near the horizon-crossing instant; \textcolor{black}{this is also not an exact moment and differs based on the coarse-graining window function on which we also elaborate in later sections}. This horizon crossing extends further into the super-Hubble scales, and the evolution of the coarse-grained curvature perturbations, also referred to as the Infra-Red (IR) component, is governed by the stochastic Langevin equation.  We later show the significance of these equations from the perspective of studying PBH formation. 

The mechanism for PBH formation has been studied quite extensively, and various approaches, like the Press-Schechter formalism, peak theory, or the recently well-focused compaction function approach, \cite{Choudhury:2023jlt,Bhattacharya:2023ysp,Choudhury:2023fwk,Choudhury:2023fjs,Choudhury:2024one,Franciolini:2023wun,DeLuca:2023tun,DeLuca:2022rfz,Musco:2021sva,Musco:2020jjb,Kalaja:2019uju,Kehagias:2019eil,Young:2019yug,Musco:2018rwt,Bardeen:1985tr,Green:2004wb,Ianniccari:2024bkh,Franciolini:2023pbf,Ferrante:2022mui,Ferrante:2023bgz} have their preferred regimes of applicability under certain conditions. In the Press-Schechter approach, the initial profile for $\zeta$ is considered to be of Gaussian nature, and PBH formation occurs when the large curvature perturbations exceed a certain threshold, $\zeta_{\rm th}\sim {\cal O}(1)$ and as a result the quantum diffusion effects in the USR region cannot be ignored during the process. \textcolor{black}{Thus, since the use of perturbation theory is best justified when non-Gaussianity is strictly very small, this assumption quickly breaks down during PBH formation where the non-Gaussian statistics, most prominent at the tail end of the perturbation distribution, become too significant to ignore for a robust understanding of the whole process.} To consider this development, many works have suggested use of the non-perturbative, stochastic-$\delta N$ formalism \cite{Enqvist:2008kt,Fujita:2013cna,Fujita:2014tja,Vennin:2015hra} which successfully allows us to relate the statistics of the curvature perturbation distribution to that of the amount of integrated expansion, $N$, for which we solve in this paper considering perturbative corrections from the different regimes (classical and quantum). \textcolor{black}{We take on this approach and show how the Fokker-Planck equation for the probability distribution function (PDF) of the e-folds of expansion variable in the USR phase receives modifications from the EFT framework following the version of the Langevin equations derived before. This approach of dealing with the stochastic effects from a USR regime has also been adopted in some recent works \cite{Firouzjahi:2018vet,Ballesteros:2020sre} and we also perform the present study with the USR by utilizing the stochastic-$\delta N$ formalism.}

Since we are not considering any specific model of a scalar field driven by a potential, the weight of realizing any particular theory falls on the parameters that characterize the slow-roll conditions necessary for inflation. In the EFT of inflation, the Hubble rate is an important parameter through which we can write the various perturbative correction terms coming from higher-derivative operators as part of the general theory. To make this realization explicit, we show how the conditions on the slow-roll parameters and the Hubble rate be realized in our three-phase scenario consisting of a USR phase in between two slow-roll phases. The information regarding mode solutions of the curvature perturbation can also be extracted within the EFT framework by working with the second-order perturbed action. This solution can then be worked out individually for the three-phase scenario, involving a specific parameterization of the transition between each phase, which is chosen here to be of a sharp nature. \textcolor{black}{However, we must emphasize here that the transition, whether sharp or smooth, does not alter our conclusions and later discuss more on such features inside the relevant section.} The EFT description also has an effective sound speed parameter $c_{s}$, where $c_{s}=1$ refers to the canonical single-field models, while $c_{s}\ne 1$ to the non-canonical single-field models of inflation. This parameter is also crucial to realise the three-phase setup used in this work. In the present context, we elaborate on the impact of the effective sound speed by, first, specifying its exact parameterization used and, secondly, discussing the changes brought by its possible values to the various auto-correlation and cross-correlation elements of the power spectrum, keeping satisfied the causality and unitarity constraints from experiments. We find that no violation of the mentioned physical conditions is needed, and perturbativity arguments remain intact, giving rise to enough amplitude necessary for PBH production. We also elaborate on the other features coming from the stochastic effects present in the system, especially near the transition between phases of the set up.

This paper is organised as follows: In section \ref{s2}, we begin with elaborating on the underlying motivation for stochastic EFT formulation given in this work.
In section \ref{s3}, a brief outline of the EFT of inflation is given followed by the second-order perturbed action for the curvature perturbation which later gets used to derive the mode solutions featuring stochasticity. In section \ref{s4}, we implement the stochastic feature in to the EFT set up by showing derivation of the Hamilton's equation of motion for the coarse-grained curvature perturbation  In section \ref{s5}, we show the explicit realisation of the slow-roll parameters and the Hubble expansion rate during inflation in our three phase set up SRI-USR-SRII. In section \ref{s6}, we present a overview of the stochastic-$\delta N$ formalism and how we utilise this in the stochastic EFT formulation. In section \ref{s7}, we analyse the Fokker-Planck equation further so as to obtain its modified version which helps to study the evolution of probability distribution function in the coming sections. In section \ref{s8}, we introduce the characteristic function and its benefits for solving the Fokker-Planck equation. Section \ref{s9} focuses on the diffusion-dominated regime analysis, where we first outline the relevant method to obtain the PDF at a specified order in the perturbative treatment, and in the rest of the section, extend this treatment up to the next-to-next-to-next-to-leading order (NNNLO). \textcolor{black}{In section \ref{s10}, we outline the relevant methods to solve for the PDF, this time in the drift-dominated regime, with a perturbative treatment also extending up to the NNNLO.} Also in the same section, along with the PDF, explicit expressions for the non-Gaussianity parameters and features of the PDF profile are discussed. In section \ref{s11}, we briefly talk about the spectral distortion effects related to the PBH formation mechanism and utilise the scalar power spectrum derived earlier to compute the magnitude of such effects. In section \ref{s12}, we outline the PBH formation mechanism in the context of present stochastic EFT picture, and discuss the PBH mass fraction and present-day abundance with giving expressions for the mass fraction at each order till NNNLO. In section \ref{s13}, we present results of the numerical outcomes starting with the scalar power spectrum, followed by  results for spectral distortion, followed by PDF profiles in the diffusion-dominated limit for each order, and ending with results for the PBH mass fraction and abundance. 
We finally conclude our work with summarizing our findings in section \ref{s15}.

\section{Underlying physical motivation and approach}
\label{s2}

This section addresses the motivation behind seeking an EFT generalization of the stochastic single-field inflation paradigm. We aim to present the procedure by which such a generalization can take place in the present context and highlight the reasons behind the higher-order (NNNLO) perturbative treatment.

The understanding of super-Hubble dynamics of the scalar perturbations in dS space and tackling the appearance of secular logarithmic IR divergences have a rich history of interest \cite{Enqvist:2008kt,Podolsky:2008qq,Finelli:2008zg,Seery:2010kh,Garbrecht:2014dca,Burgess:2015ajz,Gorbenko:2019rza,Baumgart:2019clc,Mirbabayi:2019qtx,Cohen:2020php}. A majority of these investigations center around the theory of stochastic inflation. A recent example of this is the study in \cite{Cohen:2020php}, where the authors explicitly detail the higher-derivative corrections to the evolution equation of the PDF of the long-wavelength perturbations by deriving their results with the SdSET approach. It also highlights a crucial fact regarding the corrections talked about to stochastic inflation, in that a proper Dynamical Renormalization Group (DRG) analysis, involving resummation of the logarithmic IR divergences at all orders in the loop calculations, can effectively provide the said corrections. In the present work, we focus on the scalar field perturbations and incorporate an effective field theory approach for the dynamics of the scalar metric perturbations, $\zeta(t,{\bf x})$. This approach renders our analysis as fully gauge-invariant which would not have been the case have we chosen to work with the scalar field perturbations $\delta\phi(t,{\bf x})$ instead. 

Let us first understand the idea behind the EFT of single-field stochastic inflation that is of interest here, what problems may arise concerning the quantum loop corrections, and how we aim to tackle them. We start with splitting of the gauge-invariant curvature perturbation field into its long (soft/IR) and short (hard/UV) wavelength components, as encountered in the stochastic inflation theory. After this, a perturbative expansion of the EFT action to include higher-derivative operators will be composed of various spatial and temporal derivatives combinations of $\zeta$, each accompanied with the stochastic effects. This will have major implications if one wishes to evaluate higher-point correlation functions using the Schwinger-Keldysh (\textit{in-in}) technique, which will now be described in terms of the stochastic parameter $\sigma$ for each higher-order interactions after their quantization. \textcolor{black}{The parameter $\sigma$ acts as a cut-off scale to filter out the UV modes and we are left to consider the dynamics of a system of IR modes where the UV modes act as classical noise terms coming from a separate environment.} One would eventually require the need of regularization and renormalization techniques to proceed with this modified procedure of calculating cosmological correlations with the stochastic effects, followed by a DRG analysis to remove the unwanted UV and soften the logarithmic IR divergences from quantum loops. The final step of performing the DRG, accounting for stochasticity, becomes the most challenging aspect of the above procedure. Using this method, one can capture the quantum effects of the diagrams up to all orders in the perturbative expansion that leads to the softening of the logarithmic IR divergences present at the super-hubble scales in cosmological correlations.

Now, with the inclusion of stochastic effects through $\sigma$, each interaction involves such effects in the calculations, and keeping track of all orders of the possible loop diagrams becomes cumbersome. Here, $\sigma$ plays the role of a regulator whose purpose is to coarse-grain the contributions and the softening of the logarithmic IR divergences at all orders in the SdSET. The $\sigma$ parameter remains the same throughout each diagram and demands controlling its impact such that it does not propagate to all other diagrams and results in a meaningless outcome. 
Situations like these are examples that we can avoid with the stochastic-$\delta N$ formalism since it can save us from the DRG analysis with stochastic effects, and instead, the stochastic-$\delta N$ similarly mimics the DRG analysis for the scalar perturbation modes.  
In \cite{Dias:2012qy}, the equivalence between the classical $\delta N$ formalism and the DRG method is worked out in detail. Further, following the developments in \cite{Cohen:2021fzf, Burgess:2015ajz, Burgess:2014eoa, Burgess:2009bs}, it is expected that the DRG analysis in the presence of stochastic effects and separately performing of the stochastic-$\delta N$ analysis will converge to give similar results as discussed above.

The other aspect of this work concerns why a careful analysis dealing with a next-to-next-to-next leading order (NNNLO) treatment is of concern here. To make the idea behind the analysis more transparent, we mention that in order to generate the PDF of the long-wavelength component, we deal with the drift and diffusion dominated regimes separately. For each regime, the task undertaken is such that one particular phenomenon, whether diffusion or drift, adds its corrections perturbatively to the other that dominates that regime. By performing a more rigorous study, we test for the perturbativity conditions and check the extent of validity of the analytical methods for each scenario. Also, with the EFT formulation as our backdrop, we aim to elevate our analysis further by generalizing it to include the canonical and non-canonical models of single-field inflation. Later in this work, we show how various statistical measures of the PDF, like its variance, skewness, and kurtosis, can be calculated through its higher-order moments when focusing on the drift-dominated regime and the information that they can provide on variety of non-gaussianity parameters, namely $f_{\rm NL}, g_{\rm NL}, \tau_{\rm NL}$, as they continue to receive corrections from each order treatment. Following this, since we also focus on PBH production, we carefully analyze the PDF in the diffusion-dominated regime. We later demonstrate that with each order of corrections included, the perturbative nature of the overall analysis remains robust regardless of the choice of canonical or non-canonical features. We also observe interesting features when estimating the PBH mass fraction and further elaborate on the PBH formation process with its dependence on the multiple parameters seen when working out the Fokker-Planck equation. In the course of this study, we show the robustness of the analytical methods via our results and find no instances violating any of the perturbativity conditions.

\section{The Stochastic EFT of single field inflation}
\label{s3}

\subsection{The underlying EFT setup}
\label{s3a}

The Effective Field Theory (EFT) setup involves the construction of an effective action which is valid below some extremely high energy or UV cut-off scale. This cut-off determines the scale of energy above which the effective description of the underlying theory breaks down. As of now, not much is known about the possible physics which can characterize a UV complete theory. However, whatever the physics might be present above such a  UV cut-off $\Lambda$, it should become manifest itself in the effective action used for conducting any analysis at the lower energies.
The construction of this action involves terms which are built so as to respect the underlying symmetries of the theory that survive at these low energies.

We would like to use this EFT framework to study the theory of fluctuations around a time-dependent background. To this effect, we begin with a scalar field $\phi(t,\bf{x})$ that is driving inflation. The scalar field itself obeys full diffeomorphism symmetry, however, its perturbations $\delta\phi$ are such that they they exhibit broken time diffeomorphism symmetry but remain as scalars under the spatial diffeomorphisms. 
The non-linear transformation for the scalar perturbations under time-diffeomorphisms can be written as follows:
\bea
t \rightarrow t+\xi^0(t, {\bf x}), \quad x^i \rightarrow x^i \quad \forall\;i=1,2,3 \implies
\del\phi \rightarrow \del\phi+\dot{\phi}_0(t) \xi^0(t, {\bf x})
\eea
where $\xi^0(t,{\bf x})$ is the time-diffeomorphism parameter, $\phi_{0}(t)$ represents the time-dependent background scalar field in a homogeneous isotropic FLRW space-time, \textcolor{black}{and a "dot" represents a derivative with respect to the cosmic time $t$}. To proceed further, we choose to work with the unitary gauge which brings the condition $\phi(t,{\bf x}) = \phi_0 (t)$, setting the inflationary perturbations to zero. By doing so, the scalar perturbations variable gets eaten by the metric which leads to an increased $3$ physical degrees of freedom: $1$ for the scalar mode and $2$ helicities, and this mimics exactly the phenomenon in spontaneous symmetry breaking
in the $SU(N)$ non-abelian gauge theory.
Our theory involves a quasi de Sitter solution for the spatially flat FLRW space-time having the metric:
\bea 
ds^2 = a^2(\tau) ( -d\tau^2 + d{\bf x}^2)\quad\quad {\rm where}\quad\quad a(\tau)=-\frac{1}{H\tau}\quad\quad -\infty<\tau<0.
\eea 
The next step into the construction of the effective action in the present context requires using the metric $g_{\mu \nu}$ and its derivatives, which include the Riemann tensor ($ R_{\mu\nu\alpha\beta}$), the Ricci tensor ($R_{\mu\nu}$), and the Ricci scalar $(R)$. Another important quantity needed is the temporally perturbed component $\del g^{00} = (g^{00} +  1)$, of the metric. This term \textcolor{black}{respects} the spatial diffeomorphisms and so is crucial for the construction. For the quasi de Sitter background solution, a useful identity remains the conversion between the conformal time and the physical time 
\bea
t=\frac{1}{H}\ln{\Big(-\frac{1}{H\tau }\Big)}, \quad 0 < t < \infty \quad {\rm and}\quad -\infty<\tau<0
\eea
where the scale factor $a(\tau) =- 1/H\tau$ for de Sitter is used. The extrinsic curvature of the constant time-slice surfaces is another example of a quantity invariant under the spatial diffeomorphisms. The action organized by including an expansion in powers of its temporally perturbed component at constant time slice i.e. \bea \del K_{\mu\nu}= (K_{\mu\nu}-a^2 H h_{\mu\nu}),\eea is shown to provide the most general effective Lagrangian. Here $K_{\mu\nu}$ is the extrinsic curvature, $n_\mu $ is the unit normal and $h_{\mu\nu}$ is the induced metric on the three-dimensional hyper-surface which are defined in this context as: 
\bea 
&& h_{\mu\nu}=g_{\mu\nu}+n_\mu n_\nu ,\quad 
n_\mu = \frac{\partial_\mu t}{\sqrt{-g_{\mu\nu}\partial_\mu t  \partial_\nu t}}, \quad K_{\mu\nu} = h_\mu ^ \sigma \grad _\sigma n_\nu.
\eea
We can now present the most general effective action that will help us in computing the total power spectrum including the one-loop quantum corrections:
\bea 
S &=&\int d^4x \sqrt{-g}\Bigg[ \frac{M_p^2}{2} R  -c(t) g^{00} - \Lambda(t) + \frac{M_2^4(t)}{2!} (g^{00}+1)^2
+\frac{M_3^4(t)}{3!} (g^{00}+1)^3 \nonumber \\
&& \quad\quad\quad\quad\quad\quad\quad\quad\quad\quad\quad\quad\quad -
\frac{\Bar{M}_1^{3}(t)}{2}(g^{00}+1)\del K_\mu ^\mu 
-\frac{\Bar{M}_2^{2}(t)}{2}(\del K_\mu ^\mu )^{2} 
-\frac{\Bar{M}_3^{2}(t)}{2}\del K_\mu ^\nu \del K_\nu ^\mu  \Bigg], \eea 
here 
$M_2(t) , M_3(t) ,\Bar{M_1}(t), \Bar{M_2}(t) , \Bar{M_3}(t) $ are playing the role of  the Wilson coefficients which need to be fixed by further analysis.

The corresponding Friedmann Equations after considering  only the background contributions in the action are :
\bea
&& H^2=\Bigg[\frac{\dot{a}}{a}\Bigg]^2=\frac{1}{3M_p ^2}\Big[c(t) + \Lambda(t)\Big]=\frac{\mathcal{H}^2}{a^2} \\
&& \frac{\ddot{a}}{a}=\dot{H}+H^2 = -\frac{1}{3M_p^2} \Big[2c(t)-\Lambda(t) \Big]=\frac{\mathcal{H}'}{a^2}
\eea
where the $'$ notation denotes derivative with respect to conformal time and the conformal Hubble parameter is defined as $\mathcal{H}=\displaystyle{\frac{a'}{a}}= aH$. By solving the above equations one can determine the time dependent parameters $\Lambda(t)$ and $c(t)$ for the background as follows:
\bea
&& c(t) = -M_{p}^{2}\dot{H} = -\frac{M^{2}_{p}}{a^{2}}({\cal H}' - {\cal H}^{2}), \\
&& \Lambda(t) = M_{p}^{2}(3H^{2} + \dot{H}) = \frac{M_{p}^{2}}{a^{2}}(2{\cal H}^{2} + {\cal H}'),
\eea
these solutions \textcolor{black}{fix} the coefficients $c(t),\;\Lambda(t)$ in terms of the background expansion history and through their use gives rise to the following form of the EFT action:
\bea 
\label{eftaction}
S &=&\int d^4x \sqrt{-g}\Bigg[ \frac{M_p^2}{2} R  + M_{p}^{2}\dot{H}g^{00} - M_{p}^{2}(3H^{2} + \dot{H}) + \frac{M_2^4(t)}{2!} (g^{00}+1)^2
+\frac{M_3^4(t)}{3!} (g^{00}+1)^3 \nonumber \\
&& \quad\quad\quad\quad\quad\quad\quad\quad\quad\quad\quad\quad\quad\quad -
\frac{\Bar{M}_1^{3}(t)}{2}(g^{00}+1)\del K_\mu ^\mu 
-\frac{\Bar{M}_2^{2}(t)}{2}(\del K_\mu ^\mu )^{2} 
-\frac{\Bar{M}_3^{2}(t)}{2}\del K_\mu ^\nu \del K_\nu ^\mu  \Bigg]. \eea 
where the rest of the Wilson coefficients contain information about the various inflationary models. Lastly, we mention the form of the slow-roll parameters which helps to characterize the deviation from an exact de Sitter background: 
\bea &&\epsilon = \Big[1-\displaystyle{\frac{\mathcal{H}'}{\mathcal{H}^2}}\Big],\\
&& \eta = \displaystyle{\frac{\epsilon'}{\epsilon \mathcal{H}}}.\eea and the necessary conditions required for inflation are $\epsilon \ll 1 $ and $|\eta| \ll 1 $.

\subsection{Second order perturbed action from Goldstone EFT in decoupling limit}
\label{s3b}

Studying the perturbations or the Goldstone modes in the present context of EFT in inflation becomes much easier when working in the limit of low energy scales where gravity tends to decouple. Here, we use this specific limiting procedure to study the scalar perturbations, or the comoving curvature perturbation, as we will show, in the three regions involved in our overall setup during inflation. The decoupling limit is defined where, $E_{mix}=\sqrt{\dot{H}}$, and specifies neglecting the mixing contributions in action between the Goldstone mode and the metric fluctuations. Applying this limit enables writing the second-order Goldstone action to be written as, 
\bea \label{EFTpi}
S_{\pi}^{(2)} \approx \int d^4x \,  
  a^3\Bigg\{ -M_p^2\dot{H} \Bigg( \dot{\pi}^2  - \frac{1}{a^2}(\partial_i \pi )^2 \Bigg) + 2M_2 ^4 \dot{\pi}^2 \Bigg\} = \int d^4 x \,  a^3 \Bigg( \frac{-M_p ^2 \dot{H}}{c_s ^2}\Bigg)\Bigg[ \dot{\pi}^2 - c_s ^2 \frac{(\partial_i \pi)^2 }{a^2}\Bigg]
\eea
where the effective sound speed parameter, $c_s$, can be defined in terms of EFT Wilson coefficient,
\bea
c_s = \frac{1}{\sqrt{1-\displaystyle{\frac{2M_2 ^4 }{\dot{H}M_p^2 }}}}.
\eea 
Now, we understand that the spatial part $g_{ij}$ of the perturbed metric is defined as, 
\bea
g_{ij}\sim a^2(t)\{(1+2\zeta (t,{\bf x }))\del_{ij}\} \quad  \forall \quad  i=1,2,3,
\eea
which includes the scale factor for a quasi de Sitter space-time as $a(t)=e^{H t}$. Under the conditions of broken time diffeomorphisms  $a(t)$ now transforms as, 
\bea
a(t)\rightarrow a(t-\pi(t,{\bf x })) = a(t) - H \pi(t,{\bf x})a(t)+.....\approx a(t)(1-H\pi(t,{\bf x})),
\eea 
which when modified gives us,  
\bea
a^2(t)(1-H\pi(t,{\bf x}))^2 \approx  a^2(t)(1-2H\pi(t,{\bf x})) = a^2(t)(1+2\zeta(t,{\bf x})),
\eea 
and this suggests that the comoving scalar curvature perturbation $\zeta(t,{\bf x})$ can be written in terms of Goldstone mode $\pi(t,{\bf x})$ as follows, 
\bea
\zeta(t,{\bf x}) \approx -H \pi(t,{\bf x}).
\eea 
We can now write the second-order 
perturbed action in terms of the variable $\zeta(t,{\bf x})$ with the help of eqn.(\ref{EFTpi}) and it  
remains convenient to work in conformal time $(\tau)$ rather than the physical time $(t)$,
\bea  \label{s2zeta}
S_\zeta ^ {(2)} = \int {\cal L}_{\zeta}^{(2)}d\tau = M_p ^2 \int d\tau  \, d^3x  \,  a^2 \Big( \frac{\epsilon}{c_s ^2}\Big)\Bigg[ \zeta'\,^2 - c_{s}^2(\partial_{i}\zeta)^{2} \Bigg].
\eea

\section{Implementing stochasticity within EFT setup}\label{s4}

In the previous section, we briefly presented the general EFT setup of inflation, which will form the underlying basis of our future analysis, focusing primarily on the dynamics of perturbations using stochastic inflation formalism.  
The path of implementing stochasticity within the EFT formalism involves the analysis of the stochastic nature of fluctuations using Hamilton's equations or, more precisely, the Langevin equations, followed by the evolution of the distribution function of the curvature perturbations using the Fokker-Planck equation, from the Langevin equations. In the upcoming sections, we will deduce the Fokker-Planck equation in the EFT formalism.

\subsection{Hamilton's equation: The way towards formulating Langevin equation}
\label{s4a}
\begin{figure*}[htb!]
    	\centering
    {
       \includegraphics[width=19cm,height=10.5cm]{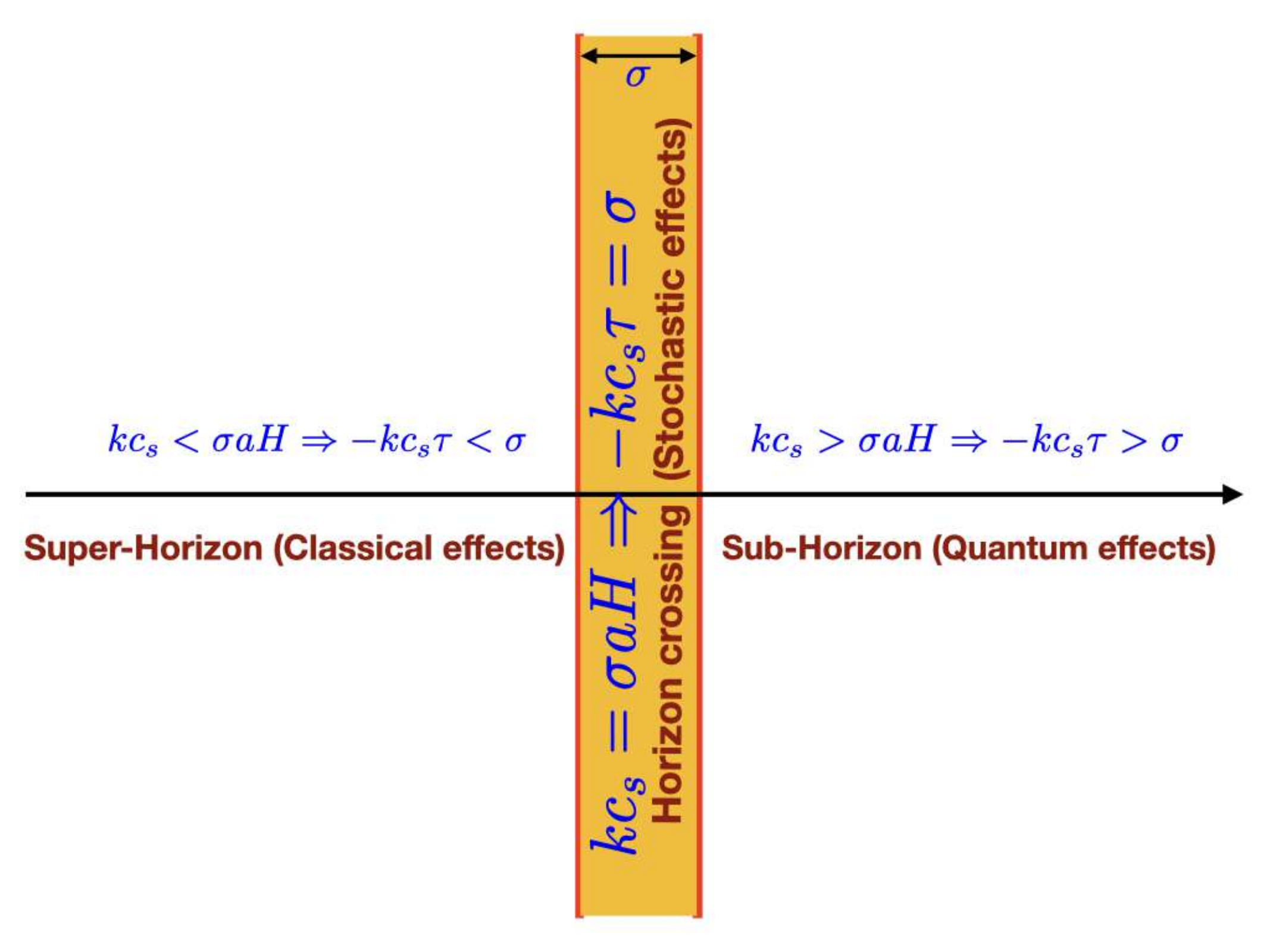}
        \label{SS}
    } 
    \caption[Optional caption for list of figures]{Schematic diagram of stochastic effects taking place for a mode during Horizon crossing. The Horizon is now represented using a region of finite width $\sigma$ where the modes classicalize and become part of the infrared regime after Horizon crossing. As $\sigma$ increases close to $1$ the stochastic effects decrease quickly and so does the width of the orange band. Conversely, for $\sigma\ll 1$, the stochastic effects also increase and similarly the width of orange band increases.  }
\label{stochasticdiag1}
    \end{figure*}

\begin{figure*}[htb!]
    	\centering
    {
       \includegraphics[width=19cm,height=12.5cm]{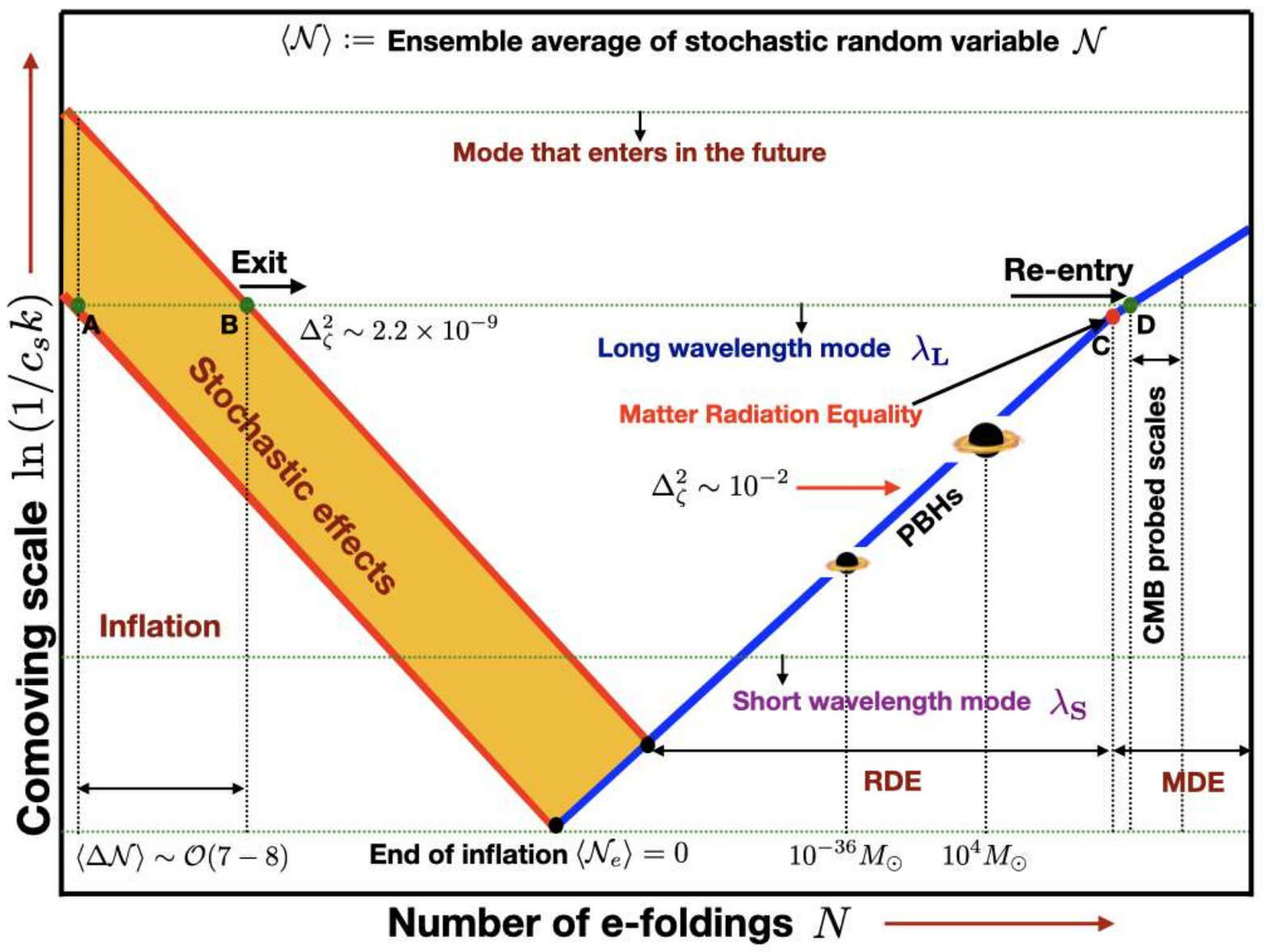}
        \label{CS}
    } 
    \caption[Optional caption for list of figures]{Schematic diagram featuring modes which travel during inflation from the Sub-Horizon, encounter stochastic effects at the Horizon crossing, and later re-enter the Horizon. The long wavelength $\lambda_{L}$ modes associated with the CMB scales exit at the position marked $A$ and $B$ and later re-enter at the instant marked as $D$. The radiation-matter equality is labelled as $C$ which is very close to the CMB re-entry scale. The short wavelength $\lambda_{s}$ are associated with the small length scales near the end of inflation which are responsible for later collapse into PBHs in the radiation-dominated era (RDE) of varied masses. the number of e-folds elapsed between an initial and final set of conditions acts as a stochastic variable ${\cal N}$. }
\label{stochasticdiag2}
    \end{figure*}

Stochastic inflation is an effective theory approach to studying the long-wavelength components of the inflationary quantum fluctuations. It involves coarse-graining these fluctuations over a fixed scale, just larger than the Hubble radius during the period of inflation. The fixed scale in the Fourier space reads as:
\bea k_{\sigma}c_{s}=\sigma aH,
\eea
where the stochastic parameter satisfies $\sigma \ll 1$, and this scale behaves as a cut-off for the modes $k$ for which, $- kc_{s}\tau \ll \sigma$ (with $\sigma \ll 1$), that contribute to the coarse-grained sector of the quantum fluctuations. As inflation proceeds, the fluctuations from the small-wavelength or UV (ultraviolet) sector exit the Hubble radius and \textcolor{black}{join} the long-wavelength or infrared (IR) sector increasing its size. The dynamics of the resulting coarse-grained sector is described under the classical stochastic theory by the Langevin equation. \textcolor{black}{It is} a stochastic differential equation that includes a classical drift term and the effects from the quantum noise which after horizon exit becomes part of the classical noise term. The fig. (\ref{stochasticdiag1}) describes the different regimes in the stochastic inflation picture. The quantum effects are most dominant for modes with wavenumber greater than the cut-off scale while the classical effects are understood for the modes with wavenumber smaller than the cut-off. The Horizon crossing boundary \textcolor{black}{becomes} an ill-defined concept due to stochastic effects continuously at play. 

We refer to the work \cite{Grain:2017dqa,Vennin:2020kng} for discussions on the Hamiltonian formulation in stochastic inflation. To develop the Langevin equations from the Hamilton's equations, we would first need to identify the UV and IR sectors of the quantum field driving inflation. In order to obtain the Langevin equations in the EFT setup where our variable of interest remains the comoving curvature perturbation $\zeta(t,{\bf x})$, we begin with a similar approach as detailed in \cite{Grain:2017dqa,Vennin:2020kng}. Our starting point would be the second-order perturbed action as mentioned in eqn. (\ref{s2zeta}) for $\zeta(t,{\bf x})$, from which one can calculate the conjugate momentum variable as:
\bea
\tilde{\Pi}_{\zeta} = \frac{\partial {\cal L}_{\zeta}^{(2)} }{\partial \zeta'} = \frac{2M_{p}^{2}a^{2}\epsilon}{c_{s}^{2}}\zeta',
\eea
using which the corresponding Hamiltonian density after a Legendre transformation can be evaluated as follows,
\bea
{\cal H}_{\zeta}^{(2)} &=& \tilde{\Pi}_{\zeta}\zeta' - {\cal L}^{(2)}_{\zeta} = M_{p}^{2}\int d^{3}x\;a^2 \Big( \frac{\epsilon}{c_s ^2}\Big) \Bigg[\frac{c_{s}^{4}}{4M_{p}^{4}a^{4}\epsilon^{2}}\tilde{\Pi}^{2}_{\zeta} + c_{s}^2(\partial_{i}\zeta)^{2} \Bigg].
\eea
Now we can analyze the following Hamilton's equations of motion using this Hamiltonian density: 
\bea
\label{Ham1}
-\tilde{\Pi}_{\zeta}' &=& \frac{\partial{\cal H}_{\zeta}^{(2)}}{\partial\zeta} = 0,\\
\label{Ham2}
\zeta' &=& \frac{\partial{\cal H}_{\zeta}^{(2)}}{\partial\tilde{\Pi}_{\zeta}} = \frac{c_{s}^2}{2M_{p}^{2}a^{2}\epsilon}\tilde{\Pi}_{\zeta},
\eea
where the prime denotes derivative with respect to the conformal time and the variable $\zeta$ turns out as a cyclic co-ordinate from the Lagrangian ${\cal L}^{(2)}_{\zeta}$. We make change in our choice of time variable from the conformal time to the e-folds $N$ and this leads to the eqn. (\ref{Ham2}) have the following form:
\bea \label{Langevin1}
\frac{d\zeta}{dN} &=& \frac{c_{s}^2}{2M_{p}^{2}Ha^{3}\epsilon}\tilde{\Pi}_{\zeta} = \Pi_{\zeta}, \\
\label{Langevin2}
\frac{d\Pi_{\zeta}}{dN} &=& \frac{1}{2M_{p}^{2}}\frac{d}{dN}\Bigg[\frac{c_{s}^2}{Ha^{3}\epsilon}\tilde{\Pi}_{\zeta}\Bigg] 
= \frac{c_{s}^2}{2M_{p}^{2}Ha^{3}\epsilon}\tilde{\Pi}_{\zeta}\Bigg[2s + \epsilon -3 - \eta \Bigg]= -(3-\epsilon)\Bigg[1 - \frac{2(s-\frac{\eta }{2})}{(3-\epsilon)} \Bigg]\Pi_{\zeta},
\eea
here conversion between the two time variables, $dN/d\tau = Ha$, is used and in the last equality of eqn. (\ref{Langevin1}) we rename the conjugate momentum and remove the tilde. To obtain the other eqn. (\ref{Langevin2}), we perform the derivative of the conjugate momentum with the e-folds $N$ where the third and last lines use the eqn. (\ref{Ham1}) and definition of $\Pi_{\zeta}$, respectively, and the following definitions for the slow-roll parameters:
\bea \label{slowrollparams}
s = \frac{d\ln c_{s}}{dN},\quad\quad \epsilon = -\frac{d\ln{H}}{dN},\quad\quad \eta = \epsilon-\frac{1}{2}\frac{d\ln{\epsilon}}{dN}.
\eea
The above two equations, (\ref{Langevin1}) and (\ref{Langevin2}), will further enable us to determine the Langevin equation governing the dynamics of the coarse-grained \textcolor{black}{super}-Hubble fields in the presence of white noises.
The fig. \ref{stochasticdiag2} describes the evolution history of the short and long-wavelength modes starting inside the Horizon at very early times, experiencing stochastic effects up to the instant they become \textit{classicalized} and continue their journey to the super-Hubble scales. Different wavelength modes re-enter at different instances in e-foldings. The short-wavelength modes can contribute to the PBH formation while the long-wavelength modes contribute to the CMB observations. The value of the stochastic parameter $\sigma$ characterizes the orange-coloured region describing the stochastic effects. With increasing $\sigma\sim 1$, the stochastic or coarse-graining effects tend to decrease quickly and the band shortens, while, on keeping $\sigma\ll 1$, the band increases in size and similarly points to increase in the duration of stochastic effects. \textcolor{black}{The choice of the coarse-graining window function also determines the nature of the noises generated. For this case involving a Heaviside Theta as our window function, the horizon crossing is marked by a fixed instant in time. The noise terms resulting from this window function are referred to as \textit{white noise}, leading to a Markovian description of the systems' future evolution; this is a signature of the memoryless property of the stochastic process. On the contrary, a window function based on a specific profile leads to what gets classified as being a \textit{coloured noise}, characteristic of a non-Markovian system.}

We start with the quantum operator picture and later present the equations in their classical version on the super-Hubble scales. Now, to obtain the Langevin equations in terms of the curvature perturbations for the coarse-grained components of the initial quantum fields, a decomposition of the said perturbations into its UV and IR components proves beneficial and can be done in the manner:
\bea \label{modedecomp}
\Hat{\Gamma} = \Hat{\Gamma}_{\bf IR} + \Hat{\Gamma}_{\bf UV}\quad\quad {\rm where}\quad \Hat{\Gamma}_{\bf IR}= \{\Hat{\zeta},\Hat{\Pi}_{\zeta}\}\quad {\rm and} \quad \Hat{\Gamma}_{\bf UV}= \{\Hat{\zeta}_{s},\Hat{\Pi}_{\zeta_{s}}\},
\eea
 where the subscript $s$ denotes small-wavelength components. These UV components can be expanded in the Fourier modes and get selected upon satisfying $k > k_{\sigma}$ or smaller than the cut-off scale:
\bea \label{uvmodesfourier}
\Hat{\zeta_{s}} &=& \int_\mathbb{R^3}\frac{d^3 k }{(2\pi)^3 }W\Bigg(\frac{k}{k_\sigma}\Bigg)\Bigg[ \Hat{a}_{{\bf k}}\zeta_k(\tau) e^{-i {\bf k}.{\bf x}} + \Hat{a}_{{\bf k}}^{\dagger}\zeta_k^*(\tau) e^{i{\bf k}.{\bf x}}\Bigg],\\
\Hat{\Pi}_{\zeta_{s}} &=& \int_\mathbb{R^3}\frac{d^3 k}{(2\pi)^3}W\Bigg(\frac{k}{k_\sigma}\Bigg)\Bigg[ \Hat{a}_{{\bf k}}\Pi_{\zeta_k}(\tau) e^{-i {\bf k}.{\bf x}} + \Hat{a}_{{\bf k}}^{\dagger}\Pi_{\zeta_k}^*(\tau) e^{i{\bf k}.{\bf x}}\Bigg].
\eea 
\textcolor{black}{where $k_{\sigma}=\sigma aH$.} These contain the annihilation and creation operators $a_{k}$ and $a_{k}^{\dagger}$ that satisfy the usual canonical commutation relations:
\bea [\Hat{a_{\bf k}}, \Hat{a_{\bf k'}}^{\dagger}] =\del^3({\bf k}-{\bf k'}), \quad\quad [\Hat{a_{\bf k}}, \Hat{a_{\bf k'}}] = [\Hat{a_{\bf k}^{\dagger}}, \Hat{a_{\bf k'}^{\dagger}}] = 0.\eea 
The window function $W$ here operates such that: 
\begin{equation}
 W = \begin{cases}
  1  & k > k_{\sigma} \\
0                                                        & k < k_{\sigma},
  \end{cases} \\
\end{equation}
\textcolor{black}{and so, it only selects modes that contribute to the small-wavelength or UV sector. Following \cite{Grain:2017dqa, Vennin:2020kng}, the Langevin equations come from using Hamilton's equation after utilizing the UV/IR field decomposition in eqn. (\ref{modedecomp}) where the UV modes, as given in eqn. (\ref{uvmodesfourier}), contribute later towards the noise terms as follows:}

\bea \label{eftlangevin}
\frac{d\hat{\zeta}}{dN} &=&  \hat{\Pi}_{\zeta} + 
 \hat{\xi}_\zeta(N),\\ 
\frac{d\hat{\Pi}_\zeta }{dN} &=& -(3-\epsilon)\hat{\Pi}_\zeta \Bigg[1 -\frac{2(s-\frac{\eta }{2})}{(3-\epsilon)} \Bigg] + \hat{\xi}_{\pi_\zeta}(N),
\eea 


The quantities $\hat{\xi}_\zeta(N)$ and $\hat{\xi}_{\pi_\zeta}(N)$ denote the quantum white noise terms, sourced by the constant outflow of UV modes into the IR modes, and they read as:
\bea 
\Hat{\xi}_\zeta &=& -\int_\mathbb{R^3}\frac{d^3 k }{(2\pi)^3 }\frac{d}{dN}W\Bigg(\frac{k}{k_\sigma}\Bigg)\Bigg[ \Hat{a}_{{\bf k}}\zeta_k(\tau) e^{-i {\bf k}.{\bf x}} + \Hat{a}_{{\bf k}}^{\dagger}\zeta_k^*(\tau) e^{i{\bf k}.{\bf x}}\Bigg],\\
\Hat{\xi}_{\pi_{\zeta}} &=& -\int_\mathbb{R^3}\frac{d^3 k }{(2\pi)^3 }\frac{d}{dN}W\Bigg(\frac{k}{k_\sigma}\Bigg)\Bigg[ \Hat{a}_{{\bf k}}\Pi_{\zeta_k}(\tau) e^{-i {\bf k}.{\bf x}} + \Hat{a}_{{\bf k}}^{\dagger}\Pi_{\zeta_k}^*(\tau) e^{i{\bf k}.{\bf x}}\Bigg],
\eea 
and the form of the window function $W$ is for convenience chosen to be a Heaviside function:
\bea 
W\Bigg(\frac{k}{k_\sigma}\Bigg) = \Theta\Bigg(\frac{k}{k_\sigma} - 1\Bigg)=\Theta\Bigg(\frac{k}{\sigma aH}-1\Bigg).
\eea
Here $\sigma$ physically represents the stochastic \textcolor{black}{coarse} graining parameter. Next, we compute the expression for the derivative of the window function with respect to the number of e-foldings which is going to be extremely useful to determine the quantized version of the white noise as explicitly mentioned above. Here we have:
\bea \frac{d}{dN}W\Bigg(\frac{k}{k_\sigma}\Bigg)=\frac{k}{k_{\sigma}}\left(\epsilon-1\right) \frac{d}{d\tau}W\Bigg(\frac{k}{k_\sigma}\Bigg)=\frac{k}{k_{\sigma}}\left(\epsilon-1\right)k_{\sigma}\delta(k-k_{\sigma})=k\left(\epsilon-1\right)\delta(k-k_{\sigma}).\eea

This function imposes a sharp cut off between IR and UV modes and its derivative with e-folds produces a Dirac Delta distribution $\delta(k-k_{\sigma}(N))$ giving us contributions to the noises after integrating over the various cut-off scales $k_{\sigma}$.
Since the noise terms are stochastic in nature this leads to a probabilistic description of the system, which can be analysed in terms of the Langevin Equations. The equations can be solved analytically by a transition into the corresponding Fokker-Planck equation, a second-order partial differential equation, whose construction we review in the appendix \ref{app:A}.


\section{Realising ultra slow-roll phase within the framework of EFT of Stochastic Single Field Inflation}
\label{s5}

\begin{figure*}[ht!]
    	\centering
    \subfigure[]{
      	\includegraphics[width=8.5cm,height=7.5cm]{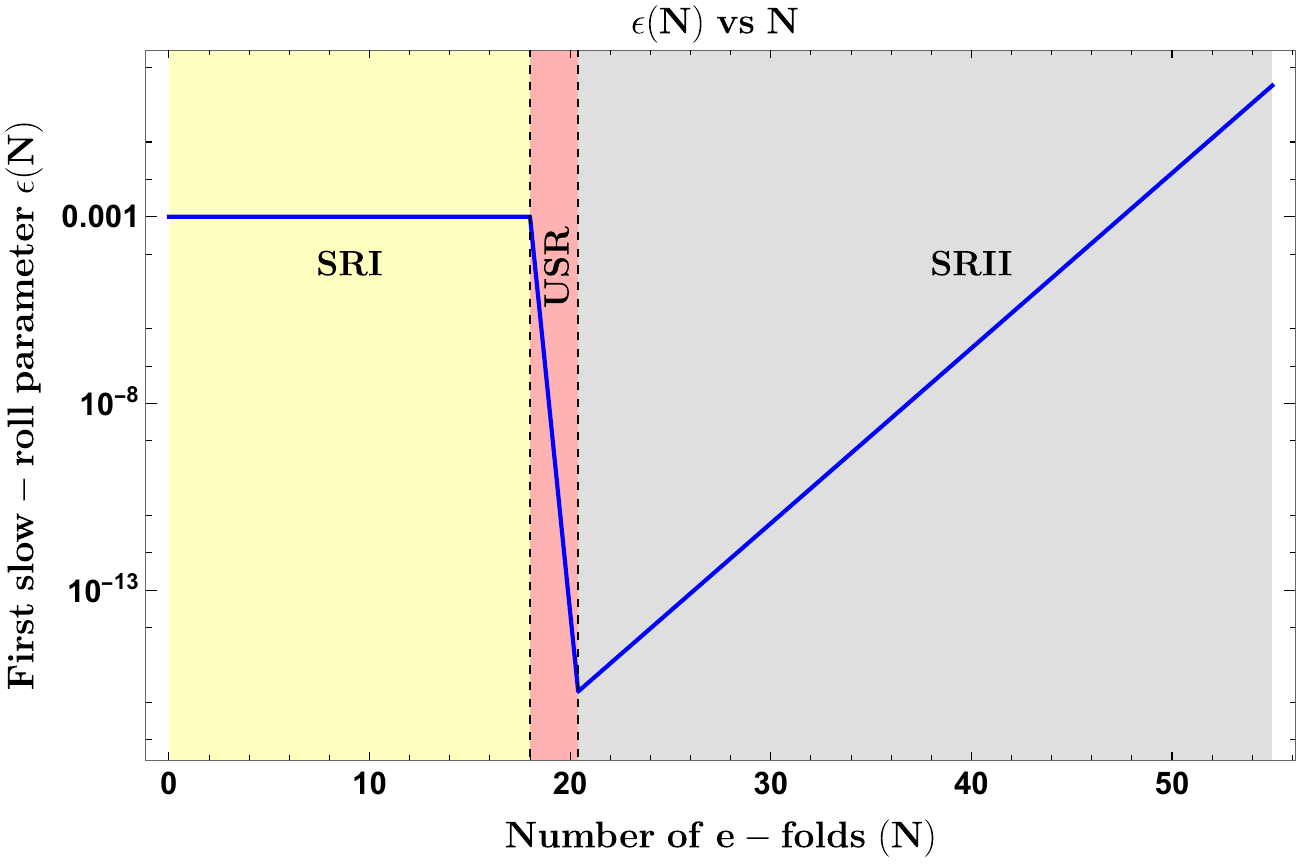}
        \label{epsilon}
    }
    \subfigure[]{
        \includegraphics[width=8.5cm,height=7.5cm]{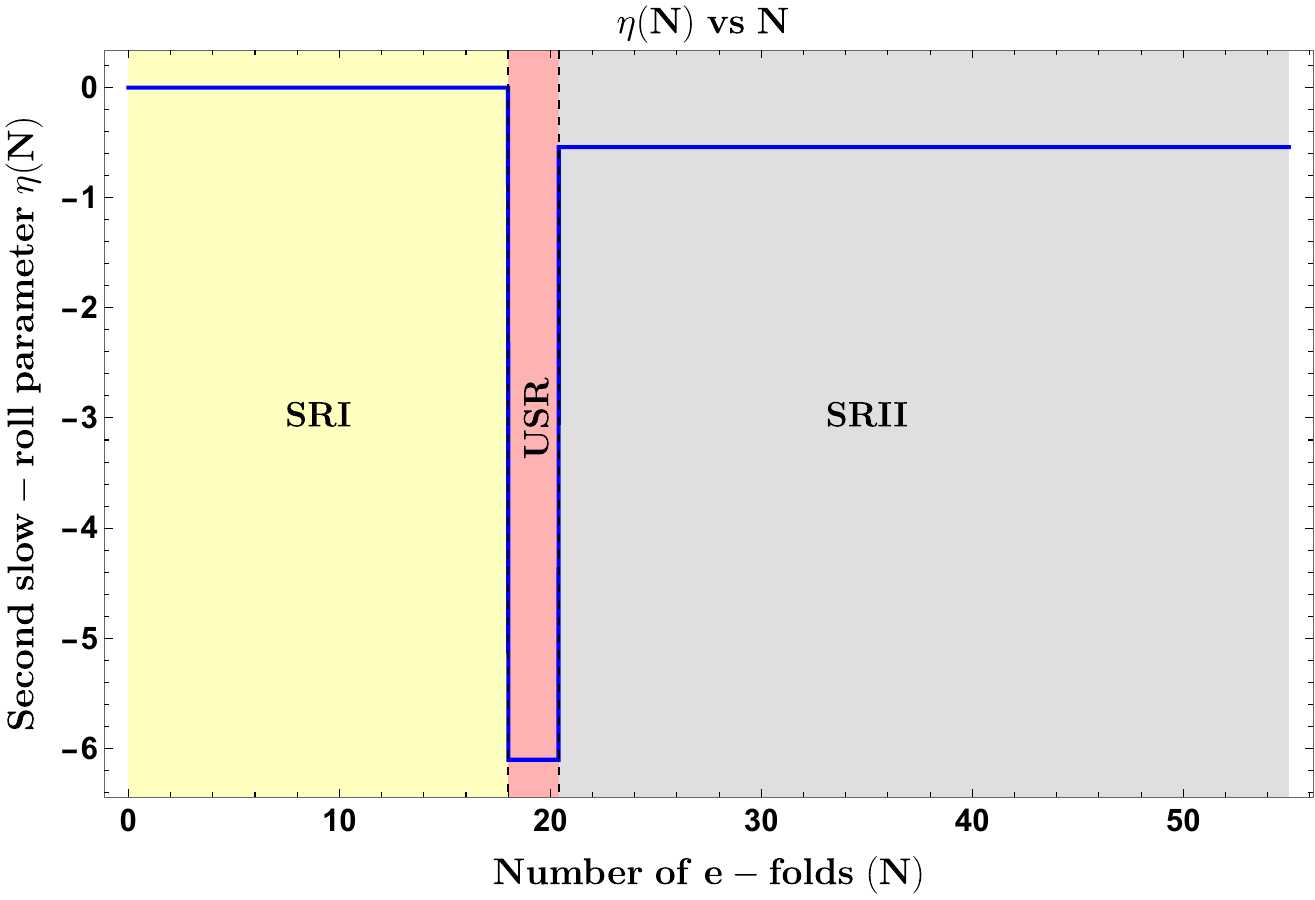}
        \label{eta}
    }
    \subfigure[]{
        \includegraphics[width=8.5cm,height=7.5cm]{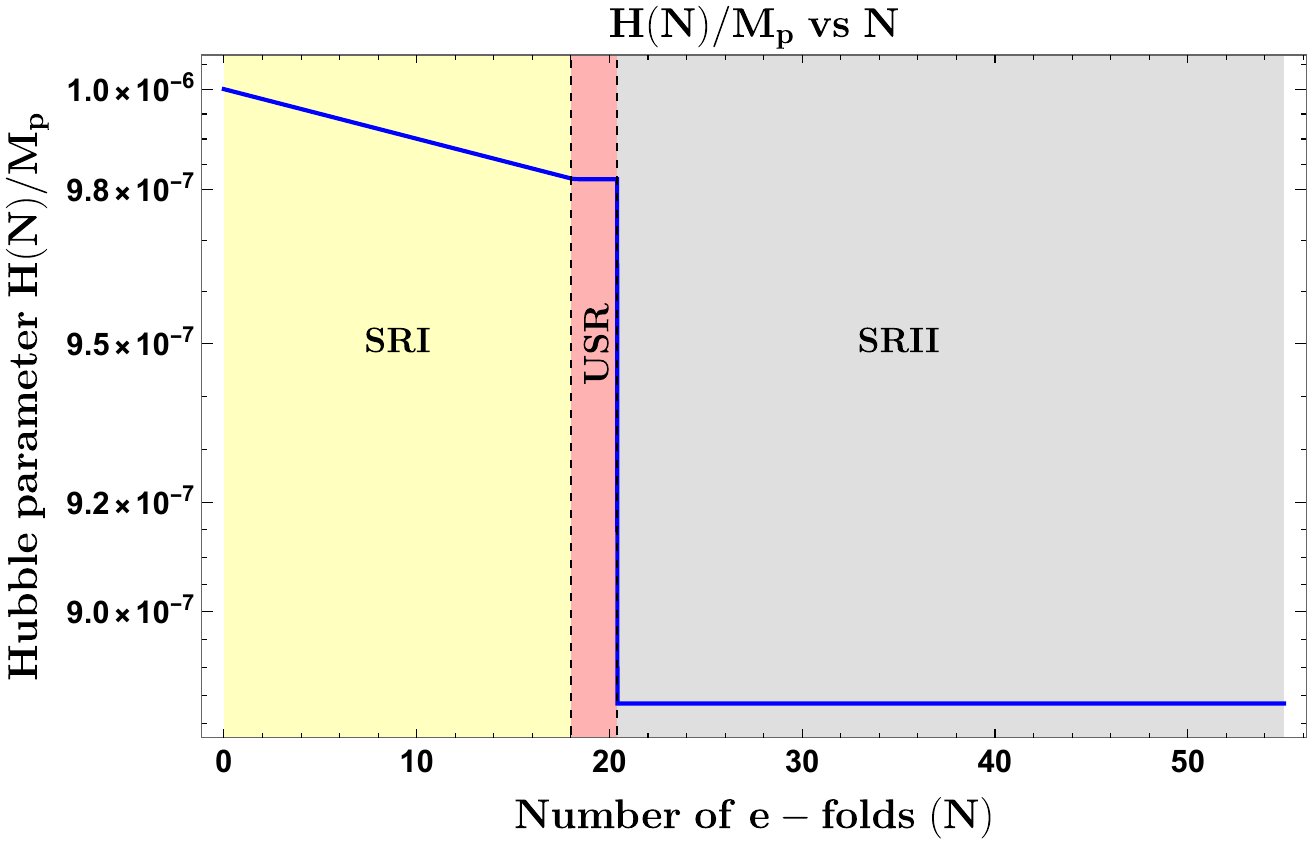}
        \label{hubble}
    }
    \subfigure[]{
        \includegraphics[width=8.5cm,height=7.5cm]{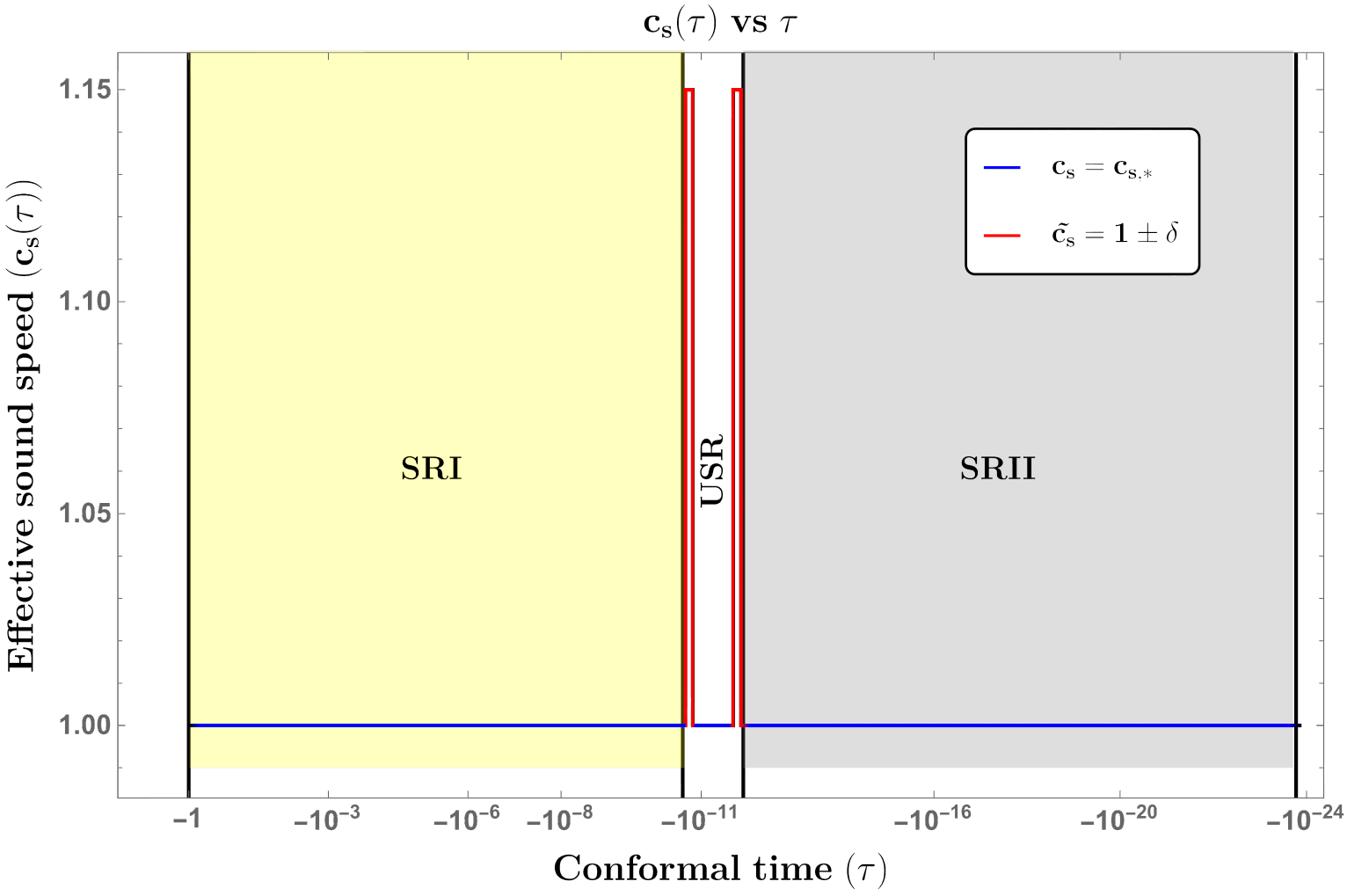}
        \label{sound}
        }
    	\caption[Optional caption for list of figures]{Realisation of the slow-roll and Hubble parameters in the presence of a USR phase during inflation as a function of the e-foldings $N$. The \textit{top-left} panel shows the $\epsilon(N)$ parameter changing through each phase of the setup, \textit{top-right} panel for the $\eta(N)$ parameter changing across the same phases, and the \textit{bottom-left} panel for the Hubble parameter $H(N)$ through each of the respective phases. The \textit{bottom-right} panel presents a schematic for the effective sound speed $c_{s}$ across the three phases. } 
    	\label{realisation}
    \end{figure*}
In this section we investigate the realisation of the ultra slow-roll phase sandwiched between the slow-roll, SRI and SRII phases, for a stochastic single field inflation framework. We follow the approach of beginning with \textcolor{black}{a particular} paramterization of the second slow-roll parameter $\eta$ across the three phases of interest and through that derive the behaviour of the other two parameters, namely $\epsilon$ and the Hubble expansion $H$. This will give us some insights into how a USR phase can be visualised from its effects on the parameters which characterize and are involved in the observables coming from inflation. 

The formulas for the slow-roll parameters with respect to the conformal time are defined previously in eqn. (\ref{slowrollparams}). Assuming constant $\eta$, the definition for $\epsilon$ gives us:
\bea
\eta = \epsilon-\frac{1}{2}\frac{d\ln{\epsilon}}{dN} \implies \int_{\epsilon_{i}}^{\epsilon}\frac{d\epsilon}{2\epsilon(\epsilon-\eta)} = \int_{N_{i}}^N dN, 
\eea
which after integration using some initial condition values, $(\epsilon_{i},N_{i})$, leads to the following expression:
\bea \label{epsilonphases}
\epsilon(N) = \eta\bigg(1-\bigg(1-\frac{\eta}{\epsilon_{i}}\bigg)e^{2\eta\Delta N}\bigg)^{-1}\sim \left\{
	\begin{array}{ll}
		\displaystyle  \epsilon(N) & \mbox{when}\quad  N_{*} \leq N \leq N_{s}  \;(\rm SRI),  \\  
			\displaystyle 
			\displaystyle \epsilon(N_{s})e^{-2\eta(N)(N-N_{s})} & \mbox{when}\quad  N_{s} \leq N \leq N_{e}  \;(\rm USR), \\ 
   \displaystyle 
			\displaystyle \epsilon(N_{s})e^{-2\eta(N_{e})(\Delta N_{es})}e^{-2\eta(N)(N-N_{e})} & \mbox{when}\quad  N_{e} \leq N \leq N_{\rm end}  \;(\rm SRII), 
	\end{array} \right.
\eea
The above equations result from taking appropriate limiting behaviour of the values for the slow-roll parameters in each phase. There complete behaviour is visualised through the fig. (\ref{slowrollparams}). The values $\epsilon(N_{s})$ and $\epsilon(N_{s})e^{-2\eta(N_{e})(\Delta N_{es})}$, where $\Delta N_{es}=N_{e}-N_{s}$, act as the initial condition values for the USR and SRII,  respectively, when joining them together. The final $\epsilon$ remains just the sum of its values in the respective regions:
\bea
\epsilon(N)= \epsilon(N\leq N_{s}) + \epsilon(N_{s}\leq N\leq N_{e}) + \epsilon(N_{e}\leq N\leq N_{\rm end}),
\eea
In the present work, the $\eta$ parameter remains as a constant value throughout each phase and using the same we can write:
\bea
\eta(N) = \left\{
	\begin{array}{ll}
		\displaystyle  \eta_{\rm I}(N) & \mbox{when}\quad  N_{*} \leq N \leq N_{s}  \;(\rm SRI),  \\  
			\displaystyle 
			\displaystyle \eta_{\rm II}(N) & \mbox{when}\quad  N_{s} \leq N \leq N_{e}  \;(\rm USR), \\ 
   \displaystyle 
			\displaystyle \eta_{\rm III}(N) & \mbox{when}\quad  N_{e} \leq N \leq N_{\rm end}  \;(\rm SRII), 
	\end{array} \right.
\eea
where $\eta_{\rm I},\;\eta_{\rm II},\;\eta_{\rm III}$ are different constants. The above can be expressed using a Heaviside Theta function to join for the three phases as follows:
\bea
\eta(N) = \eta_{\rm I}(N\leq N_{s}) + \Theta(N-N_{s})\eta_{\rm II}(N_{s}\leq N\leq N_{e}) + \Theta(N-N_{e})\eta_{\rm III}(N_{e}\leq N\leq N_{\rm end}),
\eea
The Hubble parameter is similarly obtained using the definition of $\epsilon$ as follows:
\bea
\epsilon(N) = -\frac{d\ln{H}}{dN} \implies H(N) = H_{i}\exp{\bigg(-\int_{N_{i}}^{N}\epsilon(N)dN\bigg)},
\eea
where $H_{i}\equiv H(N_{i})$ is the initial condition value of the Hubble expansion during each phase.
For each phase the Hubble can then be expressed as follows:
\bea
H(N) = \left\{
	\begin{array}{ll}
		\displaystyle  H(N_{*})\exp{\bigg(-\int_{N_{*}}^{N}\epsilon(N)dN\bigg)} & \mbox{when}\quad  N_{*} \leq N \leq N_{s}  \;(\rm SRI),  \\  
			\displaystyle 
			\displaystyle H(N_{s})\exp{\bigg(-\int_{N_{s}}^{N}\epsilon(N)dN\bigg)} & \mbox{when}\quad  N_{s} \leq N \leq N_{e}  \;(\rm USR), \\ 
   \displaystyle 
			\displaystyle H(N_{e})\exp{\bigg(-\int_{N_{e}}^{N}\epsilon(N)dN\bigg)} & \mbox{when}\quad  N_{e} \leq N \leq N_{\rm end}  \;(\rm SRII), 
	\end{array} \right.
\eea
and the final expression for the Hubble expansion as function of the e-folds after adding contributions for each phase separately becomes as:
\bea
H(N) = H(N_{*})\exp{\bigg(-\int_{N_{*}}^{N}\epsilon(N)dN\bigg)} + H(N_{s})\exp{\bigg(-\int_{N_{s}}^{N}\epsilon(N)dN\bigg)} + H(N_{e})\exp{\bigg(-\int_{N_{e}}^{N}\epsilon(N)dN\bigg)},
\eea
where $H(N_{*}),\;H(N_{s}),\;H(N_{e})$ comes as initial conditions from the continuity of  values between the different phases.

We now follow the plots in fig. \ref{eta}, \ref{epsilon}. Initially in the SRI, which starts from the instant $N_{*}$ and where slow-roll conditions are necessarily met, $\eta$ behaves almost as a negatively small and constant quantity and similarly $\epsilon$ also stays as a small constant quantity. This happens until a deviation from standard slow-roll arguments is encountered after a certain number of e-folds $N_{s}$. This instant marks the transition into the next phase where a significant violation of slow-roll condition can take place. The nature of the transition at $N_{s}$ is chosen for the present work as a sharp transition, modelled by a Heaviside Theta function $\Theta(k-k_{s})$, where $k_{s}$ is the corresponding transition wavenumber. The magnitude of $\eta$ greatly increases in the USR, as we will also see in the later eqn. (\ref{etausr}), and in a similar manner $\epsilon$ suffers a drastic decrease in magnitude, see eqn. (\ref{epsilonusr}). Such behaviour as mentioned persists till the duration of USR comes to an end at the instant $N_{e}$, with the corresponding wavenumber $k_{e}$. Soon after exiting from the USR, another sharp transition of the form $\Theta(k-k_{e})$ is found to take effect and the slow-roll conditions \textcolor{black}{start} to restore. The parameters $\epsilon$ and $\eta$ in the SRII start to increase throughout the whole phase till they achieve values of magnitude ${\cal O}(1)$ and the associated instant in e-folds $N_{\rm end}$ marks the end of inflation. For successful inflation in the set up one requires for the total change in e-foldings from the initial moment $N_{*}$ to lie within $\Delta N_{\rm Total}=N_{\rm end}-N_{*}\sim {\cal O}(55-60).$ 

The plot in fig. \ref{sound} describes the effective sound speed $c_{s}$ parameterization for the present set up. Here we choose to have $c_{s}(\tau=\tau_{*})=c_{s,*}=1$ in the initial slow-roll phase, where $c_{s,*}$ is the corresponding value at the pivot scale. The condition $c_{s}=1$ refers to a canonical stochastic single-field inflation model. When near the instant of sharp transition, at conformal time $\tau=\tau_{s}$ with $N=N_{s}$, the effective sound speed changes from its previous to a new value labelled here as $\tilde{c_{s}}=1\pm \delta$, where $\delta\ll 1$ is a fine-tuning variable. This sudden jump in its magnitude is what forces the sharp nature of the transition in between the SRI and the USR phase. Again throughout the USR ($\tau_{s}\leq \tau \leq\tau_{e}$) , the sound speed changes back to $c_{s}$. At the instant of the second sharp transition, $\tau=\tau_{e}$, we once again observe the value to experience sudden change into $c_{s}=\tilde{c_{s}}=1\pm \delta$, after which into the SRII phase the sound speed reverts back to $c_{s}$ till we encounter the end of inflation at the conformal time instant $\tau=\tau_{\rm end}$ or $N=N_{\rm end}$.  

The aforementioned parameterization of the effective sound speed can also be written for the three regions, including at instances of the sharp transitions, in the following manner:
\bea
c_{s}(\tau) = \left\{
	\begin{array}{ll}
		\displaystyle  c_{s,*} & \mbox{when}\quad  \tau_{*} \leq \tau < \tau_{s}  \;(\rm SRI),  \\
        \displaystyle 
			\displaystyle \tilde{c_{s}} = 1\pm \delta & \mbox{when}\quad  \tau = \tau_{s}  \;(\rm USR),\\
			\displaystyle 
			\displaystyle c_{s,*} & \mbox{when}\quad  \tau_{s} < \tau < \tau_{e}  \;(\rm USR), \\
        \displaystyle 
			\displaystyle \tilde{c_{s}} = 1\pm \delta & \mbox{when}\quad  \tau = \tau_{e}  \;(\rm SRII),\\
        \displaystyle 
			\displaystyle c_{s,*} & \mbox{when}\quad  \tau_{e} < N < \tau_{\rm end}  \;(\rm SRII), 
	\end{array} \right.
\eea
where $c_{s,*}=c_{s}(\tau=\tau_{*})$ is the value at the instance of pivot scale as mentioned in the above explanation of the parameterization.

\textcolor{black}{Before moving forward, we briefly digress here to highlight the importance of the transition properties in PBH formation. The exact nature of the transition, either sharp or smooth, has suffered much debate from various authors in refs.\cite{Choudhury:2023vuj,Choudhury:2023jlt,Choudhury:2023rks,Choudhury:2023hvf,Bhattacharya:2023ysp,Kristiano:2022maq,Kristiano:2023scm,Riotto:2023gpm,Firouzjahi:2023ahg,Firouzjahi:2023aum,Franciolini:2023lgy,Taoso:2021uvl}; this remains unconcluded. Also see, for example, the recent review \cite{Choudhury:2024ybk}, which discusses the equivalence between a smooth and sharp transition during the SR/USR/SR scenario, along with shedding light on why the conclusions remain the same irrespective of such transitions. It suggests that the results may reflect more coarse-graining or smoothed behaviour, but the overall result we quote here will remain unchanged. The above debate concerns the supposed role of the transition in controlling the large perturbation enhancement, like here in the USR, and how it gets reflected later in the scalar power spectrum. The consequences of this have led to a divisive stand when eventually considering the quantum loop corrections, as a result of this enhancement, to the overall power spectrum in the three phases and their role in the PBH formation.}

\section{Stochastic-$\delta N$ formalism and its applications in the Stochastic Single Field Inflation}
\label{s6}

\begin{figure*}[htb!]
    	\centering
    {
       \includegraphics[width=17cm,height=12cm]{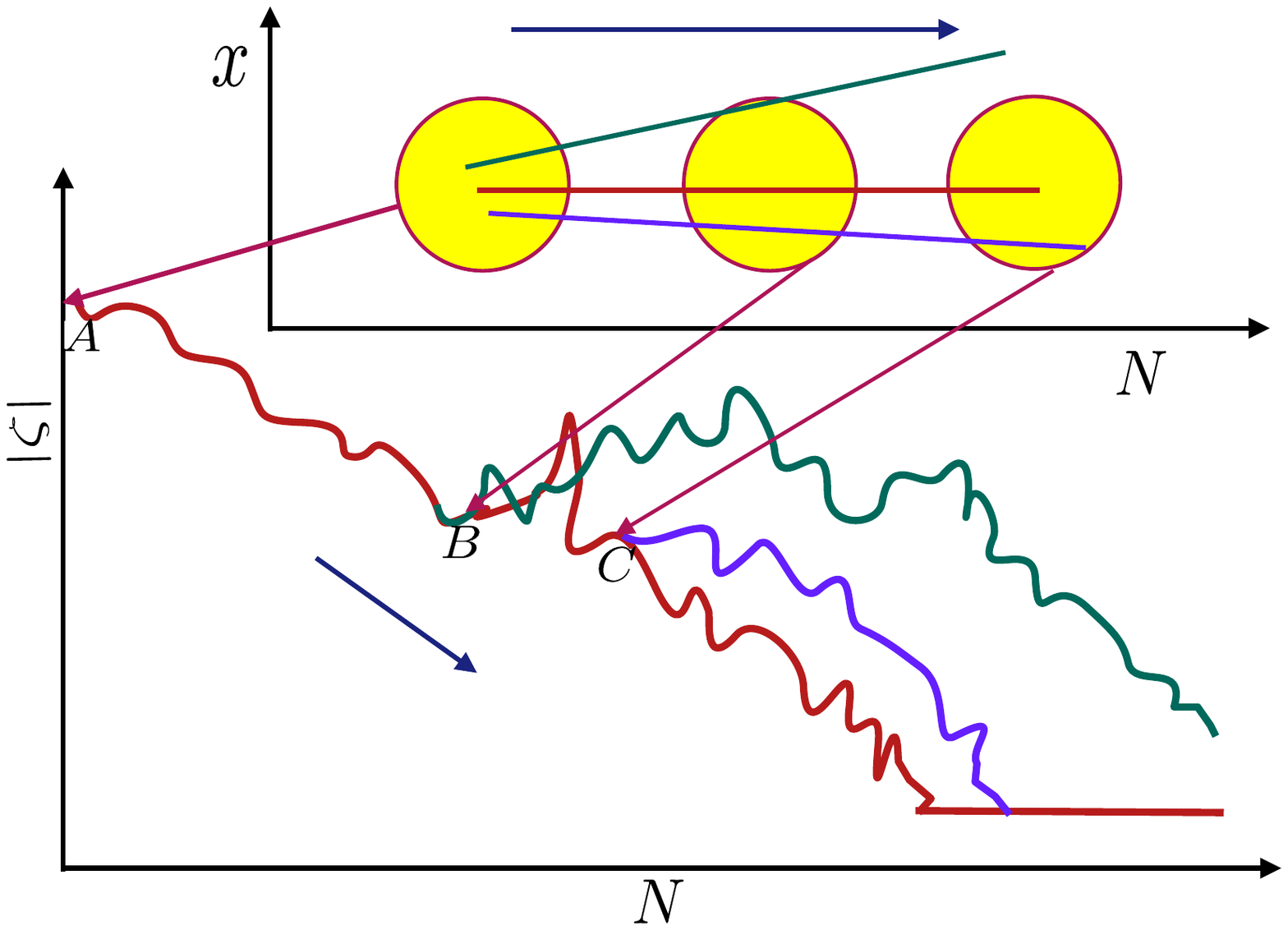}
        \label{CGCP}
    } 
    \caption[Optional caption for list of figures]{A schematic visualisation of change in coarse-grained curvature perturbation with e-folds $N$. The bottom graphs represents value of coarse-grained $\zeta$ at different spatial points. Initially, at $A$, a set of spatial points lie inside same the Hubble radius of the central point, whose evolution with time is red, inside which the white noises are correlated. The green and blue lines show branching evolution of the other points with time. After a certain instant, at points $B$ and $C$ different points move out the initial Hubble radius and experience the white noises to act differently in the future. }
\label{deltaNschematic1}
    \end{figure*}
The $\delta N$-formalism \cite{Sugiyama:2012tj,Dias:2012qy,Naruko:2012fe,Takamizu:2013gy,Abolhasani:2013zya,Clesse:2013jra,Chen:2013eea,Choudhury:2014uxa,vandeBruck:2014ata,Dias:2014msa,Garriga:2015tea,Choudhury:2015hvr,Choudhury:2016wlj,Choudhury:2017cos,Choudhury:2018glz, Starobinsky:1985ibc,Sasaki:1995aw,Sasaki:1998ug,Lyth:2005fi,Lyth:2004gb,Abolhasani:2018gyz,Passaglia:2018ixg} has emerged as an important method to compute cosmological correlations in the super-Hubble scales and has found interesting applications in the framework of stochastic inflation \cite{Cruces:2021iwq,Prokopec:2019srf,PerreaultLevasseur:2014ziv,Cruces:2022imf,Fujita:2013cna,Fujita:2014tja,Tada:2023fvd,Honda:2023unh,Asadi:2023flu,Mishra:2023lhe,Rigopoulos:2022gso,Escriva:2022duf,Tomberg:2022mkt,Ahmadi:2022lsm,Talebian:2022jkb,Figueroa:2021zah,Rigopoulos:2021nhv,De:2020hdo,Firouzjahi:2020jrj,Vennin:2020kng,Bounakis:2020jdx,Firouzjahi:2018vet,Tada:2016pmk,Assadullahi:2016gkk,PerreaultLevasseur:2014ziv} since it is able to directly relate the probability distribution of the stochastic e-folding variable ${\cal N}$ to the statistics of the curvature perturbation variable $\zeta$. The fluctuations in the e-folds across different patches of homogeneous FLRW Universes are, at leading order, related to the amount of curvature perturbations generated at some final time, usually chosen late in the super-Hubble scales. 

Generally, we start with the assumption of an unperturbed, locally homogeneous, and isotropic spacetime governed by the metric:
\bea
ds^{2} = -dt^{2} + a^{2}(t)\delta_{ij}dx^{i}dx^{j},
\eea
where $t$ is the cosmic time and $a(t)\sim \exp(Ht)$ (where $\dot{H}\ne 0$ so that $\epsilon=-\dot{H}/H^2$ exists). In the unperturbed Universes case, the Hubble parameter remains the only scale of interest to work in. In the presence of some kind of smoothing, on the scales of order $k^{-1}$, any quantity of interest is assumed to be sufficiently smooth. The perturbations come into the picture when the associated coordinate scale $k^{-1}$ becomes close to the Hubble radius, $k \sim aH$, and thereby the resulting anisotropies can appear perturbatively in powers of ${\cal O}(k/aH)$.

Next, we consider removing the gauge redundancies by the choice of a certain gauge. An example of this is choosing the uniform density gauge where the spatial slices of fixed time $t$ have uniform energy density. This choice then allows one to write the perturbed metric, considering only the scalar perturbations at this level, as follows:
\bea
ds^{2} = -dt^{2} + a^{2}(t)e^{2\zeta(t,{\bf x})}dx^{i}dx^{j},
\eea
where the new variable $\zeta(t,{\bf x})$ is the curvature perturbation present throughout our observable Universe. The form of the above-perturbed metric allows one to introduce a local scale factor consisting of a global time-dependent part and the perturbation as $\tilde{a}(t,{\bf x}) = a(t)e^{2\zeta(t,{\bf x})}$. Using this new scale factor, we can introduce the amount of expansion realised when going from an initially flat, constant $t=t_{i}$ hypersurface ($\Omega_{i}$) to a final, constant $t$ hypersurface ($\Omega_{f}$) assigned with some uniform-density as:
\bea
N(t,{\bf x}) = \ln{\bigg(\frac{\tilde{a}(t,{\bf x})}{a(t_{i})}\bigg)}.
\eea
The above can then be used to write down the amount of curvature perturbation experienced at a spatial point $x_{i}$, at instant $t$ in relation with the e-foldings elapsed \textcolor{black}{up to} that instant as follows:
\bea
\zeta(t,{\bf x}_{i}) = N(t,{\bf x}_{i}) - \bar{N}(t),\quad\quad \text{where}\;\bar{N}(t) = \ln{\bigg(\frac{a(t)}{a(t_{i})}\bigg)},
\eea
where the bar notation denotes the unperturbed amount of expansion, while $N(t,{\bf x}_{i})$ denotes the amount of expansion for unperturbed, FLRW Universes.

We now briefly talk about the stochastic formalism in inflation and the benefits of integrating it with the aforementioned $\delta N$-formalism. Here we consider not one but instead a family of FLRW Universes evolving from a certain initial condition on the phase space variables. These variables can be combined in form of a phase space vector ${\Gamma}_{i}=\{\zeta_{i},\Pi_{\zeta,i}\}$ where the index $i$ labels the various components. In the language of curvature perturbation, the stochastic formalism consists of constructing a low-energy effective theory of the long-wavelength or IR part of the initial primordial fluctuations. These fluctuations are coarse-grained over a certain fixed scale, close to the Hubble radius and defined as $k_{\sigma}=\sigma aH$, where $\sigma\ll 1$, and the resulting scalar curvature perturbation reads: 
\bea
\zeta_{\rm cg}({\bf x}) = \int_\mathbb{R^3}\frac{d^3 k }{(2\pi)^3 }\zeta_{k}e^{-i{\bf k}.{\bf x}},
\eea
where integration consider modes with wavenumber $k<k_{\sigma}$, or the IR modes in the coarse-grained curvature perturbation.
As the horizon size continues to decrease, more and more short-scale modes participate in the region of stochastic effects, getting `classicalized', and finally entering into the IR sector. The result is the appearance of classical noises, which then act on the dynamics of the super-Hubble modes, which are described by the Langevin equation; see section \ref{s4a} for a discussion on the UV and IR separation of the modes and derivation of the Langevin equation in the present context. \textcolor{black}{In the super-horizon regime, these noises remain correlated for spatial points with separations inside a Hubble patch. Past a specific instant during expansion, owing to the spatial separation between any two points exceeding the Hubble radius, the Markovian nature of the noises renders the remaining evolution of those points statistically independent. A direct consequence of this fact is the appearance of fluctuations in the total amount of e-folds realized along such trajectories until they reach the end of inflation. }

For now, we focus on a single spatial point. The amount of \textcolor{black}{expansion} realised along the worldline trajectory of the single point, starting from some initial condition to a final hypersurface, becomes a stochastic variable, represented using ${\cal N}$. From the $\delta N$-formalism mentioned earlier in this section, one can relate the curvature perturbation produced at some spatial point, coarse-grained between the scale crossing at some initial conformal time $\tau_{i}$ and the scale crossing out at the final conformal time $\tau_{f}$, to the  perturbation in the e-folding realised along the same worldline in between $\tau_{i}$ and $\tau_{f}$ as follows:
\bea \label{coarsezeta}
\zeta_{\rm cg}({\bf x}) = {\cal N}({\bf x})-\langle{\cal N}\rangle = \delta {\cal N},
\eea
where the angle brackets denote the statistical average after solving the Langevin equation for multiple realisations at a given spatial point. We reiterate that due to random noises, the e-folds realised automatically receive fluctuations, leaving ${\cal N}$ as a stochastic variable whose statistical properties can later be estimated. These fluctuations are nothing but the comoving curvature perturbation produced at the final hypersurface as consequence of the $\delta N$-formalism. The fig. \ref{deltaNschematic1} describes the evolution of the spatial points during inflation and the instances where the Gaussian random noises start to affect the points separately, at $B$ and $C$ in the figure. The coarse-grained value of the curvature perturbation remains the same at all spatial points inside a Hubble patch (yellow circles) starting from points $A$ up to $B$ and $C$ after which they evolve in a statistically independent manner (shown in green and blue lines).

\begin{table}[H]
\centering
\begin{tabular}{|c|c|c|}

\hline\hline
\multicolumn{3}{|c|}{\normalsize \textbf{Comparison between the $\delta N$ (without stochasticity) and Stochastic-$\delta N$ formalism }} \\

\hline

 & \bf{$\delta N$-formalism} & \bf{Stochastic-$\delta N$ formalism} \\
\hline
$1).$ & e-folds of evolution vary ($\delta N$) for each point $(\bf{x})$  &  e-folds for a specific point $(\bf{x})$ on the final slice \\
& on the final hypersurface slice. & receive quantum fluctuations throughout evolution.\\ \hline
$2).$ & Curvature perturbations on the final slice  &  Curvature perturbations requires coarse-graining\\
& directly related to $\delta N$ in the super-Horizon. & before relating to e-folds statistics in the super-Horizon. \\ \hline
$3).$ & $\zeta(t,{\bf x}_{i}) = N(t,{\bf x}_{i}) - \bar{N}(t) = \delta N$,  &   $\zeta_{\rm cg}({\bf x}) = {\cal N}({\bf x})-\langle{\cal N}\rangle = \delta {\cal N}$,   \\
& $N$: e-fold amount for expanding FLRW Universes  &  ${\cal N}$: stochastic variable under Langevin equations for $\zeta_{\rm cg}(\bf{x})$ \\ \hline
$4).$ & No influence of quantum noises on evolution history. & Quantum noises \textit{classicalize} after cut-off and affect $\zeta_{\rm cg}$ evolution.                              \\ \hline
\hline

\end{tabular}
\caption{Comparative study between the two frameworks: $\delta N$ (without stochasticity) and stochastic-$\delta N$. }
\label{tabcomparison}

\end{table}

Table \ref{tabcomparison} summarizes the key differences between the usual $\delta N$ (without stochasticity) and the stochastic-$\delta N$ formalism. 
From the above discussion, we outline how the power of the $\delta N$-formalism can be utilised with stochastic formalism instead of solving the Langevin equation over multiple spatial points. The power spectrum $\Delta^{2}_{f,g}$ is an example of a quantity easily estimated from this stochastic-$\delta N$ formalism which we now mention. Here $\{f,g\}$ refer to the phase space variables $\{\zeta,\Pi_{\zeta}\}$, and using Fourier mode decomposition, the dimensionless power spectrum for the e-folding fluctuation $\delta N$ can be written down as:
\bea
\Delta^{2}_{\delta N}(k) = \frac{k^{3}}{2\pi^{2}}\int\;d^{3}{\bf x}\int\;d^{3}{\bf x'}\langle\delta N({\bf x'})\delta N({\bf x})\rangle e^{-i{\bf k}.({\bf x}-{\bf x'})},
\eea
which can be inverted to give the variance in e-folds as:
\bea \label{varPspec}
\langle\delta{\cal N}^{2}\rangle = \langle({\cal N}-\langle{\cal N}\rangle)^{2}\rangle = \langle{\cal N}^{2}\rangle - \langle{\cal N}\rangle^{2}
&=& \int\frac{d^{3}{\bf k}}{(2\pi)^{3}}\int\frac{d^{3}{\bf k'}}{(2\pi)^{3}}\langle\delta N_{\bf k}\delta N_{\bf k'}\rangle \nonumber\\
&=& \int\frac{d^{3}{\bf k}}{(2\pi)^{3}}\int\frac{dk'}{k'}\frac{k'^{3}}{2\pi^{2}}\langle\delta N_{\bf k}\delta N_{\bf k'}\rangle \nonumber\\
&=& \int\frac{d^{3}{\bf k}}{(2\pi)^{3}}\int\frac{dk'}{k'}\Delta^{2}_{\delta N}(k')\delta^{(3)}({\bf k}+{\bf k'}) \nonumber\\
&=& \int_{k}^{k_{f}}\frac{dk}{k}\Delta^{2}_{\delta N}(k) \nonumber\\
&=& \int_{\ln{k_{f}}-\langle{\cal N}\rangle}^{\ln{k_{f}}}\Delta^{2}_{\delta N}(N)dN,
\eea
where we implement the definition of the power spectrum in the second line and in the last equality conversion is done using $\ln{(k)}=\ln{(aH)}$, for some wavenumber $k=|{\bf k}|$, and the fact that the average e-foldings is equal to $\langle{\cal N}\rangle=\ln{(a_{f}H/aH)}=\ln{(k_{f}/k)}$ assuming constant Hubble parameter $H$. The variance on the left is a statistical quantity and involves the average of the stochastic variable, $\langle{\cal N}\rangle$, in the right. From the above one can finally write down the dimensionless power spectrum for the curvature perturbation as:
\bea \label{pspecdeltaN}
\Delta^{2}_{\zeta\zeta}(k) = \Delta^{2}_{\delta N}(k) = \frac{d}{d\langle{\cal N}\rangle }\langle\delta{\cal N}^{2}\rangle \big|_{\langle{\cal N}\rangle=\ln{(k_{f}/k)}},
\eea
where the first equality follows from eqn. (\ref{coarsezeta}), and the above uses all the modes $k$ crossing out the Hubble radius between some initial instant and the end of inflation. In a similar fashion, we can further write down, using the higher-order statistical moments of $\langle\delta{\cal N}\rangle$, various other higher-order correlation functions. The local non-gaussianity is generally described in form of an expansion ansatz for the curvature perturbation $\zeta({\bf x})$ in the position space and around a Gaussian component:
\bea
\zeta({\bf x}) = \zeta_{g}({\bf x}) + \frac{3}{5}f_{\rm NL}\zeta^{2}_{g}({\bf x}) + \frac{9}{25}g_{\rm NL}\zeta^{3}_{g}({\bf x}) + {\cal O}(\zeta^{4}_{g}({\bf x})) + \cdots,
\eea
$\zeta_{g}({\bf x})$ where is the perturbation component obeying Gaussian statistics. Our interest lies in estimating the above introduced various non-Gaussian parameters, mainly $f_{\rm NL},g_{\rm NL},\tau_{\rm NL}$, using statistics of the e-folds ${\cal N}$ and the stochastic-$\delta N$ formalism. 

Consider the case of the bispectrum and non-Gaussianity parameter $f_{\rm NL}$. The bispectrum is defined in relation with the three-point correlation function $\langle\zeta_{\bf k_1}\zeta_{\bf k_2}\zeta_{\bf k_3}\rangle$. Just like the inverse Fourier mode of the power spectrum is related to the variance, as in eqn. (\ref{varPspec}), a similar treatment for the bispectrum, this time doubly integrated, results in the third-moment of the e-folds $\langle\delta{\cal N}^{3}\rangle$. To show this, we can write:
\bea \label{thirdmoment}
\langle\delta{\cal N}^{3}\rangle = \langle({\cal N}-\langle{\cal N}\rangle)^{3}\rangle &=& \langle{\cal N}^{3}\rangle - 3\langle{\cal N}\rangle\langle{\cal N}^2\rangle + 2\langle{\cal N}\rangle^{3}\nonumber\\
&=& \int\frac{d^{3}{\bf k_1}}{(2\pi)^{3}}\int\frac{d^{3}{\bf k_2}}{(2\pi)^{3}}
\int\frac{d^{3}{\bf k_3}}{(2\pi)^{3}}\langle\delta N_{\bf k_1}\delta N_{\bf k_2}\delta N_{\bf k_3}\rangle, \nonumber\\
&=& \int\frac{d^{3}{\bf k_1}}{(2\pi)^{3}}\int\frac{d^{3}{\bf k_2}}{(2\pi)^{3}}\int\frac{d^{3}{\bf k_3}}{(2\pi)^{3}}(2\pi)^{3}{\cal B}_{\delta N}(k_1,k_2,k_3)\delta^{(3)}({\bf k_1}+{\bf k_2}+{\bf k_3}),\nonumber\\
&=& \int\frac{d^{3}{\bf k_1}}{(2\pi)^{3}}\int\frac{d^{3}{\bf k_2}}{(2\pi)^{3}}{\cal B}_{\delta N}(k_1,k_2,k_3)\big|_{{\bf k_1}+{\bf k_2}+{\bf k_3}=0},
\eea
which involves further volume integral over a tetrahedral region due to triangular constraint from the delta function. However, for our current purpose, we observe that the third moment of e-folds $\langle\delta{\cal N}^{3}\rangle$ receives twice the integrated contribution from the bispectrum. Thus, again using eqn. (\ref{coarsezeta}), we can write the expression for the Bispectrum:
\bea
{\cal B}_{\zeta\zeta\zeta}(k_1,k_2,k_3) \propto \frac{d\langle\delta{\cal N}^{3}\rangle^2}{d\langle{\cal N}\rangle^2},
\eea
Now, the parameter $f_{\rm NL}$ and the bispectrum are related to each other as follows:
\bea \label{bispecPower}
{\cal B}_{\zeta\zeta\zeta}(k_1,k_2,k_3) = \frac{6}{5}f_{\rm NL}\left[\Delta^{2}_{\zeta\zeta}(k_1)\Delta^{2}_{\zeta\zeta}(k_2) + \Delta^{2}_{\zeta\zeta}(k_1)\Delta^{2}_{\zeta\zeta}(k_3) + \Delta^{2}_{\zeta\zeta}(k_2)\Delta^{2}_{\zeta\zeta}(k_3)\right],
\eea
from where one can consider for $f_{\rm NL}$ using eqn. (\ref{pspecdeltaN}) the relation:
\bea
f_{\rm NL} = \frac{5}{36}\frac{d\langle\delta{\cal N}^{3}\rangle^2}{d\langle{\cal N}\rangle^2}\big(\Delta^{2}_{\zeta\zeta}(k)\big)^{-2} = \frac{5}{36}\frac{d\langle\delta{\cal N}^{3}\rangle^2}{d\langle{\cal N}\rangle^2}\bigg(\frac{d\langle\delta{\cal N}^{2}\rangle}{d\langle{\cal N}\rangle}\bigg)^{-2},
\eea
which is a ratio of the bispectrum and the power spectrum squared, with the outside factor of $5/36$ coming from convention. The definition of the trispectrum involves, in a similar manner, the fourth moment of the e-folds $\langle\delta{\cal N}^{4}\rangle$ and taking its third derivative with respect to $\langle{\cal N}\rangle$, since it now receives thrice the integrated contributions. It is related to the four-point correlation function, $\langle\zeta_{\bf k_1}\zeta_{\bf k_2}\zeta_{\bf k_3}\zeta_{\bf k_4}\rangle$, and for this we start with the fourth moment of e-folds:
\bea \label{fourthmoment}
\langle\delta{\cal N}^{4}\rangle = \langle({\cal N}-\langle{\cal N}\rangle)^{4}\rangle &=& \langle{\cal N}^{4}\rangle - 4\langle{\cal N}\rangle\langle{\cal N}^3\rangle + 6\langle{\cal N}\rangle^{2}\langle{\cal N}^{2}\rangle - 3\langle{\cal N}\rangle^{4},\nonumber\\
&=& \int\frac{d^{3}{\bf k_1}}{(2\pi)^{3}}\int\frac{d^{3}{\bf k_2}}{(2\pi)^{3}}
\int\frac{d^{3}{\bf k_3}}{(2\pi)^{3}}\int\frac{d^{3}{\bf k_4}}{(2\pi)^{3}}\langle\delta N_{\bf k_1}\delta N_{\bf k_2}\delta N_{\bf k_3}\delta N_{\bf k_4}\rangle,\nonumber\\
&=& \int\frac{d^{3}{\bf k_1}}{(2\pi)^{3}}\int\frac{d^{3}{\bf k_2}}{(2\pi)^{3}}\int\frac{d^{3}{\bf k_3}}{(2\pi)^{3}}\int\frac{d^{3}{\bf k_4}}{(2\pi)^{3}}(2\pi)^{3}T_{\delta N}(k_1,k_2,k_3,k_4)\delta^{(3)}({\bf k_1}+{\bf k_2}+{\bf k_3}+{\bf k_4}),\quad\quad\quad\nonumber\\
&=& \int\frac{d^{3}{\bf k_1}}{(2\pi)^{3}}\int\frac{d^{3}{\bf k_2}}{(2\pi)^{3}}\int\frac{d^{3}{\bf k_3}}{(2\pi)^{3}}T_{\delta N}(k_1,k_2,k_3,k_4)\big|_{{\bf k_1}+{\bf k_2}+{\bf k_3}+{\bf k_4}=0},
\eea
where have used the relation for the connected part of the four-point function. Now, the connected part of the trispectrum for the curvature perturbations is defined in combination of the power spectrum as follows:
\bea \label{trispecPower}
T_{\zeta\zeta\zeta\zeta}(k_1,k_2,k_3,k_4) &=& \tau_{\rm NL}\big[\Delta^{2}_{\zeta\zeta}(k_{13})\Delta^{2}_{\zeta\zeta}(k_{3})\Delta^{2}_{\zeta\zeta}(k_{4}) + 11\;\text{perms.} \big]\nonumber\\
&& + \frac{54}{25}g_{\rm NL}\big[\Delta^{2}_{\zeta\zeta}(k_{2})\Delta^{2}_{\zeta\zeta}(k_{3})\Delta^{2}_{\zeta\zeta}(k_{4}) + 3\;\text{perms.} \big],
\eea
with the notation for magnitude $k_{ij} = |{\bf k}_i + {\bf k}_j|$ and the cubic dependence on the power spectrum being a reason behind its name as the trispectrum. We notice that there emerge two distinct non-linearity parameters from the connected part of the trispectrum, $\tau_{\rm NL}$ and $g_{\rm NL}$, based on their $k$ dependence. The $12$ permutations comes from having $3$ distinct choices for the indices in $k_{ij}$, as the magnitude remains same, $k_{ij}=k_{ji}$, and from the momentum conserving delta function we have $k_{12}=k_{34},\;k_{23}=k_{14},\;k_{13}=k_{24}$. The $4$ permutations are again result of the conservation of momentum. From eqns. (\ref{coarsezeta},\ref{pspecdeltaN},\ref{trispecPower}), we can ultimately write the following relations for the non-Gaussianity parameters in terms of the derivatives of the fourth moment:
\bea
\tau_{\rm NL} &=& \frac{1}{36}\bigg(\frac{d^2\langle\delta{\cal N}^3\rangle}{d\langle{\cal N}\rangle^2}\bigg)^{2}\big(\Delta^{2}_{\zeta\zeta}(k)\big)^{-4} = \frac{1}{36}\bigg(\frac{d^2\langle\delta{\cal N}^3\rangle}{d\langle{\cal N}\rangle^2}\bigg)^{2}\bigg(\frac{d\langle\delta{\cal N}^{2}\rangle}{d\langle{\cal N}\rangle}\bigg)^{-4},\\
g_{\rm NL} &=& \frac{d\langle\delta{\cal N}^{4}\rangle^3}{d\langle{\cal N}\rangle^3}\big(\Delta^{2}_{\zeta\zeta}(k)\big)^{-3} = \frac{d\langle\delta{\cal N}^{4}\rangle^3}{d\langle{\cal N}\rangle^3}\bigg(\frac{d\langle\delta{\cal N}^{2}\rangle}{d\langle{\cal N}\rangle}\bigg)^{-3},
\eea
where the factor $1/36$ again results from convention purposes.

\textcolor{black}{Now that we have laid out the general features of the stochastic-$\delta N$ formalism, that remain applicable in the current EFT treatment for stochastic single field inflation, and discussed} its applications into calculating higher-order correlation functions, we make use of the above developments to understand the probability distribution driving the duration of inflation. Computation of this distribution has major advantages in terms of knowing the mass fraction of the PBH when focusing on the diffusion dominated regime of inflation and identifying the non-Gaussianity parameters when focusing on the drift-dominated regime during inflation. We aim to perform these calculations in the EFT picture coupled with the stochastic-$\delta N$ formalism.
 
\section{Probability Distribution Function from Fokker-Planck equation }
\label{s7}

In this section, we derive the adjoint Fokker-Planck equation in its complete form which will later help us to analyze the probability distribution functions. The adjoint Fokker-Planck equation allows us to study the evolution of the probability density function from one point to another in the field space and here it is defined for the stochastic e-folds variable ${\cal N}$ elapsed between the start and end of the evolution. It involves the adjoint Fokker-Planck operator acting on the probability distribution function $P_{\bf \Gamma}({\cal N})$ in the following manner:
\bea 
\frac{\partial }{\partial {\cal N}}P_{\bf \Gamma}({\cal N})=\Bigg(F_i \frac{\partial }{\partial \Phi_i}+\frac{1}{2}\Sigma_{ij}\frac{\partial ^2}{\partial \Gamma_i \partial \Gamma_j}\Bigg) P_{\bf \Gamma}({\cal N}) , 
\eea
\textcolor{black}{where the quantity $\Sigma_{ij}$ represents the various noise elements as part of the correlation matrix discussed in detail in appendix \ref{app:C}} and where the adjoint Fokker-Planck operator in the parenthesis can be expanded as follows:
\bea
F_i \frac{\partial }{\partial \Gamma_i}+\frac{1}{2}\Sigma_{ij}\frac{\partial ^2}{\partial \Gamma_i \partial \Gamma_j} &=& \Pi_{\zeta}\frac{\partial}{\partial \zeta} - (3-\epsilon)\Pi_{\zeta}\bigg[1-\frac{2(s-\frac{\eta}{2})}{3-\epsilon}\bigg]\frac{\partial}{\partial\Pi_{\zeta}}  \nonumber\\
&& \quad\quad\quad\quad\quad\quad\quad + \frac{1}{2}\bigg(\Sigma_{\zeta\zeta}\frac{\partial^{2}}{\partial\zeta^{2}}  + \Sigma_{\zeta\Pi_{\zeta}}\frac{\partial^{2}}{\partial\zeta\partial\Pi_{\zeta}} + \Sigma_{\Pi_{\zeta}\zeta}\frac{\partial^{2}}{\partial\Pi_{\zeta}\partial\zeta} + \Sigma_{\Pi_{\zeta}\Pi_{\zeta}}\frac{\partial^{2}}{\partial\Pi_{\zeta}\partial\Pi_{\zeta}}\bigg).
\eea
\textcolor{black}{At this stage we implement the super-Hubble condition, $\sigma=-kc_{s}\tau \ll 1$, to examine this equation which renders the noise matrix elements $\Sigma_{\zeta\Pi_{\zeta}},\;\Sigma_{\Pi_{\zeta}\zeta}, \text{and}\; \Sigma_{\Pi_{\zeta}\Pi_{\zeta}}$, almost negligible; one can infer this from their explicit expressions provided in the appendices \ref{appCa1},\ref{appCb1}, \text{and} \ref{appCc1}}. Next, to further simplify the representation of this adjoint Fokker-Planck operator, we employ the following re-scaling of the coarse-grained variables $\{\zeta,\Pi_{\zeta}\}$ into the new variables:
\bea \label{newphasevars}
\zeta = f,\quad\quad \Pi_{\zeta} = -3y,
\eea
using this we observe the following transformation of various other differential operators present:
\bea
&&\frac{1}{2}\Sigma_{\zeta\zeta}\frac{\partial^{2}}{\partial\zeta^{2}} = \frac{1}{\mu^{2}}\frac{\partial^{2}}{\partial f^{2}},\\
&&\Pi_{\zeta}\frac{\partial}{\partial\Pi_{\zeta}} = y\frac{\partial}{\partial y},\\
&&\Pi_{\zeta}\frac{\partial}{\partial\zeta} = -3y\frac{\partial}{\partial f}.
\eea
With the new re-scaled variables in hand, we rewrite the adjoint Fokker-Planck operator as:
\bea
\frac{\partial }{\partial \mathcal{N} }P_{\bf \Gamma}(\mathcal{N})&=& \bigg[\frac{1}{\mu^{2}}\frac{\partial^{2}}{\partial f^{2}} - 3\bigg\{y\frac{\partial}{\partial f} + \bigg(1-\frac{\epsilon}{3}\bigg)\bigg(1-\frac{2(s-\frac{\eta}{2})}{3-\epsilon}\bigg)y\frac{\partial}{\partial y} \bigg\}\bigg]P_{\bf \Gamma}({\cal N}),
\eea
in which, for clarity, we further perform another set of following re-definitions:
\bea \label{EFTparam}
C = \bigg(1-\frac{\epsilon}{3}\bigg)\bigg(1-\frac{2(s-\frac{\eta}{2})}{3-\epsilon}\bigg),\quad\quad Cf = x,\quad\quad \left(\frac{C}{\mu}\right)^{2} = \frac{1}{\tilde{\mu}^{2}}
\eea
and these changes ultimately \textcolor{black}{give} rise to the adjoint Fokker-Planck equation version of our interest:
\bea \label{FPE-EFT}
\frac{\partial }{\partial \mathcal{N} }P_{\bf \Gamma}(\mathcal{N})&=& \bigg[\frac{1}{\tilde{\mu}^{2}}\frac{\partial^{2}}{\partial x^{2}} - 3Cy\bigg\{\frac{\partial}{\partial x} + \frac{\partial}{\partial y}\bigg\} \bigg]P_{\bf \Gamma}({\cal N}). 
\eea
Notice that the changes coming from an underlying EFT setup are contained within the coefficient $C$, which we now refer to as the characteristic parameter, and it will have noticeable changes when we understand the probability distribution evolution in the next section. Also, the change from $C$ is felt by the coefficient $\mu$ containing the auto-correlated power spectrum element for $\zeta$. In the case where $C=1$ reduces to the canonical single-field inflation scenario while for $C\ne 1$ we are working with the non-canonical single-field models of inflation. In the next section, we explore the method that will enable us to calculate this PDF depending on the interested conditions for a general EFT setup. 

\section{Characteristic function and Initial Conditions}
\label{s8}

In the stochastic inflation framework, our stochastic variable of interest is the e-folds ${\cal N}$, and for this, we have seen the evolution of the corresponding PDF governed by the adjoint Fokker-Planck equation. To fully determine the nature of this PDF, one requires information on its associated moments as they dictate the statistical properties of this PDF. Here, we discuss the method of the characteristic function that will let us evaluate the moments of arbitrary degree for a given PDF, $P_{\Gamma_i}({\cal N})$, and would also manage to reconstruct the entire PDF. 

The characteristic function converts the probability distribution from a function of $3$ variables, in the phase space ${\Gamma_i} = \{\zeta,\Pi_{\zeta}\}$ and the variable ${\cal N}$, to a function of just the phase space variables ${\Gamma_i}$. It is defined as the Fourier transform of the original PDF in the ${\cal N}$ co-ordinate as:
\bea \label{characterDefn}
\chi(t;x,y) = \int_{-\infty}^{+\infty}e^{it{\cal N}}P_{{\Gamma_i}}({\cal N})d{\cal N},
\eea
where $t$ acts as a dummy variable which is a complex quantity in general. This function $\chi(t;{\Gamma_i})$ is useful to provide the various statistical moments associated with the full PDF $P_{\Gamma_i}({\cal N})$. Inversely, one can obtain the PDF from the characteristic function as follows:
\bea \label{characterFourier}
P_{{\Gamma_i}}({\cal N}) = \frac{1}{2\pi}\int_{-\infty}^{+\infty}e^{-it{\cal N}}\chi(t;{\Gamma_i})dt.
\eea
The corresponding information for the moments can be extracted by Taylor expanding $\chi(t;{\Gamma_i})$ around $t=0$ and observing that:
\bea \label{chimoments}
\chi(t;{\Gamma_i}) = \sum_{n=0}^{\infty}\frac{(it)^{n}}{n!}\langle{\cal N}^{n}({\Gamma_i})\rangle \implies \langle{\cal N}^{n}({\Gamma_i})\rangle = i^{-n}\frac{\partial^{n}}{\partial t^{n}}\chi(t;{\Gamma_i})|_{t=0},
\eea
where the index $n$ denotes the $n$th moment, $\langle{\cal N}^{n}({\Gamma_i})\rangle$, of the full PDF. By using the definition of PDF from eqn. (\ref{characterFourier}), into the adjoint Fokker-Planck eqn. (\ref{FPE-EFT}), one also notices that the characteristic function obeys:
\bea \label{FPEcharacter}
\bigg[\frac{1}{\tilde{\mu}^{2}}\frac{\partial^{2}}{\partial x^{2}} - 3Cy\bigg\{\frac{\partial}{\partial x} + \frac{\partial}{\partial y}\bigg\} + it\bigg]\chi(t;{\Gamma_i}) = 0.
\eea
The solution of this equation leads us to identify the function $\chi(t;{\Gamma_i})$. In the next sections, we will analyze the characteristic function solutions under specific conditions for the $y$ parameter and in the presence of certain boundary conditions needed to sufficiently handle the above two-dimensional partial differential equation.


\section{Diffusion Dominated Regime: Methods to solve for the Characteristic function}
\label{s9}

In this section, we concern ourselves with the regime where quantum diffusion effects dominate the dynamics. This regime is also where the conjugate momentum variable, see eqn. (\ref{newphasevars}), takes on very small values, $0< y\ll 1$. This regime is significant from the perspective of analyzing PBH more accurately since we will observe distinctive features in the upper tail of the PDF, which is precisely the region of importance for the formation of PBH. Also, it can be seen from the perspective of the conjugate field momenta since it decreases in this regime, and thus, diffusion effects primarily govern the behaviour of the scalar perturbations and features of the PDF.


To generate the PDF requires solving the partial differential equation in eqn.(\ref{FPEcharacter}). We solve first for the characteristic function $(\chi)$ which will then help us to build the PDF and the related methodology is the subject of this section. 

To this effect, we start with a particular variable re-definition:
\bea
u= x-y,\quad\quad\quad v=y,
\eea
that simplifies the equation for the characteristic function as follows:
\bea \label{FPEcharacterDiff}
\bigg[\frac{1}{\tilde{\mu}^{2}}\frac{\partial^{2}}{\partial u^{2}} - 3Cv\frac{\partial}{\partial v} + it\bigg]\chi(t;u,v)=0.
\eea
The above equation is now made separable in its variables and thus accepts an ansatz of the form:
\bea \label{chiansatz}
\chi(t;u,v) = \sum_{n=0}^{\infty}v^{n}U_{n}(t;u),
\eea
and here we further utilize the fact visible from the structure of the new partial differential equation, eqn. (\ref{FPEcharacterDiff}), that the \textcolor{black}{mode function $U(t;r)$} admits oscillatory solutions that follow:
\bea
\frac{\partial^{2}}{\partial r^{2}}U_{n}(t;u) + \omega_{n}^{2}(t)U_{n}(t;u) = 0,
\eea
where the frequency contains the effects from the EFT parameter as:
\bea \label{omega1}
\omega_{n}^{2}(t) = (it-3Cn)\tilde{\mu}^{2}.
\eea
Now, in order to fully utilize the ansatz and determine $\chi(t;u,v)$, we implement the appropriate boundary conditions that can describe the behaviour of the perturbations as they enter and exit the diffusion-dominated regime. The first case corresponds to the condition where the perturbations are almost unaffected by quantum diffusion processes, with the perturbations still not present under the influence of the USR. Conversely, the second case is where the diffusion effects dominate the overall dynamics and dictate the behaviuor of the PDF for large values of the perturbations. The above described conditions for the transformed phase space variables ${\Gamma_i} = \{u,v\}$ are written using the characteristic function as follows:
\bea 
&& \label{bdrycondn1}\underline{\bf Boundary \;Condition\; I:}\quad\quad\quad\quad\chi(t;\Gamma_i)|_{u+v=0} = 1,\\
&& \label{bdrycondn2}\underline{\bf Boundary \;Condition\; II:}\quad\quad\quad \left.\frac{\partial\chi(t;\Gamma_i)}{\partial u}\right\vert_{u+v=1} = 0,
\eea
\begin{figure*}[ht!]
    	\centering
    {
       \includegraphics[width=18cm,height=12cm]{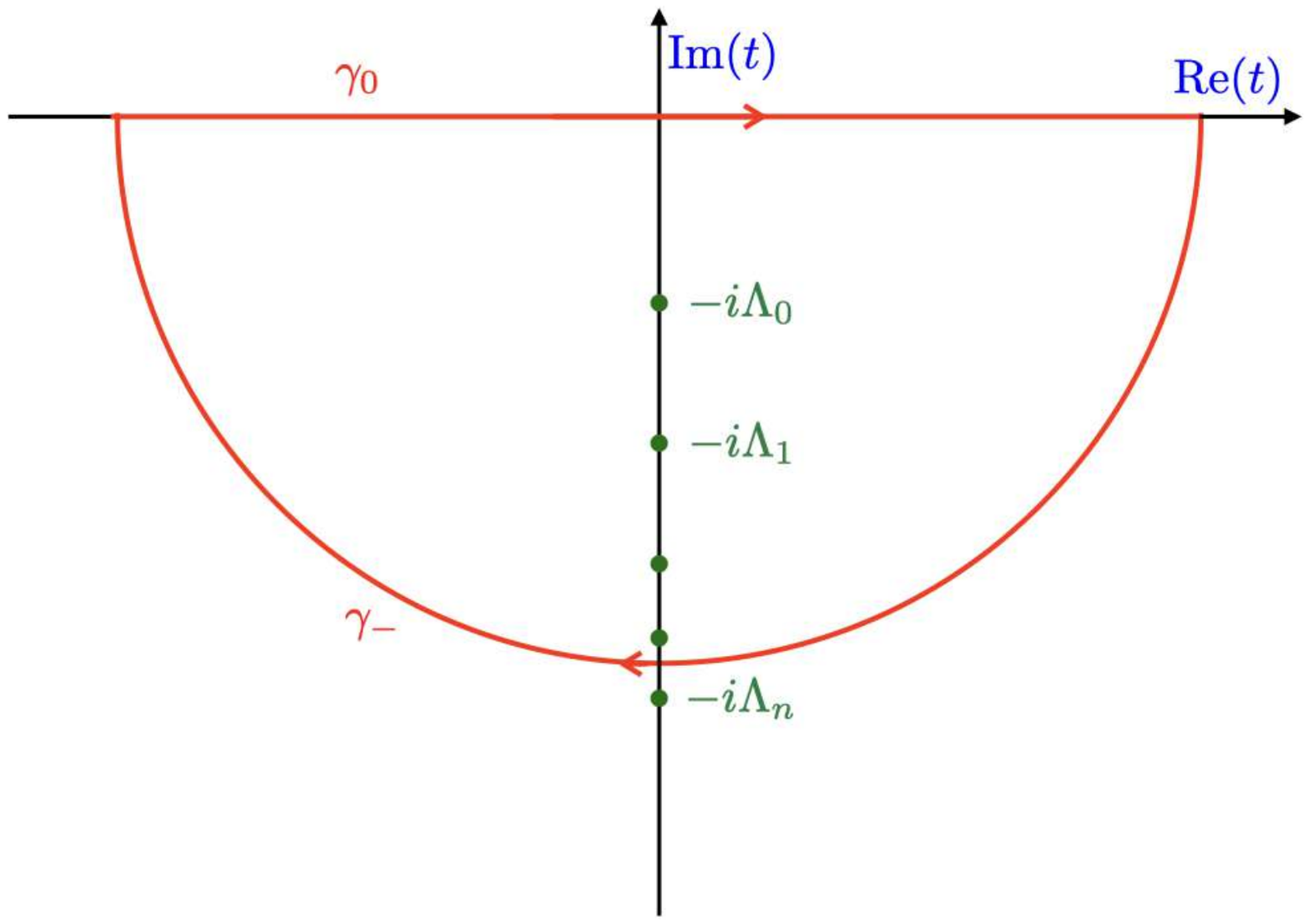}
        \label{CS}
    } 
    \caption[Optional caption for list of figures]{Closed contour $\gamma_{0}\cup\gamma_{-}$ in the complex plane containing poles of the characteristic function on the imaginary $t$ axis. The residue theorem when applied at these poles results into the PDF $P_{\Gamma}({\cal N})$.  }
\label{complecontour}
    \end{figure*}
\textcolor{black}{where, although written using the new variables, $(u,v)$, these conditions at a given $x$ also concern with the value of $\zeta$, as is evident from the original definition of $\zeta$ from eqs. (\ref{newphasevars}) and (\ref{EFTparam}).} The first condition refers to smaller values of $\zeta$ and conversely, the second condition refers to larger values of $\zeta$. However, the solution now involves mode mixing due to the nature of the conditions involving both new variables and we examine how these affect the solution from the series ansatz. Since it follows oscillatory behaviour, we can write the solution for the function $U_{n}(t;u)$ as:
\bea
U_{n}(t;u) = R_{n}(t)\cos{((1-u)\omega_{n}(t))} + Q_{n}(t)\sin{((1-u)\omega_{n}(t))},
\eea
and using this solution with the conditions from eqn.(\ref{bdrycondn1}) and eqn.(\ref{bdrycondn2}), we get the following constraints on the series:
\bea
&&\sum_{n=0}^{\infty}v^{n}\big[R_{n}\cos{((1+v)\omega_{n}(t))} + Q_{n}(t)\sin{((1+v)\omega_{n}(t))}\big] = 1,\nonumber\\
&&\sum_{n=0}^{\infty}\omega_{n}v^{n}\big[R_{n}\sin{(\omega_{n}v)} - Q_{n}\cos{(\omega_{n}v)}\big] = 0.
\eea
From this, \textcolor{black}{by series expanding the oscillatory functions in small velocities ($y=v\ll 1$)}, one can generate recurrence relations for the various coefficients $R_{n},\;Q_{n}$ for each order in index $n$. We conduct our analysis for each order in the series till we reach $n=3$ which allows us to construct the next-to-next-to-next leading order (NNNLO) version of the PDF and for that, we require the explicit form of these coefficients using the mentioned recurrence relations. We mention the coefficients up to $n=2$, or to construct the second order (NNLO) contribution, here as:
\bea \label{recurrNNLO}
R_{0} &=& \frac{1}{\cos{(\omega_{0})}},
\nonumber\\
R_{1} &=& \frac{\omega_{0}}{\omega_{1}}\frac{\omega_{1}\sin{(\omega_{0})}-\omega_{0}\sin{(\omega_{1})}}{\cos{(\omega_{0})}\cos{(\omega_{1})}}, \nonumber\\
R_{2} &=& \frac{\omega_{1}\sin{(\omega_{1})}P_{1}-\omega_{1}\cos{(\omega_{1})}Q_{1}-\sin{(\omega_{2})}Q_{2}+\frac{\omega_{0}^{2}}{2}\cos{(\omega_{0})}P_{0}}{\cos{(\omega_{2}})}, 
\eea
and 
\bea  Q_{0} &=& 0,\nonumber\\
    Q_{1} &=& \frac{\omega_{0}^{2}}{\omega_{1}}P_{0},\nonumber\\
    Q_{2} &=& \frac{\omega_{1}^{2}}{\omega_{2}}P_{1},\eea
and those for the third order are mentioned in the next section dedicated to NNNLO.

We follow the approach in \cite{Ezquiaga:2019ftu} of writing the characteristic function in a series expansion\footnote{Another approach towards solving for the tail expansion of the PDF $P_{\Gamma_i}({\cal N})$, is called the eigenvalue problem method, detailed discussions on which can be found in \cite{Ezquiaga:2019ftu}. It treats the PDE in eqn. (\ref{FPE-EFT}) as a heat equation which allows to incorporate methods designed to solve for such diffusion-like differential equations. It involves introducing a set of eigenfunctions for the adjoint Fokker-Planck operator which remains Hermitian under a newly defined scalar-product definition that accounts for the form of the model potential chosen. The resulting eigenfunction obey the equation similar to that for the characteristic function eqn. (\ref{FPEcharacter}) but with a different set of boundary conditions. } of the following form:
\bea \label{chiexpand}
\chi(t;{\Gamma_i}) = \sum_{m}\sum_{n}\frac{r_{n}^{(m)}({\Gamma_i})}{\Lambda_{n}^{(m)}-it} + h(t;{\Gamma_i}),
\eea
where the use of ${\Gamma_i}$ highlights the dependence of the residues on the phase space variables, or the initial conditions, and the poles, $\Lambda_{n}^{(m)}$, always remain independent of the set of values in ${\Gamma_i}$, which we will show in the next step. The function $h(t;{\Gamma_i})$ is the regular part of the expansion and would not contribute in the later discussions as we will show. The fig. \ref{complecontour} illustrates the application of residue theorem at the poles located on the imaginary $t$-axis enclosed by a contour $\gamma_{0}\cup\gamma_{-}$. As seen from their expressions above, the terms $R_{n},\;Q_{n}$ admits simple poles in their structure and thus, for the characteristic function $\chi(t;{\Gamma_i})$, an expansion of the type in eqn.(\ref{chiexpand}) works for our purpose. The poles correspond to the condition:
\bea
\cos{(\omega_{m}(t))}=0,\quad\quad \text{hence,}\quad \omega_{m}(t) = \left(n+\frac{1}{2}\right)\pi,
\eea
\textcolor{black}{where, from now onwards, the integer $n$ labels the number of distinct positions we have a pole for the $m$th perturbative order.} The above equation from the use of eqn. (\ref{omega1}) can be converted to give the pole expression as:
\bea \label{poles}
\Lambda_{n}^{(m)} = 3Cm + \bigg[\frac{\pi}{\tilde{\mu}}\left(n+\frac{1}{2}\right)\bigg]^{2},
\eea
and here we can notice the impact of the characteristic parameter, see eqn.(\ref{EFTparam}). Another quantity of importance are the residues, represented here by $r_{n}^{(m)}({\Gamma_i})$, and these are evaluated by inverting the expression in eqn.(\ref{chiexpand}) and taking the time-derivative in the following manner:
\bea \label{residue}
r_{n}^{(m)}({\Gamma_i}) = -i\bigg[\left.\frac{\partial}{\partial t}\chi^{-1}(t;{\Gamma_i})\right\vert_{t=-i\Lambda_{n}^{(m)}}\bigg]^{-1},
\eea
The above information regarding the poles and the residues combine together to construct the PDF if we use the eqn.(\ref{chiexpand}) with the Fourier transform expression in eqn. (\ref{characterFourier}), which results in:
\bea \label{PDFfinal}
P_{\Gamma_i}({\cal N}) = \sum_{m=0,1,\cdots}\sum_{n=0}^{\infty}r_{n}^{(m)}({\Gamma_i})\exp{(-\Lambda_{n}^{(m)}{\cal N})}.
\eea
with the outside sum going up to the desired order $m$th order. The exponential decay of the PDF at each order in the expansion is determined by the value of the poles $\Lambda_{n}^{(m)}$ and, for large values of the e-folds ${\cal N}$ the dominant contribution solely comes from the lowest pole $\Lambda_{0}$ and it corresponding residue $r_{0}^{(m)}$. As mentioned earlier, the regular function in the expansion, eqn. (\ref{chiexpand}), plays no role in determining the final PDF expansion. We can thus use the characteristic function and find its residues corresponding to the simple poles at each subsequent order in the expansion to construct, using the above, the full PDF at the $m$th order.

\subsection{Calculation of PDF at LO}

In this section, we discuss the calculation of the probability distribution at the leading order (LO) for the diffusion-dominated regime. We follow the approach described in the previous section based on the characteristic function and use its series expansion to find the residues and the corresponding decay factors involving the poles.

For the leading order, we only require to use the first set of coefficients, $(R_{0},Q_{0})$, in eqn. (\ref{recurrNNLO}). These imply that the drift contributions are completely ignored since we are working in the limit where $(v\rightarrow 0,\;u\rightarrow x)$ and using the corresponding LO characteristic function we can proceed to find its PDF. The first job is to use the following function at LO:
\bea \label{chiLO1}
\chi^{\rm LO}(t;x,0) = \frac{\cos{((1-x)\omega_{0})}}{\cos{(\omega_{0})}},
\eea
and find the residues using eqn. (\ref{residue}) using the $m=0$ poles, $\Lambda_{n}^{(0)}$, in eqn. (\ref{poles}). Notice here that the characteristic parameter $C$ for the EFT does not directly come into the picture at LO. The residues then obtained are given by,
\bea \label{resLO}
r_{n,0}^{(m=0)} = (2n+1)\frac{\pi}{\tilde{\mu}^{2}}\sin{\bigg((2n+1)\frac{\pi x}{2}\bigg)},
\eea
which combines into the eqn. (\ref{PDFfinal}) with the poles to give for the PDF,
\bea \label{pdfLO}
P_{\Gamma_{i}}^{\rm LO}({\cal N}) &=& \sum_{n=0}^{\infty}r_{n,0}^{(m=0)}\exp{(-\Lambda_{n}^{(0)}{\cal N})}= -\frac{\pi}{2\tilde{\mu}^2}{\cal V}'_{2}\bigg(\frac{\pi}{2}x,\exp{\bigg(-\frac{\pi^2}{\tilde{\mu}^2}{\cal N}\bigg)}\bigg),
\eea
where ${\cal V}_2$ represents the elliptic theta function,
\bea
{\cal V}_{2}(\alpha,x)=2\sum_{n=0}^{\infty}x^{(n+\frac{1}{2})^2}\cos{((2n+1)\alpha)},
\eea
and the notation, ${\cal V}'_{2}$, refers to the derivative of the function with respect to its first argument. 
Here, the contribution coming from $n=0$ dominates the most when ${\cal N}$ is taken very large. In similar fashion, we develop the PDF for the next-to-leading order (NLO). The first and second moments from the above PDF at the leading order can also be calculated using the relation in eqn. (\ref{chimoments}). The knowledge of $\langle {\cal N}\rangle$ will prove necessary when we concern ourselves with the primordial black hole abundance in the coming sections. The expression for these moments of the number of e-folds ${\cal N}$ at LO is given by
\bea \label{meanLO}
\langle{\cal N}(x,y=0)\rangle_{\rm LO} &=& \tilde{\mu}^{2}x\bigg(1-\frac{x}{2}\bigg),\\
\langle{\cal N}^{2}(x,y=0)\rangle_{\rm LO} &=& \frac{1}{3}\tilde{\mu}^{4}x(2-x^{2}+x^{3}),
\eea
One immediate effect is clear that the mean value is more sensitive to changes in $\tilde{\mu}$. To get a better picture of this sensitivity let us evaluate the variance in the leading order, which gives us:
\bea
\sigma_{\rm LO}^{2} = \langle{\cal N}^{2}(x,y=0)\rangle_{\rm LO} - \langle{\cal N}(x,y=0)\rangle_{\rm LO}^{2} = \frac{1}{6}\tilde{\mu}^{4}x(2+x^{2}-2x)(2-x),
\eea
The variance experiences increased sensitivity as $\tilde{\mu}$ is increased slightly.

\subsection{Calculation of PDF at NLO}
   
Here we determine the first sub-dominant correction to the pure diffusion limit result in the previous section. We require here to use the second set of coefficients, $(R_{1},Q_{1})$, in eqn. (\ref{recurrNNLO}) to properly determine the characteristic function and subsequent residues at the next-to-leading order in our $y$-expansion. We have at NLO the following function:
\bea \label{chiNLO1}
\chi^{\rm NLO}(t;\Gamma_{i}) = U_{0}(t;u) + vU_{1}(t;u),
\eea
and further utilising the formulas in eqns. (\ref{poles},\ref{residue}) using the above characteristic function, we now mention the residues at each order with information up to the first order ${\cal O}(y)$, as follows:
\bea \label{resNLO0}
r_{n,1}^{(m=0)} &=& (2n+1)\frac{\pi}{\tilde{\mu}^{2}}\bigg[\sin{\left((2n+1)\frac{\pi x}{2}\right)} - y(2n+1)\frac{\pi}{2}\cos{\left((2n+1)\frac{\pi x}{2}\right)}\bigg] \nonumber\\
&+& y(2n+1)\frac{\pi}{2}\frac{1}{\omega^{(0)}_{n,1}\cos{(\omega^{(0)}_{n,1})}}\bigg[\omega^{(0)}_{n,1}\cos{\omega^{(0)}_{n,1}(1-x)}-(-1)^{n}(2n+1)\frac{\pi}{2}\sin{(x\; \omega^{(0)}_{n,1})}\bigg],
\eea
where the notation for the zero-order frequency at NLO means:
\bea
\omega^{(0)}_{n,1} = \sqrt{\left(n+\frac{1}{2}\right)^{2}\pi^{2}-3C\tilde{\mu}^{2}},
\eea
the above residues come associated with the zero-order pole at NLO, $\Lambda_{n}^{(0)}$. Similarly, for the first order poles, $\Lambda_{n}^{(1)}$, the residues have the following structure:
\bea \label{resNLO1}
r_{n,1}^{(m=1)} &=& \frac{(-1)^{n+1}2y}{\tilde{\mu}^{2}}\frac{\sin{\left((n+\frac{1}{2})\pi x\right)}}{\cos{\left(\sqrt{\left(n+\frac{1}{2}\right)^{2}\pi^{2}+3C\tilde{\mu}^{2}}\right)}}\bigg[\left(n+\frac{1}{2}\right)^{2}\pi^{2}+3C\tilde{\mu}^{2}\nonumber\\
&& \quad\quad\quad\quad\quad -(-1)^{n}\frac{\pi}{2}(2n+1)\sqrt{\left(n+\frac{1}{2}\right)^{2}\pi^{2}+3C\tilde{\mu}^{2}}\sin{\bigg(\sqrt{\left(n+\frac{1}{2}\right)^{2}\pi^{2}+3C\tilde{\mu}^{2}}\bigg)}\bigg],
\eea
Combining together eqns.(\ref{resNLO0},\ref{resNLO1}) into the final PDF from eqn.(\ref{PDFfinal}) gives
\bea \label{PDFN1}
P_{\Gamma_{i}}^{\rm NLO}({\cal N}) = \sum_{n=0}^{\infty}\big[r_{n,1}^{(m=0)}+r_{n,1}^{(m=1)}e^{-3C{\cal N}}\big]e^{-\Lambda_{n}^{(0)}{\cal N}},
\eea
where in addition to the set of poles for  $m=0$, we have another version of poles, \bea \Lambda_{n}^{(1)} = 3C + \Lambda_{n}^{(0)},\eea contributing to the decay rate. The presence of characteristic parameter $C$ also alters the decay feature depending on its value, which remains close to $1$. Similar to the case of LO, here also one can evaluate the first moment or mean number $\langle{\cal N}\rangle$ to get the following:
\bea \label{meanNLO}
\langle{\cal N}(\Gamma)\rangle_{\rm NLO} = \tilde{\mu}^{2}x\bigg(1-\frac{x}{2}\bigg) + \tilde{\mu}^{2}y\bigg(x-1 + \frac{\cosh{[\sqrt{3C}\tilde{\mu}(1-x)]}}{\cosh{[\sqrt{3C}\tilde{\mu}]}} - \frac{\sinh{[\sqrt{3C}\tilde{\mu}x]}}{\sqrt{3C}\tilde{\mu}\cosh{[\sqrt{3C}\tilde{\mu}]}} \bigg),
\eea
which contains the linear correction terms, that is up to ${\cal O}(y)$. We will use this expression in the upcoming sections for PBH analysis.

\subsection{Calculation of PDF at NNLO}

Here we discuss the calculation of the probability distribution at the next-to-next-leading order (NNLO) for the diffusion-dominated regime. This means we work with the series ansatz in eqn. (\ref{chiansatz}) truncated up to the quadratic terms to find those respective corrections. We would require knowledge of the residues and the corresponding poles to begin. The last set of coefficients in eqn. (\ref{recurrNNLO}), $(R_{2},Q_{2})$, will be of use here. The characteristic function at NNLO now has the form:
\bea \label{chiNNLO1}
\chi^{\rm NLO}(t;\Gamma_{i}) = U_{0}(t;u) + vU_{1}(t;u) + v^{2}U_{2}(t;u),
\eea
following this, the residues from eqn. (\ref{residue}) at orders $m=0,1,2$, each expanded up to the second order ${\cal O}(y^{2})$, is required to eventually construct the PDF. We do not explicitly mention the analytic form of each residue due to its overall complexity. However, the figures obtained for $P_{\Gamma_{i}}^{\rm NNLO}(\cal N)$ will be discussed in good detail in the numerical analysis section.
The final form of the PDF can be combined into the following expression:
\bea \label{PDFN2}
P_{\Gamma_i}^{\rm NNLO}({\cal N}) = \sum_{n=0}^{\infty}\big[r_{n,2}^{(0)}+ r_{n,2}^{(1)}e^{-3C{\cal N}} + r_{n,2}^{(2)}e^{-6C{\cal N}} \big]e^{-\Lambda_{n}^{(0)}{\cal N}},
\eea

Here also, we can calculate the corrections to the mean number of e-folds at NNLO following the eqn. (\ref{chimoments}) along with the characteristic function in the above eqn.(\ref{chiNNLO1}). This gives us the following expression for the mean:
\bea \label{meanNNLO}
\langle{\cal N}(\Gamma)\rangle_{\rm NNLO} &=& \tilde{\mu}^{2}x\bigg(1-\frac{x}{2}\bigg) + \tilde{\mu}^{2}y(-1+x-y) + \frac{\tilde{\mu}^{2}y^2}{2}\bigg(\frac{\cosh{[\sqrt{6C}\tilde{\mu}(1-x)]}}{\cosh{[\sqrt{3C}\tilde{\mu}]}\cosh{[\sqrt{6C}\tilde{\mu}]}}\bigg)\nonumber\\
&&\times\bigg(-2+\cosh{[\sqrt{3C}\tilde{\mu}]}+\sqrt{6C}\tilde{\mu}\sinh{[\sqrt{6C}\tilde{\mu}]} - (2\sqrt{3C}\tilde{\mu}^{2}+\sqrt{2}\sinh{[\sqrt{6C}\tilde{\mu}]})\sinh{[\sqrt{3C}\tilde{\mu}]}\bigg)\nonumber\\
&&+\frac{\tilde{\mu}^{2}y^2}{\sqrt{2}}\frac{\sinh{[\sqrt{6C}\tilde{\mu}(1-x)]}}{\sqrt{3C}\tilde{\mu}^2} - \tilde{\mu}y\sinh{[\sqrt{3C}\tilde{\mu}(1-x+y)]} -\frac{\tilde{\mu}^{2}y^2}{\sqrt{2}}\sinh{[\sqrt{6C}\tilde{\mu}(1-x)]}\tanh{[\sqrt{3C}\tilde{\mu}]} \nonumber\\
&& + \tilde{\mu}^{2}y\cosh{[\sqrt{3C}\tilde{\mu}(1-x+y)]}\bigg(\frac{1}{\cosh{[\sqrt{3C}\tilde{\mu}]}}-\frac{\tanh{[\sqrt{3C}\tilde{\mu}]}}{\sqrt{3C}\tilde{\mu}}\bigg),
\eea 
where we consider terms only up to ${\cal O}(y^{2})$.

\subsection{Calculation of PDF at NNNLO}

To completely determine the PDF we would require the recurrence relation between the coefficients, which we now mention:
\bea
R_{3} &=& \frac{1}{\cos{(\omega_{3})}}\bigg[\frac{1}{2}\omega_{1}^{2}(R_{1}\cos{(\omega_{1})} + Q_{1}\sin{(\omega_{1})}) + \omega_{2}R_{2}\sin{(\omega_{2})}-\frac{1}{6}\omega_{0}^{3}R_{0}\sin{(\omega_{0})}-\omega_{2}Q_{2}\cos{(\omega_{2})}-Q_{3}\sin{(\omega_{3})}\bigg],\nonumber\\
Q_{3} &=& \frac{1}{\cos{(\omega_{3})}}\bigg[\omega_{2}^{2}R_{2}+\frac{1}{2}\omega_{1}^{3}Q_{1} - \frac{1}{6}\omega_{0}^{4}R_{0}\bigg],
\eea
where the simple pole due to $\cos{(\omega_{3})}$ at $t=-i\Lambda_{n}^{(3)}$ is explicit. Already we notice that the expressions quickly become more cumbersome as we go beyond NNLO and these together with the previous values of the coefficients in eqn.(\ref{recurrNNLO}) make up the characteristic function that will be needed to build the complete NNNLO PDF utilizing eqs.(\ref{chiansatz}), (\ref{residue}) and (\ref{PDFfinal}). 

Regarding the poles, the index $m$ now runs from $m=0,1,2,3$ with the last $m=3$, or \bea \Lambda_{n}^{(3)} = \Lambda_{n}^{(0)} + 3Cm,\eea providing the smallest order of contribution while the $m=0$ provides the dominant contribution, which seems more important as we approach large values of ${\cal N}$. We do not mention here the complete form of the residues, $r_{n}^{(m)}\;\forall\;m=0,1,2,3$, as they are extremely complicated and not illuminating by themselves, and thus, we only later show their contribution in the form of the PDF realized after estimating them. The final NNNLO PDF using eqn.(\ref{PDFfinal}) has the following form:
\bea \label{PDFN3}
P_{\Gamma_i}^{\rm NNNLO}({\cal N}) = \sum_{n=0}^{\infty}\big[r_{n,3}^{(0)}+ r_{n,3}^{(1)}e^{-3C{\cal N}} + r_{n,3}^{(2)}e^{-6C{\cal N}} + r_{n,3}^{(3)}e^{-9C{\cal N}} \big]e^{-\Lambda_{n}^{(0)}{\cal N}}
\eea
The residues, especially the term $r_{n}^{(0)}$ controlling the amplitude, do depend on the initial phase space variables, ${\Gamma_i}$, while the decay from the dominant exponential, $\exp{(-\Lambda_{n}^{0}{\cal N})}$, is independent of the same. 
The formula in eqn. (\ref{poles}) also tells us that when $\tilde{\mu} \gtrsim 1$, the poles remain well-separated along the imaginary $t$-axis. Expression for the mean number of e-folds now becomes much lengthy and complicated, and it is represented by the following relation:
\bea \label{meanNNNLO}
\langle{\cal N}(\Gamma)\rangle_{\rm NNNLO} &=& \frac{\tilde{\mu}^{2}}{2}\bigg\{1-(1-x+y)^{2} + \frac{12\tilde{\mu} ^2 y\left(\sqrt{3C\tilde{\mu} ^2}-\sinh \big[\sqrt{3C\tilde{\mu} ^2}\big]\right) 
   \cosh \big[\sqrt{3C\tilde{\mu} ^2} (x-y-1)\big]}{\sqrt{3C\tilde{\mu} ^2}\cosh\big[\sqrt{3C\tilde{\mu} ^2}\big]}\nonumber\\
&-& 3\tilde{\mu}^2 y^2 \cosh \big[\sqrt{6C \tilde{\mu} ^2} (x-y-1)\big]\nonumber\\
&\times& \bigg(\frac{4-2\sqrt{6C\tilde{\mu} ^2}
   \sinh \big[\sqrt{6C \tilde{\mu} ^2}\big]+2 \sinh \big[\sqrt{c \tilde{\mu} ^2}\big] \left(2 \sqrt{3C\tilde{\mu} ^2}+\sqrt{2} \sinh
   \big[\sqrt{6C \tilde{\mu} ^2}\big]\right)-2 \cosh \big[\sqrt{3C\tilde{\mu} ^2}\big]}{\cosh\big[\sqrt{3C \tilde{\mu} ^2}\big] \cosh\big[\sqrt{6C\tilde{\mu} ^2}\big]} \bigg)\nonumber\\
&-& \frac{12 y\sqrt{3C \tilde{\mu} ^2} \sinh \big[\sqrt{3C\tilde{\mu}^2}(x-y-1)\big]}{3C}+6 \sqrt{2}\tilde{\mu} ^2 y^2 \left(\sqrt{3C \tilde{\mu} ^2}-\sinh \big[\sqrt{3C \tilde{\mu} ^2}\big]\right)
   \frac{\sinh\big[\sqrt{6C\tilde{\mu} ^2} (x-y-1)\big]}{\cosh\big[\sqrt{3C\tilde{\mu} ^2}\big]}\nonumber\\
&+& \frac{2 \tilde{\mu} ^2 y^3\cosh
   \big[\sqrt{3} (x-1) \sqrt{3C \tilde{\mu} ^2}\big]}{\sqrt{3C \tilde{\mu} ^2}\cosh\big[\sqrt{6C \tilde{\mu} ^2}\big]\cosh\big[\sqrt{9C \tilde{\mu}^2}\big]}\bigg[6 c\tilde{\mu} ^2 \left(\sqrt{3C \tilde{\mu} ^2}-\sinh \big[\sqrt{3C \tilde{\mu} ^2}\big]\right) \frac{\cosh ^2\big[\sqrt{6C \tilde{\mu} ^2}\big]}{\cosh\big[\sqrt{3C\tilde{\mu} ^2}\big]}\nonumber\\
&+& 6 \left(3\sqrt{c^2 \tilde{\mu} ^4}\sinh \big[\sqrt{3C \tilde{\mu} ^2}\big]-\left(3C \tilde{\mu} ^2\right)^{3/2}\right) \frac{\sinh ^2\big[
   \sqrt{6C \tilde{\mu} ^2}\big]}{\cosh\big[\sqrt{3C \tilde{\mu} ^2}\big]}
   -3C\tilde{\mu} ^2 \left(3 \sqrt{3C \tilde{\mu} ^2}-\sqrt{3} \sinh \big[3\sqrt{C\tilde{\mu} ^2}\big]\right)\nonumber\\
&\times&\cosh \big[\sqrt{6C
   \tilde{\mu} ^2}\big] -6\sqrt{3}C \tilde{\mu} ^2 \sinh \big[3 \sqrt{C\tilde{\mu} ^2}\big]\left(2 \sqrt{3C \tilde{\mu} ^2} \tanh \big[\sqrt{3C \tilde{\mu} ^2}\big] +\frac{2}{\cosh\big[\sqrt{3C\tilde{\mu} ^2}\big]}
-1\right)\nonumber\\
&+& \sqrt{2} \sinh \big[\sqrt{6C\tilde{\mu} ^2}\big] \bigg(6C\tilde{\mu} ^2 \left(3 \sqrt{3C\tilde{\mu} ^2}-\sqrt{3} \sinh \big[3
\sqrt{C\tilde{\mu} ^2}\big]\right)\tanh \big[\sqrt{3C \tilde{\mu} ^2}\big]\nonumber\\
&+& \frac{\left(18\sqrt{C^2 \tilde{\mu} ^4}+2 \sqrt{3} \left(3C\tilde{\mu} ^2\right)^{3/2} \sinh \big[3
\sqrt{C\tilde{\mu} ^2}\big]\right)}{\cosh\big[\sqrt{c \tilde{\mu} ^2}\big]}-9\sqrt{C^2\tilde{\mu}
   ^4}\bigg)\bigg]+ 2 \sqrt{3}\tilde{\mu}^2 y^3 \sqrt{3C\tilde{\mu} ^2} \sinh \big[3(x-1) \sqrt{C \tilde{\mu} ^2}\big] \nonumber\\
&\times& \bigg(1-2 \sqrt{2} \tanh \big[\sqrt{3C\tilde{\mu} ^2}\big] \tanh \big[\sqrt{6C \tilde{\mu} ^2}\big] +\frac{1}{\cosh\big[\sqrt{6C
   \tilde{\mu} ^2}\big]}\bigg(2-4\sqrt{3C\tilde{\mu} ^2} \tanh \big[\sqrt{3C\tilde{\mu} ^2}\big]\nonumber\\
&-& \frac{4}{\cosh\big[\sqrt{3C\tilde{\mu}
   ^2}\big]} \bigg)+ \frac{2\sqrt{6C \tilde{\mu} ^2} \tanh \big[\sqrt{6C \tilde{\mu} ^2}\big]}{\cosh\big[\sqrt{3C \tilde{\mu}
   ^2}\big]}\bigg)
   \bigg\}
\eea

Since from the eqn. (\ref{PDFN3}), we can now construct the PDF by including more of the higher-order terms in the complete expansion; we have more information on the upper tail region of the PDF, corresponding to an increase in the non-Gaussian nature and thus translating into important consequences for the PBH formation possibilities.   

\section{Drift Dominated Regime: Methods to Solve for the characteristic and calculating moments}
\label{s10}

In this section, we concern the opposite condition where the quantum diffusion effects remain subdominant, and thus, the behavior of the scalar perturbations $\zeta$ through the USR is governed mainly in a classical limit where drift effects from the Langevin equation become essential. Here, the scalar perturbations are not affected predominantly by diffusion processes, which do not contribute to most of the e-folding realizations between some initial and final conditions for a given $\Gamma_{i}$.

The conditions on the variables to solve the partial differential equation, eqn. (\ref{FPEcharacter}), involve working under the limit where $y\gg 1$. In terms of the phase space variables, eqn. (\ref{newphasevars}) also tells us that this limit implies large conjugate canonical momenta. Since the nature of the analysis here is mainly classical, in terms of the PDF, we will obtain information about features around the maximum of the distribution. Starting with the mean value of the stochastic e-folds variable ${\cal N}$, or the first moment of the PDF from eqn. (\ref{chimoments}), one can construct the higher-order moments, that includes the skewness and kurtosis, to accurately study the structure of the complete PDF away from the maximum. 


Here we discuss the general method to solve for the various moments of the PDF, $P_{\Gamma_{i}}({\cal N})$, related to the set of coarse-grained variables ${\Gamma_{i}} = \{\zeta_i,\Pi_i\}$. We start with working at the leading order in the variable $y$ and develop the PDF up to going till the next-to-next-to-next-to-leading order in $y$. The method adopted here is known as the method of characteristics. Here we choose to parameterize our phase space variables $\Gamma= \{x,y\}$ with some arbitrary parameter that introduces characteristic lines in the phase space, and for each of these lines we need to solve linear ordinary differential equation whose solutions when combined gives us the characteristic curves in the phase space. These solutions are a convenient method to solve for the original PDE in eqn. (\ref{FPEcharacter}) as we will show now starting with the solution leading order in $y$.

\subsection{Calculation of PDF at LO}

We start with writing the phase space variables as $\Gamma(p)=\{x(p),y(p)\}$, where $p$ is a parameter defined for the respective set of characteristic lines in phase space. We need to solve these lines with the help of the PDE in eqn. (\ref{FPEcharacter}). This is done as follows by writing the differential:
\bea
\frac{d\chi(t;\Gamma(p))}{dp}= \bigg(\frac{dx}{dp}\frac{\partial}{\partial x}+ \frac{dy}{dp}\frac{\partial}{\partial y}\bigg)\chi(t;\Gamma(p)),
\eea
the RHS resembles the differential operator in the PDE eqn. (\ref{FPEcharacter}), given if we are able to obtain the solution curves for the equations:
\bea \label{solutionODE}
\frac{dx(p)}{dp}=\frac{dy(p)}{dp}=-3C,
\eea
and the resulting family of curves allows us to write the leading order PDE in the ordinary differential equation form, that completely ignores any effects coming from the first diffusion term in eqn. (\ref{FPEcharacter}) as:
\bea \label{chiLO}
\bigg(\frac{d}{dp}+\frac{it}{y_{0}-3Cp}\bigg)\chi(t;\Gamma(p))= 0,
\eea
where the solution $y= y_{0}-3Cp$, for some initial condition value $y_{0}$, is used which forms the solution curve for the eqn. (\ref{solutionODE}) parameterized by $p$. The above first-order ODE can be straightforwardly solved to give:
\bea
\chi(t;\Gamma(p)) = \chi_{0}(t;\Gamma_{0})\big[y_{0}-3Cp\big]^{\frac{it}{3C}},
\eea
where the characteristic parameter $C$ appears explicitly. Re-labelling the parameterization by using the difference $x_{0}-y_{0}$, and therefore, setting one constant $y_{0}=0$, we have for the original variables, $y=-3Cp$ and $x_{0}=x-y$. This makes the initial condition solution dependent only on $y$. Now, applying the similar boundary conditions in eqn. (\ref{bdrycondn1}) and eqn. (\ref{bdrycondn2}) by converted back into $\{x,y\}$ we have from the first condition:
\bea
\chi(t;\Gamma)|_{x=0} = 1,\quad\implies\quad\;\chi_{0}(-y)=y^{-\frac{it}{3C}},
\eea
From this, the final solution can be easily written down to the following expression:
\bea \label{chiLOsol}
\chi^{\rm LO}(t;\Gamma(p)) = \bigg(1-\frac{x}{y}\bigg)^{-\frac{it}{3C}},
\eea
To further build the PDF at LO we just have to make use of the Fourier transform eqn.(\ref{characterFourier}):
\bea \label{pdfLO1}
P_{\Gamma}^{\rm LO}({\cal N})= \delta({\cal N}-\langle{\cal N}_{\rm LO}\rangle),
\eea
where the first and second moments of the stochastic variable ${\cal N}$ comes out as:
\bea
\langle{\cal N}(\Gamma)\rangle_{\rm LO} &=& -\frac{1}{3C}\ln{\bigg(1-\frac{x}{y}\bigg)}.\\
\langle{\cal N}^2(\Gamma)\rangle_{\rm LO} &=& \langle{\cal N}(\Gamma)\rangle^2_{\rm LO},
\eea
and from the above we conclude that,
\bea \sigma_{\rm LO}^{2} = \langle{\cal N}^2(\Gamma)\rangle_{\rm LO} - \langle{\cal N}(\Gamma)\rangle^2_{\rm LO} = 0.\eea
The various higher-order parameters describing deviations from Gaussianity, such as the skewness and the kurtosis also remain absent and thus, by extension, the parameters $f_{\rm NL},\;g_{\rm NL},\;\tau_{\rm NL}$ also vanish. The PDF eqn. (\ref{pdfLO}) tells us one important fact about the fluctuations in the e-folds, they completely shut off and effectively lead to zero curvature perturbations since ${\cal N}-\langle{\cal N}\rangle=\zeta_{\rm cg}$ from the stochastic-$\delta N$ formalism.

\subsection{Calculation of PDF at NLO}

In this section, we continue with the approach of the previous section to derive the characteristic function at the next-to-leading order (NLO) which will give us further corrections to properties like the mean number of e-folds and as well introduce us to more new statistical properties like the variance of the stochastic number of e-folds and many more important measures of interest. Notice as we explained at the end of the previous analysis, to incorporate the curvature perturbations we will need to go beyond the LO results and thus it starts with the current discussions of NLO corrections.  

To begin we consider eqn. (\ref{FPEcharacter}) but this time let the first term, which brings the diffusion effects, act only on the leading order solution obtained above, as for the rest terms they act trivially on $\chi^{\rm LO}$. We write 
\bea \chi = \chi^{\rm LO} + \chi^{\rm NLO},\eea 
where the term $\chi^{\rm NLO}$ represents a sub-dominant correction to the LO result. Substituting the above into eq. (\ref{FPEcharacter}), the NLO contribution obeys the equation:
\bea \label{chiNLO}
\bigg(-3Cy\bigg[\frac{\partial}{\partial x}+ \frac{\partial}{\partial y}\bigg] + it\bigg)\chi^{\rm NLO}(t;\Gamma(p)) = -\frac{1}{y^{2}}\frac{it}{3C\tilde{\mu}^{2}}\bigg(1+\frac{it}{3C}\bigg)\bigg(1-\frac{x}{y}\bigg)^{-\frac{it}{3C}-2}.
\eea
Notice the important fact that the quantum diffusion effects contribute sub-dominantly when working at the NLO, which is why the LO solution when acted by the diffusion operator, acts as the source term for the NLO drift solution of the characteristic function. The characteristic lines for $\{x(p),y(p)\}$ remain the same as before. With these we obtain a similar linear ODE for $\chi^{\rm NLO}$ of the form:
\bea
\bigg(\frac{d}{dp} + \frac{it}{y}\bigg)\chi^{\rm NLO}(t;\Gamma(p)) = -\frac{1}{y^{3}}\frac{it}{3C\tilde{\mu}^{2}}\bigg(1+\frac{it}{3C}\bigg)\bigg(1-\frac{x}{y}\bigg)^{-\frac{it}{3C}-2},
\eea
solving which gives us a general solution:
\bea \label{gensolNLO}
\chi^{\rm NLO}(t;\Gamma(p)) = \bigg(\chi_{1}(y-x) + \frac{it}{9\tilde{\mu}^{2}C^{2}}\bigg(1+\frac{it}{3C}\bigg)\ln{(y)}(y-x)^{-i\frac{it}{3C}-2}\bigg)y^{\frac{it}{3C}},
\eea
for the initial condition solution we again use the boundary condition $\chi(t;\Gamma)|_{x=0}=1$, and get:
\bea
\chi_{1}(y-x) = (y-x)^{-\frac{it}{3C}}\bigg(1-\frac{it}{9C^{2}\tilde{\mu}^{2}(y-x)^{2}}\bigg(1+\frac{it}{3C}\bigg)\ln{(y-x)}\bigg),
\eea
which when used in the general NLO solution eqn. (\ref{gensolNLO}) gives us:
\bea \label{chiNLOsol}
\chi^{\rm NLO}(t;\Gamma) = \bigg(1-\frac{x}{y}\bigg)^{-\frac{it}{3C}}\bigg[1-\frac{it}{9C^{2}\tilde{\mu}^{2}}\bigg(1+\frac{it}{3C}\bigg)\frac{\ln{\big(1-\frac{x}{y}\big)}}{(y-x)^{2}}\bigg].
\eea
The above contains corrections of ${\cal O}(\tilde{\mu}^{-2}y^{-2})$ to the LO result. This solution can further provide us with the PDF at NLO order and associated statistical moments since ${\cal N}$ is eventually a stochastic quantity. First, we take a look at the PDF coming from the inverse Fourier transform of $\chi^{\rm NLO}(t;\Gamma)$ which has a structure utilizing derivatives of the Dirac delta distribution as follows:
\bea \label{pdfNLO1}
P^{\rm NLO}_{\Gamma}({\cal N}) &=& P^{\rm LO}_{\Gamma}({\cal N}) + \frac{1}{9C^2\tilde{\mu}^2(y-x)^2}\bigg(\delta^{(1)}({\cal N}-\langle{\cal N}_{\rm LO}\rangle) - \frac{1}{3C}\delta^{(2)}({\cal N}-\langle{\cal N}_{\rm LO}\rangle)\bigg)\ln{\bigg(1-\frac{x}{y}\bigg)}\nonumber\\
&=& P^{\rm LO}_{\Gamma}({\cal N}) - \frac{1}{9C^2\tilde{\mu}^2(y-x)^2}\bigg(\frac{1}{\cal N} + \frac{2}{3C{\cal N}^2}\bigg)\delta({\cal N}-\langle{\cal N}_{\rm LO}\rangle)\ln{\bigg(1-\frac{x}{y}\bigg)},
\eea
where the superscripts $(i)$ denote the $i$th derivative of the Dirac delta with respect to the e-folds ${\cal N}$ and in the second line the derivative properties of the Dirac delta are used.
From this result, we can also quickly determine the mean value and variance in the number of e-folds. The mean value at NLO is evaluated as follows:
\bea
\langle{\cal N}(\Gamma)\rangle_{\rm NLO} = \langle{\cal N}(\Gamma)\rangle_{\rm LO}\bigg(1+\frac{1}{3C^2\tilde{\mu}^2(y-x)^2}\bigg), 
\eea
and similarly evaluating the variance at NLO requires first to know the second moment which comes out to be:
\bea
\langle{\cal N}^{2}(\Gamma)\rangle_{\rm NLO} = \frac{2\langle{\cal N}\rangle_{\rm LO}}{9C^2\tilde{\mu}^2(y-x)^2}\bigg[1-\bigg(1+\frac{3}{2}C\tilde{\mu}^2(y-x)^2\bigg)\ln{\bigg(1-\frac{x}{y}\bigg)}\bigg],
\eea
using which the variance gets evaluated to the following:
\bea \label{delta2nlo}
\sigma_{\rm NLO}^{2} = \langle\delta{\cal N}^2(\Gamma)\rangle_{\rm NLO}&=&  \langle{\cal N}^{2}(\Gamma)\rangle_{\rm NLO} - \langle{\cal N}(\Gamma)\rangle_{\rm NLO}^{2}\nonumber\\
&=& \frac{2\langle{\cal N}(\Gamma)\rangle_{\rm LO}}{9C^2\tilde{\mu}^2(y-x)^2}\bigg(1-\frac{\langle{\cal N}(\Gamma)\rangle_{\rm LO}}{2C^2\tilde{\mu}^2(y-x)^2}\bigg).
\eea
where the first term gives the leading contribution. The benefits of calculating multiple statistical moments come to aid when we want to calculate important quantities of observational significance such as the power spectrum, and other non-Gaussianity parameters. 
We now check for corrections to the third moment induced from the above PDF at NLO order. Here we obtain for the third moment at NLO the following:
\bea
\langle{\cal N}^3\rangle_{\rm NLO} = \frac{2-\ln{\big(1-\frac{x}{y}\big)}(1+C\tilde{\mu}^2(x-y)^2)}{27C^4\tilde{\mu}^2(x-y)^2}\ln^2{\bigg(1-\frac{x}{y}\bigg)},
\eea
using which the leading contribution to the third moment away from the mean is found to be:
\bea \label{delta3nlo}
\langle\delta{\cal N}^3\rangle_{\rm NLO} = \langle({\cal N}-\langle{\cal N}\rangle)^3\rangle_{\rm NLO} &=& \langle{\cal N}^3\rangle_{\rm NLO} + 2\langle{\cal N}\rangle_{\rm NLO}^3 - 3\langle{\cal N}\rangle_{\rm NLO}\langle{\cal N}^2\rangle_{\rm NLO}, \nonumber\\
&=& -\frac{2\langle{\cal N}(\Gamma)\rangle_{\rm LO}^2}{9\;C^3\tilde{\mu}^4(x-y)^4},
\eea
and the above comes out as more suppressed in the order of expansion considered for NLO, that is ${\cal O}((\tilde{\mu}y)^{-2})$. Similarly, checking for corrections to the fourth moment from the mean also leads to suppressed contributions. First we require the following fourth moment:
\bea
\langle{\cal N}^4\rangle_{\rm NLO} = \frac{-12+\ln{\big(1-\frac{x}{y}\big)}(4+3C\tilde{\mu}^2(x-y)^2)}{243C^5\tilde{\mu}^2(x-y)^2}\ln^3{\bigg(1-\frac{x}{y}\bigg)},
\eea
using which the leading contribution to the quantity of interest is calculated as:
\bea \label{delta4nlo}
\langle\delta{\cal N}^4\rangle_{\rm NLO} = \langle({\cal N}-\langle{\cal N}\rangle)^4\rangle_{\rm NLO} &=& \langle{\cal N}^4\rangle_{\rm NLO} - 4\langle{\cal N}\rangle_{\rm NLO}\langle{\cal N}^{3}\rangle_{\rm NLO} + 6\langle{\cal N}\rangle_{\rm NLO}^{2}\langle{\cal N}^2\rangle_{\rm NLO}- 3\langle{\cal N}\rangle_{\rm NLO}^{4}, \nonumber\\
&=& \frac{4\langle{\cal N}(\Gamma)\rangle_{\rm LO}^3}{27\;C^4\tilde{\mu}^6(x-y)^6}.
\eea
From these higher-order statistical moments, one can measure deviations from having a Gaussian distribution by using the properties of skewness and kurtosis of the distribution. These are defined as follows:
\bea
\gamma = \frac{\langle({\cal N}-\langle{\cal N}\rangle)^3\rangle}{\sigma^3}, \quad\quad\quad \kappa = \frac{\langle({\cal N}-\langle{\cal N}\rangle)^4\rangle}{\sigma^4},
\eea
and from the eqns. (\ref{delta2nlo},\ref{delta3nlo},\ref{delta4nlo}) we determine the skewness and kurtosis as the following at NLO:
\bea
\gamma_{\rm NLO} &=& \frac{\langle({\cal N}-\langle{\cal N}\rangle)^3\rangle_{\rm NLO}}{\sigma_{\rm NLO}^3} = -3C\sqrt{\frac{\langle{\cal N}\rangle_{\rm LO}}{2C^2(x-y)^2\tilde{\mu}^2}},\\
\kappa_{\rm NLO} &=& \frac{\langle({\cal N}-\langle{\cal N}\rangle)^4\rangle_{\rm NLO}}{\sigma_{\rm NLO}^4} = \frac{3\langle{\cal N}\rangle_{\rm LO}}{(x-y)^2\tilde{\mu}^2},
\eea

Following the above results we can now evaluate the expressions for various quantities of statistical and observational significance, starting with the power spectrum using eqn. (\ref{delta2nlo}) as:
\bea
\Delta^{2}_{\zeta\zeta,\rm NLO} = \frac{d\langle\delta{\cal N}^2\rangle}{d\langle{\cal N}\rangle} = \frac{2}{9C^2\tilde{\mu}^2(y-x)^2},
\eea
which represents the leading order contribution. Next we determine the non-Gaussianity parameter $f_{\rm NL}$ using eqn. (\ref{delta3nlo}) and eqn. (\ref{delta2nlo}) as follows:
\bea \label{fnlNLO}
f_{\rm NL} = \frac{5}{36}\frac{1}{[\Delta^{2}_{\zeta\zeta,\rm NLO}]^2}\frac{d^2\langle\delta{\cal N}^3\rangle}{d\langle{\cal N}\rangle^2} = -\frac{5C}{4},
\eea
and similarly, the other non-Gaussian parameters $\tau_{\rm NL}$ and $g_{\rm NL}$ at NLO are determined as follows:
\bea \label{tauNLO}
\tau_{\rm NL} &=& \frac{1}{36}\frac{1}{[\Delta^{2}_{\zeta\zeta,\rm NLO}]^4}\bigg(\frac{d^2\langle\delta{\cal N}^3\rangle}{d\langle{\cal N}\rangle^2}\bigg)^2 = \frac{9}{4}C^2,\\
g_{\rm NL} &=&  \frac{1}{[\Delta^{2}_{\zeta\zeta,\rm NLO}]^3}\frac{d^3\langle\delta{\cal N}^4\rangle}{d\langle{\cal N}\rangle^3} = 18C^2.
\eea
The above gives us some intriguing insights into the nature of non-Gaussianity to be expected at the NLO. See refs. \cite{Maldacena:2002vr,Alishahiha:2004eh,Mazumdar:2001mm,Choudhury:2002xu,Panda:2005sg,Chingangbam:2004ng,Armendariz-Picon:1999hyi,Garriga:1999vw,Burrage:2010cu,Choudhury:2012yh,Choudhury:2012whm,Chen:2010xka,Chen:2006nt,Chen:2009zp,Chen:2009we,Chen:2008wn,Chen:2006xjb,Chen:2013aj,Chen:2012ge,Chen:2009bc,Creminelli:2010ba,Kobayashi:2010cm,Mizuno:2010ag,Burrage:2011hd,Kobayashi:2011pc,DeFelice:2011zh,Renaux-Petel:2011lur,DeFelice:2011uc,Gao:2011qe,deRham:2012az,Ohashi:2012wf,DeFelice:2013ar,Arroja:2013dya,Choudhury:2013qza,Pirtskhalava:2015zwa,Baumann:2009ds,Senatore:2016aui,Baumann:2018muz,Das:2023cum,Choudhury:2011sq,Choudhury:2012yh,Choudhury:2012ib,Choudhury:2012whm,Choudhury:2013jya,Choudhury:2013zna,Esposito:2019jkb,Goldstein:2022hgr,Arkani-Hamed:2015bza,Arkani-Hamed:2023kig,Green:2023ids,Baumann:2021fxj,Baumann:2020dch,Baumann:2019oyu,Meerburg:2019qqi,Arkani-Hamed:2018kmz}
for more details. The $f_{\rm NL}$ comes with a negative signature due to the expression in eqn. (\ref{delta3nlo}) and thus, when working at the NLO expansion in the diffusion-dominated regime, one can expect to obtain ${\cal O}(1)$ negative non-Gaussianity dependent on the underlying EFT structure characterized by the parameter $C$. The case $C=1$ results in $f_{\rm NL}=-5/4$, which is for the canonical stochastic single-field inflation and the value violates the bound coming from Maldcena's consistency condition, $f_{\rm NL}=(5/12)(1-n_{s})$ \cite{Maldacena:2002vr} where $n_{s}\ll 1$ represents the spectral tilt. Large negative non-Gaussianity of $f_{\rm NL}\lesssim {\cal O}(-1)$ have found significant use in curing the PBH overproduction issue in light of the latest NANOGrav15 data, see \cite{Franciolini:2023pbf,Franciolini:2023wun,Inomata:2023zup, Inui:2023qsd,Chang:2023aba,Gorji:2023ziy,Li:2023xtl,Firouzjahi:2023xke,Gorji:2023sil,Raatikainen:2023bzk,Choudhury:2023fwk,Choudhury:2023fjs,Ferrante:2023bgz,Gorji:2023sil} for recent attempts to tackle this problem. The relation in eqn. (\ref{tauNLO}) maintains the Suyama-Yamaguchi relation, $\tau_{\rm NL}=(6f_{\rm NL}/5)^{2}$ \cite{Suyama:2007bg}, where the equality sign holds only for the case of single-field inflation models. In a multi-field scenario, the equality sign in the relation is replaced with an inequality, giving us, $\tau_{\rm NL}< (6f_{\rm NL}/5)^{2}$. The case $C=1$ predicts a large $\tau_{\rm NL}=9/4$ relative to $f_{\rm NL}$ and gives $g_{\rm NL}=18$.

\subsection{Calculation of PDF at NNLO}

Here we proceed to an order higher in the characteristic function, at the next-to-next-to-leading order. The approach is similar to the previous case of NLO and it is here that we observe some significant non-Gaussian features. To that effect, we write the characteristic function satisfying the eqn. (\ref{FPEcharacter}) as: \bea \chi=  \chi^{\rm LO} + \chi^{\rm NLO} + \chi^{\rm NNLO},\eea
that now includes an extra sub-dominant contribution which we aim to evaluate and using this we can obtain an equation for $\chi_{\rm NNLO}$, after not acting it with the diffusion terms, as follows:
\bea
\bigg(\frac{d}{dp} + \frac{it}{y}\bigg)\chi^{\rm NNLO}(t;\Gamma(p)) = -\frac{1}{y\tilde{\mu}^{2}}\frac{\partial^2}{\partial x^2}\chi^{\rm NLO},
\eea
where we have used the eqns. (\ref{chiLO},\ref{chiNLO}), including the previous characteristic lines which do not change. The expression on the RHS becomes more cumbersome based on the function in eqn. (\ref{chiNLOsol}). We still provide the RHS in a compact form as follows:
\bea
-\frac{1}{y\tilde{\mu}^{2}}\frac{\partial^2}{\partial x^2}\chi^{\rm NLO} &=& \frac{1}{3\tilde{\mu}^{2}}\bigg(1-\frac{x}{y}\bigg)^{-\frac{it}{3C}-2}\bigg[3b\bigg(-\frac{1}{y^{3}}-\frac{4\big(1-\frac{x}{y}\big)}{Cy^{2}}+\frac{6\big(1-\frac{x}{y}\big)^{2}\ln{\big(1-\frac{x}{y}\big)}}{yC^2}\bigg)\nonumber\\
&-& 2itb\bigg(\frac{1}{Cy^3}-\frac{2\big(1-\frac{x}{y}\big)\ln{\big(1-\frac{x}{y}\big)}}{C^{2}y^2}\bigg)-it\bigg(1+\frac{it}{3C}\bigg)\frac{1-b\ln{\big(1-\frac{x}{y}\big)}}{Cy^{3}}\bigg],
\eea
where the parameter $b$ is defined as:
\bea b=\frac{it}{9C^{4}\tilde{\mu}^{2}}\left(1+\frac{it}{3C}\right).\eea From here, we obtain a similar linear ODE for $\chi^{\rm NNLO}$. We now directly mention the PDF after performing the same procedure of solving the ODE and determining the initial condition function at NNLO, which ultimately gives us the following expression:
\bea \label{chiNNLOsol}
\chi^{\rm NNLO}(t;\Gamma) &=& \bigg(1-\frac{x}{y}\bigg)^{-\frac{it}{3c}}\bigg[1-\frac{it(1+\frac{it}{3C})\ln{\big(1-\frac{x}{y}\big)}}{9(x-y)^{2}C^{2}\tilde{\mu}^2}\nonumber\\
&&\quad\quad\quad\quad\quad\quad\quad\quad\quad- \frac{it(1+\frac{it}{3C})\ln{\big(1-\frac{x}{y}\big)^2}\big(90C^2+12iCt+(t-9Ci)(t-6Ci)\big)}{486(x-y)^{4}C^{4}\tilde{\mu}^4}\bigg], \quad
\eea
where the last term inside the second bracket comes as a new addition to the rest of the terms which already form $\chi^{\rm NLO}$, see eqn.(\ref{chiNLOsol}). Notice that the expression for the characteristic function results in a series expansion in powers of ${(C\tilde{\mu}y)^{-2}}$, where for the above $\chi^{\rm NNLO}$ the expansion gets terminated at ${\cal O}((C\tilde{\mu}y)^{-4})$. Analogously, the PDF corresponding to the inverse Fourier transform of the function $\chi^{\rm NNLO}$ can be represented using the Dirac delta derivatives, as done in the previous case with $\chi^{\rm NLO}$. The final form of the PDF turns out as follows:
\bea \label{pdfNNLO1}
P_{\Gamma}^{\rm NNLO}({\cal N}) &=& P_{\Gamma}^{\rm NLO}({\cal N}) \nonumber\\
&&- \frac{1}{486\;{C}^4\tilde{\mu}^4(y-x)^4}\ln{\bigg(1-\frac{x}{y}\bigg)}\bigg[-90\delta^{(1)}({\cal N}-\langle{\cal N}_{\rm LO}\rangle)+ 40C\delta^{(2)}({\cal N}-\langle{\cal N}_{\rm LO}\rangle)
\nonumber\\
&&- 4\delta^{(3)}({\cal N}-\langle{\cal N}_{\rm LO}\rangle)+\ln{\bigg(1-\frac{x}{y}\bigg)}\bigg(54C^2\delta^{(1)}({\cal N}-\langle{\cal N}_{\rm LO}\rangle)- 33C\delta^{(2)}({\cal N}-\langle{\cal N}_{\rm LO}\rangle)\nonumber\\
&&+ 6\delta^{(3)}({\cal N}-\langle{\cal N}_{\rm LO}\rangle)- \frac{1}{3C}\delta^{(4)}({\cal N}-\langle{\cal N}_{\rm LO}\rangle) \bigg)\bigg],\nonumber\\
&=& P_{\Gamma}^{\rm NLO}({\cal N}) \nonumber\\
&&- \frac{1}{486\;{C}^4\tilde{\mu}^4(y-x)^4}\delta({\cal N}-\langle{\cal N}_{\rm LO}\rangle)\bigg[\frac{90}{\cal N}+ \frac{80C}{{\cal N}^2}+ \frac{12C}{{\cal N}^3}+  \ln{\bigg(1-\frac{x}{y}\bigg)}\nonumber\\
&&\times \bigg(-\frac{54C^2}{\cal N}-\frac{66C}{{\cal N}^2}- \frac{18}{{\cal N}^3}-\frac{4}{3C{\cal N}^4}\bigg)\bigg]\ln{\bigg(1-\frac{x}{y}\bigg)},
\eea

Using Eqn.(\ref{chiNNLOsol}) and Eqn.(\ref{chimoments}), further higher-order corrections estimated for the mean number of ${\cal N}$ can be collected and written down as follows:
\bea
\langle{\cal N}(\Gamma)\rangle_{\rm NNLO} = \langle{\cal N}(\Gamma)\rangle_{\rm LO}\bigg(1+\frac{1}{3C^2\tilde{\mu}^2(y-x)^2} + \frac{9C^2\langle{\cal N}(\Gamma)\rangle_{\rm LO}+5}{9C^3\tilde{\mu}^4(y-x)^4} \bigg). 
\eea
Similarly, the variance estimated for the current NNLO at order ${\cal O}((\tilde{\mu}y)^{-4})$ first requires for the second moment given as follows:
\bea
\langle{\cal N}^{2}(\Gamma)\rangle_{\rm NNLO} = \frac{14C+3\langle{\cal N}\rangle_{\rm LO}(10+11C^2)}{27C^3(x-y)^4\tilde{\mu}^2}\langle{\cal N}\rangle_{\rm LO}
\eea
which in turn provides the next order correction to the variance:
\bea \label{delta2nnlo}
\sigma_{\rm NNLO}^{2} = \langle\delta{\cal N}^2(\Gamma)\rangle_{\rm NNLO}&=& -\frac{2}{27C^3\tilde{\mu}^2(y-x)^2}\ln{\bigg(1-\frac{x}{y}\bigg)} + \frac{(11C-1)}{81C^4\tilde{\mu}^4(x-y)^4}\ln^2{\bigg(1-\frac{x}{y}\bigg)},\nonumber\\
&=& \frac{2\langle{\cal N}(\Gamma)\rangle_{\rm LO}}{9C^2\tilde{\mu}^2(y-x)^2}+  \frac{(11C-1)\langle{\cal N}(\Gamma)\rangle^{2}_{\rm LO}}{9C^2\tilde{\mu}^4(x-y)^4},
\eea
For the present case, third moment from the mean of the probability distribution is calculated from eqn.(\ref{pdfNNLO1}) and eqn.(\ref{chimoments}). First evaluating the third moment itself gives us the following contribution:
\bea
\langle{\cal N}^3(\Gamma)\rangle_{\rm NNLO} = \frac{4\langle{\cal N}(\Gamma)\rangle_{\rm LO}}{27C^4\tilde{\mu^4}(x-y)^4}\bigg[1+15C^2\langle{\cal N}(\Gamma)\rangle_{\rm LO}\bigg],
\eea
and further at leading order in ${\cal O}((\tilde{\mu}y)^{-4})$, discarding other higher powers of logarithms, corrections to the third moment comes out as follows:
\bea \label{delta3nnlo}
\langle\delta{\cal N}^3(\Gamma)\rangle_{\rm NNLO} = \langle({\cal N}-\langle{\cal N}\rangle)^3\rangle_{\rm NNLO} &=& \frac{1}{81\;C^5\tilde{\mu}^4(x-y)^4}2\ln{\bigg(1-\frac{x}{y}\bigg)}\bigg[(-1+3C)\ln{\bigg(1-\frac{x}{y}\bigg)}-2\bigg],\nonumber\\
&=& \frac{4\langle{\cal N}(\Gamma)\rangle_{\rm LO}}{27\;C^4\tilde{\mu}^4(x-y)^4}\bigg[1+\frac{3C}{2}(3C-1)\langle{\cal N}(\Gamma)\rangle_{\rm LO}\bigg], 
\eea
which includes effects from the characteristic parameter $C$. Similarly, one can evaluate corrections to the fourth moment away from the mean. For the fourth moment after discarding higher multiples of logarithms we have:
\bea
\langle{\cal N}^{4}(\Gamma)\rangle_{\rm NNLO} = \frac{4(4+C)\langle{\cal N}(\Gamma)\rangle_{\rm LO}^2}{27C^4\tilde{\mu^4}(x-y)^4}
\eea
and further, to leading order in ${\cal O}((\tilde{\mu}y)^{-4})$, we have:
\bea \label{delta4nnlo}
\langle\delta{\cal N}^4(\Gamma)\rangle_{\rm NNLO} = \langle({\cal N}-\langle{\cal N}\rangle)^4\rangle_{\rm NNLO} &=& \frac{4}{243\;C^6\tilde{\mu}^4(x-y)^4}\ln^2{\bigg(1-\frac{x}{y}\bigg)}\bigg[8+C+\ln{\bigg(1-\frac{x}{y}\bigg)}(4+40C)\bigg],\nonumber\\
&=& \frac{4\langle{\cal N}(\Gamma)\rangle_{\rm LO}^2}{27\;C^4\tilde{\mu}^4(x-y)^4}\bigg[8+C+3C\langle{\cal N}(\Gamma)\rangle_{\rm LO}(4+40C)\bigg].
\eea
From here we can again move towards determining the properties of  
skewness and kurtosis for the distribution at NNLO. These are defined for the present distribution as follows:
\bea
\gamma_{\rm NNLO} &=& \frac{\langle({\cal N}-\langle{\cal N}\rangle)^3\rangle_{\rm NNLO}}{\sigma_{\rm NNLO}^3} = \frac{1}{\langle{\cal N}\rangle_{\rm LO}^2}\frac{2(2+3\langle{\cal N}\rangle_{\rm LO}C(C-1))}{C(11C-1)^{3/2}}(x-y)^{2}\tilde{\mu}^2,\\
\kappa_{\rm NNLO} &=& \frac{\langle({\cal N}-\langle{\cal N}\rangle)^4\rangle_{\rm NNLO}}{\sigma_{\rm NNLO}^4} = \frac{1}{\langle{\cal N}\rangle^{2}_{\rm LO}}\frac{12C(x-y)^4\tilde{\mu}^4}{(11C-1)^2}.
\eea

We now estimate the corrected version of the power spectrum from the variance corrections derived before at the NNLO in eqn.(\ref{delta2nnlo}):
\bea
\Delta^{2}_{\zeta\zeta,\rm NNLO} = \frac{d\langle\delta{\cal N}^2\rangle}{d\langle{\cal N}\rangle} =  \frac{2}{9C^2\tilde{\mu}^2(y-x)^2} + \frac{2(11C-1)\langle{\cal N}(\Gamma)\rangle_{\rm LO}}{9C^2\tilde{\mu}^4(x-y)^4}.
\eea
For the remaining set of non-Gaussianity parameters, we evaluate them using eqns.(\ref{delta3nnlo},\ref{delta4nnlo}) as follows:
\bea \label{fnlNNLO}
f_{\rm NL} = \frac{5}{36}\frac{1}{[\Delta^{2}_{\zeta\zeta,\rm NNLO}]^2}\frac{d^2\langle\delta{\cal N}^3\rangle}{d\langle{\cal N}\rangle^2} &=& \frac{5}{36}\frac{3C-1}{2C^3}\frac{8}{9\tilde{\mu}^4(x-y)^4}\frac{81\tilde{\mu}^4(x-y)^4}{4}\bigg(1+\frac{(11C-1)\langle{\cal N}\rangle_{\rm LO}}{\tilde{\mu}^{2}(x-y)^{2}}\bigg)^{-2},\nonumber\\
&=& \frac{5}{2}\frac{3C-1}{2C^3} - \frac{2(11C-1)\langle{\cal N}\rangle_{\rm LO}}{\tilde{\mu}^{2}(x-y)^{2}},
\eea
where the term ${\cal O}(\tilde{\mu}^{-2}y^{-2})$ comes as a suppressed correction from the power spectrum and due to having $\tilde{\mu}y \gg 1$ as the necessary condition to achieve a drift-dominated scenario, we approximate the result at leading order in the series expansion. Next we have:
\bea \label{tauNNLO}
\tau_{\rm NL} &=&  \frac{1}{36}\frac{1}{[\Delta^{2}_{\zeta\zeta,\rm NNLO}]^4}\bigg(\frac{d^2\langle\delta{\cal N}^3\rangle}{d\langle{\cal N}\rangle^2}\bigg)^{2} = \bigg(\frac{3C-1}{2C^3}\frac{8}{9\tilde{\mu}^4(x-y)^4}\bigg)^{2}\bigg(\frac{81\tilde{\mu}^4(x-y)^4}{4}\bigg)^{2}\bigg(1+\frac{(11C-1)\langle{\cal N}\rangle_{\rm LO}}{\tilde{\mu}^{2}(x-y)^{2}}\bigg)^{-4},\nonumber\\
&=& \frac{9}{4}\bigg(\frac{3C-1}{C^{3}}\bigg)^{2}\bigg(1-\frac{4(11C-1)\langle{\cal N}\rangle_{\rm LO}}{\tilde{\mu}^{2}(x-y)^{2}}\bigg),\\
g_{\rm NL} &=& \frac{1}{[\Delta^{2}_{\zeta\zeta,\rm NNLO}]^3}\frac{d^3\langle\delta{\cal N}^4\rangle}{d\langle{\cal N}\rangle^3} = \frac{24C(4+40C)}{9}\frac{81C^2\tilde{\mu}^2(x-y)^2}{4}\bigg(1+\frac{(11C-1)\langle{\cal N}\rangle_{\rm LO}}{\tilde{\mu}^{2}(x-y)^{2}}\bigg)^{-3},\nonumber\\
&=& 54C^3(4+40C)\bigg(1- \frac{3(11C-1)\langle{\cal N}\rangle_{\rm LO}}{\tilde{\mu}^{2}(x-y)^{2}}\bigg).
\eea
where, as mentioned above, a similar term leading order in the expansion is considered. The case $C=1$ in eqn. (\ref{fnlNNLO}) gives us at leading order, $f_{\rm NL}\sim 5/2$, which is positive and larger in amplitude compared to the $C=1$ estimate at NLO, see eqn. (\ref{fnlNLO}). The current non-Gaussianity estimate also violates Maldacena's bound for canonical single-field inflation models. We must notice that for $C<1/3$ we start to see negative $f_{\rm NL}$ at NNLO, indicating possibility to achieve negative non-Gaussianity for some non-canonical stochastic single-field model giving thus a possible candidate to also tackle the PBH overproduction problem. For $C=1$ in eqn. (\ref{tauNNLO}) we get $\tau_{\rm NL} \sim 9 $ at leading order and only small corrections relative to its value from $C=1$ in eqn. (\ref{tauNNLO}). The Suyama-Yamaguchi bound, $\tau_{\rm NL}=(6f_{\rm NL}/5)^{2}$, still remains intact at NNLO.  

\subsection{Calculation of PDF at NNNLO}

We finally discuss results from the characteristic function in the next-to-next-to-next-leading order. The present characteristic function comes as another higher-order term in the series expansion satisfying the eqn.(\ref{FPEcharacter}) with the diffusion terms as: \bea \chi=  \chi^{\rm LO} + \chi^{\rm NLO} + \chi^{\rm NNLO} + \chi^{\rm NNNLO}.\eea To obtain the NNNLO contribution from the method of characteristics is not a tractable task to undertake in this section and thus we refer to \cite{Pattison:2021oen} for a general series expansion method developed to obtain the characteristic function terms to any arbitrary degree in the classical or drift-dominated regime.

Instead of fully mentioning the PDF, which is not illuminating in itself, we mention the necessary corrections brought from its structure to the statistical moments of our interest so far. We start with the mean value correction,
\bea
\langle{\cal N}(\Gamma)\rangle_{\rm NNNLO} = \langle{\cal N}(\Gamma)\rangle_{\rm NNLO}+ \langle{\cal N}(\Gamma)\rangle_{\rm LO}\frac{60\langle{\cal N}(\Gamma)\rangle^2_{\rm LO}+77\langle{\cal N}(\Gamma)\rangle_{\rm LO}+17}{9C^4\tilde{\mu}^6(y-x)^6}, 
\eea
where this time the correction happens at order ${\cal O}((\tilde{\mu}y)^{-6})$. For the next statistical quantity, the variance of the distribution, a correction at the same order in the expansion looks as follows:
\bea \label{delta2nnnlo}
\sigma_{\rm NNNLO}^{2} = \langle\delta{\cal N}^2(\Gamma)\rangle_{\rm NNNLO}= \frac{2\langle{\cal N}(\Gamma)\rangle_{\rm LO}}{81\;C^4\tilde{\mu}^6(y-x)^6}\bigg(28+143\;C\langle{\cal N}(\Gamma)\rangle_{\rm LO}+384\;C^2\langle{\cal N}(\Gamma)\rangle^{2}_{\rm LO}\bigg).
\eea
Corrections to the third moment away from the mean, at NNNLO, and in the expansion order ${\cal O}((C\tilde{\mu}y)^{-6})$ requires first the following expression for the third moment:
\bea
\langle{\cal N}^3\rangle_{\rm NNNLO} = \frac{\langle{\cal N}(\Gamma)\rangle_{\rm LO}}{81\;C^5\tilde{\mu}^6(y-x)^6}\bigg(76+1056\;C\langle{\cal N}(\Gamma)\rangle_{\rm LO}+3798\;C^2\langle{\cal N}(\Gamma)\rangle^{2}_{\rm LO}\bigg)
\eea
using which we can obtain the expression:
\bea \label{delta3nnnlo}
\langle\delta{\cal N}^3(\Gamma)\rangle_{\rm NNNLO}= \frac{4\langle{\cal N}(\Gamma)\rangle_{\rm LO}}{27\;C^4\tilde{\mu}^4(y-x)^4}\bigg[1+3C\langle{\cal N}(\Gamma)\rangle_{\rm LO} + \frac{19+ 120\;C\langle{\cal N}(\Gamma)\rangle_{\rm LO}+ 132\;C^2\langle{\cal N}(\Gamma)\rangle^{2}_{\rm LO}}{3\;C\tilde{\mu}^2(y-x)^2}\bigg],
\eea
and lastly, corrections to the fourth moment away from the mean, at NNNLO, also requires first to know the fourth moment:
\bea
\langle{\cal N}^4\rangle_{\rm NNNLO} = \frac{4\langle{\cal N}(\Gamma)\rangle_{\rm LO}}{243\;C^6\tilde{\mu}^6(y-x)^6}\bigg(10+ 366\;C\langle{\cal N}(\Gamma)\rangle_{\rm LO}+ 2667\;C^2\langle{\cal N}(\Gamma)\rangle^{2}_{\rm LO}\bigg),
\eea
having which can gives us the following form:
\bea \label{delta4nnnlo}
\langle\delta{\cal N}^4(\Gamma)\rangle_{\rm NNNLO}= \frac{4\langle{\cal N}(\Gamma)\rangle_{\rm LO}}{27\;C^4\tilde{\mu}^4(y-x)^4}\bigg[\langle{\cal N}(\Gamma)\rangle_{\rm LO} + \frac{2(5+ 63\;C\langle{\cal N}(\Gamma)\rangle_{\rm LO}+ 105\;C^2\langle{\cal N}(\Gamma)\rangle^{2}_{\rm LO})}{9\;C^2\tilde{\mu}^2(y-x)^2}\bigg].
\eea
Lastly, using the above results we further mention the
skewness and kurtosis for the distribution at NNNLO. These are as follows:
\bea
\gamma_{\rm NNNLO} &=& \frac{\langle({\cal N}-\langle{\cal N}\rangle)^3\rangle_{\rm NNNLO}}{\sigma_{\rm NNNLO}^3} = 9\sqrt{2}C^3\tilde{\mu}^3(x-y)^3 \frac{19 +120C\langle{\cal N}\rangle_{\rm LO} +132C^2\langle{\cal N}\rangle_{\rm LO}^2}{\sqrt{\langle{\cal N}\rangle_{\rm LO}}(28+ 143C\langle{\cal N}\rangle_{\rm LO} + 384C^2\langle{\cal N}\rangle_{\rm LO}^2)^{3/2}},\\
\kappa_{\rm NNNLO} &=& \frac{\langle({\cal N}-\langle{\cal N}\rangle)^4\rangle_{\rm NNNLO}}{\sigma_{\rm NNNLO}^4} = \frac{54C^2(x-y)^6\tilde{\mu}^6}{\langle{\cal N}\rangle_{\rm LO}}\frac{5+ 63C\langle{\cal N}\rangle_{\rm LO}+ 105C^2\langle{\cal N}\rangle^{2}_{\rm LO}}{28+ 143C\langle{\cal N}\rangle_{\rm LO} + 384C^2\langle{\cal N}\rangle_{\rm LO}^2}.
\eea
The corrected power spectrum after using the addition from eqn.(\ref{delta2nnnlo}) can be evaluated to give:
\bea
\Delta^{2}_{\zeta\zeta, \rm NNNLO} = \frac{d\langle\delta{\cal N}^2\rangle}{d\langle{\cal N}\rangle} &=& \frac{2}{9C^2\tilde{\mu}^2(y-x)^2} + \frac{2(11C-1)\langle{\cal N}(\Gamma)\rangle_{\rm LO}}{9C^2\tilde{\mu}^4(y-x)^4} \nonumber\\
&+& \frac{4}{81C^4\tilde{\mu}^6(y-x)^6}\bigg(14+ 143C\langle{\cal N}(\Gamma)\rangle_{\rm LO}+ 576C^2\langle{\cal N}(\Gamma)\rangle^{2}_{\rm LO}\bigg),
\eea
and the set of non-Gaussianity parameters at the NNNLO are ultimately estimated to give:
\bea
\label{fnlNNNLO}
f_{\rm NL} &=& \frac{5}{36}\frac{1}{[\Delta^{2}_{\zeta\zeta,\rm NNNLO}]^2}\frac{d^2\langle\delta{\cal N}^3\rangle}{d\langle{\cal N}\rangle^2} = \bigg[\frac{5C}{2} + \frac{10(10+33C\langle{\cal N}\rangle_{\rm LO})}{3\tilde{\mu}^{2}(y-x)^{2}}\bigg]\bigg(1+ \frac{g_{1}(C,\langle{\cal N}\rangle_{\rm LO})}{\tilde{\mu}^{2}(y-x)^{2}} + \frac{2g_{2}(C,\langle{\cal N}\rangle_{\rm LO})}{9C^{2}\tilde{\mu}^{4}(y-x)^{4}}\bigg)^{-2},\\
\label{tauNNNLO}
\tau_{\rm NL}&=& \frac{1}{36}\frac{1}{[\Delta^{2}_{\zeta\zeta,\rm NNNLO}]^4}\bigg(\frac{d^2\langle\delta{\cal N}^3\rangle}{d\langle{\cal N}\rangle^2}\bigg)^{2} = \bigg[\frac{C}{2}+\frac{2(10+33C\langle{\cal N}\rangle_{\rm LO})}{3\tilde{\mu}^{2}(y-x)^{2}}\bigg]^{2}\bigg(1+ \frac{g_{1}(C,\langle{\cal N}\rangle_{\rm LO})}{\tilde{\mu}^{2}(y-x)^{2}} + \frac{2g_{2}(C,\langle{\cal N}\rangle_{\rm LO})}{9C^{2}\tilde{\mu}^{4}(y-x)^{4}}\bigg)^{-4},\quad\quad\\
g_{\rm NL}&=& \frac{1}{[\Delta^{2}_{\zeta\zeta,\rm NNNLO}]^3}\frac{d^3\langle\delta{\cal N}^4\rangle}{d\langle{\cal N}\rangle^3} = 1890C^2\bigg(1+ \frac{g_{1}(C,\langle{\cal N}\rangle_{\rm LO})}{\tilde{\mu}^{2}(y-x)^{2}} + \frac{2g_{2}(C,\langle{\cal N}\rangle_{\rm LO})}{9C^{2}\tilde{\mu}^{4}(y-x)^{4}}\bigg)^{-3}.
\eea
where the coefficients, $g_{1}(C,\langle{\cal N}\rangle_{\rm LO})$, $g_{2}(C,\langle{\cal N}\rangle_{\rm LO})$ are defined by the following expressions:
\bea && g_{1}(C,\langle{\cal N}\rangle_{\rm LO})=(11C-1)\langle{\cal N}\rangle_{\rm LO},\\
&& g_{2}(C,\langle{\cal N}\rangle_{\rm LO})= 14+143C\langle{\cal N}\rangle_{\rm LO}+576C^{2}\langle{\cal N}\rangle_{\rm LO}^{2}.\eea
From the above expressions for the various non-Gaussianity indicators, we notice the effect of corrections to the leading order result. For $f_{\rm NL}$, at $C=1$ one notices a slightly increased magnitude, coming from the ${\cal O}(\tilde{\mu}^{-2}y^{-2})$ suppressed term with positive signature, than the leading order value of $f_{\rm NL}=5/2$. Here there does not seem to be a case for negative non-Gaussianity to appear. The above eqns. (\ref{fnlNNNLO},\ref{tauNNNLO}) are also able to maintain the Suyama-Yamaguchi relation, $\tau_{\rm NL}=(6f_{\rm NL}/5)^{2}$, at this order of analysis. The other two parameters $\tau_{\rm NL}$ and $g_{\rm NL}$ take on large positive magnitudes with $g_{\rm NL}\gg \tau_{\rm NL}$ at their leading order.

\section{Spectral distortion and its implications}
\label{s11}

In this section, we discuss a different but essential physical effect present as inhomogeneities in the cosmic microwave background distributions known as spectral distortions. In the context of PBH, these effects are related to the collapsing phase of their formation mechanism and can become visible as significant deviations in the CMB spectrum depending on the mass of the formed PBH. These distortions are classified in the literature based on the redshift intervals and the variety of mechanisms taking place in the respective history, with their ability to induce temperature changes in the standard black-body distribution of the CMB photons. Detection of these distortion effects can appear to give well-needed signals from a pre-recombination era. See refs.\cite{zeldovich1969interaction,Sunyaev:1970eu,Illarionov:1975ei,Hu:1992dc,Chluba:2012we,Khatri:2012tw,Chluba:2003exb,Stebbins:2007ve,Chluba:2011hw,Pitrou:2014ota,Hooper:2023nnl,Deng:2021edw} for detailed explanations on these effects and their connection with PBH formation.

There are mainly two types of important spectral distortions in study, the $\mu$ and $y$ type distortions. The history during formation of the CMB spectrum consists of an epoch known as the $\mu$-era when redshitfs are in between, $2\times 10^{5}\leq z\leq 2\times 10^{6}$, responsible for the $\mu$-type distortions. Beyond $z \geq 2\times 10^{6}$ lies the high-temperature, thermalization era where processes like the bremsstarhlung, double Compton scattering, and Compton scattering are able to maintain a condition of thermal equilibrium implying highly suppressed distortion effects. During the $\mu$-era, deviations from a black-body spectrum start to appear and the Compton scattering processes still function but instead create a Bose-Einstein spectrum characterized by a non-zero chemical potential $\mu\ne 0$, hence the name of the era. If we go to redshifts below, $z\leq 2\times 10^{5}$, even the Compton scattering effects start to become inefficient in maintaining a Bose-Einstein distribution. The result is a loss of equilibrium and the temperature differences become more sensitive in the late-time scales, near the CMB era. These are known as the $y$-type distortions.   

In terms of the wavenumber, these can be evaluated from the scalar power spectrum of the curvature perturbations as follows \cite{Hooper:2023nnl}:
\bea \label{mudistortion}
\mu &\simeq& 2.2\int_{k_{i}}^{\infty}\frac{dk}{k}\Delta^{2}_{\zeta\zeta}(k)\bigg\{\exp{\bigg[-\frac{k}{5400\;{\rm Mpc^{-1}}}\bigg]}-\exp{\bigg[-\bigg(\frac{k}{31.6\;{\rm Mpc^{-1}}}\bigg)^{2}\bigg]}\bigg\},\\
\label{ydistortion}
y &\simeq& 0.4\int_{k_{i}}^{\infty}\frac{dk}{k}\Delta^{2}_{\zeta\zeta}(k)\exp{\bigg[-\bigg(\frac{k}{31.6\;{\rm Mpc^{-1}}}\bigg)^{2}\bigg]},
\eea
where $k_{i}=1\;{\rm Mpc^{-1}}$ is used. The definition of the power spectrum used above, $\Delta^{2}_{\zeta\zeta}(k)$, is the total power spectrum derived for the three phases in eqns. (\ref{pspecsr1dS}, \ref{pspecusrdS}, \ref{pspecsr2dS}). The two distortion estimates from the mentioned total power spectrum gives us values, $\mu\sim 2.35\times 10^{-8}$ and $y\sim 2.78\times 10^{-9}$, when the wavenumber for PBH formation satisfies $k_{\rm PBH}\sim {\cal O}(10^{7}{\rm Mpc^{-1}})$. The PBH \textcolor{black}{mass is} related to the transition wavenumber as follows:
\bea \label{pbhmass}
\frac{M_{\rm PBH}}{M_{\odot}} = 1.13 \times 10^{15} \times \bigg(\frac{\gamma}{0.2}\bigg)\bigg(\frac{g_{*}}{106.75}\bigg)^{-1/6}\bigg(\frac{k_{*}}{k_{\rm PBH}}\bigg)^{2},
\eea
where $k_{*}\sim 0.02{\rm Mpc^{-1}}$. In order to investigate the allowed PBH mass range, we implement the above formula along with the definitions in eqn. (\ref{mudistortion},\ref{ydistortion}) to obtain $\mu$ and $y$ as function of the PBH mass. We later confront the results with the existing observational constraints on their values in a dedicated numerical outcome section. A useful quantity to remember here is the amplitude of the total scalar power spectrum which reaches sufficient values needed for PBH production, that is $A\sim {\cal O}(10^{-2})$. The expression for the amplitude using the power spectrum expression mentioned before can be extracted as follows:
\bea
\Delta^{2}_{\zeta\zeta}(k) &=& \big[\Delta^{2}_{\zeta\zeta,{\bf dS}}(k<k_{s})\big]_{\bf SRI} + \Theta(k-k_{s})\big[\Delta^{2}_{\zeta\zeta,{\bf dS}}(k_{s}\leq k < k_{e})\big]_{\bf USR} + \Theta(k-k_{e})\big[\Delta^{2}_{\zeta\zeta,{\bf dS}}(k_{e}\leq k< k_{\rm end})\big]_{\bf SRII},\nonumber\\
&=& A(k_s)\bigg[1+\bigg(\frac{k}{k_{s}}\bigg)^{2}\bigg](1+\sigma^{2})\times \bigg\{\bigg(\frac{k_{s}}{k_{e}}\bigg)^{6}\nonumber\\
&& \quad\quad+ \bigg(\Theta(k-k_{s})\Big|\alpha_{2,{\bf dS}} e^{i\sigma}-\frac{(1+i \sigma)}{(1-i \sigma)}\beta_{2,{\bf dS}} e^{-i\sigma}\Big|^2 + \Theta(k-k_{e})\Big|\alpha_{3,{\bf dS}} e^{i\sigma}-\frac{(1+i \sigma)}{(1-i \sigma)}\beta_{3,{\bf dS}} e^{-i\sigma}\Big|^2\bigg)\bigg\},
\eea
and the amplitude is written as:
\bea
A(k_{s}) = \bigg(\frac{H^2}{8\pi^2\epsilon  c_s M_p^2}\bigg)_{*}\bigg(\frac{k_{e}}{k_{s}}\bigg)^{6},
\eea
where the $*$ notation implies evaluation at the pivot scale, $k_{*}\simeq 0.02{\rm Mpc^{-1}}$ and $k_{s}$ stands for the wavenumber associated with the sharp transition into the USR leading to PBH formation, as used before in eqn. (\ref{pbhmass}). The wavenumber dependency in the total power spectrum above comes from the Bogoliubov coefficients for the USR and SRII, and the $k-$dependent factor outside the power spectrum is present for all scales, before we consider a specific regime.

\section{PBH Formation from EFT of Stochastic Single Field Inflation using stochastic-$\delta N$ formalism}
\label{s12}

This section focuses on the primordial black hole formation analysis within stochastic inflation. We first discuss the general mechanism for PBH formation in the early Universe. In the context of stochastic inflation, after we describe the formation then, we move towards the study of PBH mass fraction and how the analysis of the probability distribution done in the previous sections proves helpful in obtaining numerical values for the current density fraction of PBH and its ratio concerning the present-day dark matter density in the form of PBH abundance.  

\subsection{Formation Mechanism}

We examine the PBH formation mechanism when stochastic effects are important and tend to facilitate the increase of curvature perturbations to sufficiently large values. Mostly this happens with the perturbations at the small scales during inflation when they re-enter the Hubble radius, and if they are large enough, can force the Hubble patches to undergo gravitational collapse and generate PBHs. The standard approach to calculate the resulting PBH mass fraction goes as follows. We start with an initial Gaussian distribution $P(\zeta)$ of the curvature perturbation $\zeta$. From such a distribution, its value after integrating above some threshold magnitude for the perturbations, say $\zeta_{\rm th}$, inside a Hubble patch, provides the probability for PBH formation of mass $M_{\rm PBH}$ or its PBH mass fraction,
\bea \label{betazeta}
\beta(M_{\rm PBH}) \sim \int_{\zeta_{\rm th}}P(\zeta)d\zeta,
\eea
Now, in the case of dominant quantum diffusion processes, the curvature perturbations suffer huge enhancement until the condition on their threshold $\zeta_{\rm th}\sim {\cal O}(1)$ is met. Such large fluctuations lie at the far tail region of the probability distribution, thus making PBH formation a rare event. This feature answers why the non-Gaussian nature of the distribution affects this process significantly. During inflation, the existence of an ultra-slow roll phase for the small scales provides for such enhancements and invites the quantum diffusion effects to take over as the driving phenomenon affecting PBH formation. Since the expected nature of the distribution function for the curvature perturbations is now not Gaussian due to prominent non-Gaussian tail features, we will be using the results of the previous sections in the diffusion-dominated regime to determine how the obtained PDFs affects PBH mass fraction calculations. 

Another important element in our analysis comes from using of the stochastic variable ${\cal N}$. The various distributions as mentioned before are calculated to realise the stochastic period ${\cal N}$ given some initial values of the phase space variables $\Gamma$. The idea is to work with the stochastic-$\delta N$ formalism where one is able to directly map the statistics of ${\cal N}$ to that of the curvature perturbation $\zeta$ on the super-Hubble scales. The identification is as follows:
\bea \label{deltaN}
\zeta_{\rm cg}({\bf x}) = \delta {\cal N} =  {\cal N}({\bf x}) -\langle{\cal N}\rangle
\eea
where the subscript `cg' indicates coarse-grained value of the curvature perturbations in terms of the modes which enter and exit the Hubble radius for a given set of initial and final condition on phase space variables. The PBH mass formed is related to the scale of the mode involved in the coarse-graining and we will be using the eqns. (\ref{betazeta}, \ref{deltaN}) to make an estimate on the resulting PBH mass fraction.

\subsection{Mass Fraction of PBH $(\beta)$ }

To calculate the PBH mass fraction we make use of the results for the PDF as mentioned in eqns.(\ref{pdfLO},\ref{PDFN1},\ref{PDFN2},\ref{PDFN3}) and apply the stochastic-$\delta N$ formalism. First we notice that the above PDFs are described as functions of the stochastic e-folds ${\cal N}$. Since the mass fraction is defined from the curvature perturbation distribution as in eqn.(\ref{betazeta}), we write the same definition following eqn.(\ref{deltaN}) as:
\bea
\beta \sim \int_{\zeta_{\rm th}+\langle{\cal N}\rangle}^{\infty}P_{\Gamma}({\cal N})d{\cal N},
\eea
which requires mean value of e-folds from the various distribution functions calculated before in eqns.(\ref{meanLO},\ref{meanNLO},\ref{meanNNLO},\ref{meanNNNLO}). The value for the threshold maintains $\zeta_{\rm th}\sim {\cal O}(1)$. The above integral is better approximated with a summation version of the same due to the fact that the contributions of the set of higher-order poles in the PDF quickly decay to being negligible and thus when we use the series formula for the PDF, we find for the mass fraction the following relation:
\bea
\beta_{m} \sim \sum_{i=0}^{m}\sum_{n=0}^{\infty}\frac{r_{n,m}^{(i)}}{\Lambda_{n}^{(i)}}\exp{(-\Lambda_{n}^{(i)}[\langle{\cal N}\rangle_{m} + \zeta_{\rm th}])},
\eea
where the subscript $(m)$ signifies the order of perturbative expansion being considered and the set of poles at $i$-th order, $\Lambda_{n}^{(i)}$, is defined in eqn.(\ref{poles}). In the above, we use the analytical expressions for the residues $r_{n,m}^{(i)}$ evaluated before at each order in eqns. (\ref{resLO}, \ref{resNLO0}, \ref{resNLO1}), except for the NNLO and NNNLO residues which have not been mentioned before due to they being lengthy and not illuminating on their own. We will be using the above formula in the small $y$ limit to determine the PBH fraction and comment on its utility from the results obtained at each order in the $y$-expansion. 

The analytic result of the mass fraction $\beta$ at the leading order can still be mentioned here using the residue expression, mentioned before in eqns. (\ref{resLO}), as follows:
\bea
\beta_{\rm LO} &=& \sum_{n=0}^{\infty}\frac{r_{n,0}^{(0)}}{\Lambda_{n}^{(0)}}\exp{(-\Lambda_{n}^{(0)}[\langle{\cal N}\rangle_{\rm LO} + \zeta_{\rm th}])},\nonumber\\
&=& \frac{4}{\pi}\sum_{n=0}^{\infty}\frac{1}{2n+1}\sin{\left(\frac{(2n+1)\pi x}{2}\right)}\exp{\bigg\{-\pi^{2}\left(n+\frac{1}{2}\right)^{2}\left[x\left(1-\frac{x}{2}\right)+\frac{\zeta_{\rm th}}{\tilde{\mu}^{2}}\right]\bigg\}},
\eea
where we implement the expression for the mean e-folds from eqn. (\ref{meanLO}). Similarly, for the NLO one can write:
\bea
\beta_{\rm NLO} &=& \sum_{i=0}^{m=1}\sum_{n=0}^{\infty}\frac{r_{n,1}^{(i)}}{\Lambda_{n}^{(i)}}\exp{(-\Lambda_{n}^{(i)}[\langle{\cal N}\rangle_{\rm NLO} + \zeta_{\rm th}])},\nonumber\\
&=& \sum_{n=0}^{\infty}\bigg(\frac{r_{n,1}^{(0)}}{\Lambda_{n}^{(0)}} + \frac{r_{n,1}^{(1)}\exp{(-3C[\langle{\cal N}\rangle_{\rm NLO} + \zeta_{\rm th}])}}{\Lambda_{n}^{(0)}+3C} \bigg)\exp{(-\Lambda_{n}^{(0)}[\langle{\cal N}\rangle_{\rm NLO} +\zeta_{\rm th}])},
\eea
which includes the mean at NLO as derived in eqn. (\ref{meanNLO}). Likewise, we can go on to evaluate the mass fraction using the subsequent higher-order contributions in the perturbative expansion. For NNLO, the result is then written as:
\bea
\beta_{\rm NNLO} &=& \sum_{n=0}^{\infty}\bigg(\frac{r_{n,2}^{(0)}}{\Lambda_{n}^{(0)}} + \frac{r_{n,2}^{(1)}\exp{(-3C[\langle{\cal N}\rangle_{\rm NNLO} + \zeta_{\rm th}])}}{\Lambda_{n}^{(0)}+3C} \nonumber\\
&&\quad\quad\quad\quad\quad\quad\quad\quad\quad\quad\quad\quad + \frac{r_{n,2}^{(2)}\exp{(-6C[\langle{\cal N}\rangle_{\rm NNLO} + \zeta_{\rm th}])}}{\Lambda_{n}^{(0)}+6C} \bigg)\exp{(-\Lambda_{n}^{(0)}[\langle{\cal N}\rangle_{\rm NNLO} +\zeta_{\rm th}])},
\eea
and lastly for the NNNLO, we can write the PBH mass fraction as:
\bea
\beta_{\rm NNNLO} &=& \sum_{n=0}^{\infty}\bigg(\frac{r_{n,3}^{(0)}}{\Lambda_{n}^{(0)}} + \frac{r_{n,3}^{(1)}\exp{(-3C[\langle{\cal N}\rangle_{\rm NNNLO} + \zeta_{\rm th}])}}{\Lambda_{n}^{(0)}+3C} + \frac{r_{n,3}^{(2)}\exp{(-6C[\langle{\cal N}\rangle_{\rm NNNLO} + \zeta_{\rm th}])}}{\Lambda_{n}^{(0)}+6C}\nonumber\\
&&\quad\quad\quad\quad\quad\quad\quad\quad\quad\quad\quad + \frac{r_{n,3}^{(3)}\exp{(-9C[\langle{\cal N}\rangle_{\rm NNNLO} + \zeta_{\rm th}])}}{\Lambda_{n}^{(0)}+9C} \bigg)\exp{(-\Lambda_{n}^{(0)}[\langle{\cal N}\rangle_{\rm NNNLO} +\zeta_{\rm th}])},
\eea

\subsection{Abundance of PBH $(f_{PBH})$}

The current day fraction of PBH density relative to the current day dark matter density is expressed using the fraction $f_{\rm PBH}$. During inflation, the presence of diffusion effects dominating the short scale regime provides the right situation for the production of PBH when these enhanced fluctuations re-enter the Horizon. The abundance of these PBHs is an important quantity to study which can tell us the percent of current total dark matter present in this form. From the previous section, the concept of the PBH mass fraction was detailed in the context of stochastic inflation and the subtleties involved regarding the distribution function of curvature perturbations. The same mass fraction $\beta(M_{\rm PBH})$ can be further used to evaluate the abundance $f_{\rm PBH}$ using the relation:
\bea
\label{fPbh}
f_{\rm PBH}&=&1.68\times 10^8 \bigg(\frac{\gamma}{0.2}\bigg)^{\frac{1}{2}}\bigg(\frac{g_{*}}{106.75}\bigg)^{-\frac{1}{4}}\bigg(\frac{M_{\rm PBH}}{M_{\odot}}\bigg)^{-\frac{1}{2}}\times \beta(M_{\rm PBH}).
\eea
where the mass of PBH in solar mass units, $M_{\odot}$, is expressed in terms of the  wavenumber of PBH formation $k_{\rm PBH}$ through the relation:
\bea
\frac{M_{\rm PBH}}{M_{\odot}} = 1.13 \times 10^{15} \times \bigg(\frac{\gamma}{0.2}\bigg)\bigg(\frac{g_{*}}{106.75}\bigg)^{-1/6}\bigg(\frac{k_{*}}{k_{\rm PBH}}\bigg)^{2},
\eea
and $k_{*}\simeq 0.02\;{\rm Mpc^{-1}}$ is the pivot scale.
The threshold $\zeta_{\rm th}\sim {\cal O}(1)$ plays a crucial role in determining the outcome of PBH abundance in a sizeable window $f_{\rm PBH}\leq 1$. In the numerical outcome section we will show the behaviour of $f_{\rm PBH}$ as function of the PBH masses calculated from the mass fraction at each order in the perturbative expansion for diffusion. 

\section{Numerical Outcomes}
\label{s13}

This section focuses on various results from the numerical analysis of important observational features related to the production of PBH and the scalar power spectrum studied before for each phase in our set up. We start by discussing the tree-level power spectrum across the three phases of interest and point out their connection with the noise matrix element. Utilising this power spectrum we evaluate the spectral distortion effects for a wide range of PBH masses and study the implications of observational constraints on the amplitude of the power spectrum.  We follow this with studying the behaviour of the probability distribution function with the number of e-folds, ${\cal N}$, at each order in the perturbative expansion. Later, we study results for the PBH mass fraction and present-day PBH abundance in the similar perturbative study carried out for the diffusion-dominated regime as before.

\subsection{Outcomes of tree-level power spectrum and noise matrix elements}
\begin{figure*}[ht!]
    	\centering
    \subfigure[]{
      	\includegraphics[width=8.5cm,height=7.5cm]{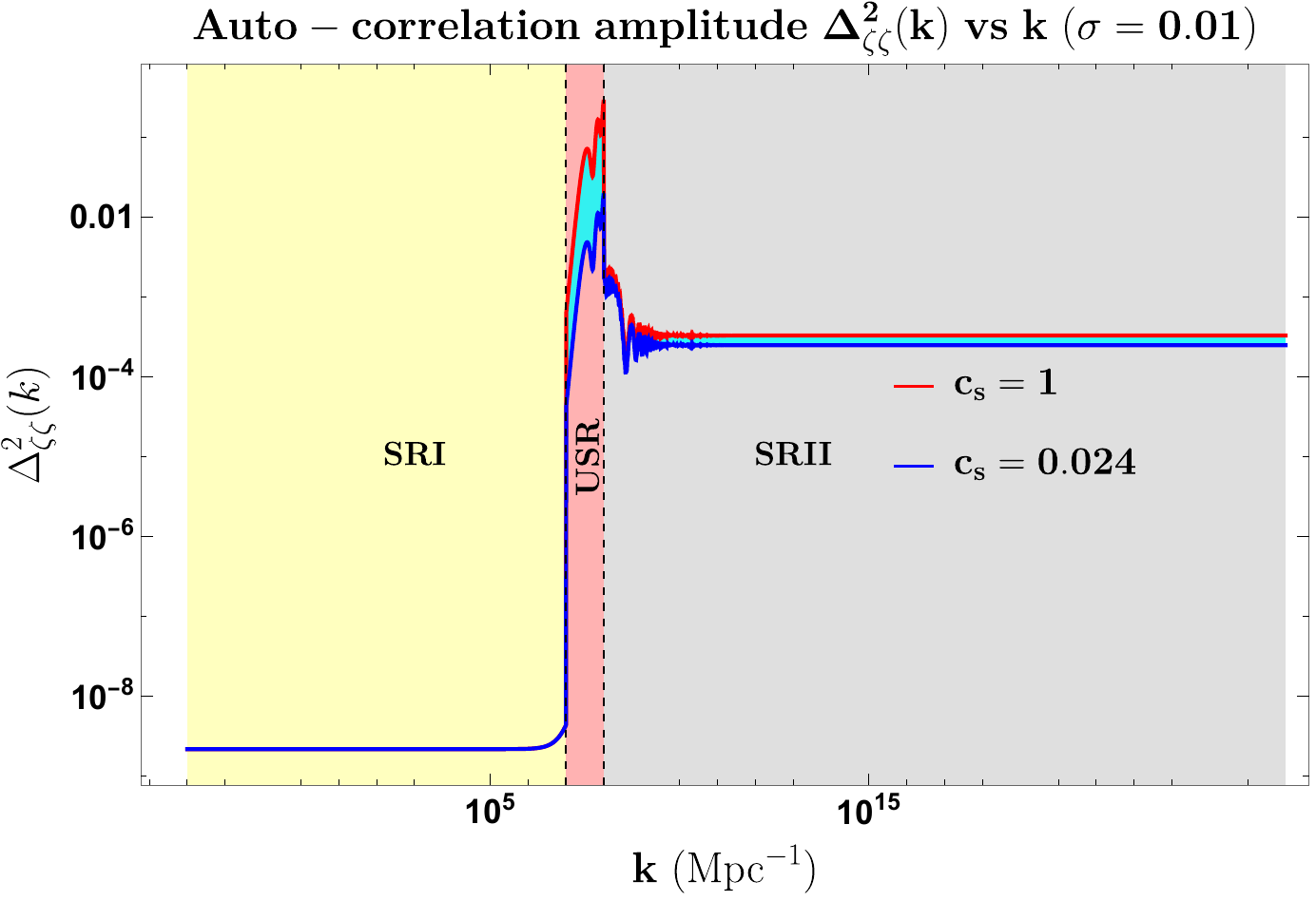}
        \label{pspeczz1}
    }
    \subfigure[]{
        \includegraphics[width=8.5cm,height=7.5cm]{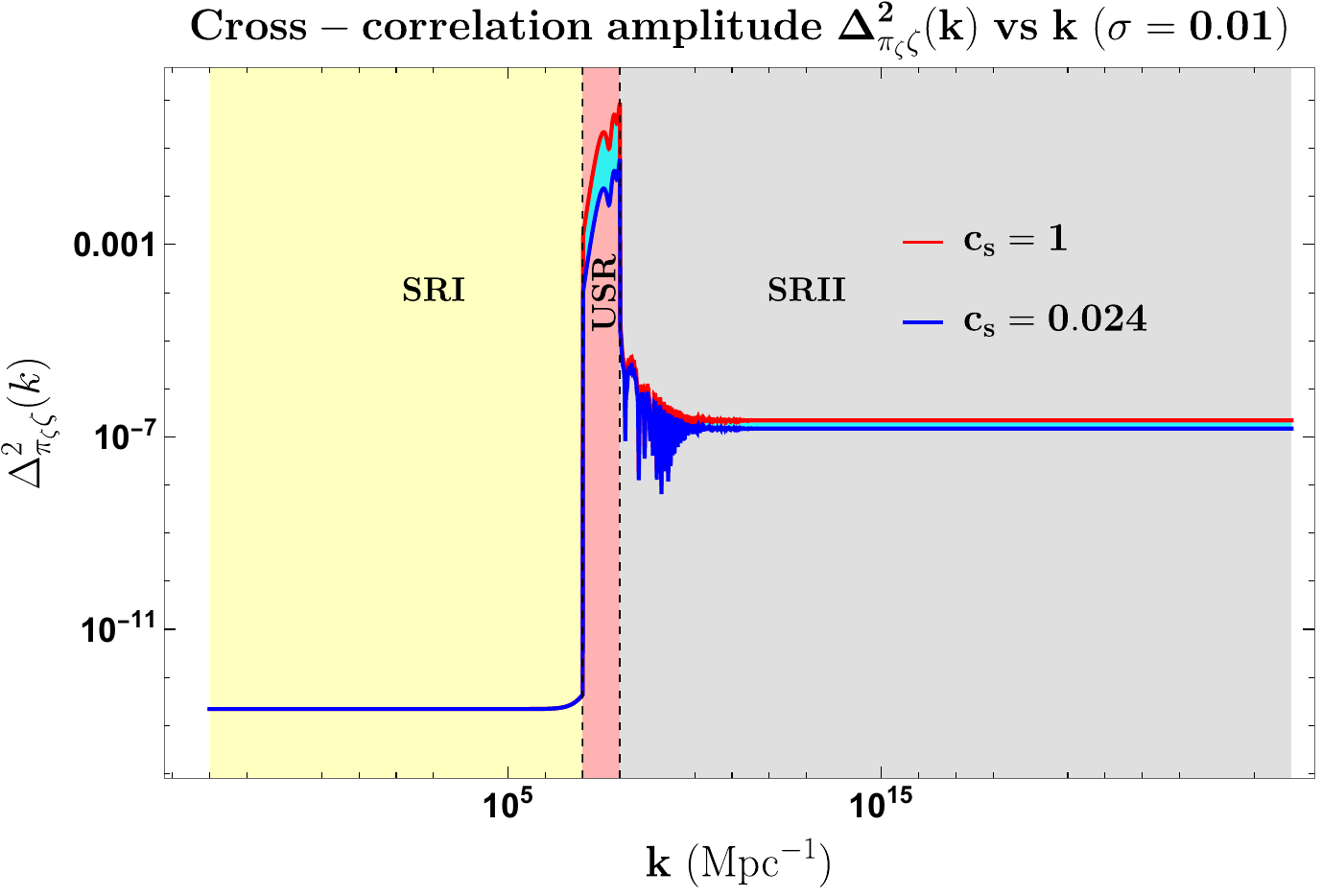}
        \label{pspecpz1}
    }
       \subfigure[]{
        \includegraphics[width=8.5cm,height=7.5cm]{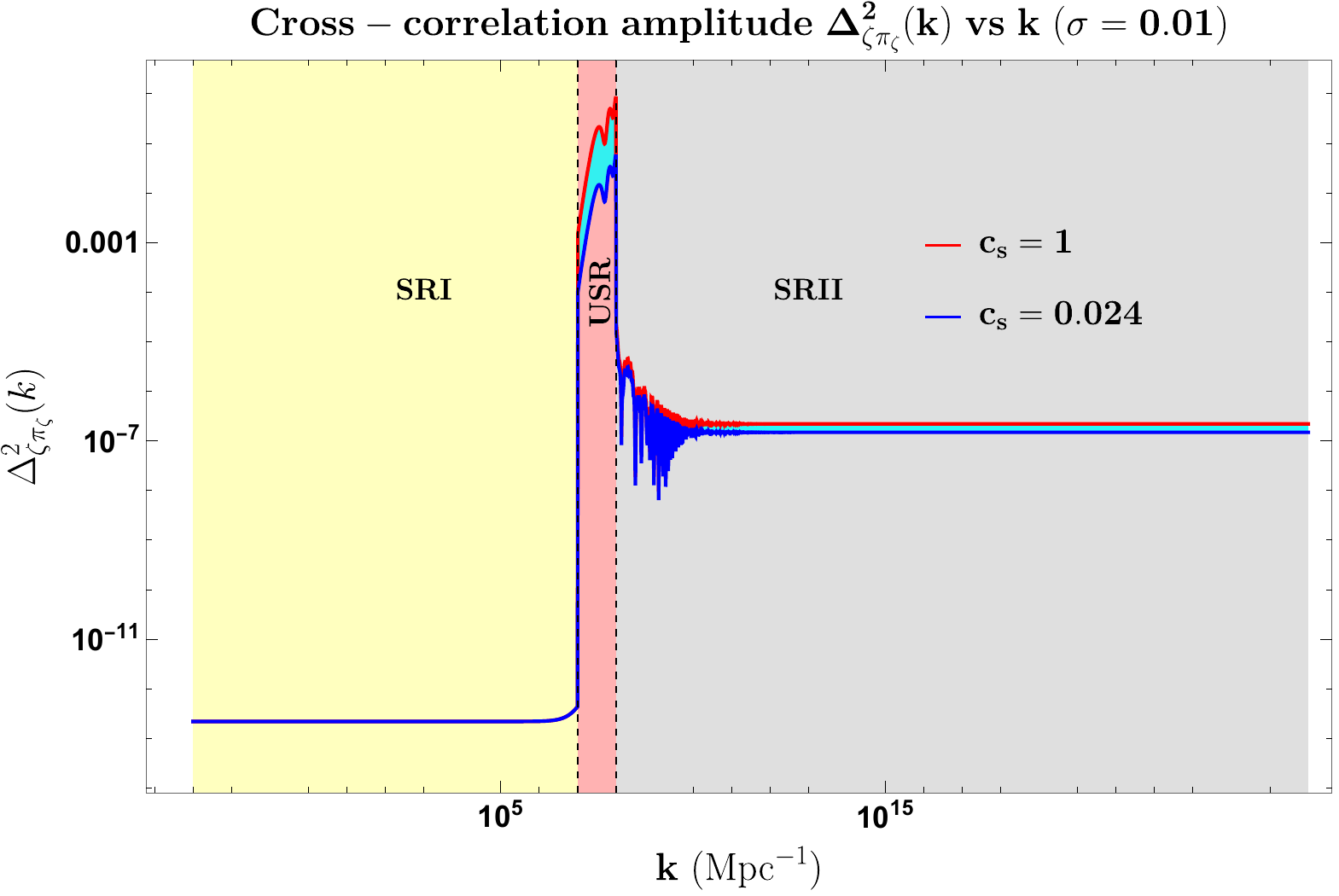}
        \label{pspeczp1}
    }
    \subfigure[]{
        \includegraphics[width=8.5cm,height=7.5cm]{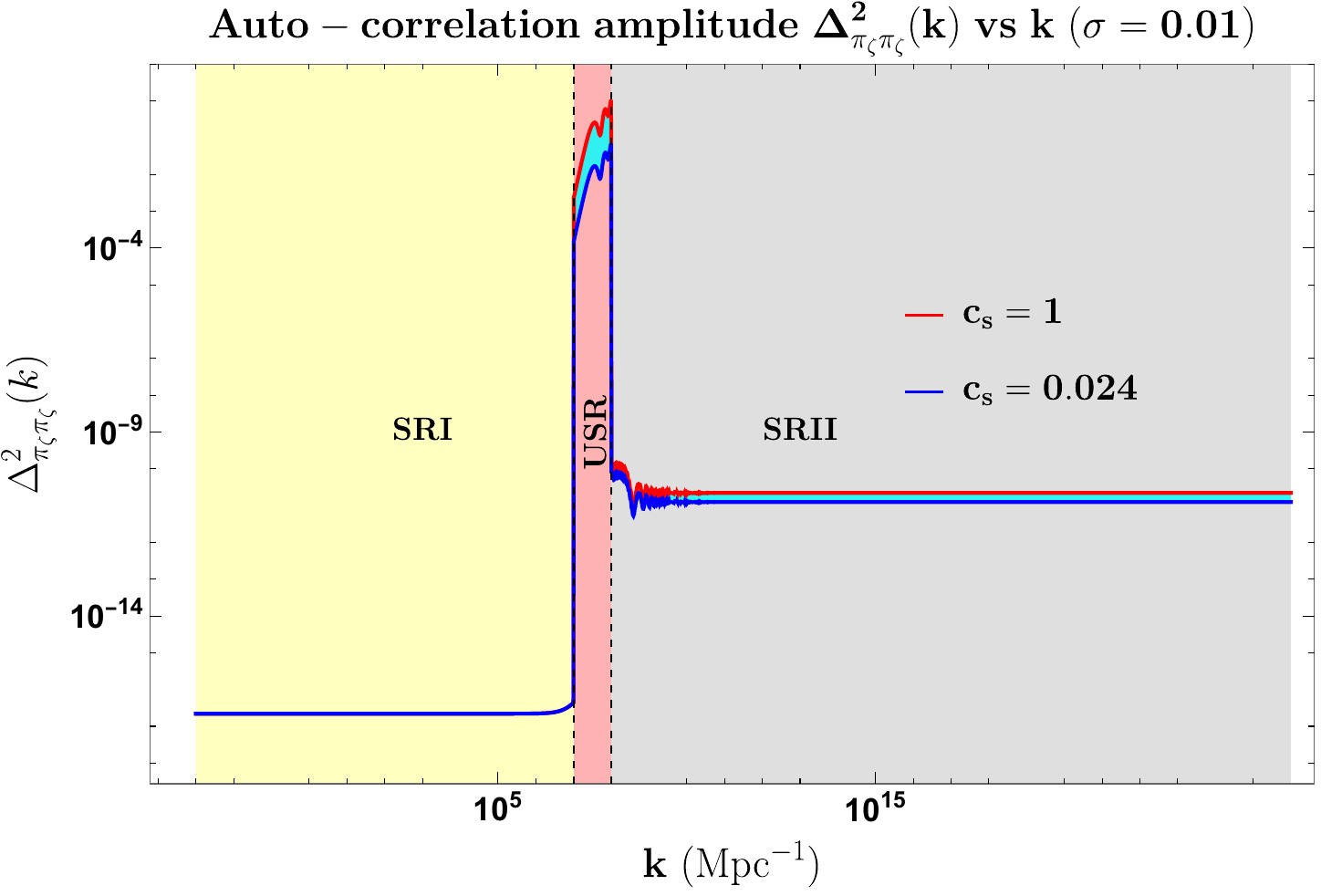}
        \label{pspecpp1}
       }
    	\caption[Optional caption for list of figures]{Plot of various power spectrum elements $\Delta^{2}_{fg}$ as function of the wavenumber $k$, where $f,g\;\in\{\zeta,\Pi_{\zeta}\}$. The \textit{top-left} panel shows the element $\Delta^{2}_{\zeta\zeta}$, the \textit{top-right} and \textit{bottom-left} panels show the elements $\Delta^{2}_{\zeta\Pi_{\zeta}}$,$\Delta^{2}_{\Pi_{\zeta}\zeta}$, and the \textit{bottom-right} panel shows the element $\Delta^{2}_{\Pi_{\zeta}\Pi_{\zeta}}$. The stochastic parameter is fixed to $\sigma=0.01$. The effective sound speed parameter is, $c_{s}=c_{s,*}=\{0.024,1\}$, which is its value fixed at the pivot scale and we plot each correlation for these two values in blue ($c_{s}=0.024$) and red ($c_{s}=1$). All the elements have their behaviour in the three phases, SRI, USR, and SRII, indicated by distinct shading. } 
    	\label{spectrumplots}
    \end{figure*}
In this section, we present with behaviour of the various power spectrum elements as function of the wavenumber across the three phases of interest in our set up, namely SRI, USR, and SRII.

From the fig. (\ref{spectrumplots}), we find that the auto-correlation $\Delta^{2}_{\Pi_{\zeta}\Pi_{\zeta}}$, see \ref{pspecpp1}, is the most suppressed in magnitude until slow-roll conditions remain satisfied. This happens during the SRI phase as well as in the SRII phase. In a similar sense, the auto-correlation $\Delta^{2}_{\zeta\zeta}$, see \ref{pspeczz1}, is the most dominant in both the SRI and SRII phases with an amplitude of ${\cal O}(10^{-9})$ in the SRI and ${\cal O}(10^{-3})$ in the SRII phases. The behaviour of all the auto and cross-correlations in the USR duration is more important to note here. All the correlation amplitudes in the USR experience large enhancements to approach magnitudes of order ${\cal O}(10^{-1}-1)$ with the correlation amplitude $\Delta^{2}_{\zeta\zeta}$ being relatively smaller than the other two elements. This fact of the USR have important consequences if we wish to examine the effects from such large enhancements getting carried into the quantum correction calculations.   
Soon after a sharp exit from the USR occurs at $k=k_{e}$, the amplitude features rapid oscillations till they die out to give a constant magnitude, and these oscillations remain larger in the cross-correlations, see \ref{pspecpz1} and \ref{pspeczp1} for $\Delta^{2}_{\zeta\Pi_{\zeta}}$ and $\Delta^{2}_{\Pi_{\zeta}\zeta}$, than in the auto-correlations, $\Delta^{2}_{\zeta\zeta}$ and $\Delta^{2}_{\Pi_{\zeta}\Pi_{\zeta}}$. We put emphasis on the fact that the oscillatory nature of the power spectrum in the USR and after a while into the SRII is a result of the new set of Bogoliubov coefficients, $(\alpha_{2},\beta_{2})$ and $(\alpha_{3},\beta_{3})$ in eqn. (\ref{alpha2qds},\ref{beta2qds},\ref{alpha3qds},\ref{beta3qds}), which contains complex phase factors after solving for the boundary conditions at each sharp transition $k=k_{s}$ and $k=k_{e}$. 

\begin{figure*}[ht!]
    	\centering
    \subfigure[]{
      	\includegraphics[width=8.5cm,height=7.5cm]{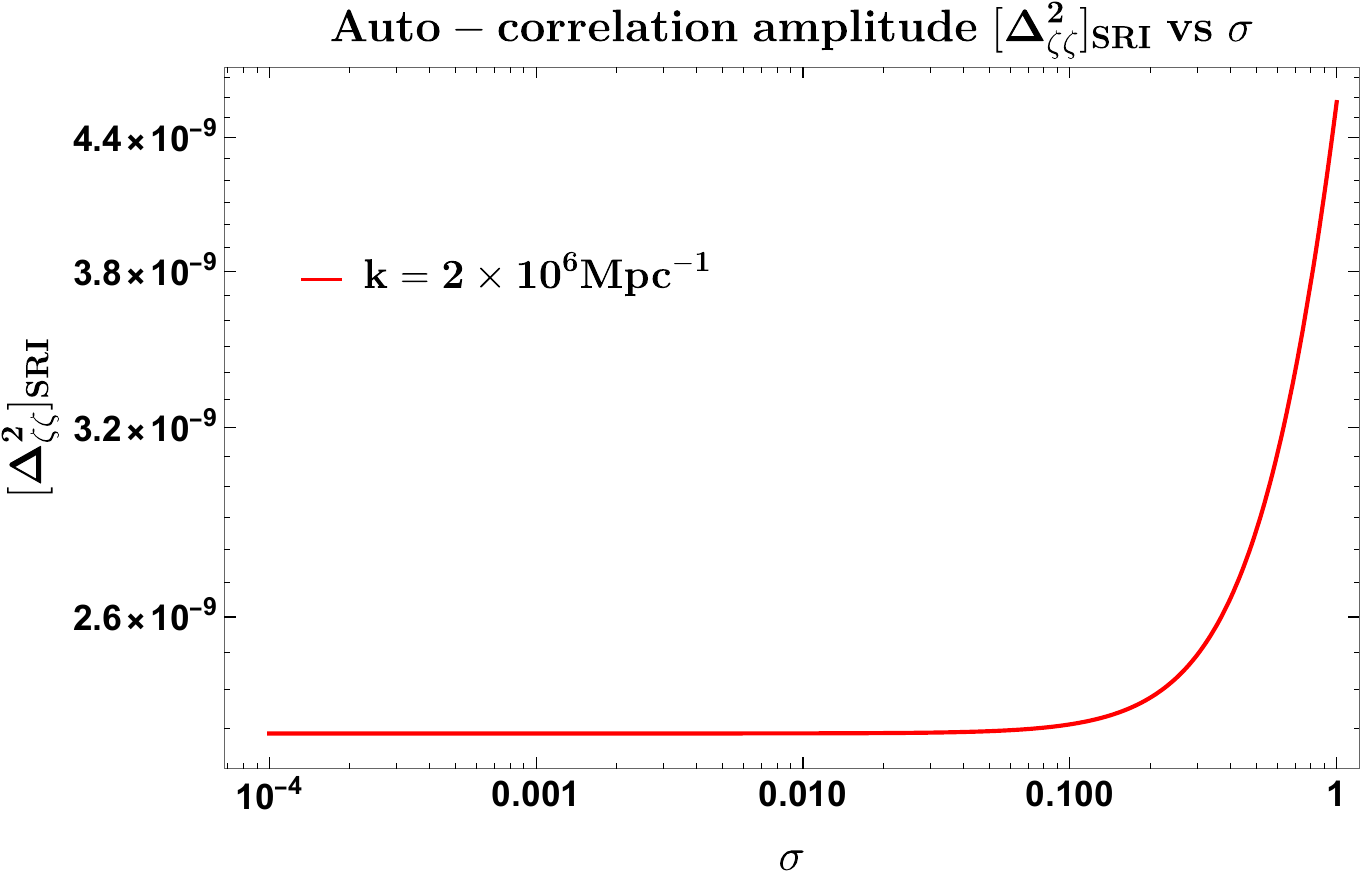}
        \label{sr1pspecsigma}
    }
    \subfigure[]{
        \includegraphics[width=8.5cm,height=7.5cm]{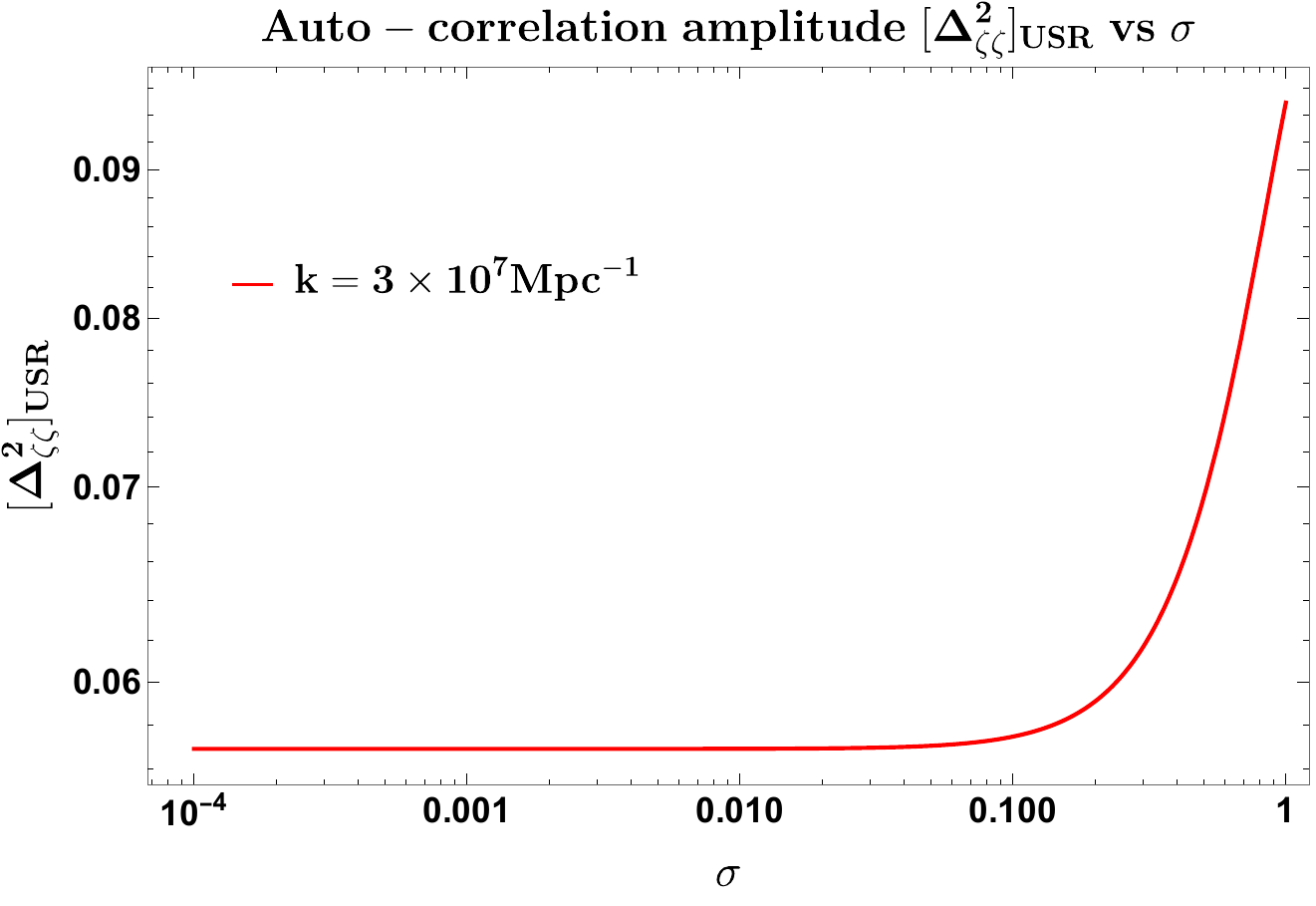}
        \label{usrpspecsigma}
    }
       \subfigure[]{
        \includegraphics[width=8.5cm,height=7.5cm]{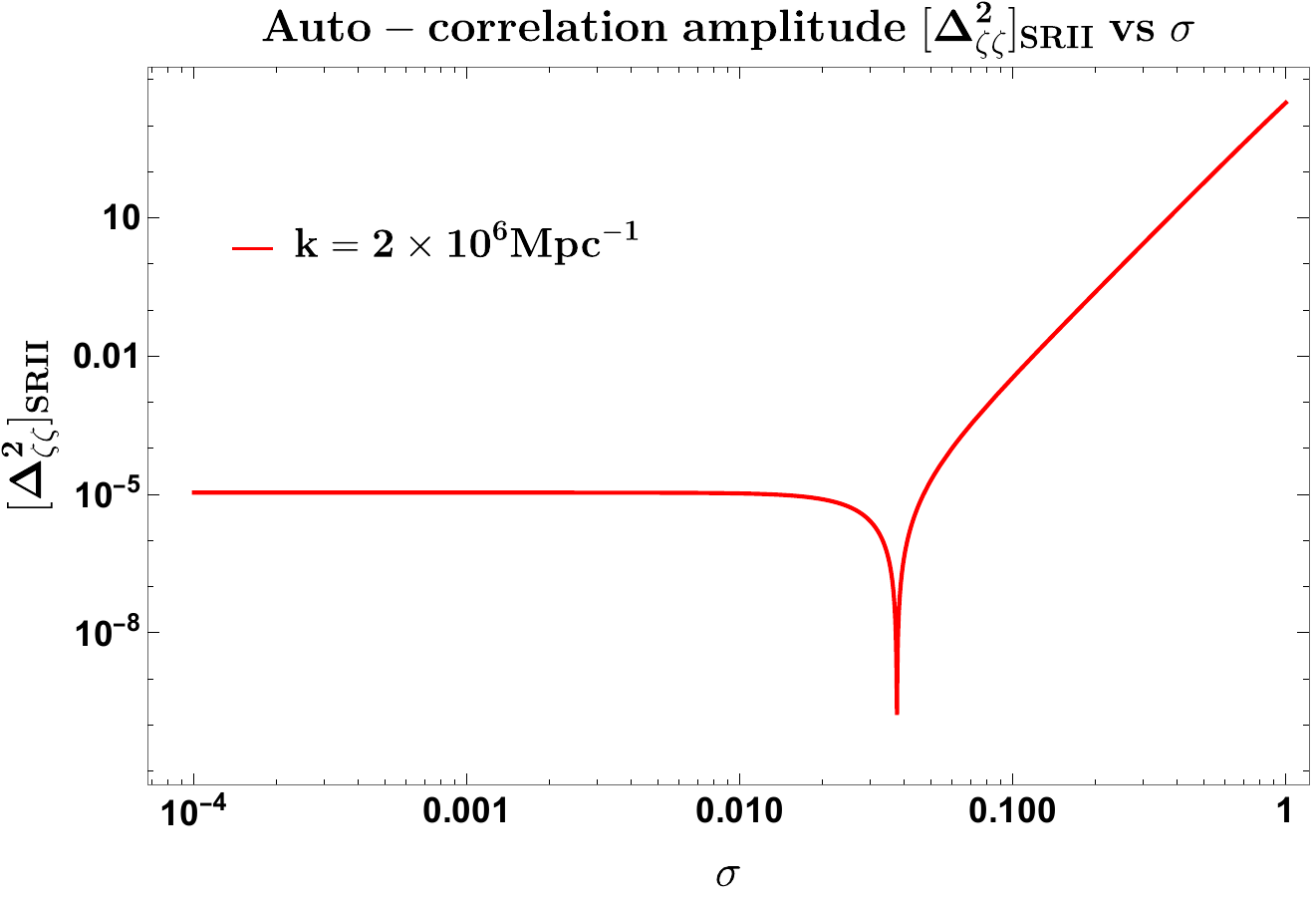}
        \label{sr2pspecsigma}
    }
    	\caption[Optional caption for list of figures]{Behaviour of the auto-correlation amplitude $\Delta^{2}_{\zeta\zeta}(k)$ for each phase with respect to change in stochastic parameter $\sigma$. The \textit{top row} shows for the SRI and USR phase and the \textit{bottom} for the SRII phase. For $\sigma\in (10^{-4},0.1)$, amplitudes of the correlation in their respective phases remains satisfied and quickly change once $\sigma> 0.1$ is considered.  } 
    	\label{spectrumplotsigma}
    \end{figure*}

The role played by the effective sound speed, $c_{s}$, is also on display in fig. (\ref{spectrumplots}). We show each correlation for two values of $c_{s}=0.024,1$, specifying the lower and upper bounds from observational constraints for causality and unitarity \cite{Planck:2018jri}. The red curve corresponds to the case of canonical stochastic single-field inflation, where $c_{s}=c_{s,*}=1$, and here $c_{s,*}$ refers to its value fixed at the pivot scale, see discussions on fig. \ref{sound}. The value $c_{s,*}=1$ signifies the maximum allowed value, and beyond that, the causality and unitarity constraints get violated. The power spectrum amplitude achieves $\Delta^{2}_{\zeta\zeta}\gtrsim {\cal O}(10^{-2})$ that is more than sufficient to produce PBH. As we decrease $c_{s,*}$ to its lowest value, we get the blue curve where $c_{s,*}=0.024$ and the cyan shaded region in between the two curves represent the possible space of values coming from different $c_{s}$ existing for other non-canonical stochastic single-field models. An important conclusion can be drawn from this is that decreasing sound speed, which keeps the causality and unitarity constraints intact, lowers the amplitude of both the auto-correlations and cross-correlations in the USR phase, thereby keeping the perturbativity assumptions intact. The stochastic parameter value, $\sigma=0.01$, is kept fixed. The lowest amplitude in the USR can reach within $\Delta^{2}_{\zeta\zeta}\sim {\cal O}(10^{-3})$, which is still sufficient from the perspective of PBH production. Also, we highlight here that it is only in the USR where we observe the most significant impact coming from changes in $c_{s}$, while in the two SRI and SRII phases, the changes are minimal, with the least significant being in SRI.    

We would now like to underscore the phenomenon of out-of-equilibrium features and first-order phase transitions within the present cosmological set up of stochastic inflation. In the first-order phase transitions, the free energy exhibits a discontinuity in its first derivative, and in the present context, the primordial curvature perturbations and its conjugate momenta play the role of free energies and experience the need to satisfy certain boundary conditions. The Israel junction conditions are the continuity and differentiability requirements imposed during moments of each sharp transition, and this mimics a similar role observed for the phase transitions in statistical physics. As a result, the mode solutions and their conjugate momenta incorporate a different set of Bogoliubov coefficients coming from each subsequent phase transition starting from the SRI phase. The framework of stochastic inflation invites the presence of stochastic effects at the Horizon crossing instant, which comes with a stochastic coarse-graining parameter $\sigma$ and where this parameter facilitates the joining of the different phases near the sharp transitions.
Under the influence of stochasticity, at the encounter of each sharp transition,  the possible auto-correlations and cross-correlations between the primordial fluctuations become significant and individually bring in out-of-equilibrium effects for the different phases in the vicinity of their Horizon crossing. The net effect is the loss of a precise instance of the quantum to classical phase transition. The stochastic parameter $\sigma$ must always satisfy a magnitude of order $\sigma \ll 1$ to preserve perturbativity in the underlying theoretical framework. The role of $\sigma$ as a coarse-graining factor or a regulator becomes increasingly essential when quantum loop corrections are the subject of interest, and where they can appear in smoothing out the structure of logarithmic IR divergences. The stochastic inflation theory has also received interest in the past to tackle the IR-associated issues in dS space \cite{Podolsky:2008qq,Finelli:2008zg,Seery:2010kh,Garbrecht:2014dca,Burgess:2015ajz,Gorbenko:2019rza,Baumgart:2019clc,Mirbabayi:2019qtx,Cohen:2020php}. In the absence of any stochastic features, the dominant contribution comes from the observationally significant auto-correlation $\Delta^{2}_{\zeta\zeta}$ and the other auto-correlation $\Delta^{2}_{\Pi_{\zeta}\Pi_{\zeta}}$ and cross-correlations $\Delta^{2}_{\Pi_{\zeta}\zeta}$ and $\Delta^{2}_{\zeta\Pi_{\zeta}}$ turn out to be highly suppressed.

The fig. \ref{spectrumplotsigma} show behaviour of the $\Delta^{2}_{\zeta\zeta}$ auto-correlation as a function of the stochastic parameter $\sigma$ for each phase in our set up. Below we append pointwise the interpretation regarding the value of $\sigma$ in each phase:
\begin{itemize}
    \item \underline{\textbf{In SRI:}}
        We notice that for a given wavenumber mode, $k<{\cal O}(10^{7}{\rm Mpc^{-1}})$, the amplitude of the power spectrum in the SRI stays within its magnitude found at the pivot scale, $[\Delta^{2}_{\zeta\zeta}(k)]_{\rm SRI}\sim 2.2\times 10^{-9}$, when we consider the interval of the parameter, $\sigma\in (10^{-4},0.1)$. Beyond the values $\sigma>0.1$, the power spectrum amplitude starts to increase as the contribution coming from the presence of $\sigma$ starts to become increasingly relevant. However, here we must take caution, in that there does not exist a smooth crossover between $\sigma\ll 1$ and $\sigma\sim 1$, as it clearly gives the wrong amplitude for the pivot scale value seen from eqn. (\ref{pspecsr1dS}).
    \item \underline{\textbf{In USR:}}
        We keep fixed a wavenumber $k$ near the transition scale of $k_{s}\sim {\cal O}(10^{7}){\rm Mpc^{-1}}$ as it provides us with the production of large mass PBH, $M_{\rm PBH}\sim {\cal O}(10^{-2}-1\;M_{\odot})$. Having this satisfied, the power spectrum amplitude in the USR reaches $[\Delta^{2}_{\zeta\zeta}(k)]_{\rm USR}\sim{\cal O}(10^{-2})$ which is sufficient to generate PBHs. We notice that for $\sigma\in (10^{-4},0.1)$, the amplitude remains almost a constant value until it rises to reach magnitude of $\sim{\cal O}(10^{-1})$ when $\sigma>0.1$. This interval of the stochastic parameter $\sigma$, prevents amplitude from breaking the perturbativity constraint of reaching $\sim {\cal O}(1)$, and does not greatly enhance the amplitude which can easily lead to overproduction of PBH. 
    \item \underline{\textbf{In SRII:}}
        Similar to the two previous phases, in the SRII for a given mode, the respective power spectrum amplitude maintains a value of $[\Delta^{2}_{\zeta\zeta}(k)]_{\rm SRII}\sim{\cal O}(10^{-5})$ till the interval of interest remains $\sigma\in (10^{-4},1)$. The overall amplitude increases at a larger rate afterwards for $\sigma>0.1$, but we again remind of the fact that, within the same setting, a limiting case of $\sigma\sim 1$ cannot be physically interpreted with the behaviour at lower values of $\sigma$. The amplitude quickly breaks perturbativity once $\sigma>0.1$ is considered. We conclude that $\sigma<0.1$ remains a good interval to observe effects of coarse-graining in the power spectrum.  
\end{itemize}
From the above discussions, we find that $\sigma> 0.1$ does not lead to meaningful conclusions when comparing with estimates of theoretical and observable importance, such as amplitude of power spectrum at pivot scale, the necessary enhancement for PBH production without overproducing them, and not breaking the perturbativity constraints. On the other hand, any changes coming from $\sigma\ll 1$ remain completely negligible. Hence, we can say that any value of $\sigma<0.1$ suffices to work with and we cannot interpret this scenario with the case of $\sigma>0.1$ together.

A crucial identification can be made directly in reference to the connection between the noise matrix elements and the power spectrum elements as seen in the appendix with eqn. (\ref{noisepower}). As a result of the same formula, in sections \ref{appCa2}, \ref{appCb2}, and \ref{appCc2}, we obtain the noise matrix auto-correlations and cross-correlations corresponding to the respective power spectrum correlations after multiplying with a factor of $(1-\epsilon)$. The value of the slow-roll parameter $\epsilon$ is vanishingly small in the USR while it is still much less than unity in the SRI and SRII, thus indicating that the plots in fig. (\ref{spectrumplots}) also represent the behaviour of associated noise auto and cross-correlations. These noise correlations represent effects of the \textit{quantum kicks} coming from the UV modes after they classicalize in the far super-Horizon (or IR) regions. The $\Sigma_{\zeta\zeta}$ correlation corresponds to most significant of the noise contributions while the correlations involving the conjugate momentum fluctuations, $\Sigma_{\Pi_{\zeta}\zeta}$, $\Sigma_{\zeta\Pi_{\zeta}}$, $\Sigma_{\Pi_{\zeta}\Pi_{\zeta}}$, generate sub-dominant noise amplitudes in the SRI and SRII regime. In the USR, however, all mentioned noise correlations are equally important, implying significant quantum kicks affecting the IR dynamics.

Without including any stochastic features, quantum loop effects have been previously studied by authors in \cite{Choudhury:2023vuj,Choudhury:2023jlt,Choudhury:2023rks,Choudhury:2023hvf}. The existence of a stochastic parameter $\sigma$ serves the purpose of a coarse-graining factor or a regulator in the theory. If one performs the one-loop calculations correctly, then it may happen that in the final result, one can observe the underlying logarithmic IR divergences to be smoothed out more effectively after proper regularization and renormalization techniques than in the calculations done with the absence of such stochastic regulator. We wish to venture towards this goal in the future with great detail. From the perspective of PBHs, their formation faces restrictions in the framework of single-field inflation, where a single-sharp transition occurs. The calculations disallow the formation of any $M_{\rm PBH}\sim {\cal O}(M_{\odot})$ due to solid constraints on the total number of e-folds coming from the necessary procedures of renormalization and a resummation of logarithmic IR divergences at all orders in the loop calculations via the Dynamical Renormalization Group (DRG) analysis \cite{Chen:2016nrs,Baumann:2019ghk,Boyanovsky:1998aa,Boyanovsky:2001ty,Boyanovsky:2003ui,Burgess:2015ajz,Burgess:2014eoa,Burgess:2009bs,Dias:2012qy,Chaykov:2022zro,Chaykov:2022pwd}, $\Delta{\cal N}\sim {\cal O}(20-25)$ (unsuccessful inflation). Some previous attempts, including the presence of multiple sharp transitions (MSTs) \cite{Bhattacharya:2023ysp,Choudhury:2023fjs} and within the framework of single-field Galileon theory with a single sharp transition \cite{Choudhury:2023hvf,Choudhury:2023kdb,Choudhury:2023hfm,Choudhury:2024one} where the non-renormalization theorem plays the most critical role in tackling the PBH mass constraints, it is shown to avoid this strong restriction and eventually make possible the generation of large solar mass PBHs. The present formalism of stochastic inflation serves as a much better alternative in that with just a single transition, and without bringing in any additional symmetry effects, the stochastic regulator $\sigma$ contains the power to avoid the mentioned restrictions on PBH mass and leads to the formation of $M_{\rm PBH}\sim {\cal O}(M_{\odot})$.

\subsection{Outcomes of spectral distortions}
\begin{figure*}[htb!]
    	\centering
    \subfigure[]{
      	\includegraphics[width=8.5cm,height=7.5cm]{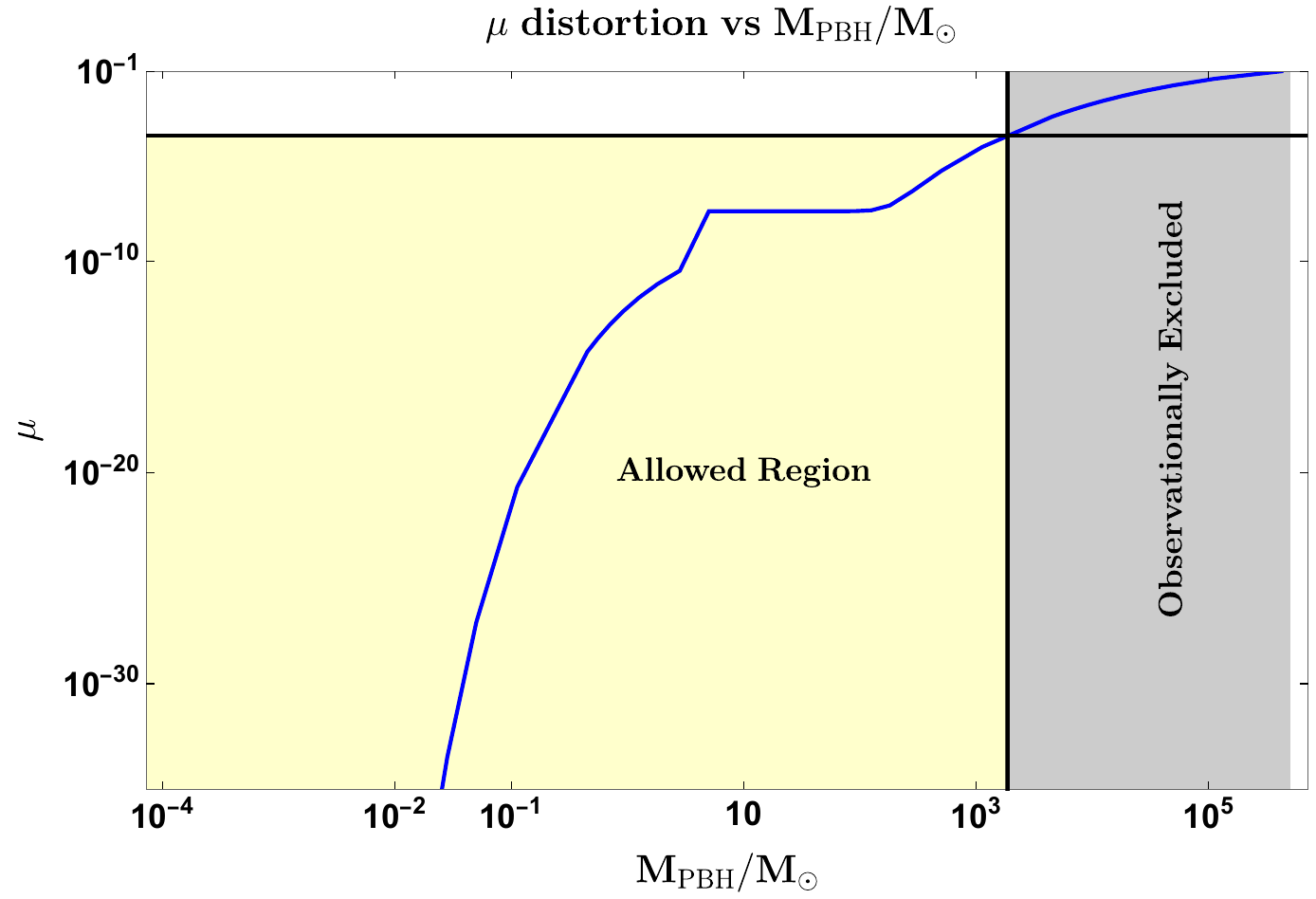}
        \label{mudistort}
    }
    \subfigure[]{
        \includegraphics[width=8.5cm,height=7.5cm]{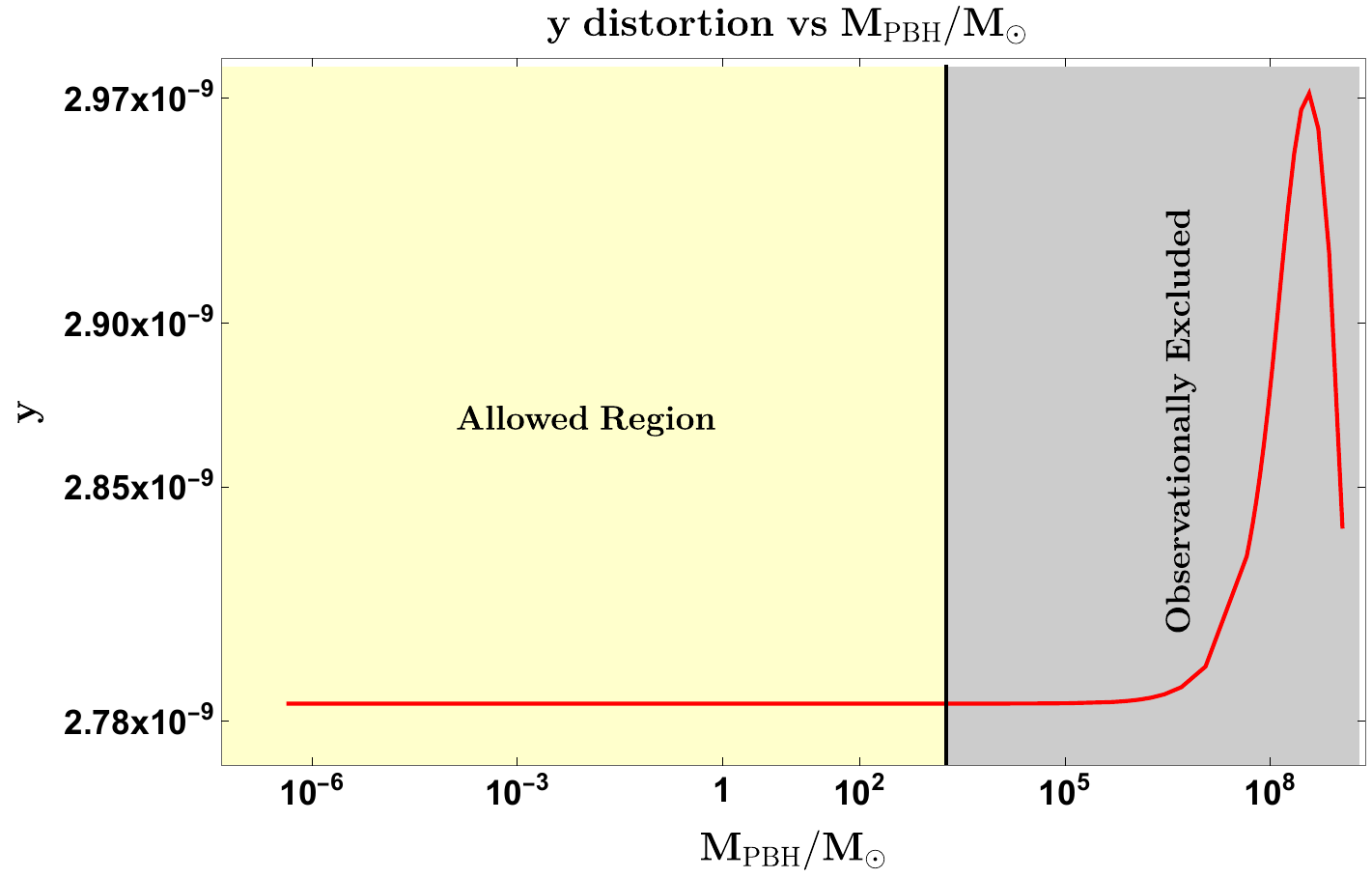}
        \label{ydistort}
       }
    	\caption[Optional caption for list of figures]{Figure displays the $\mu$-distortion (\textit{left-panel}), and the y-distortion (\textit{right-panel}), as function of the PBH mass $M_{\rm PBH}$ (in $M_{\odot}$). The yellow shaded region represents allowed values of the distortion effects after constraints from the COBE/FIRAS observations. The gray shaded regions represents the values disallowed by the same observations. } 
    	\label{distortion}
    \end{figure*}

\begin{figure*}[ht!]
    	\centering
    {
       \includegraphics[width=17cm,height=11cm]{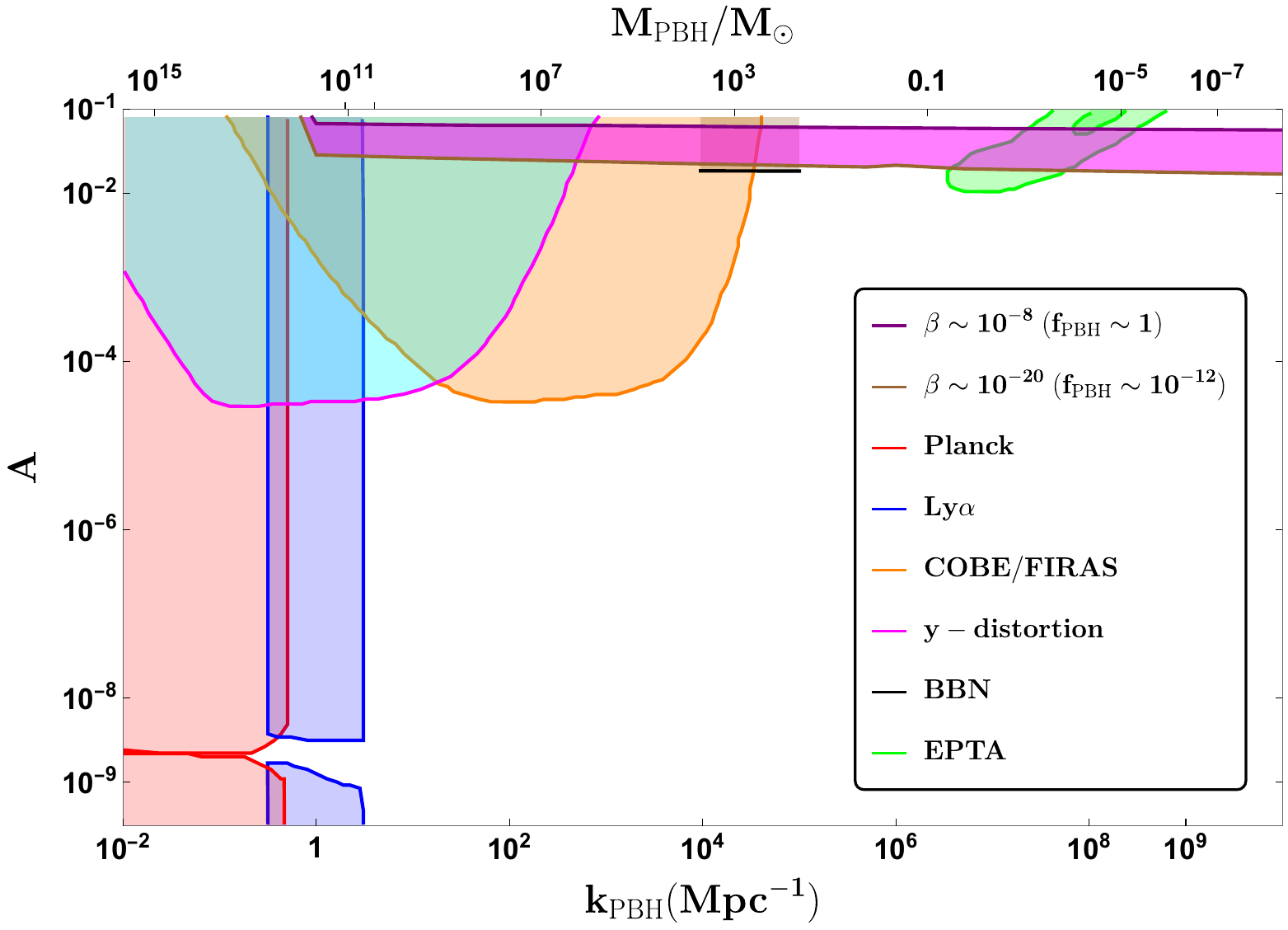}
        \label{ampdistortion}
    } 
    \caption[Optional caption for list of figures]{Amplitude of the scalar power spectrum required to achieve non-negligible PBH mass fraction as function of the PBH forming wavenumber $k_{s}$. The background involves constraints on the amplitude coming from the CMB temperature anisotropies (red) \cite{Planck:2018jri} at the large scales, Lyman-$\alpha$ forest (blue) \cite{bird2011minimally}, COBE/FIRAS (orange) and the y-distortion (cyan) effects \cite{Cyr:2023pgw}, BBN (black)\cite{Jeong:2014gna}, and the $1$ and $2\sigma$ contours reported by the pulsar timing array collaboration (green) \cite{EPTA:2023xxk}. The magenta-coloured band, bounded by the brown and purple lines, highlights the region of amplitude where the initial PBH mass fraction from the Press-Schechter formalism lies in the window $\beta\sim {\cal O}(10^{-20}-10^{-8})$. }
\label{amplitudedistortion}
    \end{figure*}

This section discusses the numerical outcomes for the estimates of the two $\mu$ and $y$ distortion effects. We follow the expressions contained in eqns. (\ref{mudistortion}, \ref{ydistortion}) to obtain various estimates as function of the PBH mass $M_{\rm PBH}$ formed corresponding to the wavenumber, $k_{\rm PBH}.$ 

The fig. \ref{mudistort} displays the $\mu$ distortion estimates as function of $M_{\rm PBH}$. Based on the current strongest constraints coming from the COBE/FIRAS observations, the $\mu$ distortion must satisfy, $|\mu|\simeq 9\times 10^{-5}$. The gray shaded region in the plot highlights the excluded mass range as a result of the disallowed $\mu$ values. We notice that the $\mu$ values decrease much rapidly for the masses lower than, $M_{\rm PBH}\lesssim {\cal O}(M_{}\odot)$ and thus make those masses fall in the allowed region. The horizontal line demarcates the above mentioned upper bound on $\mu$ and the black vertical line corresponds to the respective mass value indicating that PBHs with $M_{\rm PBH}\gtrsim 1.8\times 10^{3}\;M_{\odot}$, are equivalently removed from the $\mu$ constraints. The fig. \ref{ydistort} displays the behaviour of the $y$ distortion values as function of the PBH masses. The $y$-distortion does not come out as a significant effect for $M_{\rm PBH}\lesssim {\cal O}(10^{7}M_{\odot})$ after which its behaviour can change in an unpredictable manner. The constraints on this distortion from the COBE/FIRAS observations are $|y|\simeq 1.5\times 10^{-5}$. 

To achieve PBH formation requires the scalar power spectrum amplitude to lie within the order of magnitude $A\sim {\cal O}(10^{-2})$. We utilise the same formulas, in eqns. (\ref{mudistortion}, \ref{ydistortion}), to determine the necessary amplitude such that existing bounds on distortion effects remain satisfied. At the same time, the resulting initial PBH mass fraction should also remain non-negligible and for the same purpose we choose the window of $\beta$ to lie within $\beta\sim {\cal O}(10^{-20}-10^{-8})$. 

The figure (\ref{amplitudedistortion}) provides the results for the power spectrum amplitude $A$ as function of the PBH forming wavenumber $k_{s}$. The magenta coloured band highlights the region of initial mass fraction, $\beta\sim {\cal O}(10^{-20}-10^{-8})$, with the interval bounded between purple and brown lines, respectively. The background contains various contours of observational constraint on the amplitude $A$. CMB anisotropy already strictly constraints the primordial power spectrum amplitude at larger scales. We also notice that constraints from the distortion effects, captured by the COBE/FIRAS (in orange) and the regime from y-distortion constraint (in magenta), disfavours PBHs with masses $M_{\rm PBH}\gtrsim 1.8\times 10^{3}M_{\odot}$ upon using the amplitude of $A\sim {\cal O}(10^{-2})$. The earliest studies properly establishing the idea of constraining the primordial power spectrum amplitude by observation of CMB spectral distortions on the larger scales was carried out in \cite{Chluba:2012gq,Chluba:2012we}. The orange and cyan curves showcase the updated constraints from $\mu-$ and $y-$ type distortions on the power spectrum amplitude, and a detailed analysis on improving these distortion effects constraints and their connections in light of the recently observed stochastic gravitational wave background signal can be found in \cite{Cyr:2023pgw}. The Big Bang Nucleosynthesis (BBN) also limits the amplitude with constraints at the right end of the distortion contour in orange. At scales after BBN but before spectral-distortion era, $k\simeq 10^{4}-10^{5}\;{\rm Mpc^{-1}}$, the energy dissipation of the associated waves ultimately leads to constraints on the primordial power spectrum from overproduction of primordial deuterium and helium estimates. The smaller wavenumbers or larger mass PBH require larger amplitudes, $A\sim {\cal O}(10^{-2}-10^{-1})$, to achieve substantial initial abundance, and that amplitude only tends to increase even faster for $M_{\rm PBH}\gtrsim {\cal O}(10^{11}M_{\odot})$. Going beyond the purple line on top leads to overproduction of PBHs; however, here we remind that the calculations involve the assumption of using an initial Gaussian profile for the curvature perturbation $\zeta$ when examining the distortion estimates with the total power spectrum from eqns. (\ref{pspecsr1dS}, \ref{pspecusrdS}, \ref{pspecsr2dS}).

\subsection{Outcomes of PDF}

\begin{figure*}[ht!]
    	\centering
    \subfigure[]{
      	\includegraphics[width=8.5cm,height=7.5cm]{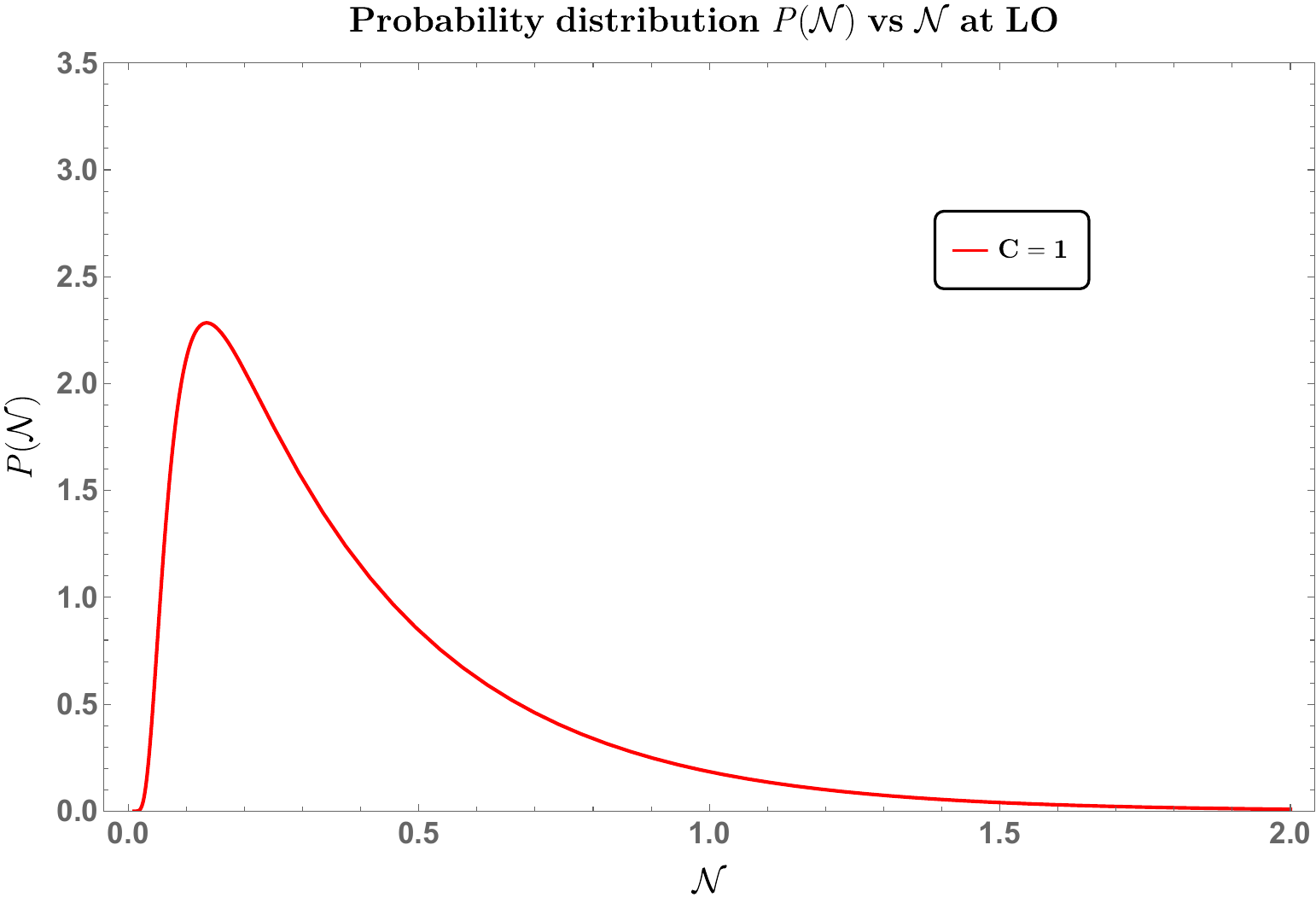}
        \label{N0PDF}
    }
    \subfigure[]{
        \includegraphics[width=8.5cm,height=7.5cm]{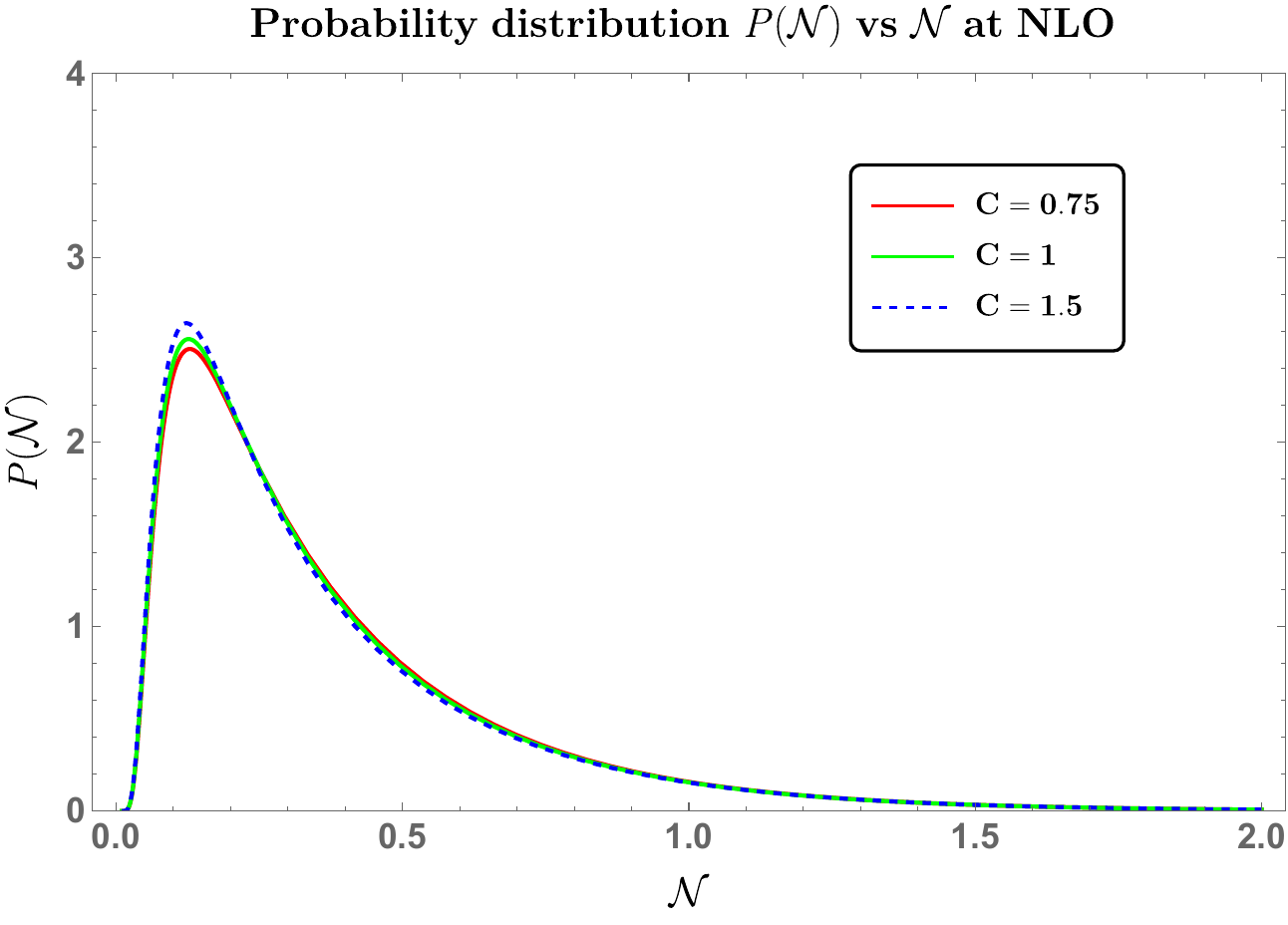}
        \label{N1PDF}
       }
    \subfigure[]{
      	\includegraphics[width=8.5cm,height=7.5cm]{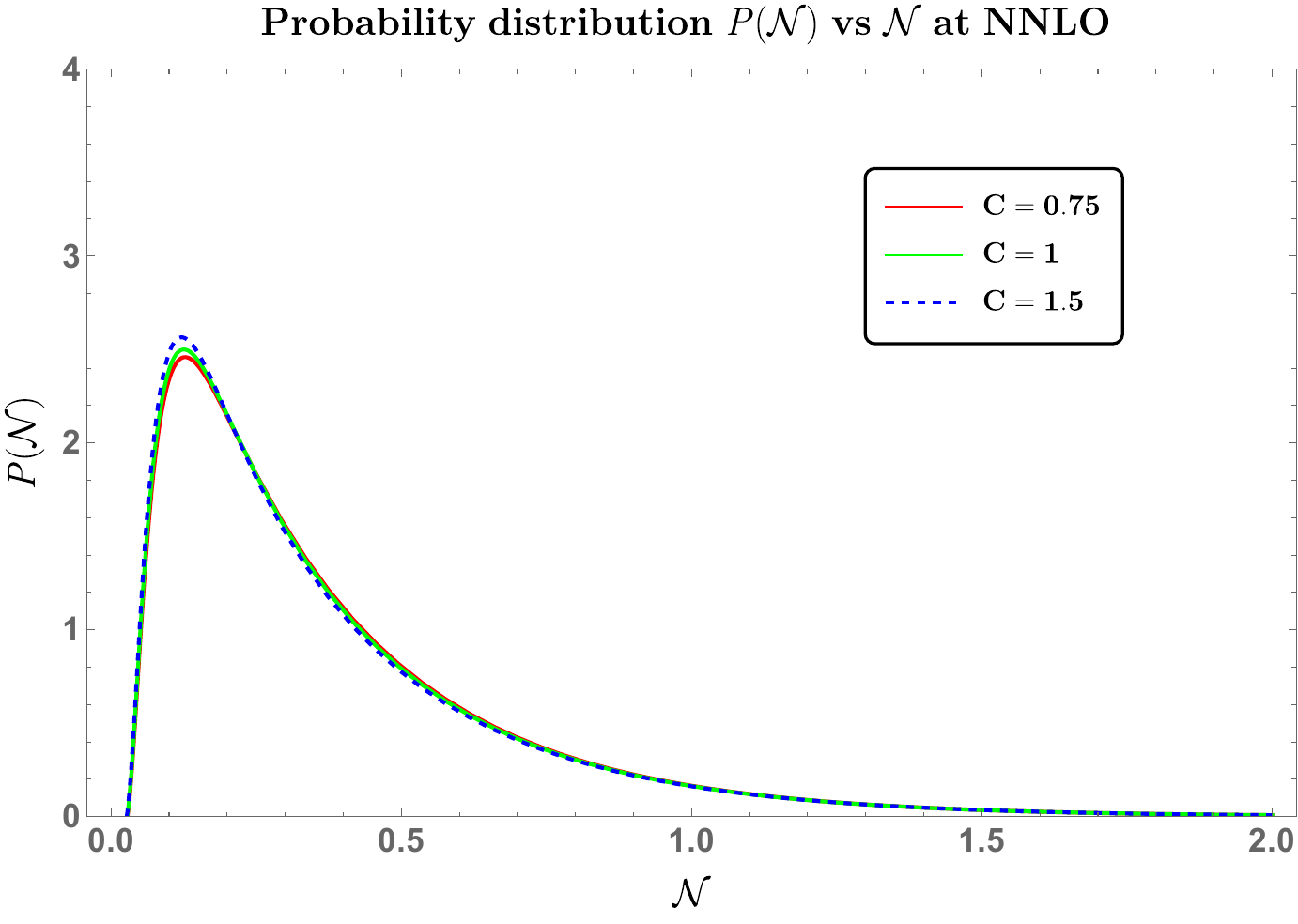}
        \label{N2PDF}
    }
    \subfigure[]{
        \includegraphics[width=8.5cm,height=7.5cm]{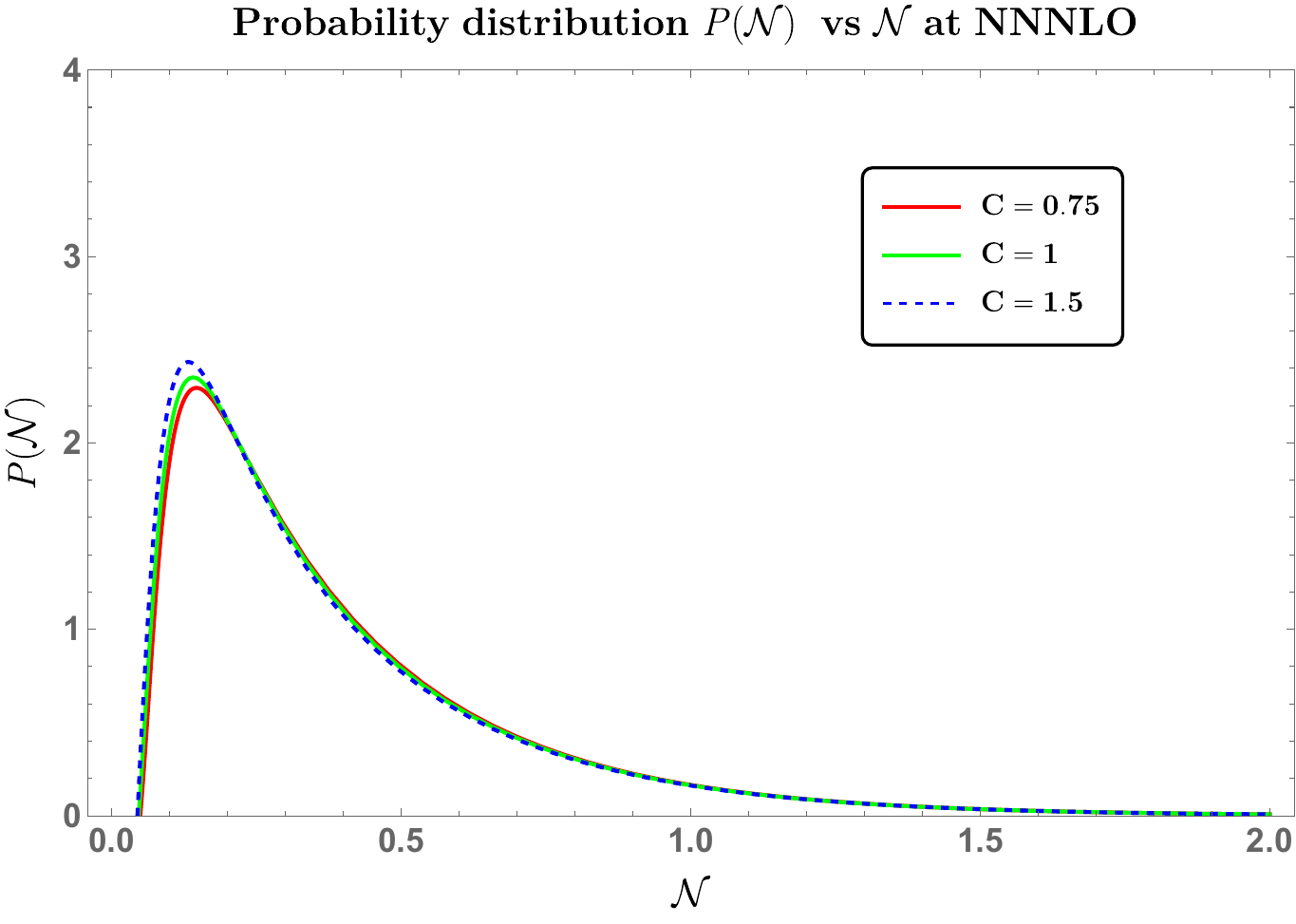}
        \label{N3PDF}
       }
    	\caption[Optional caption for list of figures]{Behaviour of the Probability Distribution Function with changing number of e-folds ${\cal N}$. The \textit{top-left} panel shows PDF at the leading order (LO) along with different $y$ values. The \textit{top-right} panel shows the PDF at the next-to-leading order (NLO), the \textit{bottom-left} panel shows the PDF at the next-to-next-to-leading order (NNLO), and the \textit{bottom-right} panel shows the PDF at the next-to-next-to-next-to-leading order (NNNLO), all for different values of the characteristic parameter $C=\{0.75,1,1.5\}$ in red, green, blue, respectively.  } 
    	\label{diffPDF}
    \end{figure*}

\begin{figure*}[ht!]
    	\centering
    \subfigure[]{
      	\includegraphics[width=8.5cm,height=7.5cm]{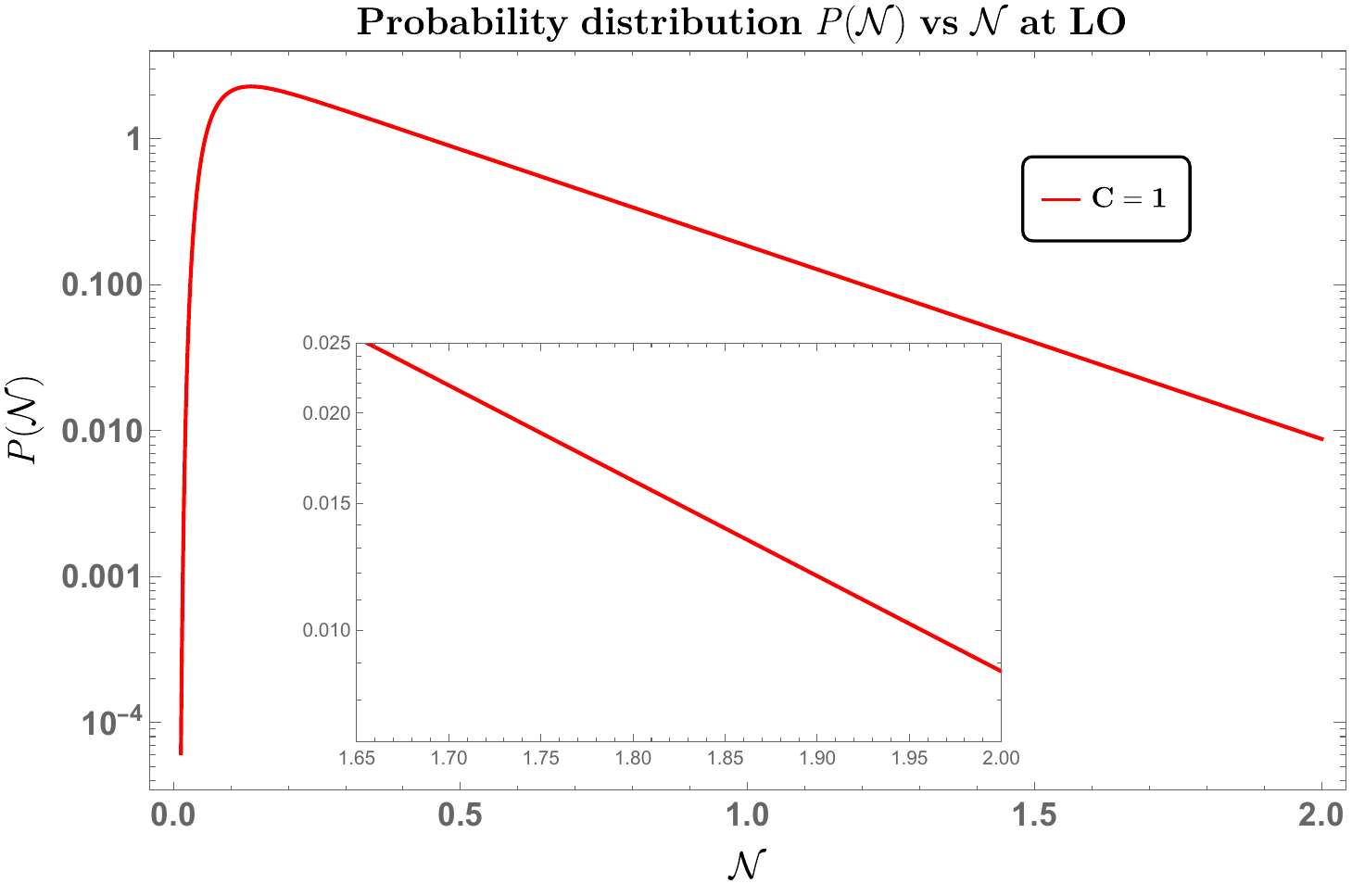}
        \label{N0PDFlog}
    }
    \subfigure[]{
      	\includegraphics[width=8.5cm,height=7.5cm]{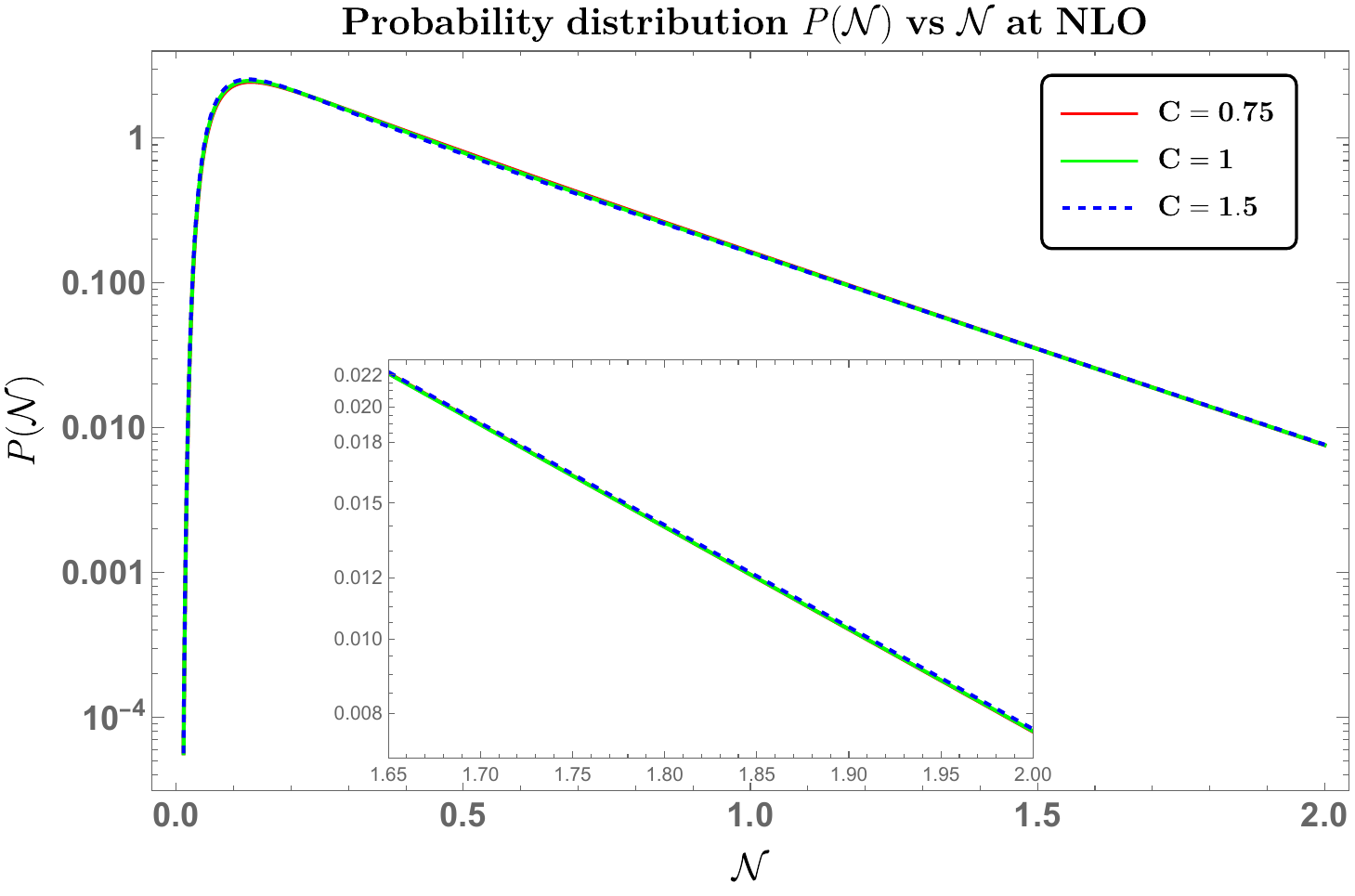}
        \label{N1PDFlog}
    }
    \subfigure[]{
        \includegraphics[width=8.5cm,height=7.5cm]{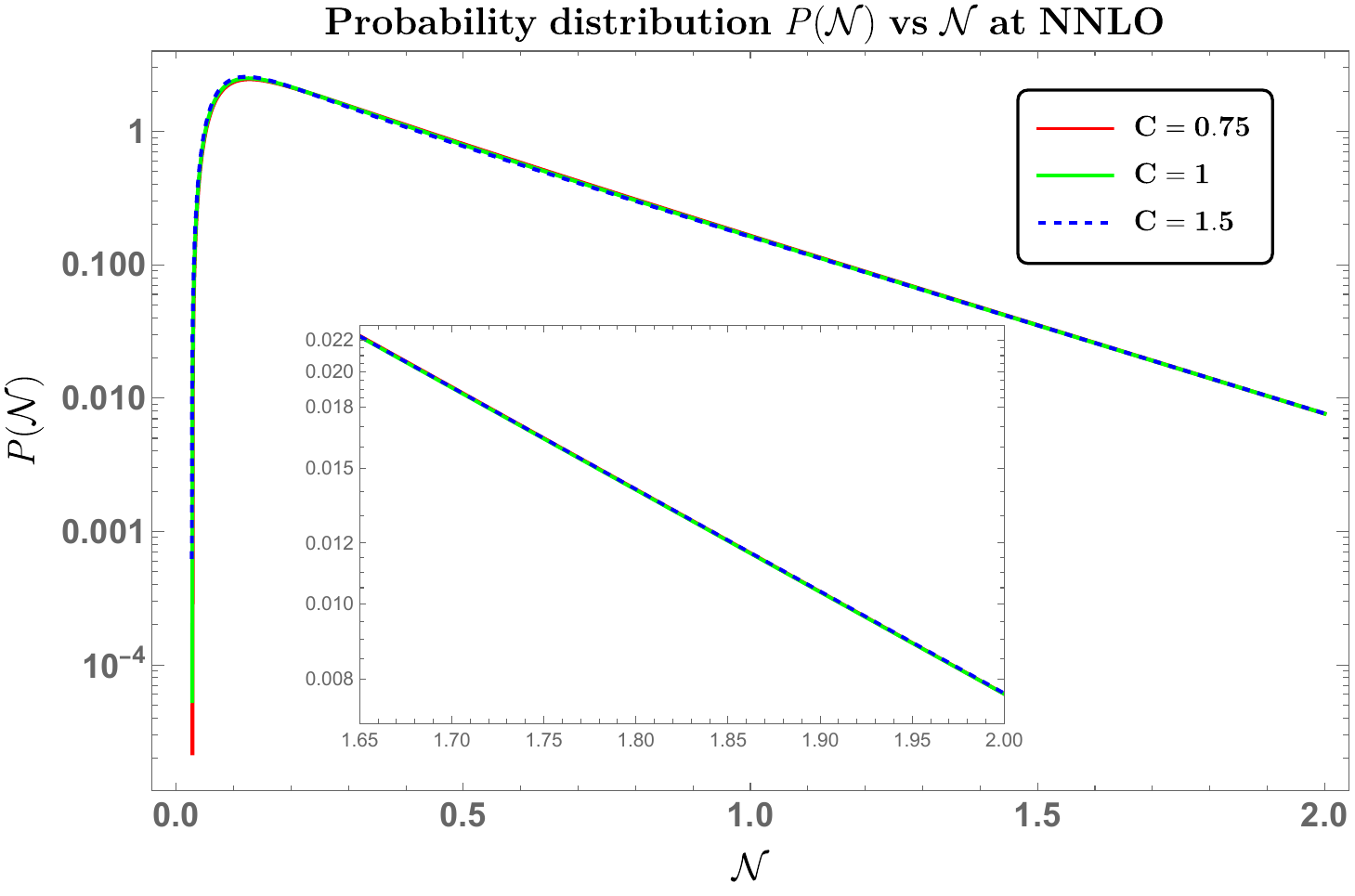}
        \label{N2PDFlog}
       }
    \subfigure[]{
      	\includegraphics[width=8.5cm,height=7.5cm]{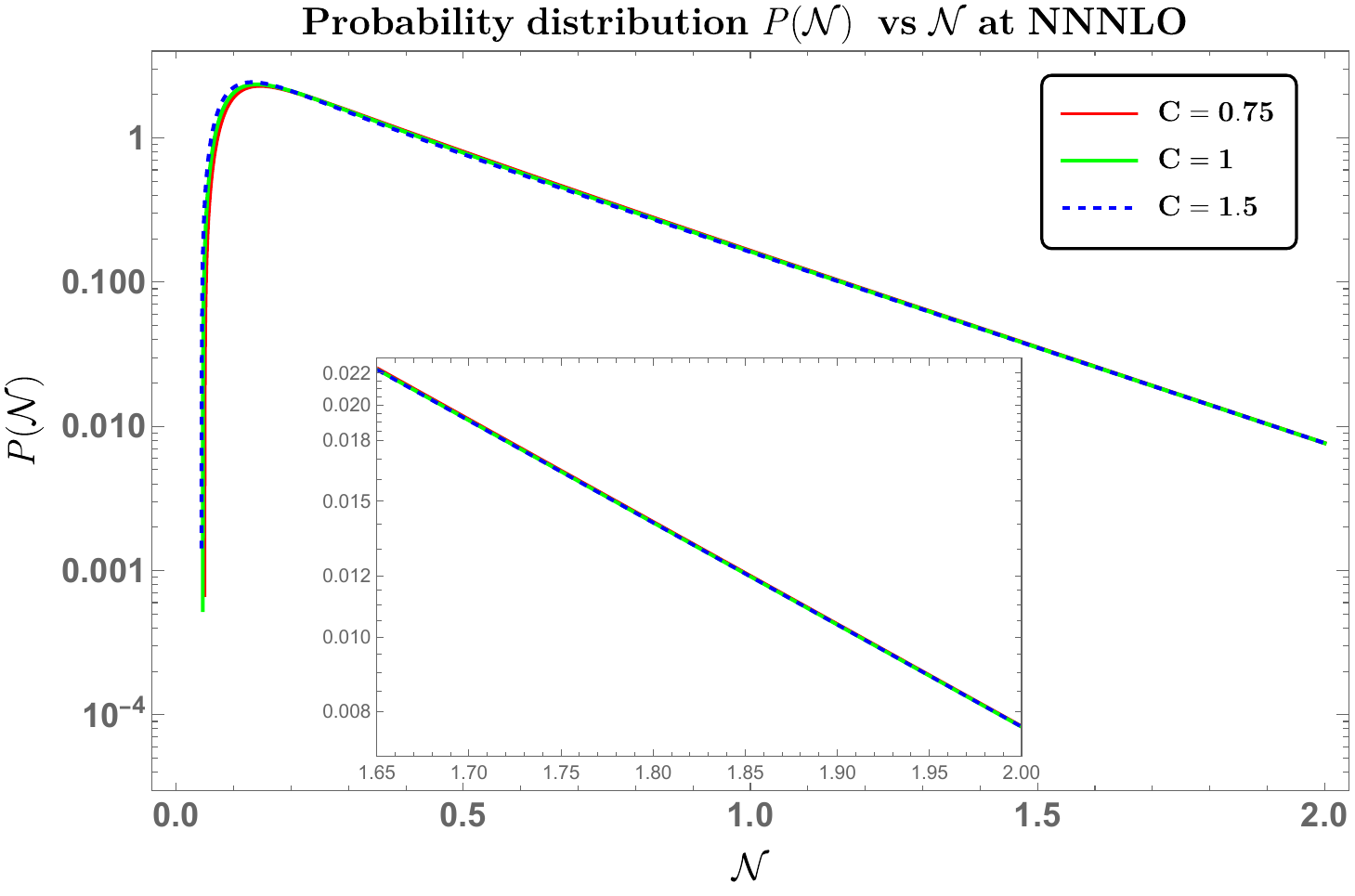}
        \label{N3PDFlog}
    }
    	\caption[Optional caption for list of figures]{\textcolor{black}{Logarithmic plots for the probability distribution function of the stochastic variable, ${\cal N}$, at different orders in perturbations theory. The inset figures in each plot provide a magnified view of the amplitude $P({\cal N})$ for the e-folds, $1.65 \leq {\cal N}\leq 2$. The values of $x=1$, $y=0.15$, and $\tilde{\mu}=0.9$, are kept fixed. } } 
    	\label{diffPDFlog}
    \end{figure*}

\begin{figure*}[ht!]
    	\centering
    \subfigure[]{
      	\includegraphics[width=8.5cm,height=7.5cm]{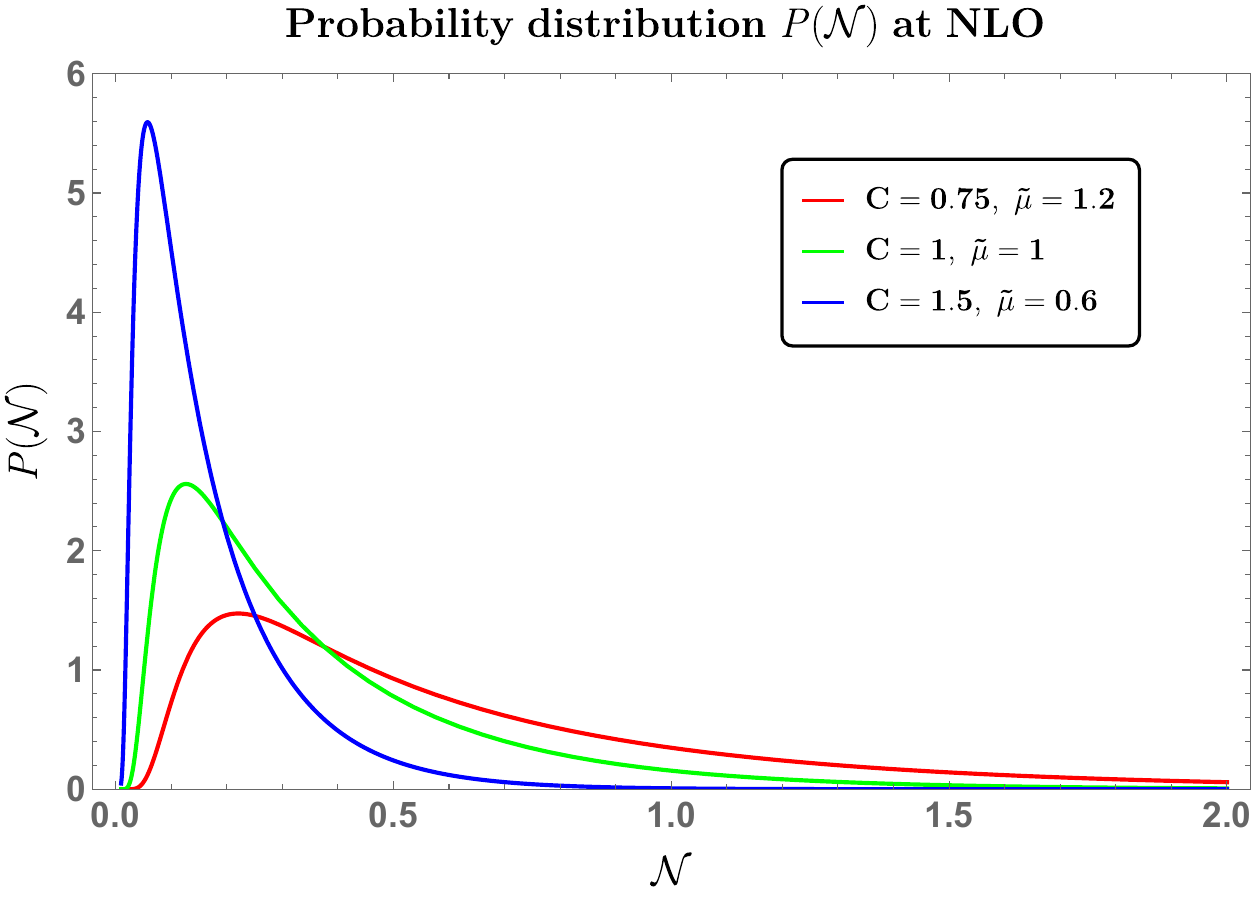}
        \label{N1PDFdiffmu}
    }
    \subfigure[]{
        \includegraphics[width=8.5cm,height=7.5cm]{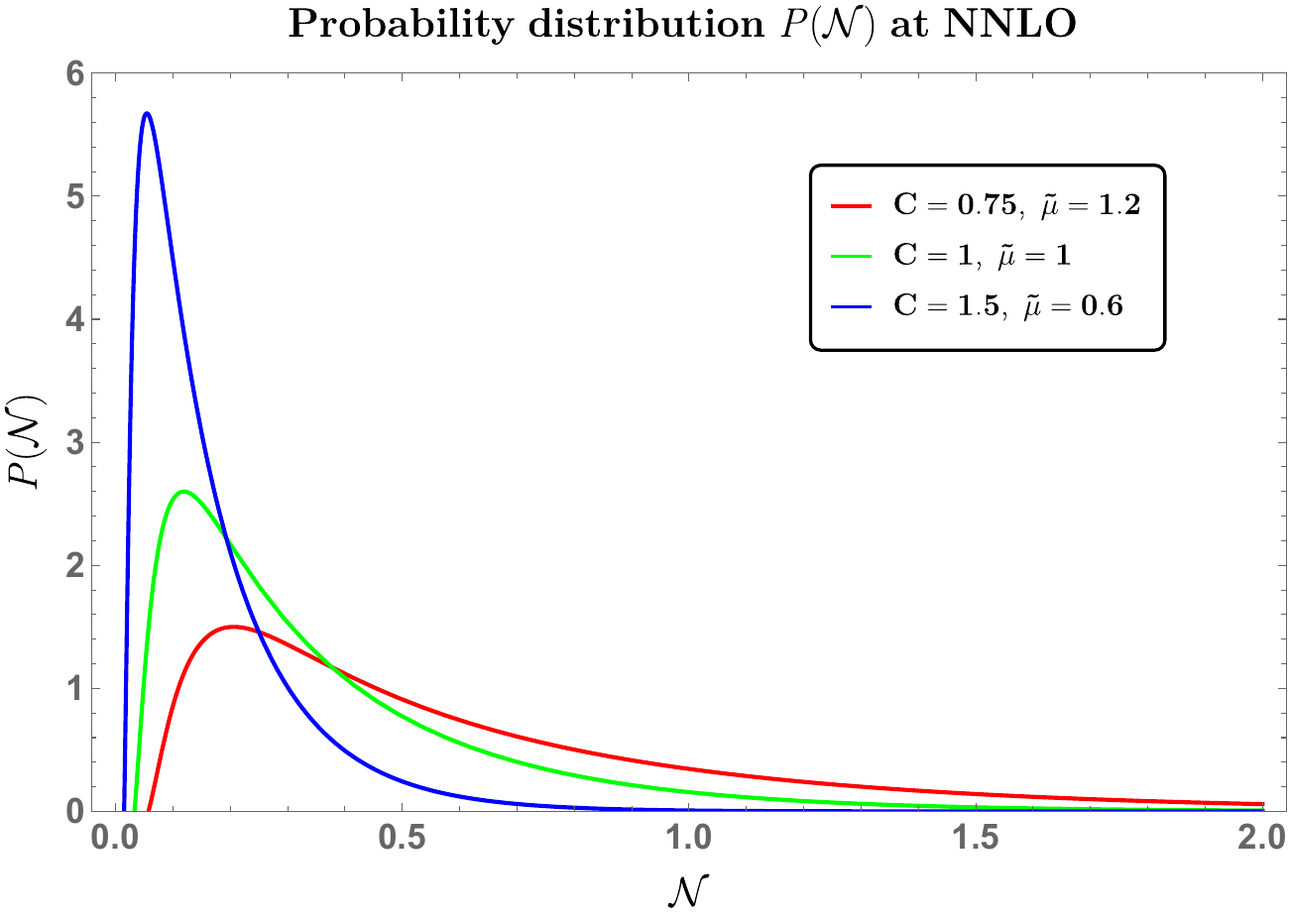}
        \label{N2PDFdiffmu}
       }
    \subfigure[]{
      	\includegraphics[width=8.5cm,height=7.5cm]{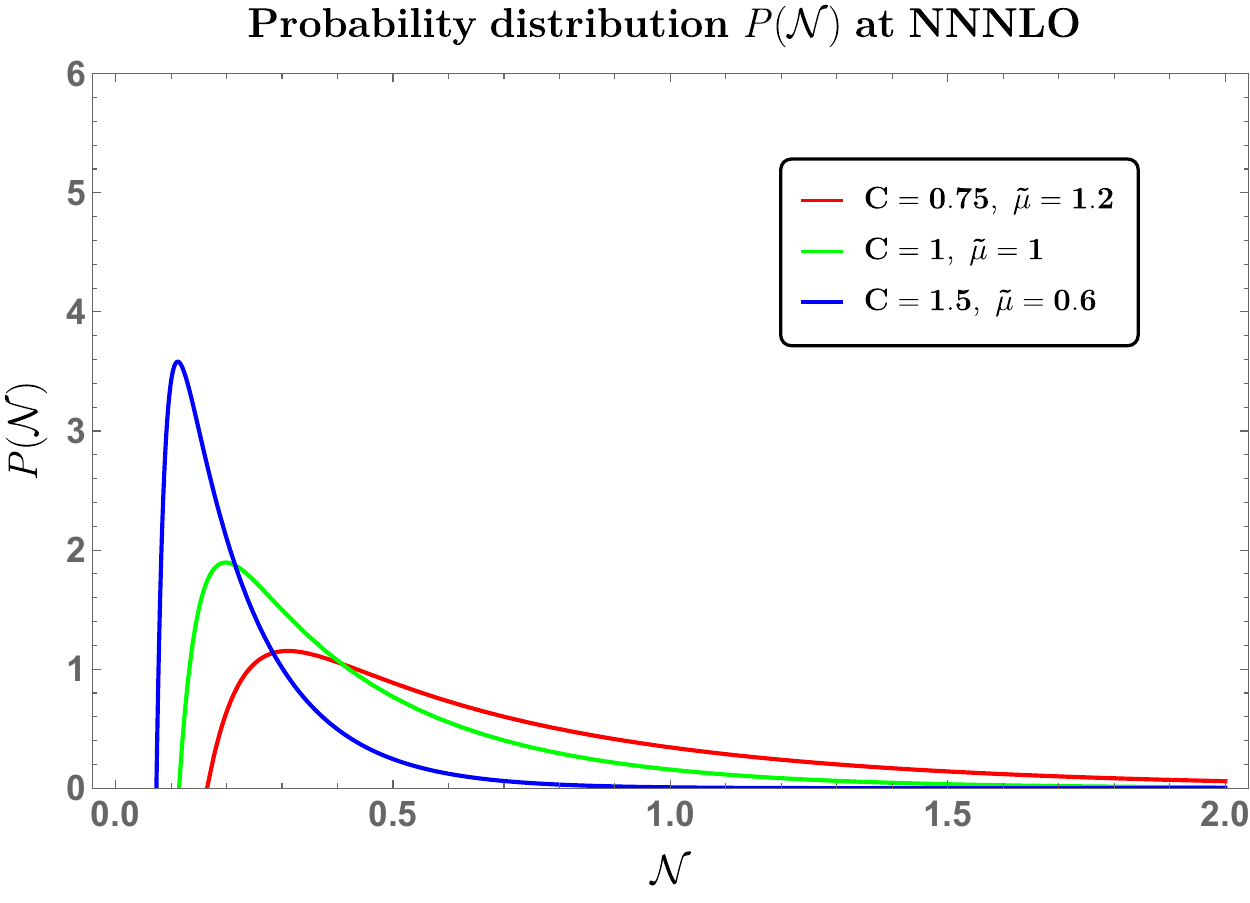}
        \label{N3PDFdiffmu}
    }
    	\caption[Optional caption for list of figures]{\textcolor{black}{The probability distribution function of the stochastic variable, ${\cal N}$, at different orders in perturbations theory. The \textit{top-row} shows the PDF at NLO (left) and NNLO (right), while at the \textit{bottom}, we have the PDF after including the NNNLO corrections. The colors denote the cases where, in contrast to fig. (\ref{diffPDF}), now $\tilde{\mu}$ changes between $\{0.6,1,1.2\}$ in red, green, and blue, respectively. Notice the drastic change in PDF shapes for ${\cal N}$ close to $0$ and at the right tail regions.} } 
    	\label{diffPDFdiffmu}
    \end{figure*}

\begin{figure*}[ht!]
    	\centering
    \subfigure[]{
      	\includegraphics[width=8.5cm,height=7.5cm]{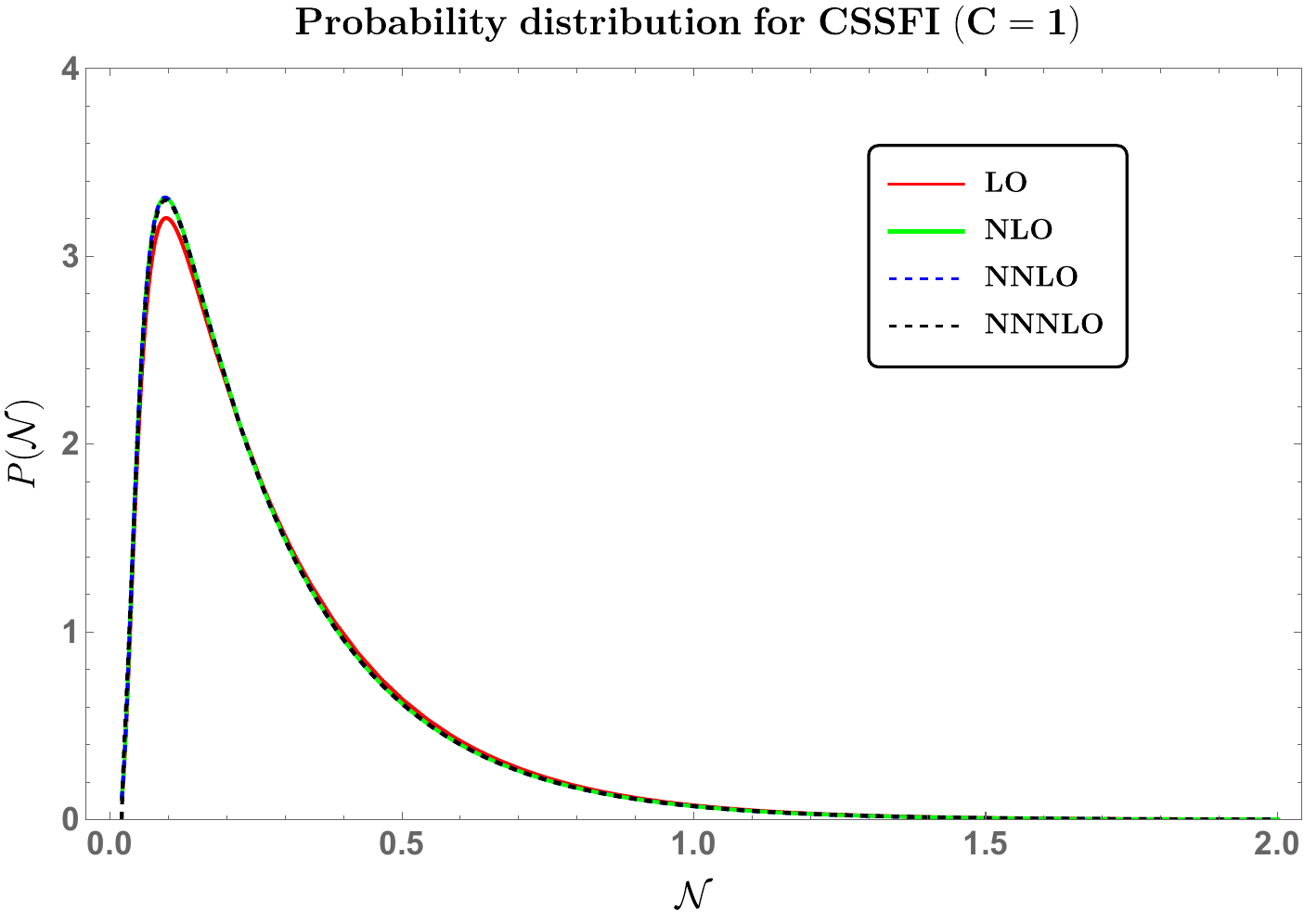}
        \label{cssfi}
    }
    \subfigure[]{
        \includegraphics[width=8.5cm,height=7.5cm]{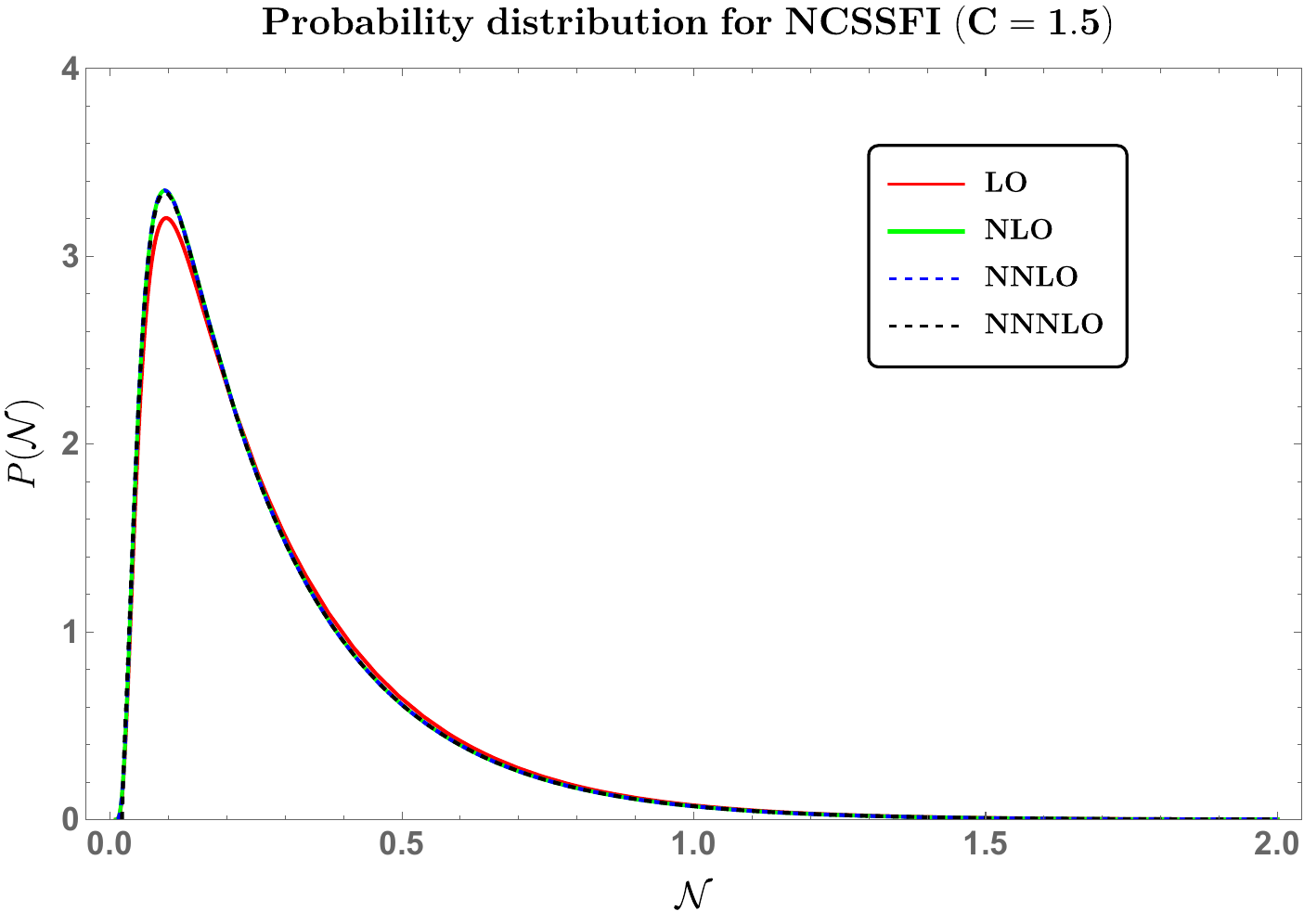}
        \label{ncssfi}
    }
    \subfigure[]{
      	\includegraphics[width=8.5cm,height=7.5cm]{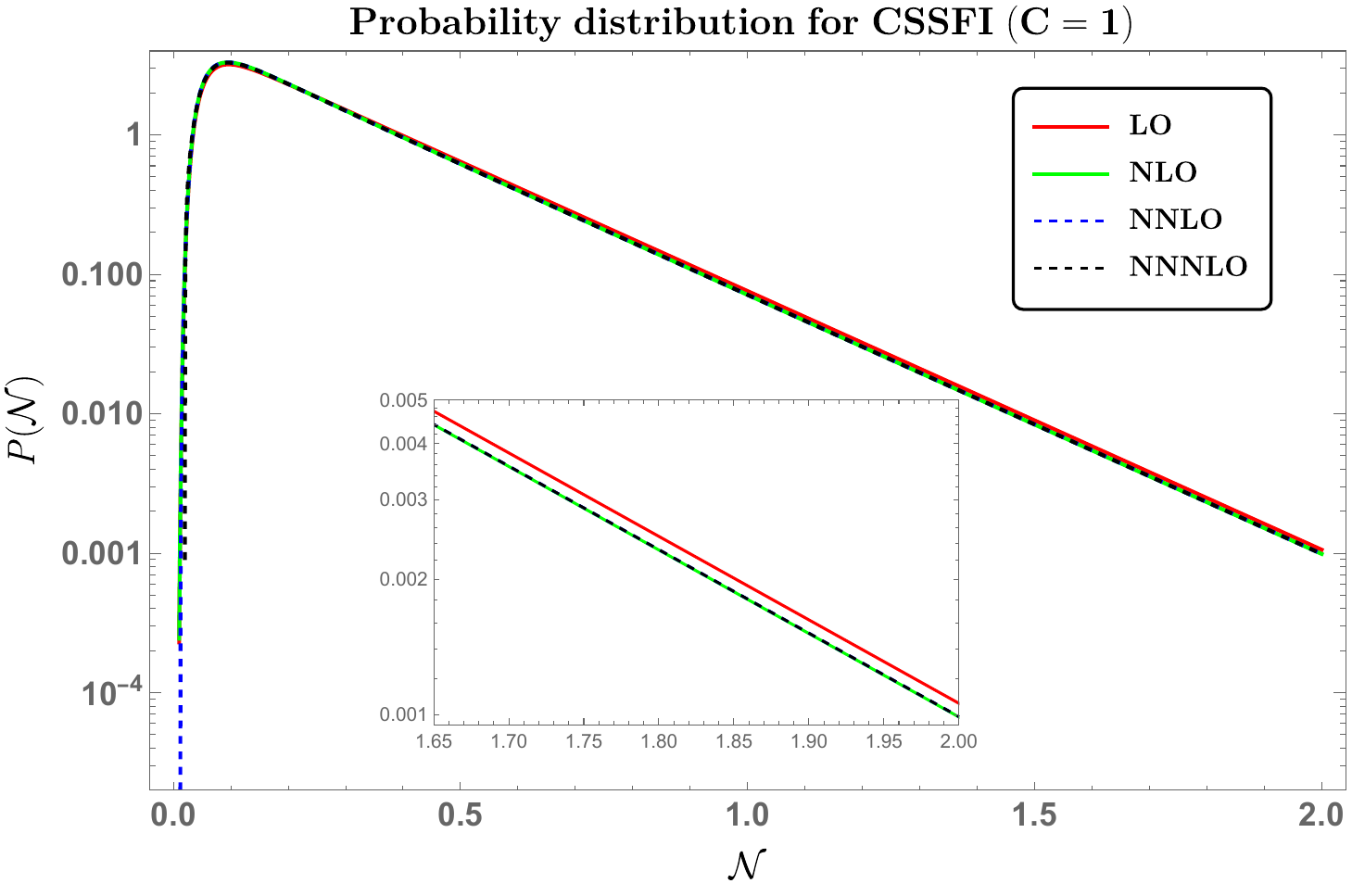}
        \label{cssfilog}
    }
    \subfigure[]{
        \includegraphics[width=8.5cm,height=7.5cm]{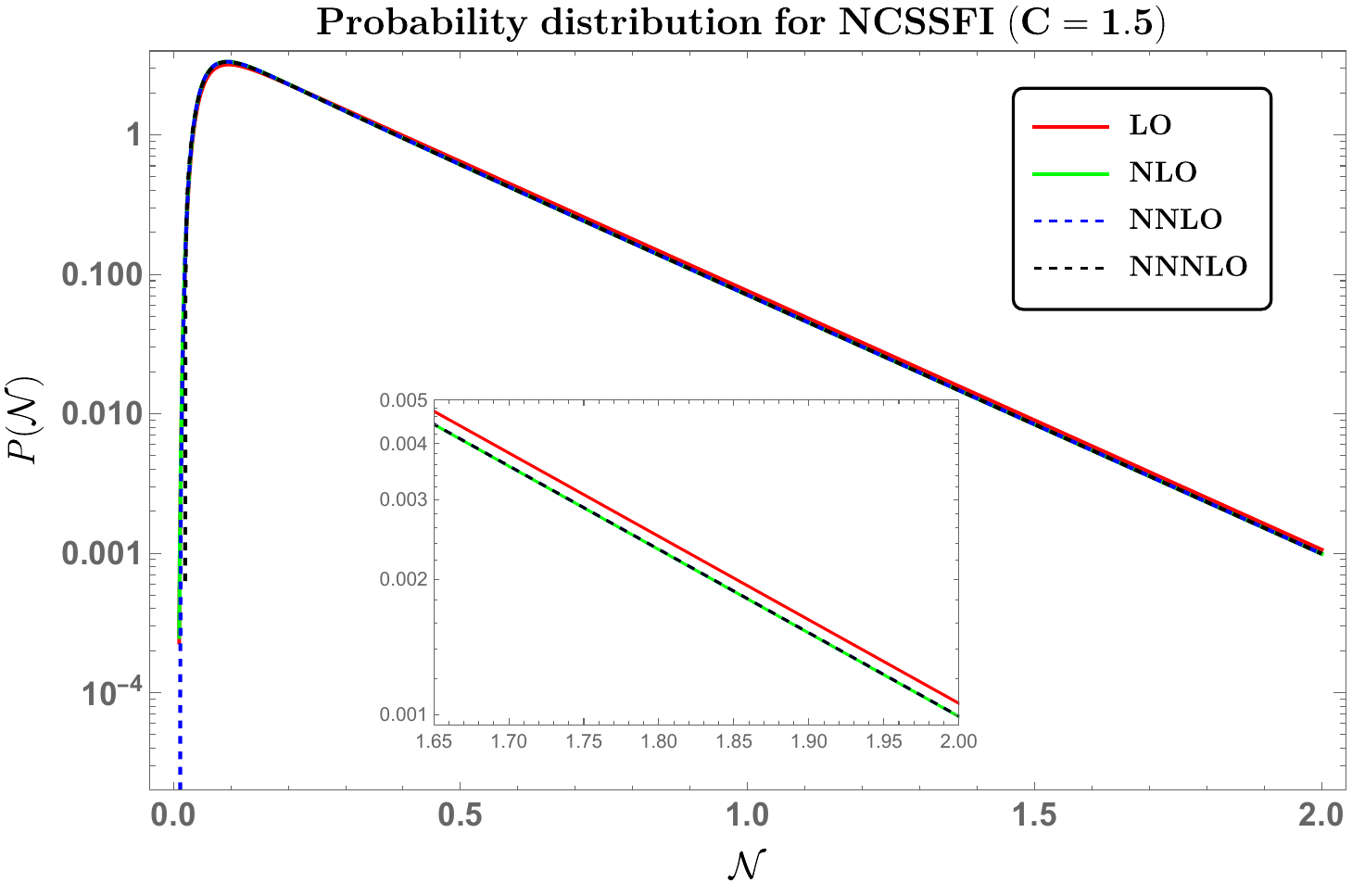}
        \label{ncssfilog}
    }
    	\caption[Optional caption for list of figures]{Plot shows PDF as a function of the stochastic variable ${\cal N}$ including all order of the perturbative expansion up to NNNLO. The \textit{top-row} considers the case of canonical stochastic single-field inflation $(C=1)$ superposed with all the distinct PDF from the perturbative expansion and, similarly, for the case of non-canonical single-field inflation $(C=1.5)$. \textcolor{black}{The \textit{bottom-row} shows the same PDFs as present above now scaled logarithmically along the vertical axis. The two inset plots provide a magnified view of the far-right tail region of $P({\cal N})$ for, $1.65 \leq {\cal N}\leq 2$. Notice the clear separation in between the LO value and other higher-order corrections which will later become crucial in understanding the PBH abundance results. }
        } 
    	\label{allSSFI}
    \end{figure*}
In this section, we study the outcomes of our PDF calculations up to the order NNNLO in the diffusion-dominated regime. The expressions for the PDF at the NNLO and NNNLO have not been discussed explicitly but here we provide with their graphical representations and discuss their features. The initial phase space variables $(x,y)$ are defined previously in eqn. (\ref{newphasevars}), as being related to the coarse-grained curvature perturbation and its conjugate momentum variable.

We begin with the analysis at the leading order (LO). Earlier eqn. (\ref{pdfLO}) did not saw any involvement of the characteristic parameter $C$ in the residues, as well as the poles. From eqn. (\ref{poles}), it is clear that in order to explicitly observe $C$ requires the existence of higher-order poles, at least where $m=1$. The fig. \ref{N0PDF} shows the distribution from eqn. (\ref{pdfLO}), plotted against the stochastic e-folding number, ${\cal N}$, for a fixed $C=1$. Since this is at the leading order in $y$ expansion, no change in the features for changing $y$ is observed. The remaining parameters are fixed here to have $x=1$ and $\tilde{\mu}= 0.9$.  
Going to the next order in the expansion, we now have two sets of poles that combine to give the full PDF at NLO, displayed in the fig. \ref{N1PDF}. The analytic expression is also available from eqn. (\ref{pdfNLO1}). Notice the difference in curves associated with different scenarios of non-canonical single-field inflation, characterized by a finite and changing $C$. The relative difference is not significant enough to render the EFT description undesirable. We have chosen here to keep $C$ close to $1$, as seen from the red and dotted blue curves. This time, the amplitude has risen compared to the LO case, with the difference being in the first decimal place. Here $y=0.15$ is fixed while $x=1$ and $\tilde{\mu}=0.9$ remain the same. In this regime of $y$ values, most of the PDF gets well approximated by the leading terms in the series expansion of eqn. (\ref{pdfLO1}) as the rest decay very quickly. Further in the bottom-left fig. \ref{N2PDF} we have the PDF at the NNLO. The analytic expression for this PDF is not made explicit before but we can see from the distribution that its behaviour remains almost similar to the previous PDFs with small relative changes at the maximum when different values of $C$ are considered. The peak amplitude is also close to the PDF at NLO with differences now pushed to the second decimal place. The parameter set, $x=1$, $y=0.15$, and $\tilde{\mu}=1$, is kept fixed. In the bottom-right panel of fig. \ref{N2PDF}, we finally consider the distribution at NNNLO. The peak amplitude shows clear changes for various $C$ values in the allowed range for a valid EFT formulation and the amplitude change relative to NNLO now occurs at the first decimal place and lowers the estimate. This also signals preserving the perturbativity argument in the analysis after keeping fixed the same set of values for the other parameters, $x=1$, $y=0.15$, and $\tilde{\mu}=1$. 

\textcolor{black}{From fig. (\ref{diffPDFlog}), we can observe similar plots in fig. (\ref{diffPDF}) but logarithmically scaled along the vertical axis. Inside each plot, the insets look closer at the far-right tail region, $1.65\leq {\cal N}\leq 2$. At the NLO, we infer from the inset plots that the PDF amplitude suppresses in value at the third decimal place compared to the PDF at LO. Also, for the PDF at NLO, upon closer examination, we find that the amplitude of PDF for large ${\cal N}$ descends with $C=1.5$ at the top and $C=0.75$ at the bottom with differences pushed to the fourth decimal place. Such differences in amplitude become even smaller when considering higher-order corrections like NNLO and NNNLO PDFs in the bottom row. Likewise, we expect from such behaviour of these PDFs that the resulting PBH mass function should also exhibit almost indistinguishable deviations for the $C$ values and other parameters as chosen here. In the upcoming mass fraction section, we will analyze such changes, including the effect of changing the $x$ and $\tilde{\mu}$ variables. }

In the diffusion-dominated regime, we conclude that the distribution functions at subsequent orders maintain the perturbativity conditions satisfactorily. The observed changes for various EFT descriptions are characterized by $C$ value of interest only near the peak amplitude regions, and the size of the differences does not drastically change the overall features of the PDF. At all orders shown in fig. \ref{diffPDF}, the left and right tails follow similar trends and meet together while the only slight changes are visible near the peak amplitude. 

\textcolor{black}{In contrast to the conditions in fig. (\ref{diffPDF}), if we now fix the original diffusion coefficient, $\mu=0.9$, everywhere while observing changes in the EFT-modified coefficient, $\tilde{\mu}=\mu/C$, then the results are illustrated in fig. (\ref{diffPDFdiffmu}). The LO behaviour remains unchanged since any $C$-dependent factors remain absent in the corresponding PDF. As for the remaining plots, at each order, we observe a significant change in overall PDF shapes and including concerning features near small ${\cal N}$. This observation already tells us that the PDF features are quite sensitive to the coefficient $\tilde{\mu}$; the use of $\tilde{\mu}$ in the exponential decay factors for each PDF in eqns. (\ref{PDFN1}), (\ref{PDFN2}), and (\ref{PDFN3}), being the reason for such sensitivity. Apart from the pronounced shape changes, the left tail regime demonstrates their inability to converge for small ${\cal N}$ as higher-order corrections get incorporated into the PDF beyond NLO. Most drastic of such changes come at NNNLO, where for ${\cal N}<0.2$, the various PDFs start to separate more noticeably, which only worsens when compared to shapes at previous orders and exhibit large deviations (not shown here are the large negative values) that signals breaking the perturbative analysis for smaller e-folds. The heavy exponential behaviour of the large ${\cal N}$ tails after crossing the peaks remains roughly the same at each order and for each value of $C$, which can strongly alter the PBH abundance estimates as they are highly sensitive to this regime of the PDF. Moreover, we notice that including corrections at NLO raises the PDF higher in magnitude for $C=1$ but introduces a huge change in overall features when slightly changing the characteristic coefficient $C$ which then remains more or less unchanged at NNLO but suffers a noticeable suppression in amplitude after including the NNNLO corrections.
}


Figure \ref{allSSFI} displays behaviour of all the distribution functions superimposed, that captures various higher-order corrections in the diffusion-dominated regime in the small $y$ limit, keeping $y=0.073$, $x=1$, and $\tilde{\mu}=0.76$ fixed. The reason for these specific parameter values will soon become evident when we examine the related PBH abundance features we expect in the upcoming sections. After superimposing the PDFs, we notice minute changes in the PDF values for large ${\cal N}$ among different orders. This time, the expansion variable $y$ is also kept much lower than the analysis in fig. (\ref{diffPDF}), and this is reflected in the fact that relative differences in the amplitude near the peak are pushed further to the second and third decimal places. The $C=1$ in fig. \ref{cssfi} represents canonical stochastic single-field inflation scenario while the case $C=1.5$ in fig. \ref{ncssfi} represents the non-canonical stochastic single-field inflation scenario. \textcolor{black}{The peak amplitudes remain close to each other for both the cases, $C=1$ and $C=1.5$, thus keeping perturbativity intact, and to closely examine the tail features, we refer to the plots at the bottom, namely \ref{cssfilog} and \ref{ncssfilog}. The log-scaled versions in these plots provide an interesting outlook into the behaviour of PDFs for large values of ${\cal N}\geq 1$. We infer that already beyond ${\cal N}\sim 0.5$, the LO PDF keeps a higher amplitude with the other higher-order corrections combined, with differences present at the third decimal place. Such a feature remains till we reach ${\cal N}=2$, with no change in the qualitative differences whether the case with $C=1$ or $C=1.5$. The subtle tail features eluded here will prove crucial to understanding the PBH abundance features shortly since PBH formation is quite sensitive to the tail regions of the PDF above. } 
The combined combination of the dominant contribution from the leading residue, $r^{(0)}_{n,i}$ where $i$ specifies the order of expansion, and the $m=0$ poles, $\Lambda_{n}^{(0)}$, quickly give the similar dominant features in the PDF shape across all orders in the fig. \ref{allSSFI}. 
\begin{figure*}[ht!]
    	\centering
    \subfigure[]{
      	\includegraphics[width=8.5cm,height=7.5cm]{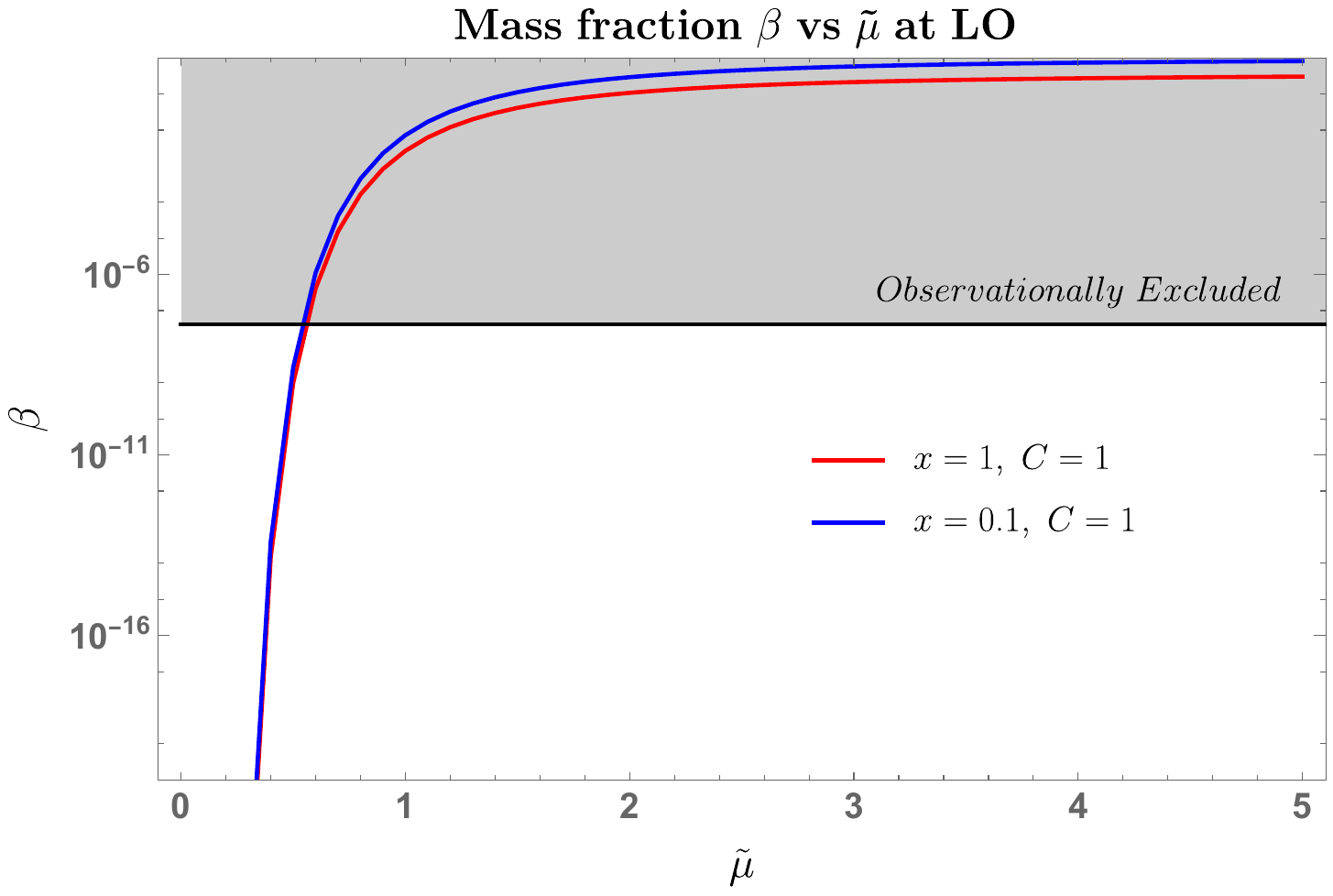}
        \label{massfracLO}
    }
    \subfigure[]{
        \includegraphics[width=8.5cm,height=7.5cm]{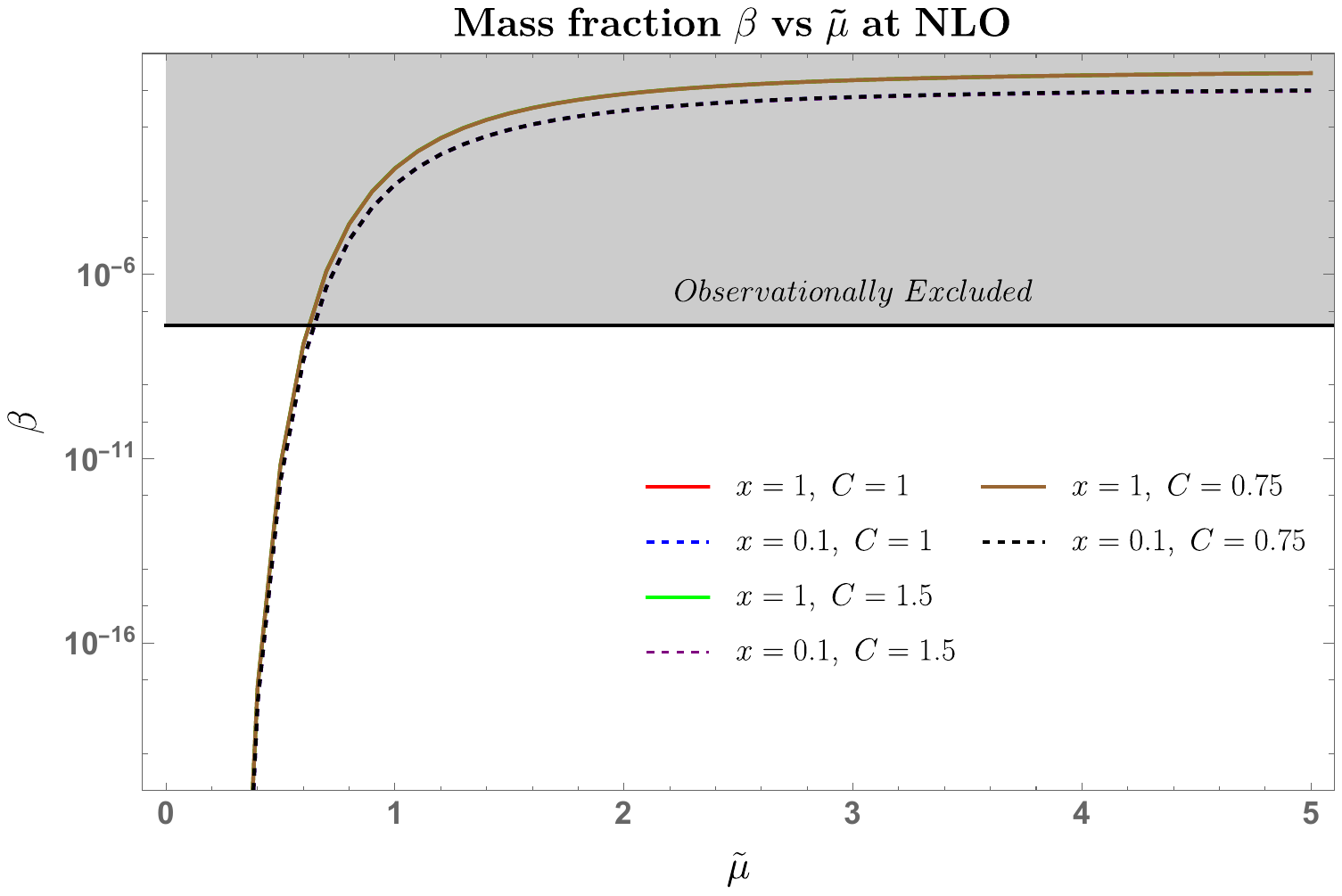}
        \label{mfracNLO}
    }
       \subfigure[]{
        \includegraphics[width=8.5cm,height=7.5cm]{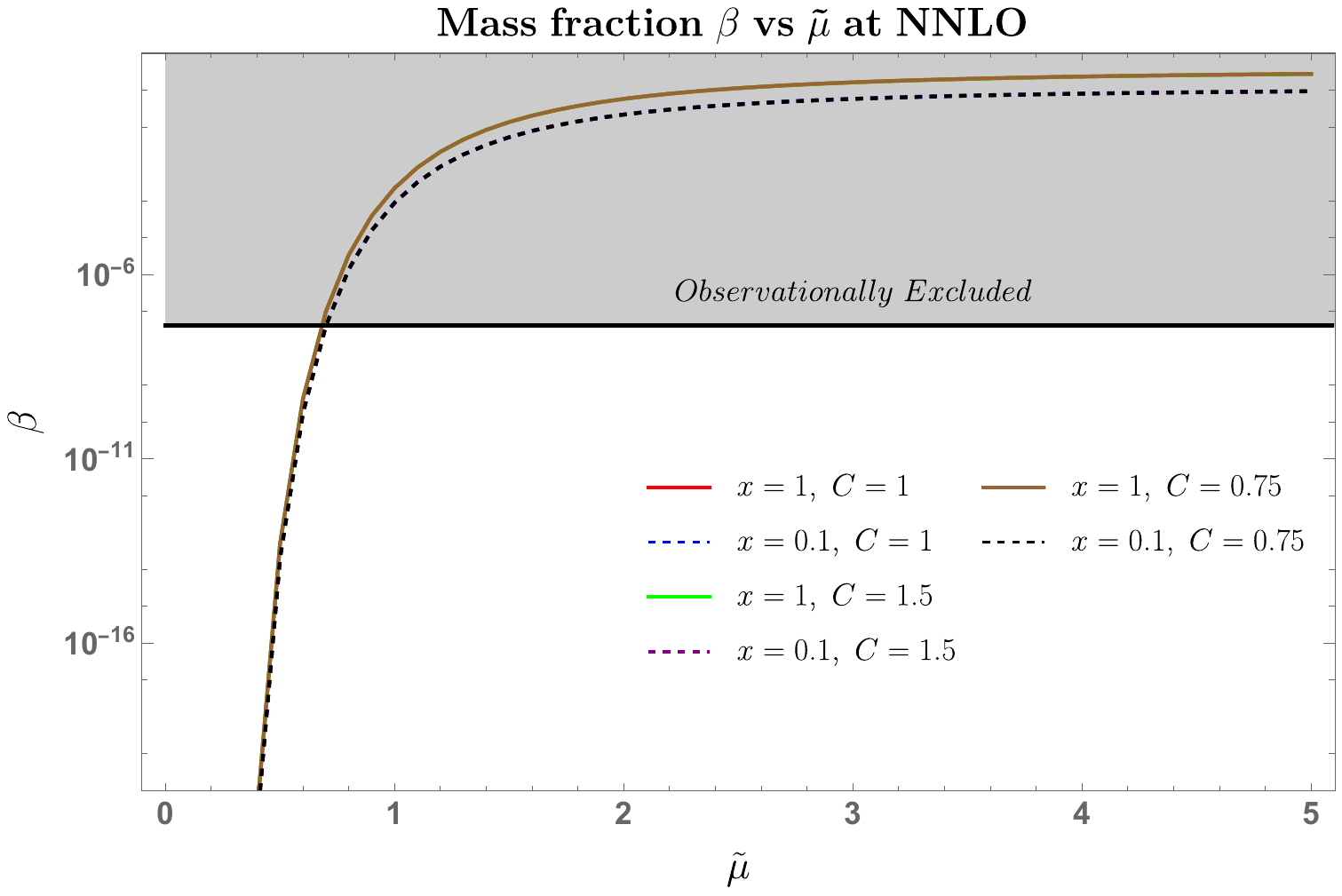}
        \label{massfracNNLO}
    }
    \subfigure[]{
        \includegraphics[width=8.5cm,height=7.5cm]{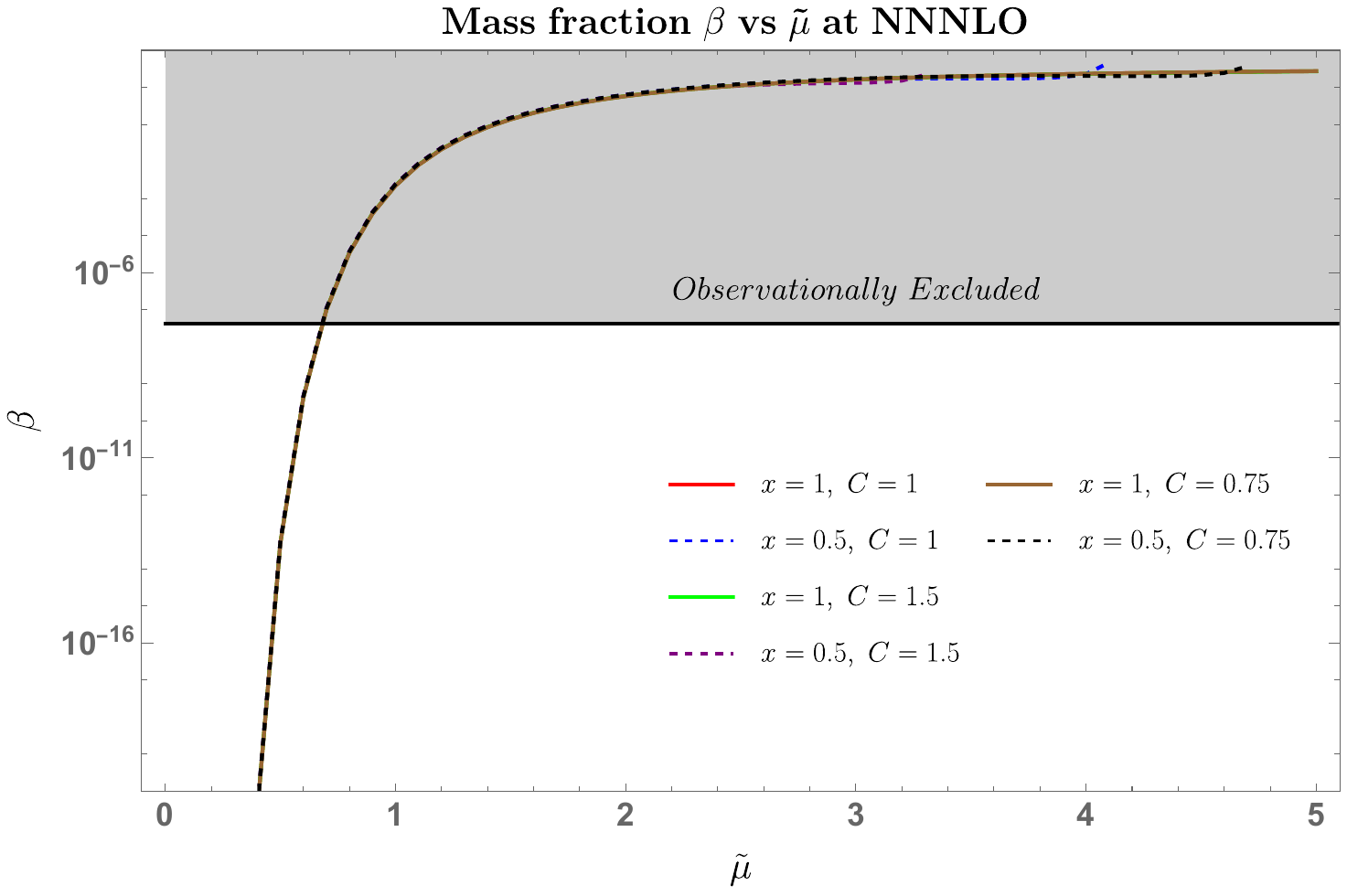}
        \label{massfracNNNLO}
       }
    	\caption[Optional caption for list of figures]{Mass fraction $\beta$ against variation in $\tilde{\mu}$. The mass fraction is extremely sensitive to $\mu$ values close to or less than $1$ and equally less sensitive to changes in $x$. The \textit{top-left} panel displays behaviour of $\beta$ for PDF at LO, $P_{\Gamma}^{\rm LO}({\cal N})$ and does not incorporate a variable $C$. The \textit{top-right} panel displays behaviour of $\beta$ for PDF at NLO, $P_{\Gamma}^{\rm NLO}({\cal N})$, the \textit{bottom-left} panel displays behaviour of $\beta$ for PDF at NNLO, $P_{\Gamma}^{\rm NNLO}({\cal N})$, and the \textit{bottom-right} panel displays behaviour of $\beta$ for PDF at NNNLO, $P_{\Gamma}^{\rm NNNLO}({\cal N})$. Multiple cases with values of $C \in\{0.75,1,1.5\}$ are shown. The gray shaded region highlights the values of $\beta\gtrsim {\cal O}(10^{-8})$ excluded by current observations for heavy mass PBH, in the range $10^{16}$g to $10^{50}$g (or $M_{\rm PBH}\sim {\cal O}(10^{-17}-10^{16})M_{\odot}$). } 
    	\label{betaplot0123}
    \end{figure*}

\begin{figure*}[ht!]
    	\centering
    \subfigure[]{
        \includegraphics[width=8.5cm,height=7.5cm]{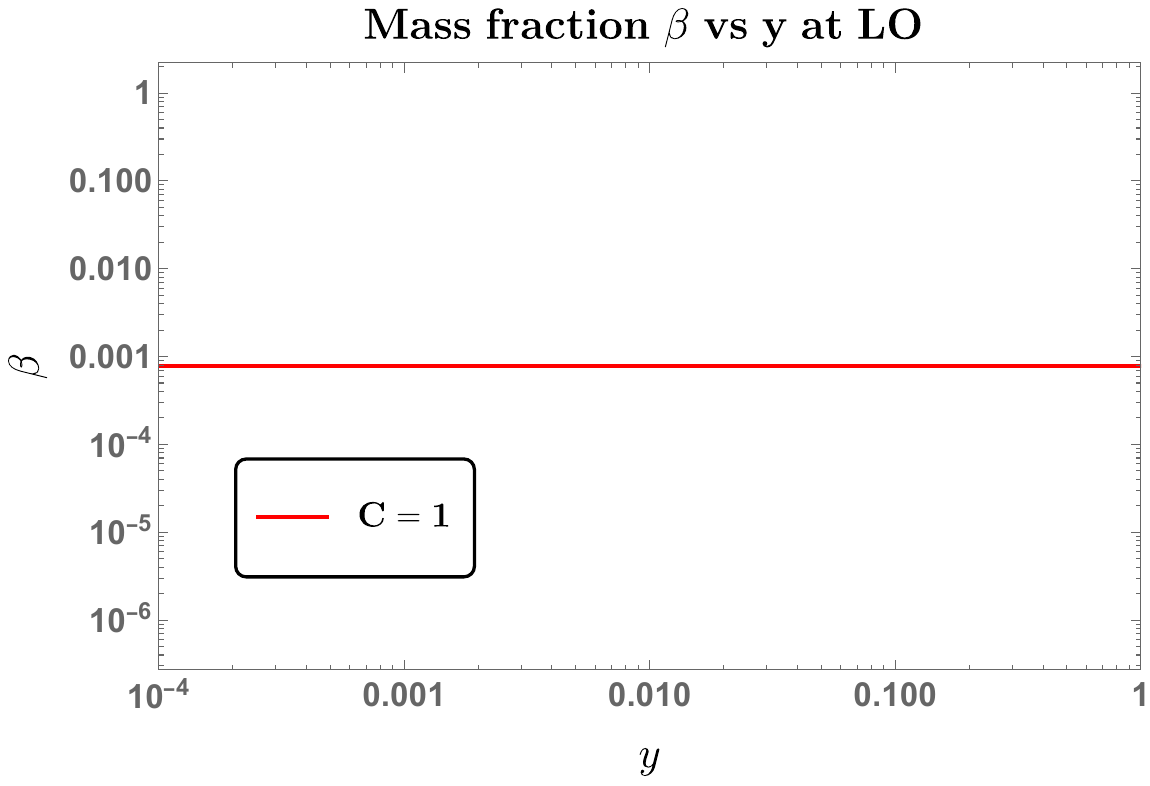}
        \label{massfracLOy}
    }
    \subfigure[]{
        \includegraphics[width=8.5cm,height=7.5cm]{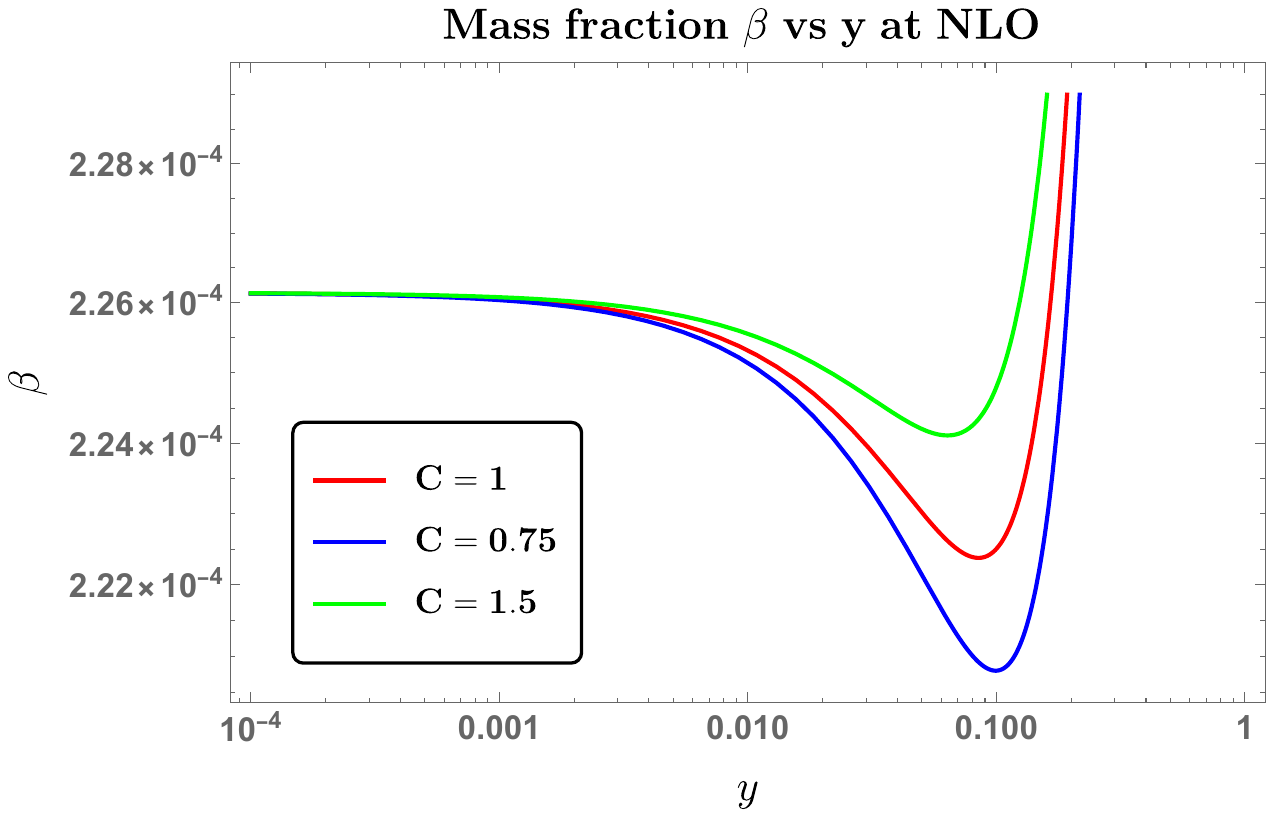}
        \label{mfracNLOy}
    }
       \subfigure[]{
        \includegraphics[width=8.5cm,height=7.5cm]{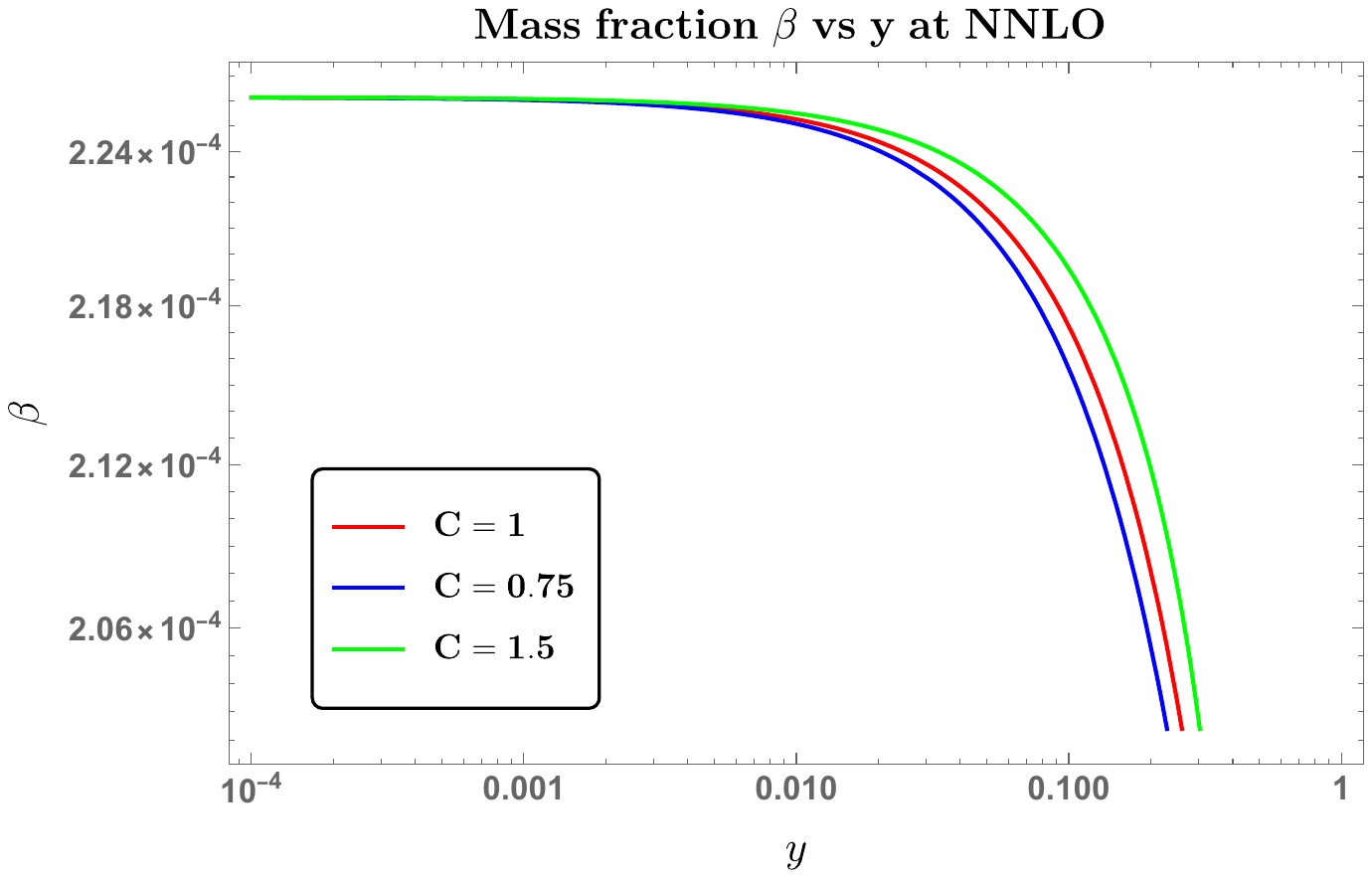}
        \label{massfracNNLOy}
    }
    \subfigure[]{
        \includegraphics[width=8.5cm,height=7.5cm]{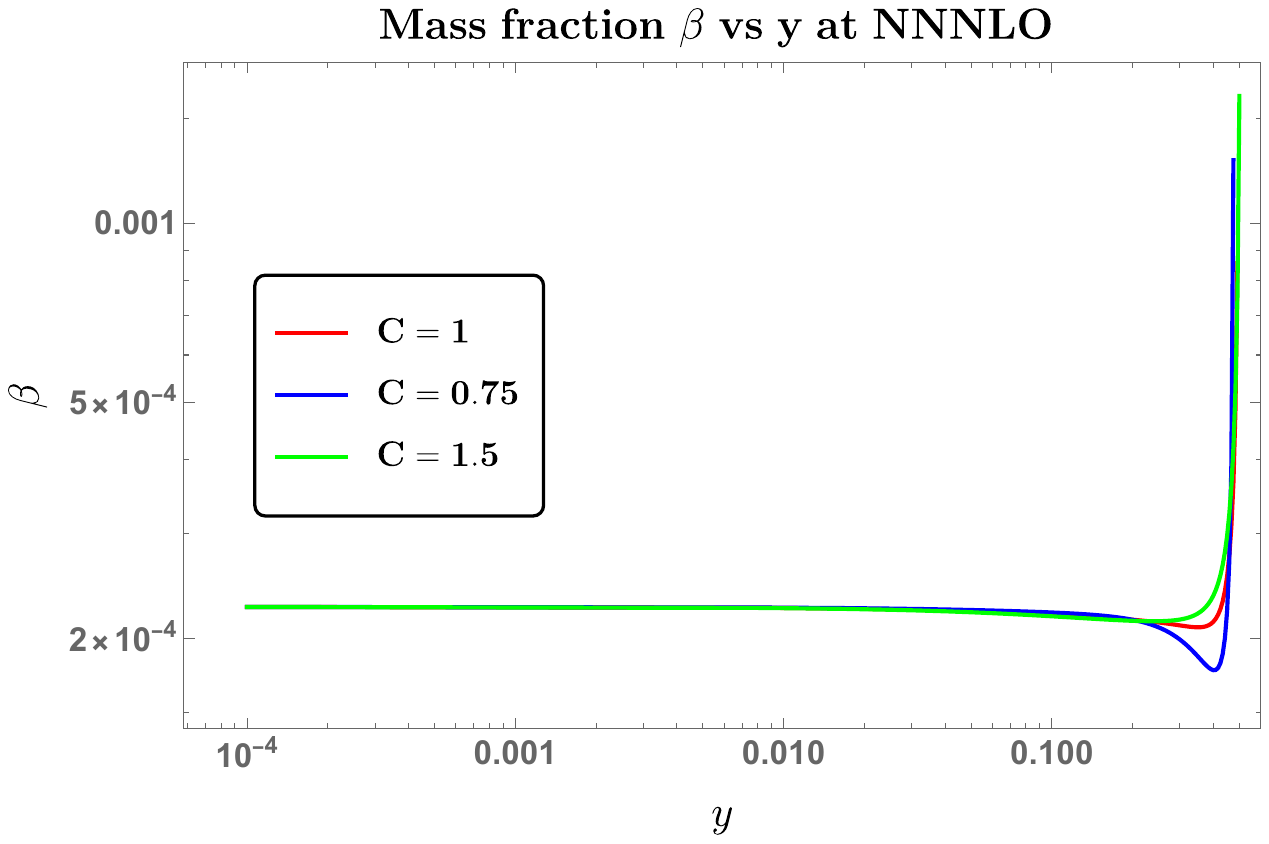}
        \label{massfracNNNLOy}
       }
    	\caption[Optional caption for list of figures]{Mass fraction $\beta$ against variation in $y$.  The \textit{top-left} panel displays behaviour of $\beta$ at LO with changing $y$ and does not incorporate variable $C\ne 1$. The \textit{top-right} panel displays behaviour of $\beta$ at NLO with $y$, the \textit{bottom-left} panel displays behaviour of $\beta$ at NNLO with changing $y$, and the \textit{bottom-right} panel displays behaviour of $\beta$ at NNNLO with changing $y$. Multiple cases with values of $C=1$ (red), $C=0.75$ (blue), and $C=1.5$ (green) are shown and $x=1$ with $\tilde{\mu}=1$ and $\zeta_{\rm th}\sim {\cal O}(1)$ are kept fixed. } 
    	\label{betaplot0123y}
    \end{figure*}
\subsection{Outcomes from PBH mass fraction $\beta$}

\begin{figure*}[ht!]
    	\centering
    \subfigure[]{
      	\includegraphics[width=8.5cm,height=7.5cm]{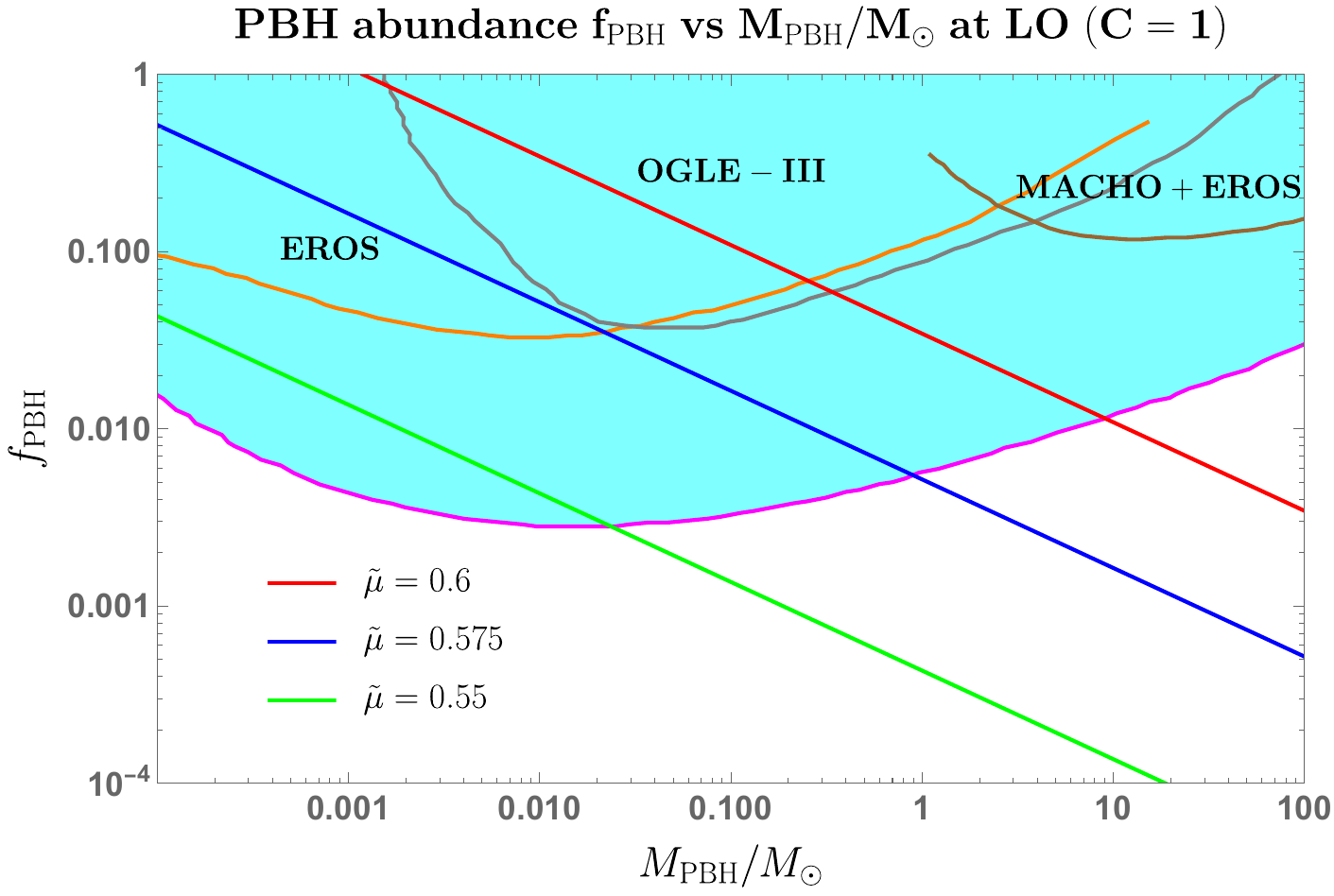}
        \label{fpbhLO}
    }
    \subfigure[]{
        \includegraphics[width=8.5cm,height=7.5cm]{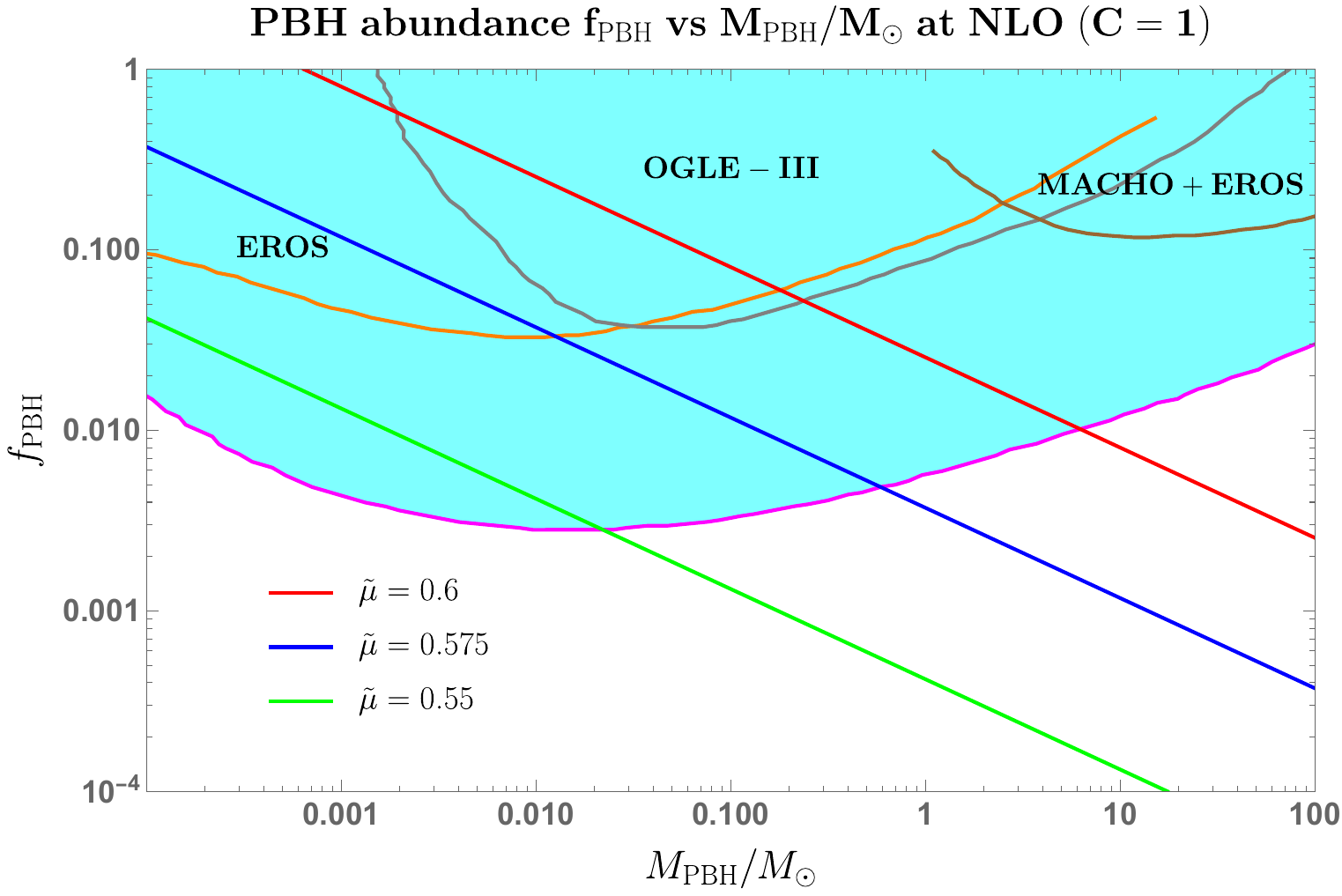}
        \label{fpbhNLO}
    }
       \subfigure[]{
        \includegraphics[width=8.5cm,height=7.5cm]{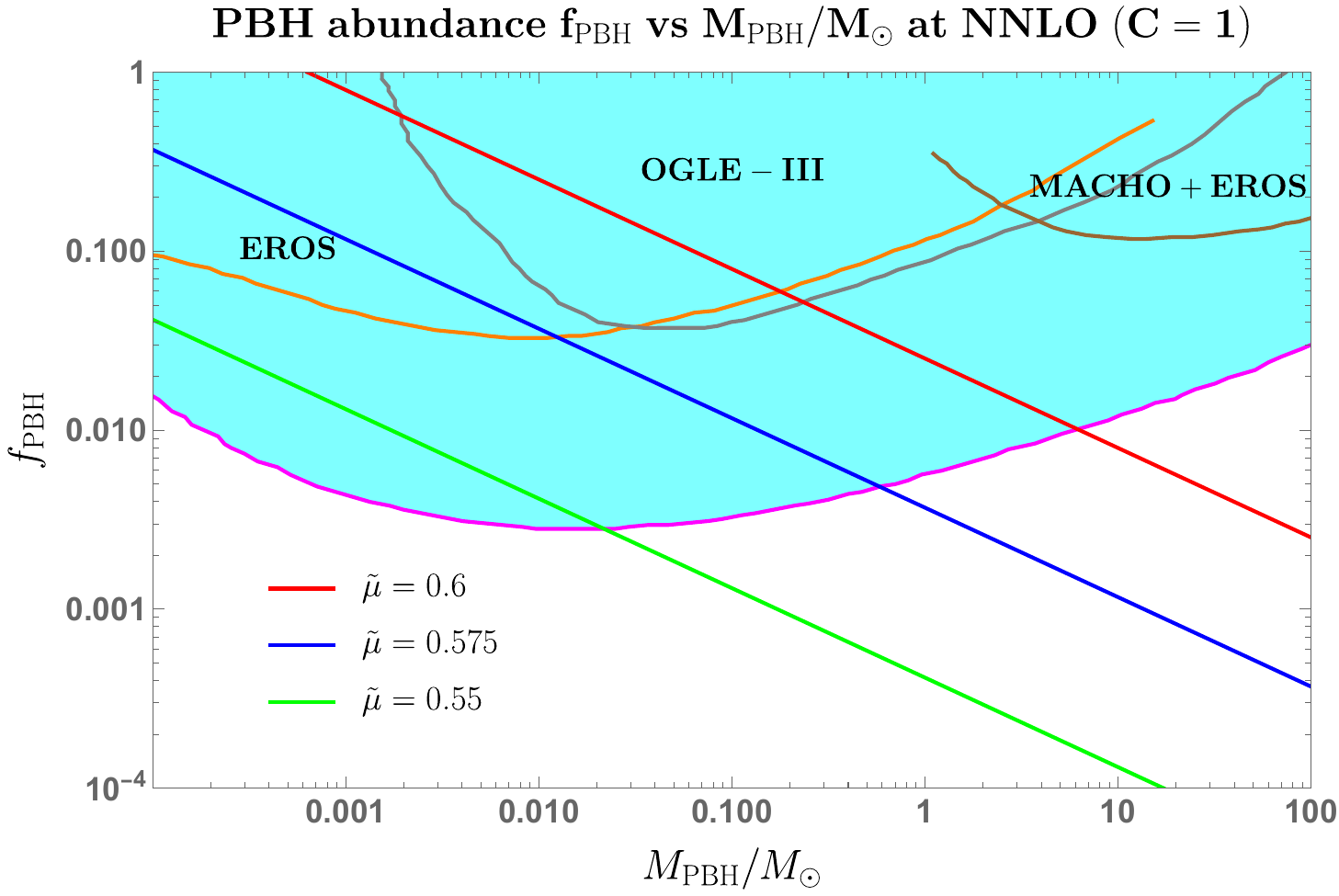}
        \label{fpbhNNLO}
    }
    \subfigure[]{
        \includegraphics[width=8.5cm,height=7.5cm]{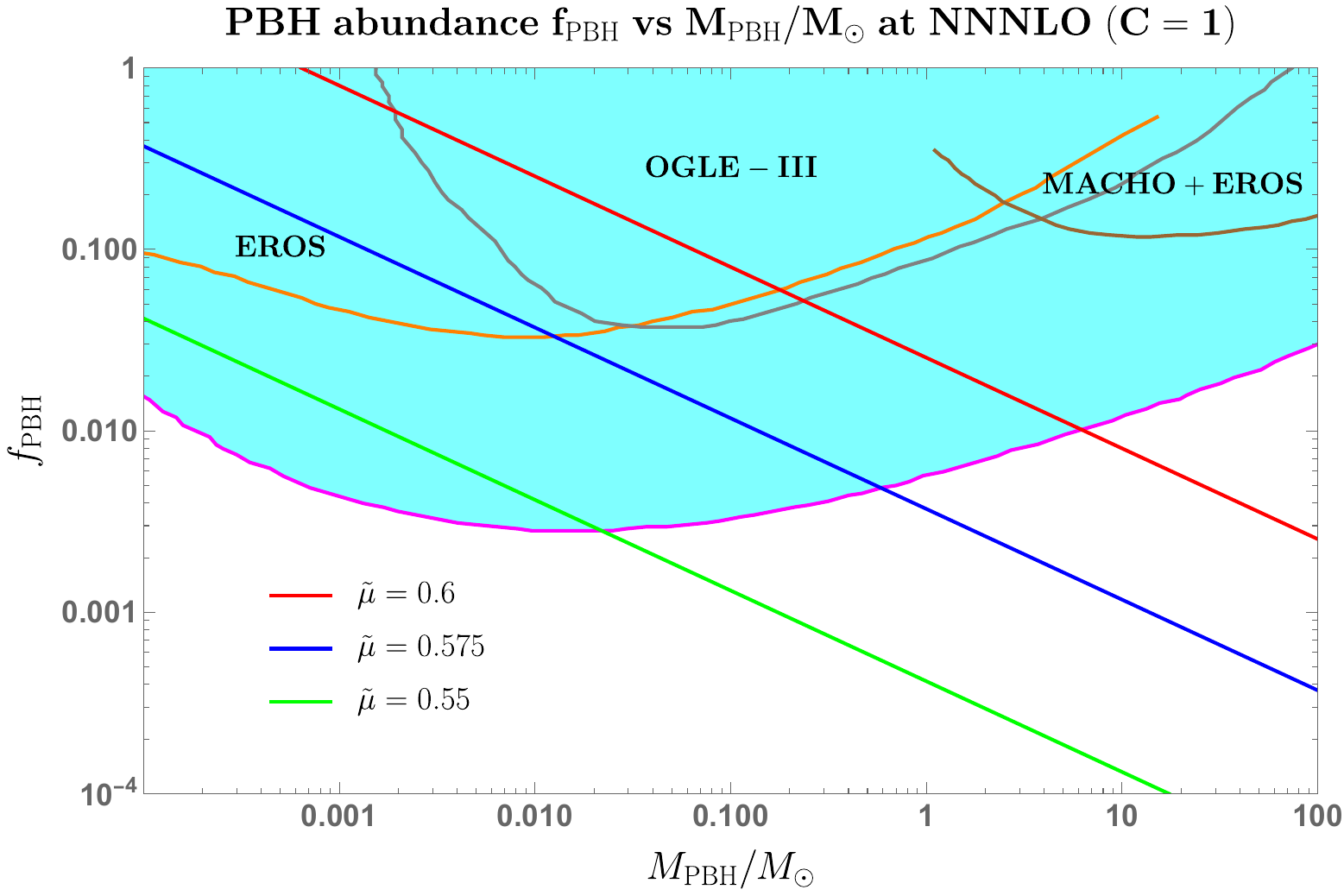}
        \label{fpbhNNNLO}
       }
    	\caption[Optional caption for list of figures]{PBH abundance $f_{\rm PBH}$ as a function of the mass (in $M_{\odot}$). The fixed values of different parameters are $x=1$, $y=0.06$, and $C=1$. Each plot features different values of $\mu\in \{0.55,0.575,0.6\}$. The cyan-coloured region highlights the recently obtained $95\%$ upper limits on PBH abundance from microlensing events. The magenta boundary marks the strict limits on $f_{\rm PBH}$ \cite{Mroz:2024mse} and also includes limits from other dark matter surveys: EROS (orange) \cite{EROS-2:2006ryy}, OGLE-III (gray) \cite{wyrzykowski2011ogle}, MACHO+EROS (brown) \cite{Blaineau:2022nhy}. } 
    	\label{fpbhplot0123}
    \end{figure*}

In this section we analyze the results for the PBH mass fraction and abundance based on the analytical treatment of the distribution functions for the diffusion-dominated regime. The fig. \ref{betaplot0123} features behaviour of the PBH mass fraction for changing values of $\tilde{\mu}$. Here, $\beta$ experiences steep change in its value for $\tilde{\mu}\leq 1$ and after $\tilde{\mu}$ becomes greater than $1$, the mass fraction quickly saturates to values close to but less than $1$. This time changes in $x$ value have a small but noticeable impact on $\beta$ and the effects only become distinguishable once $\tilde{\mu}\geq 1$. The left-panel (\ref{massfracLO}) shows $\beta$ when PDF at leading order is considered. It displays the scenario of canonical stochastic single-field inflation. The right-panel (\ref{mfracNLO}) focuses on $\beta$ from the PDF at the next-to-leading order. The effects of various $C$ are not significant enough to distinguish but we notice that $\beta$ at NLO, for $\tilde{\mu}\geq 1$, changes by an order of magnitude amount less as $x$ is decreased compared to the changes in $\beta$ at LO. Also, the PDF at NLO predicts less mass fraction compared to the PDF at LO for the same $\tilde{\mu}$ value. 

Focusing on the next order in the perturbative expansion, we get the mass fraction behaviour from the PDF at NNLO in the bottom-left, \ref{massfracNNLO}. The mass fraction $\beta$ suffers a further decrease of more than an order of magnitude as $\tilde{\mu}\leq 1$; with this rate of decline improved compared to NLO for smaller $\tilde{\mu}$ values. Such effects will also contribute towards the resulting PBH abundance from the mass fraction. Similar to the previous case at NLO, the effects coming from the characteristic parameter $C$ in the mass fraction are not significant enough to distinguish properly. Lastly, some interesting changes happen with the PDF at NNNLO and its derived $\beta$ behaviour. From figure \ref{massfracNNNLO} for $\beta$ at NNNLO, we notice that for $x=0.5$, a positive magnitude of $\beta$ lasts up to a certain value of $\tilde{\mu}>1$ after which the numerical values return negative results. Thus, $x$ cannot be decreased to smaller values close to zero as doing so would only push $\beta$ to negative values faster and for values of $\mu\sim 1$. The relative change after decreasing $x$ from $x=1$ has also almost vanished. The effects of the characteristic parameter $C$ is more interesting to notice here. Increasing $C$ beyond $C=1$ along with $x<1$, results with the corresponding $\beta$ going negative for lower values of $\tilde{\mu}$ as compared to lowering $C$ than $C=1$, in which case the $\beta$ extends to larger $\tilde{\mu}$ above which it again drops to negative values which are not visible. No such issue arises when $x\gtrsim 1$ is considered.        

In the fig. \ref{betaplot0123y}, we explore the behaviour of the mass fraction $\beta$, at each order in the perturbative expansion, with changing values of $y$ or the coarse-grained conjugate momentum variable as defined before in eqn. (\ref{newphasevars}). Since this behaviour is studied using the features of diffusion-dominated regime, see section \ref{s11}, it follows that the limit of $y\ll 1$ would provide for much suitable interval from where PBH mass fraction can get estimated accurately, including higher-order corrections of NNNLO analysis. The leading order (LO) scenario is an expansion independent of $y$, hence, any change with $y$ in the top-left panel \ref{massfracLOy} would not give any useful insights into the mass fraction behaviour. Going to the NLO scenario, here we get additional terms linear in $y$ to our PDF and as a consequence we observe, from the top-right panel, fig. \ref{mfracNLOy}, small changes to $\beta$. In terms of different $C$ values, the relative separation becomes larger as we set $y> 0.01$. The mass fraction then changes quickly by at least an order of magnitude, increasing till we reach $y\sim 1$. The effect of sub-dominant corrections changes as we consider the NNLO case, fig. \ref{massfracNNLOy}. Here, the mass fraction $\beta$ continues to move further lower in magnitude as $y>0.01$ is considered. Mass fraction for the lower values, $C<1$, fall early as compared to the higher values, $C\geq 1$. At last we consider the correction up to NNNLO to $\beta$ and analyse their effects from fig. \ref{massfracNNNLOy}. This time, the mass fraction stays the same for a large interval of $y<0.1$. Even while taking multiple $C$ into account, the relative difference only becomes significant if we choose to have $y>0.5$. The case of canonical stochastic single-field $(C=1)$ does not show much differences and only changes sharply by an order of magnitude when we reach $y\sim 1$. The mass fraction converges to have similar behaviour for different $C$ values in the $(y\ll 1)$ and $y\sim 1$ limits. 

We conclude by saying that the choice of interval $y\sim {\cal O}(0.01)$ agrees well after comparison between each higher-order correction to the mass fraction $\beta$. For $y\geq 0.1$, the behaviour can change non-trivially and depending on the order of expansion considered.

\subsection{Outcomes from PBH abundance $f_{PBH}$}

In this section we discuss the observed outcomes for the PBH abundance as function of the PBH mass after the magnitude of curvature perturbations exceed a threshold value of $\zeta_{\rm th}\sim {\cal O}(1).$

\begin{figure*}[htb!]
    	\centering
    \subfigure[]{
      	\includegraphics[width=8.5cm,height=7.5cm]{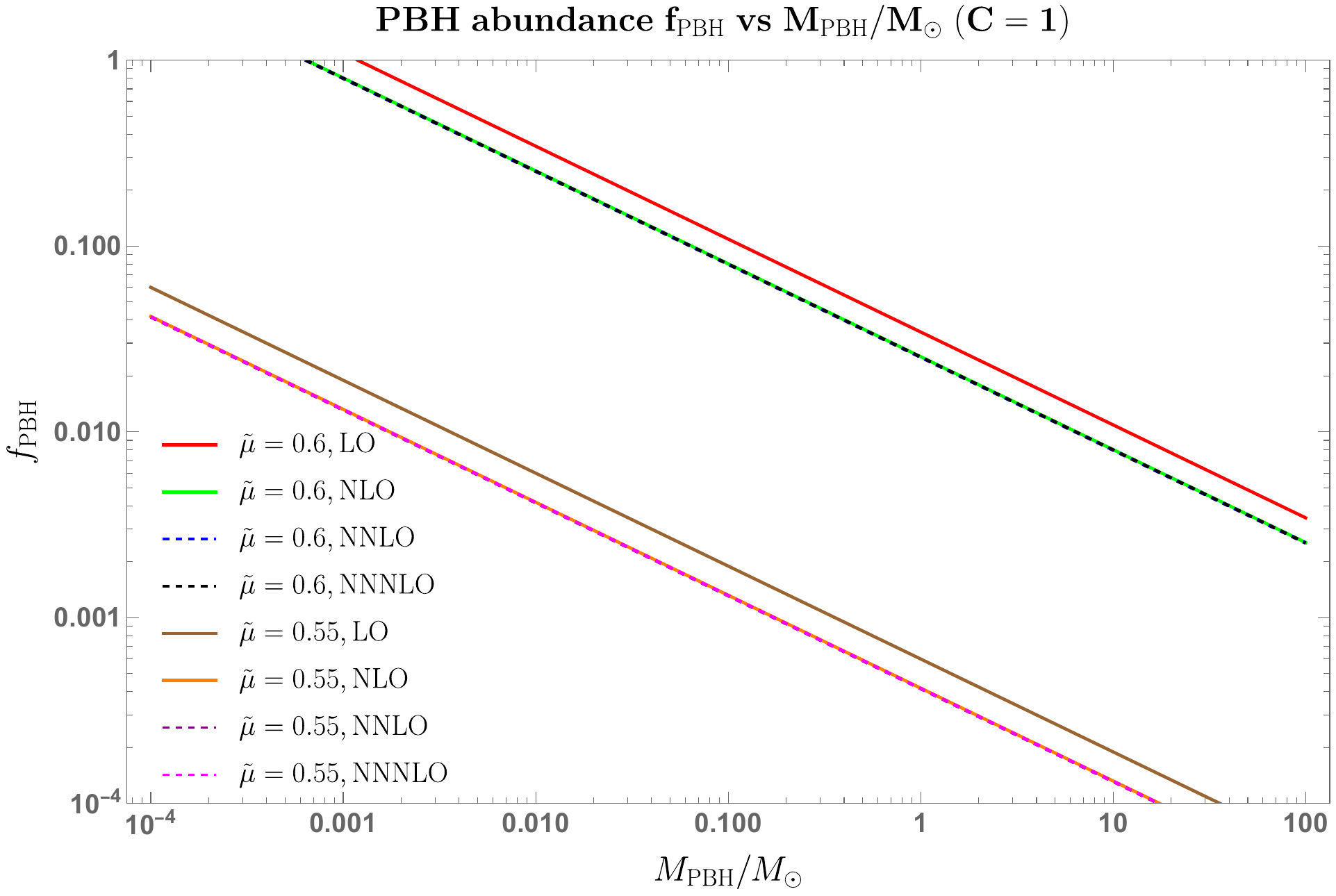}
        \label{fpbhALL}
    }
    \subfigure[]{
        \includegraphics[width=8.5cm,height=7.5cm]{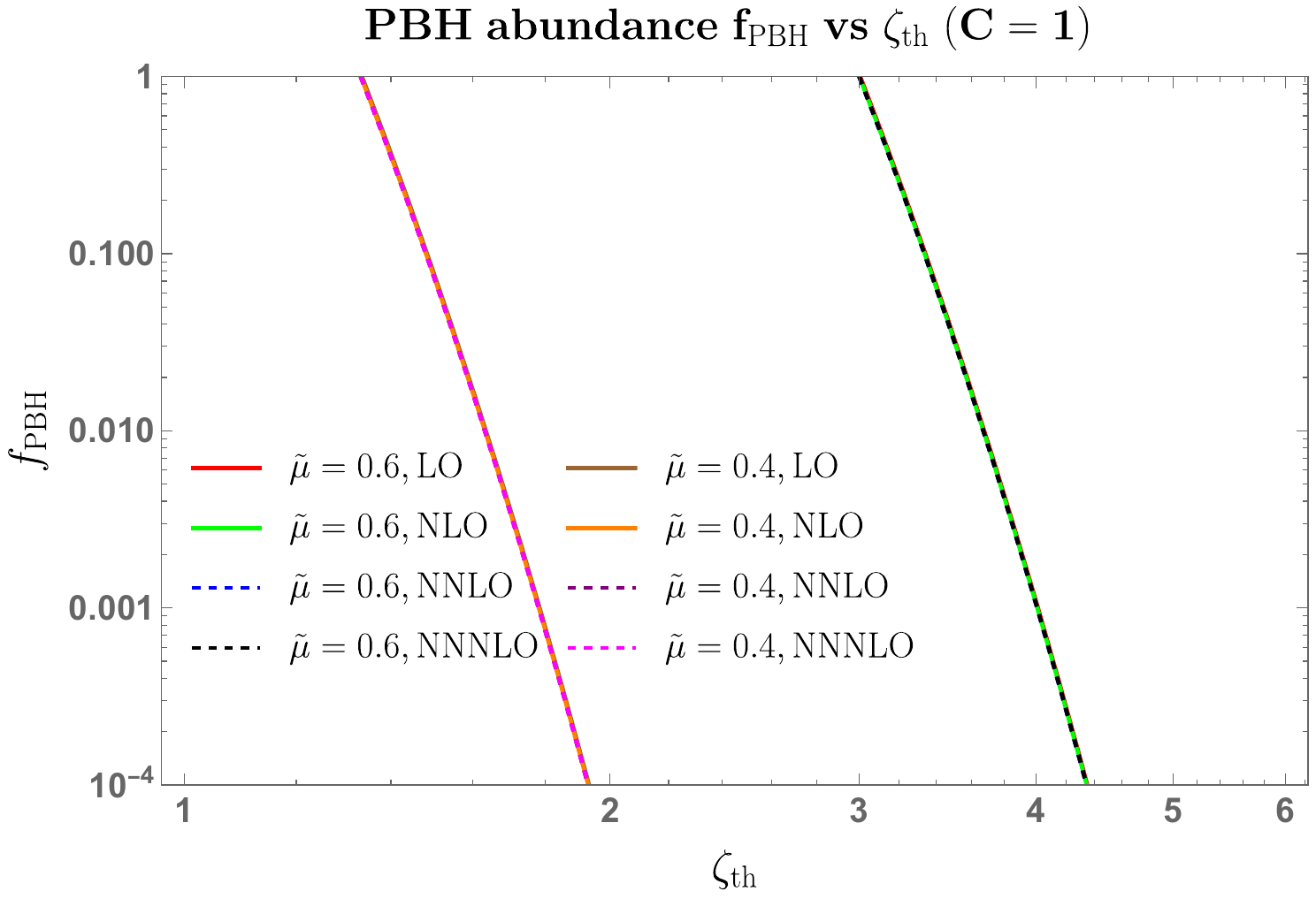}
        \label{fpbhALL2}
    }
    	\caption[Optional caption for list of figures]{\textit{left-panel} displays PBH abundance $f_{\rm PBH}$ as function of the mass (in $M_{\odot}$) including all orders of the perturbative expansion. \textit{right-panel} shows $f_{\rm PBH}$ as function of the threshold $\zeta_{\rm th}$ including all order of the perturbative expansion. The figures are for a fixed set of $\tilde{\mu} \in \{0.55,0.6\}$ and $C=1$. } 
    	\label{fpbhplotab}
    \end{figure*}

The figure \ref{fpbhplot0123} features in it the abundance of a spectrum of PBH masses ranging from $M_{\rm PBH}\sim {\cal O}(10^{-4}M_{\odot})$ to $M_{\rm PBH}\sim {\cal O}(10^{2}M_{\odot})$. The plots shows that the abundance is extremely sensitive to the values of $\tilde{\mu}$ for a fixed $x=1$, $y=0.06$, $C=1$ and keeping $\zeta_{\rm th}\sim {\cal O}(1)$ satisfied throughout. With the distribution function at different orders we observe from the various sub-figures, \ref{fpbhLO}, \ref{fpbhNLO}, \ref{fpbhNNLO}, \ref{fpbhNNNLO}, the scenario of a large enough mass range that can achieve sizeable PBH abundance as a result. The higher we go in the perturbation expansion, their influence on the abundance remains the same with relatively small deviations throughout the mass range being explored. We include the results obtained from different microlensing surveys on the fraction of total dark matter distributed in form of PBHs. The cyan coloured region with solid magenta line marks the region constrained from the data of Optical Gravitational Lensing Experiment (OGLE) in their $20$ years run which comprises of the  OGLE-III $(2001-2009)$ and OGLE-IV $(2010-2020)$ runs. For more details on the observation setup and proper data generation analysis consider studies in \cite{Mroz:2024wag,Mroz:2024mse}. From the strict constraints, we infer that large $M_{\rm PBH}\sim {\cal O}(1-100)M_{\odot}$ can compose at most $10\%$ fraction of the total dark matter.    
The sensitivity to the corrections in PDF at each order can be seen more clearly from the plot in fig. \ref{fpbhALL}. The leading order (LO) contribution predicts a larger magnitude of abundance relative to the next-to-leading order (NLO) corrections coming from the associated PDF, and the subsequent corrections do not bring much noticeable changes. 

The threshold curvature perturbation $\zeta_{\rm th}$ is another important parameter to manipulate. Changes in its value show high sensitivity to the possible $\tilde{\mu}$ values, with the larger $\tilde{\mu}\geq 0.5$ being more sensitive to $\zeta_{\rm th}$ resulting in a sizeable abundance than compared to lower values of $\tilde{\mu}\leq 0.5$. 
The right-panel, fig. \ref{fpbhALL2}, features the effect of changing $\zeta_{\rm th}$ for a fixed $\tilde{\mu}$ and fixed PBH mass, $M_{\rm PBH}\sim {\cal O}(10^{-3}M_{\odot})$, including corrections from all cases in the perturbative analysis of the PDF. Effects of threshold change is found to be independent of the order of correction considered in the PDF expansion and likewise in the PBH abundance which follows, as a result, similar decreasing trend upon increasing the threshold. From the slope of \ref{fpbhALL}, we can also interpret that changing $\tilde{\mu}$ affects the abundance for different masses in the same manner for a given $C$ value. We conclude with the fact that additional perturbative corrections to the PDF bring in almost negligible changes into the abundance estimates for a large spectrum of PBH masses and the quantity $f_{\rm PBH}$ is highly sensitive to changes in $\tilde{\mu}$ which affects all PBH masses equally for a certain threshold $\zeta_{\rm th}$.

\section{Conclusion}
\label{s15}

\begin{figure*}[htb!]
    	\centering
    {
       \includegraphics[width=19cm,height=12.5cm]{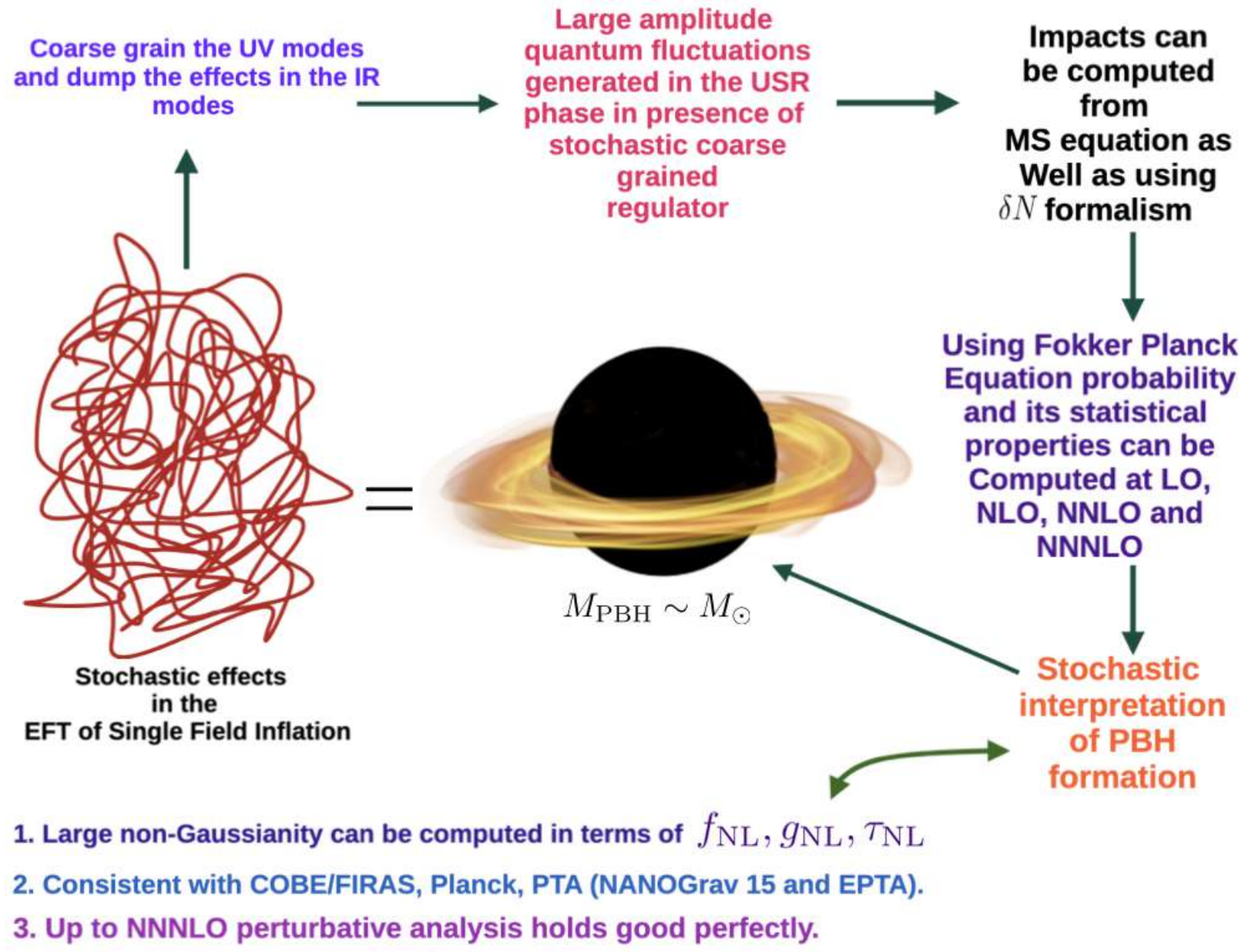}
        \label{conclusion}
    } 
    \caption[Optional caption for list of figures]{Illustrative diagram describing the main features and results from this work. The results from this work include computation of various non-Gaussianity parameters, $f_{\rm NL}, g_{\rm NL}, \tau_{\rm NL}$, following perturbative corrections coming from each order in the diffusion-dominated regime.   }
\label{stochasticdiag3}
    \end{figure*}

The focus of this work is threefold. We attempt to develop an Effective Field Theory (EFT) understanding of stochastic single-field inflation. In parallel with the stochastic-$\delta N$ formalism, we perform a detailed analysis of the non-Gaussianity features in the drift-dominated regime and the tail features responsible for PBH formation when in the diffusion-dominated regime. Finally, we study the tree-level scalar power spectrum, both auto-correlated and cross-correlated amplitudes, employing the stochastic inflation formalism.


We construct the EFT description for the long-wavelength part of the curvature perturbations, following closely the analysis in refs.\cite{Grain:2017dqa, Vennin:2020kng}, and solve for Hamilton's equations of motion for the coarse-grained variables, $\{\zeta,\Pi_{\zeta}\}$. Following earlier developments, we also discuss the essential modifications in the adjoint Fokker-Planck equation driving the probability distribution of the stochastic duration variable ${\cal N}$. The consequence of working in the EFT language comes with a new characteristic parameter, $C$, in the adjoint Fokker-Planck equation, whose values describe a class of models, either canonical $(C=1)$ or non-canonical $(C\ne 1)$, in stochastic single-field inflation. The parameter $C$ is shown to be determined by a combination of slow-roll parameters, which ultimately carry the EFT picture into the remainder of our analysis concerning both non-Gaussianity and PBH formation. Before continuing with this detailed analysis, we outline a comparative study focusing on the merits of the stochastic-$\delta N$ formalism and how this new integrated version of the two different frameworks, \textcolor{black}{that was initially developed in refs.\cite{Fujita:2013cna,Fujita:2014tja,Vennin:2015hra} to solve for multiple realizations of the Langevin equation,} will help us in our later computations in the drift and diffusion-dominated regimes. 

\textcolor{black}{We follow the analysis of ref.\cite{Pattison:2021oen}, that studied the implications of higher-order corrections on PBH mass fraction, and further highlighted the results on the stochastic variable, ${\cal N}$, statistics till the NLO.} In the drift-dominated regime, we carried out the analysis up to the next-to-next-to-next-to-leading order (NNNLO) that showed $C$ dependent corrections coming into each order version of the non-Gaussianity parameters, $f_{\rm NL}, g_{\rm NL}, \tau_{\rm NL}$, and equivalently into the corresponding statistical moments of the stochastic e-folds ${\cal N}$. The probability distribution function (PDF) of the e-folds ${\cal N}$ was provided for each order of the analysis, and the effects of the parameter $C$ on the PDF shape at each order via measures such as skewness and kurtosis are also provided. 

In the diffusion-dominated regime, \textcolor{black}{a similar but more sophisticated perturbative analysis inspired by the refs.\cite{Ezquiaga:2019ftu,Pattison:2021oen}, was presented up to the NNNLO for the probability distribution function}. Though explicit expressions become highly cumbersome beyond the next-to-leading order (NLO), numerical results are possible to design. The perturbative expansion is realised through the figure \ref{allSSFI} and holds perfectly for this regime. The analytic expressions for the mean number of e-folds are given at each order, which later appears to contribute to the PBH mass fraction using stochastic-$\delta N$ formalism. 

The implications of corrections up to NNNLO,
in the diffusion-dominated scenario, on PBH formation, shows an order of decrease in the magnitude of mass fraction $\beta$ when the other parameter $\tilde{\mu}$, containing the tree-level scalar power spectrum, remains of $\tilde{\mu}\gtrsim 1$. However, for $\tilde{\mu}< 1$, the mass fraction suffers a steep decrease, a change of at least two orders and more in magnitude, as we go towards the lower values for $\tilde{\mu}$. The above observation shows that taking into account the higher-order corrections can reduce the mass fraction to favourable estimates in a controllable manner, see figure \ref{betaplot0123} for respective details. Changes from different $C$ are not discernible from the numerical outcomes at each order as they remain very small. The current-day PBH abundance outcomes also reflect the similar effect of taking higher-order corrections into account. The abundance reduces equally for a range of PBH masses, $M_{\rm PBH}\sim {\cal O}(10^{-4}-10^{2} M_{\odot})$, and further corrections from NNLO and NNNLO do not bring in  visible changes; see figure \ref{fpbhplotab} for the same. 

We also provide explicit expressions for the tree-level scalar power spectrum after including stochastic effects for each possible auto and cross-correlation amplitude. Here, the stochastic parameter $\sigma$, where $\sigma\ll 1$, is shown to play a crucial role by acting as a regulator in the presence of large quantum fluctuations and promoting controlled behaviour of the correlations at the transition junctions from SRI to USR and USR to SRII, which we outline. The impact of different effective sound speed values, $c_{s}$, is also studied for each correlation. We find that $c_{s}=1$ manages to provide a correlation amplitude of, $\Delta^{2}_{\zeta\zeta}\sim {\cal O}(10^{-2})$, in the USR that also enables PBH production, and upon lowering $c_{s}$ within the constraint coming from experimental analysis, $0.024 \leq c_{s}< 1$, reduces the amplitude to, $\Delta^{2}_{\zeta\zeta}\sim {\cal O}(10^{-3})$. The above allows us to infer that by keeping the causality and unitarity constraints satisfied, with the value of $c_{s}$ in the mentioned interval, the correlation amplitude always tends to decrease, which also implies preserving the underlying perturbativity assumptions. As an aside, we discuss some important features of the $\sigma$ parameter when quantum loop corrections get treated as a genuine concern from each of the mentioned possible power spectrum correlations. On the observational side, we also explore the spectral distortion effects, which require the use of the tree-level scalar power spectrum. As a result of the strong constraints on the power spectrum amplitude from such effects, the case of PBH mass with $M_{\rm PBH}\gtrsim 1.8\times10^{3}M_{\odot}$ remains excluded.

Figure \ref{stochasticdiag3} pictorially summarizes the results of this work and lists out the key findings from observations in relation to PBHs and the non-Gaussianity parameters. We have a plan to carry forward our analysis by incorporating the quantum loop effects in the presence of stochasticity in the soft de Sitter EFT framework using both Schwinger Keldysh (in-in) and stochastic-$\delta {\cal N}$ formalism. In this extended version of the work, we have a further plan to study the issues related to renormalization, DRG resummation, and its connection to the stochastic-$\delta {\cal N}$ formalism.

\section*{Acknowledgement}
SC would like to thank the National Academy of Sciences (NASI), Prayagraj, India, for being elected as a member of the academy. SC would like to especially thank Soumitra SenGupta and Supratik Pal for inviting to IACS, Kolkata, and ISI, Kolkata, during the work. Additionally, SC thanks Supratik Pal and his students for inviting to give an inaugural plenary talk at the discussion meeting titled, {\it Cosmo Mingle}, where part of the work was presented. SC would also like to thank all the members of Quantum Aspects of the Space-Time \& Matter
(QASTM) for elaborative discussions. MS is supported by Science
and Engineering Research Board (SERB), DST, Government of India under the Grant Agreement number CRG/2022/004120 (Core Research Grant). MS is
also partially supported by the Ministry of Education
and Science of the Republic of Kazakhstan, Grant No.
0118RK00935, and CAS President’s International Fellowship Initiative (PIFI). We would like to thank 
Jens Chluba, Przemek Mroz, and Jun'ichi Yokoyama for useful comments and suggestions. Last but not least, we would like to acknowledge our debt to the people
belonging to the various parts of the world for their generous and steady support for research in natural sciences.

\newpage

\appendix
\section*{Appendix} 

\section{Fokker-Planck equation from the Langevin equation}   \label{app:A}

In this section of the appendix, we focus on briefly discussing the derivation of the Fokker-Planck equation from the Langevin equation. The Fokker-Planck is useful to describe the evolution of the probability distribution of the field variables in the phase space as they evolve during inflation from an initial condition at some arbitrary moment in time to a final field configuration at some another moment later, usually chosen where inflation ends. 

To initiate we require the use of a probability rate for the system to start from a given initial field configuration and land to some infinitesimal increment away into a different field configuration in a small increment in the time parameter.
To this effect, we introduce the transition probability rate $W_{\Delta {\bf \Gamma}}({\bf \Gamma},N)$ in terms of the field variables, ${\bf \Gamma} = \{\zeta , \Pi_\zeta\}$, by the expression: 
\bea
W_{\Delta {\bf \Gamma}}({\bf \Gamma},N)\del N = P({\bf \Gamma} +\Delta {\bf \Gamma},N +\delta N|{\bf \Gamma}, N),
\eea
which corresponds to the probability where the system with ${\bf \Gamma}$ at time $N$ evolves to ${\bf \Gamma}+\Delta {\bf \Gamma}$ at $ N+\del N $ in a $\del N $ infinitesimal increment in time. Using this, equation for the probability $P({\bf \Gamma},N)$ to arrive at a field configuration ${\bf \Gamma}$ starting from some initial configuration ${\bf \Gamma}_{\rm in}$ is given by: 
\bea \label{Prob1}
\frac{\partial}{\partial N}P({\bf \Gamma},N)=\int d\Delta{\bf \Gamma}\Bigg[W_{\Delta{\bf \Gamma}}({\bf \Gamma}-\Delta {\bf \Gamma},N)P({\bf \Gamma}-\Delta {\bf \Gamma},N)-W_{-\Delta {\bf \Gamma}}( {\bf \Gamma},N)P( {\bf \Gamma},N)\Bigg].
\eea
This captures the increase by the first term in going from ${\bf \Gamma}-\Delta {\bf \Gamma}$ to ${\bf \Gamma}$, and the decrease from the second term as we go from ${\bf \Gamma}$ to ${\bf \Gamma} -\Delta {\bf \Gamma} $, and later integrated over the increments $\Delta {\bf \Gamma}$. If we Taylor expand the first term in the integrand of eqn. (\ref{Prob1}), we notice: 
\bea
W_{\Delta{\bf \Gamma}}({\bf \Gamma}-\Delta {\bf \Gamma},N)P({\bf \Gamma}-\Delta {\bf \Gamma},N)=W_{\Delta{\bf \Gamma}}({\bf \Gamma},N)P({\bf \Gamma},N)+\Bigg( -\Delta \Gamma_i\frac{\partial}{\partial \Gamma_i}+ \frac{1}{2}\Delta \Gamma_i \Delta \Gamma_j \frac{\partial^2}{\partial \Gamma _i \partial \Gamma_j}\nonumber \\
+........+\frac{(-1)^l}{l!}\Delta \Gamma_i \Delta \Gamma_j......\Delta \Gamma_l\frac{\partial^l}{\partial \Gamma_i \partial \Gamma_j...\partial \Gamma _l}+.....\Bigg)[W_{\Delta{\bf \Gamma}}({\bf \Gamma},N)P({\bf \Gamma},N)].
\eea
and the dummy indices between the field variables, $\Gamma_{i}=\{\zeta_{i},\Pi_{\zeta,i}\}$, are summed over. The integration variable of the second term in eqn. (\ref{Prob1}) can be redefined by $\Delta {\bf \Gamma}\rightarrow - \Delta {\bf \Gamma}$ and this enables the following expression: 
\bea \label{Prob2}
\frac{\partial}{\partial N}P({\bf \Gamma},N)=\sum_{l=1}^\infty\frac{(-1)^l}{l!}\frac{\partial^l}{\partial \Gamma_i \partial \Gamma_j...\partial \Gamma _l}[a_{ij....l}({\bf \Gamma},N)P({\bf \Gamma},N)],
\eea
where the various field moments of ${\bf \Gamma}$ are labelled using: 
\bea \label{Probmoment}
a_{i j....l}({\bf \Gamma},N)=\int d\Delta {\bf \Gamma }\;\Delta \Gamma_i \Delta \Gamma_j....\Delta \Gamma_l W_{\Delta {\bf \Gamma}}({\bf \Gamma },N).
\eea
these moments will later prove essential to determine the Fokker-Planck differential equation. To evaluate these moments corresponding to the field increments we take help of the Langevin equation, see eqn. (\ref{eftlangevin}). Evolving $\bf{\Gamma}$ value between $N$ and $N+\delta N$ where  ${\bf\Gamma} = \{\zeta , \Pi_\zeta\}$
\bea \label{evolveGamma}
{\bf\Gamma}(N +\del N) = {\bf\Gamma}(N)+ F({\bf\Gamma})\del N +G({\bf\Gamma}).\int_N ^{N+\del N}\xi(\tilde{N})d\tilde{N}.
\eea
where $F({\bf \Gamma})$ encapsulates the classical part of the motion and $G({\bf \Gamma})$ refers to the part which upon squaring forms the noise matrix element arising from the different possible noise correlators, \bea \Sigma_{f_1,g_1}\equiv (G^{2})_{f,g}\del(\tau_1 - \tau_2), \quad\quad\quad {\rm where},\quad\quad\quad (G^{2})_{f,g}\equiv \Sigma_{f,g}(\tau_1),\eea and the function $\delta(\tau_{1}-\tau_{2})$ represents white noise condition. Further critical details on these matrix elements are expanded in the later section \ref{s7}. For the current derivation purposes, we now introduce a new notation in terms of a parameter $\alpha$
\bea \label{meanGamma}
{\bf \Gamma}_\alpha (N)=(1-\alpha){\bf \Gamma}(N)+\alpha {\bf \Gamma}(N+\del N),
\eea
which allows to parameterize or give certain weights, here $(1-\alpha)$ and $\alpha$, to the position of interest on where to evaluate the functions $F({\bf \Gamma})$ and $G({\bf \Gamma})$. The parameter $\alpha$ can range in between $\alpha\in [0,1]$. In terms of this definition the eqn.(\ref{evolveGamma}) becomes:
\bea \label{evolveGamma2}
{\bf \Gamma}(N +\del N) = {\bf \Gamma}(N)+ F({\bf \Gamma}_\alpha (N))\del N +G({\bf \Gamma}_\alpha (N)).\int_N ^{N+\del N}\xi(\tilde{N})d\tilde{N},
\eea
The use of new notation makes it clear that we can solve eqn.(\ref{evolveGamma2}) in terms of ${\bf \Gamma}(N)$ for any arbitrary $\alpha$. We invoke a perturbative approach to solve by writing \bea \del {\bf \Gamma}= {\bf \Gamma}(N+\del N)-{\bf \Gamma}(N),\eea and Taylor expanding functions $F$ and $G$. With the phase space variable in the form \bea {\bf \Gamma}_\alpha = {\bf \Gamma}+\alpha \del {\bf \Gamma},\eea we can write eqn.$(\ref{meanGamma})$ as:
\bea
F_{m}[{\bf \Gamma}_\alpha (N)]= F_{m}({\bf \Gamma})+\alpha \del \Gamma_i\frac{\partial}{\partial \Gamma_i}F_{m}({\bf \Gamma})+\frac{\alpha^2}{2}\del \Gamma_i\del \Gamma_j \frac{\partial^2}{\partial \Gamma_i \partial \Gamma_j}F_{m}({\bf \Gamma})+.....
\eea
where the subscript $m$ will be clear shortly in the next equation. Also, a similar analysis can be done for $G[{\Gamma_\alpha}(N)]$. By putting these expansions for $F$ and $G$ into eqn. (\ref{evolveGamma2}), one gets a series in powers of $\delta N$ for the $m$th component of $\delta{\bf \Gamma}$ which has to form: 
\bea 
\delta\Gamma_{m} = F_{m}({\bf \Gamma})\del N + G_{mi}({\bf \Gamma})\int _N ^{N
+\del N}\xi_i(\tilde{N})d\tilde{N}+\alpha G_{pq}({\bf \Gamma})\frac{\partial G_{mj}({\bf \Gamma})}{\partial \Gamma_p}\int_{N} ^{N
+\del N}\xi_q(\tilde{N})d\tilde{N}\int_{N} ^{N
+\del N}\xi_j(\tilde{N})d\tilde{N}+\cdots,
\eea 
where the ellipses `$\cdots$' refer to the terms higher-order in the expansion. Using this expression with the eqn.(\ref{Probmoment}) we can finally begin evaluating the moments of field displacement in the following form:
\bea
&& a_i({\bf \Gamma})=\lim_{\del N\rightarrow 0}\frac{\langle \del \Gamma_i\rangle}{\del N}=\Bigg( F_i({\bf \Gamma})+\alpha G_{mj}({\bf \Gamma})\frac{\partial G_{ij}({\bf \Gamma})}{\partial \Gamma_m}\Bigg) \\
&& a_{ij}({\bf \Gamma})=\lim_{\del N \rightarrow 0}\frac{\langle \del \Gamma_i\del \Gamma_j\rangle}{\del N} = \Bigg( G_{im}({\bf \Gamma})G_{jm}({\bf \Gamma}) \Bigg),\\
&&a_{ij.....l}({\bf \Gamma})=0.
\eea 
where the white noise condition, \bea \langle\xi_{i}(N)\xi_{j}(\tilde{N})\rangle = \delta_{ij}\delta(N-\tilde{N}),\eea is applied. It becomes clear from the above that only the first and second moments are non-vanishing and putting these into the evolution for the PDF, eqn. (\ref{Prob2}), we obtain the desired Fokker-Planck equation.

The Fokker-Planck Equation corresponding to the Langevin Equation in terms of the variables ${\bf\Gamma} = \{\zeta , \Pi_\zeta\}$ is finally written as:
\bea
\frac{\partial P({\bf \Gamma},N)}{\partial N}=\mathcal{L}_{\rm FP}({\bf \Gamma})  P({\bf \Gamma},N),
\eea 
where, $\mathcal{L}_{\rm FP}({\bf\Gamma})$ is the Fokker-Planck operator for the probability density function $P({\bf\Gamma},N)$. As a result of the non-zero moments, this Fokker-Planck operator has the form:
\bea
\mathcal{L}_{\rm FP}({\bf\Gamma})=\Bigg\{-\Bigg(F_i({\bf \Gamma})+\alpha G_{lj}({\bf \Gamma})\frac{\partial G_{ij}({\bf \Gamma})}{\partial \Gamma_l}\Bigg)\frac{\partial}{\partial \Gamma_i} 
 + \frac{1}{2}\frac{\partial^2}{\partial\Gamma_i\partial\Gamma_j}G_{il}({\bf \Gamma} )G_{jl}({\bf \Gamma})\Bigg\},
\eea
Similarly useful is the adjoint Fokker-Planck equation, which has the following differential equation:
\bea
\frac{\partial}{\partial N}P_{{\bf \Gamma}}(N)= -\mathcal{L}^\dagger_{\rm FP}({\bf \Gamma})P_{{\bf \Gamma}}(N).
\eea
The corresponding adjoint Fokker-Planck operator is defined when integrating by parts the following operation using the Fokker-Planck operator:
\bea 
\int d{\bf \Gamma} f_1({\bf \Gamma}) \Bigg[\mathcal{L}_{\rm {FP}}({\bf \Gamma}).f_2({\bf \Gamma})\Bigg] =\int d{\bf \Gamma} \Bigg[\mathcal{L}_{\rm FP}^\dagger.f_1({\bf \Gamma})\Bigg ] f_2({\bf \Gamma}).
\eea 


There are two important descriptions for interpreting and handling stochastic differential equations :
\begin{itemize}
    \item \underline{\textbf{It\^{o} prescription $(\alpha=0)$}:}
    \bea
\mathcal{L}^{\dagger , \text{It\^{o}}}_{\rm FP}({\bf\Gamma}) = \Bigg\{F_i(\Gamma)\frac{\partial}{\partial \Gamma_i}  +\frac{1}{2}G_{il}(\Gamma)G_{jl}(\Gamma)\frac{\partial^2}{\partial\Gamma_i \partial\Gamma_j}\Bigg\},
\eea
    \item  \underline{\textbf{Stratonovich prescription $(\alpha=\frac{1}{2})$ }:}
    \bea
\mathcal{L}^{\dagger ,\rm  {Stratonovich}}_{\rm {FP}}({\bf\Gamma})  = \Bigg\{ F_i(\Gamma)\frac{\partial}{\partial \Phi_i}+ \frac{1}{2}G_{ij}(\Gamma)\frac{\partial G_{lj}(\Gamma)}{\partial \Gamma _l}\frac{\partial }{\partial \Gamma_i} +\frac{1}{2}G_{il}(\Gamma)G_{jl}(\Gamma)\frac{\partial^2}{\partial\Gamma_i \partial\Gamma_j} \Bigg\},
\eea
\end{itemize}

Taking $ G=0 $ i.e. in the absence of stochastic noise the $\alpha$ term has no significance since   differential equations which are deterministic are independent of any prescription. However, the prescription parameter  $\alpha$ can become relevant with scenarios where multiple inflating fields on curved field spaces are present. For the purpose of this paper, we are choosing to work in It\^{o}'s prescription, $(\alpha=0)$. The adjoint Fokker-Planck operator can then be written as,
\bea
\mathcal{L}^{\dagger , \text{It\^{o}}}_{\rm FP}({\bf\Gamma}) = \Bigg(F_i \frac{\partial }{\partial \Gamma_i}+\frac{1}{2}\Sigma_{ij}\frac{\partial ^2}{\partial \Gamma_i \partial \Gamma_j}\Bigg)
\eea
where ${\bf \Gamma} = \{\zeta ,\Pi_\zeta \}$ is the same as before and $F_i$ represent the classical drift terms,
\bea
F_i=\Bigg\{\Pi_\zeta, -(3-\epsilon)\Pi_\zeta \Bigg[1 -\frac{2(s- \frac{\eta }{2})}{(3-\epsilon)} \Bigg]
\Bigg \}, 
\eea 
where $\Sigma_{i j}$ are the noise correlation matrix elements which we will explicitly determine for our underlying theoretical setup in the later half of this paper.

\section{Modes representing comoving  scalar curvature perturbation}
\label{app:B}

This section focuses on developing the solutions for the comoving curvature perturbation in a quasi de Sitter background from the underlying EFT in inflation formalism. We employ the decoupling limit to safely examine the behaviour of the mode solutions for the three phases comprising our setup of SR, USR, and SRII during inflation. The general formulas once understood will later help to construct the various elements of the scalar power spectrum, by finding correlations among the comoving curvature perturbation and its conjugate momentum variable, and to establish the corresponding noise matrix elements.

We start with the Mukhanov-Sasaki (MS) equation solutions for the three distinct phases in our setup.
The variation of the action eqn. (\ref{s2zeta}) allows one to obtain the MS equation and solving it in the Fourier space provides us with the curvature perturbation modes for different phases. The MS equation in the Fourier space takes on the form, 
\bea \label{MSfourier}
\bigg(\partial^2_{\tau}+2 \frac{z'(\tau)}{z(\tau)}\partial_{\tau}+c_s ^2 k^2\bigg)\zeta_{\bf k}(\tau) = 0.
\eea 
Here $z(\tau)$ is known as the {\it Mukhanov-Sasaki} variable which is given by the following expression:
\bea z(\tau):=\frac{a(\tau)\sqrt{2\epsilon}}{c_s}\quad\quad\quad {\rm where}\quad\quad \quad  a(\tau)=-\frac{1}{H\tau}\quad\quad\quad {\rm with} -\infty<\tau<0.\eea
Now we use the following results which will be extremely useful to solve the above-mentioned second order differential equation:
\bea  \frac{z'(\tau)}{z(\tau)}={\cal H}\left(1-\eta+\epsilon-s\right)=-\frac{1}{\tau}\left(1-\eta+\epsilon-s\right)\quad\quad\quad{\rm where}\quad\quad\quad s:=\frac{c^{'}_s}{{\cal H}c_s},\eea
and 
\bea  \frac{z''(\tau)}{z(\tau)}={\cal H}^2\left(1-\eta+\epsilon-s\right)^2+{\cal H}^{'}\left(1-\eta+\epsilon-s\right)=2{\cal H}^2\left(1-\frac{\eta}{2}+\frac{3}{2}\epsilon-\frac{3}{2}s\right)=\frac{1}{\tau^2}\left(\nu^2--\frac{1}{4}\right),\quad\quad\quad\eea
where the parameter $\nu$ is defined by the following expression:
\bea \nu:=\frac{3}{2}+3\epsilon-\eta-3s.\eea
Here $s$ is another slow-roll parameter in all the three phases and $\nu=3/2$ represents the de Sitter limiting solution in the present context of discussion.

\subsection{First Slow Roll phase (SRI)}
\label{appBa}

\begin{figure*}[ht!]
    	\centering
    \subfigure[]{
      	\includegraphics[width=8.5cm,height=7.5cm]{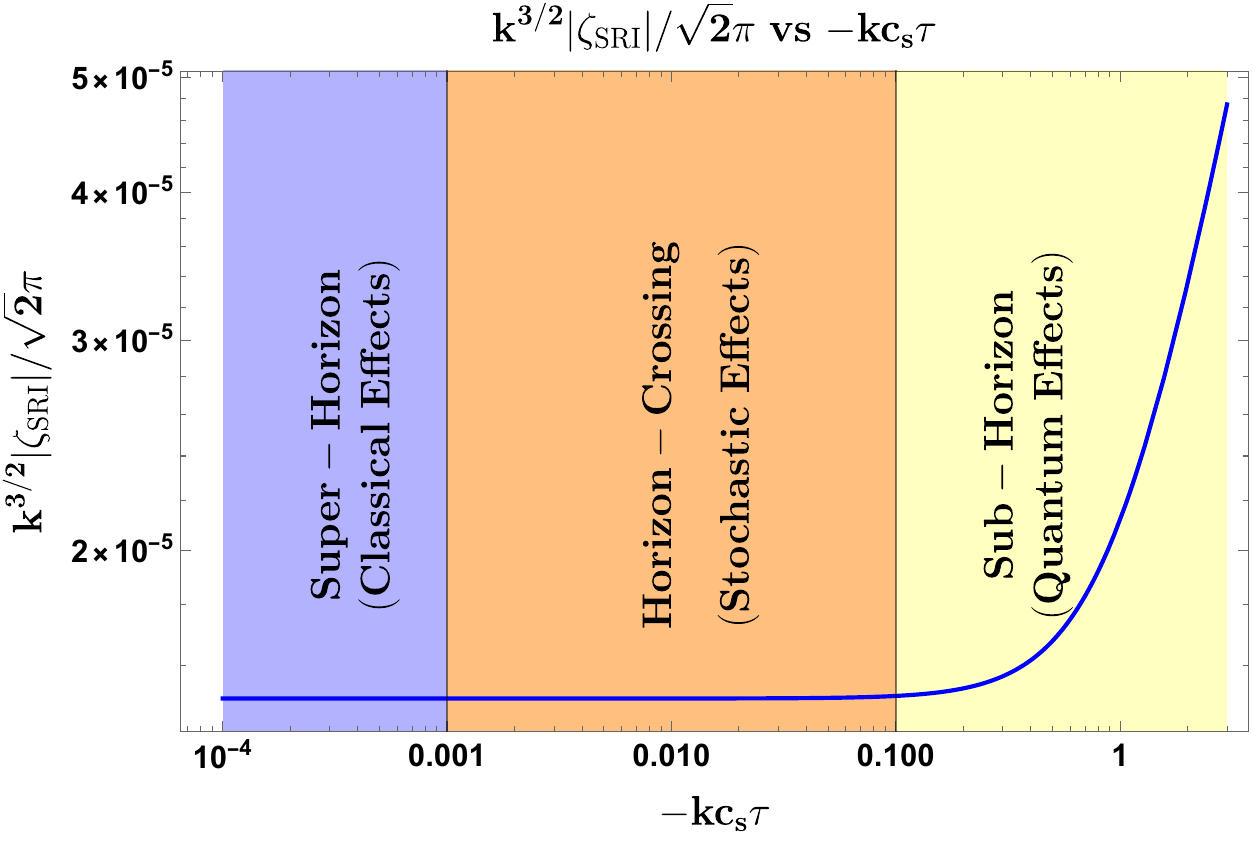}
        \label{zetasr1mode}
    }
    \subfigure[]{
        \includegraphics[width=8.5cm,height=7.5cm]{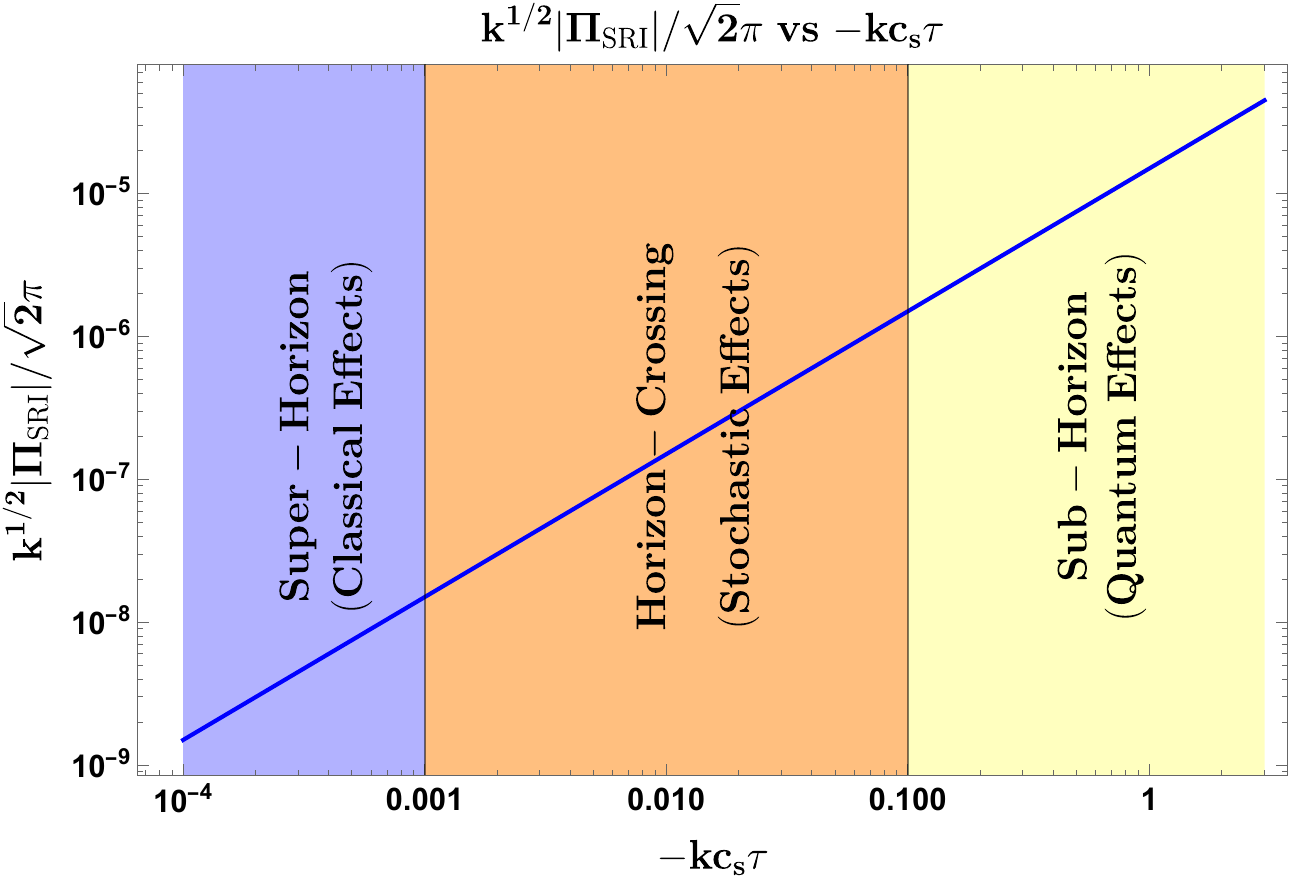}
        \label{pisr1mode}
    }
    	\caption[Optional caption for list of figures]{Behaviour of the curvature perturbation and its conjugate momentum mode in the SRI as a function of $-kc_{s}\tau$. The \textit{left-panel} shows evolution of $k^{3/2}|\zeta_{\rm SRI}|/\sqrt{2}\pi$ after choosing the Bunch-Davies initial conditions $(\alpha_{1}=1,\beta_{1}=0)$ and $\nu\sim 3/2$. Similarly, the \textit{right-panel} shows evolution of $|k^{1/2}\Pi_{\rm SRI}|/\sqrt{2}\pi$. The orange shaded region represents the stochastic effects where $-kc_{s}\tau = \sigma \in (0.001,0.1)$. The stochastic parameter $\sigma$ acts a coarse-graining factor and explains the quantum-to-classical transition of the modes. The blue and yellow shaded regions highlights the quantum and classical effects, respectively. } 
    	\label{sr1dSmodes}
    \end{figure*}

The general solution for the MS equation (\ref{MSfourier}) in the first slow-roll or SRI region for a quasi de Sitter background and arbitrary initial quantum vacuum state is as follows,
\bea \label{modezetaSRI}
{\bf \zeta}_{\bf SRI}=\frac{2^{\nu-\frac{3}{2}} c_s H (-k c_s \tau )^{\frac{3}{2}-\nu}}{i \sqrt{2 \epsilon}(k c_s)^{\frac{3}{2}}\sqrt{2} M_p}\Bigg|\frac{\Gamma(\nu)}{\Gamma(\frac{3}{2})}\Bigg |\Bigg\{\alpha_1 (1+i k c_s\tau) e^{-i(kc_s\tau+\frac{\pi}{2}(\nu+\frac{1}{2}))}-\beta_1(1-i k c_s \tau)e^{i(k c_s\tau+\frac{\pi}{2}(\nu+\frac{1}{2}))}\Bigg\}.
\eea
The mode function for the canonically conjugate momentum can be obtained by differentiating $\zeta_{\rm SRI}$,
\bea \label{modepiSRI}
{\bf \Pi_\zeta}_{\bf  SRI} = {\bf \zeta'}_{\bf SRI} = \frac{2^{\nu-\frac{3}{2}} c_s H (-k c_s \tau )^{\frac{3}{2}-\nu}}{i \tau  \sqrt{2 \epsilon}(k c_s)^{\frac{3}{2}}\sqrt{2} M_p }\Bigg|\frac{\Gamma(\nu)}{\Gamma(\frac{3}{2})}\Bigg|\Bigg [\alpha_1 \Bigg\{\Bigg(\frac{3}{2}-\nu \Bigg)(1+i k c_s \tau ) + k^2 c_s^2 \tau^2 \Bigg\}e^{-i(k c_s\tau+\frac{\pi}{2}(\nu+\frac{1}{2}))}\nonumber\\
-\beta_1 \Bigg\{\Bigg(\frac{3}{2}-\nu \Bigg)(1 - i k c_s \tau ) + k^2 c_s^2 \tau^2 \Bigg\} e^{i(k c_s\tau+\frac{\pi}{2}(\nu+\frac{1}{2}))}\Bigg],
\eea

where in the above solutions $\alpha_1$ and $\beta_1$ are the Bogoliubov coefficients in the SRI region which can be fixed using initial conditions in form of choosing a suitable quantum vacuum state. 
The SRI phase persists for the interval $\tau < \tau_s $. At $\tau = \tau_s$, the SRI transits to the Ultra Slow Roll (USR) and the nature of this transition will be of utter importance for rest of the analysis. During SRI, $\epsilon$ is approximately a constant quantity that varies very slowly with time scale whereas the second slow roll parameter, $\eta$, is very small and can be treated almost as a constant,
\bea
&&\epsilon = -\frac{\dot{H}}{H^2} = \Bigg(1-\frac{\mathcal{H}'}{\mathcal{H}^2}\Bigg),\\
&& \eta = \epsilon-\frac{1}{2}\frac{\epsilon'}{\epsilon\mathcal{H}}.
\eea
If we choose the well-known Euclidean quantum vacuum state i.e. the Bunch-Davies vacuum state, then in the SRI period the corresponding Bogoliubov coefficients are given by the following expression:
    \bea
    \alpha_{1} = 1 ,\quad  \beta_{1} = 0.
    \eea 
After substituting the mentioned values of the Bogoliubov coefficients for SRI the expression for the comoving curvature perturbation in the case of Bunch Davies initial vacuum state can be further recast in the following form:
\bea
{\bf \zeta}_{{\bf SRI},{\bf BD}}=\frac{2^{\nu-\frac{3}{2}} \, c_s \, H (-k c_s \tau )^{\frac{3}{2}-\nu}}{i \sqrt{2 \epsilon}(k c_s)^{3/2}\sqrt{2} M_p}\Bigg|\frac{\Gamma(\nu)}{\Gamma(\frac{3}{2})}\Bigg |(1+i k c_s\tau) e^{-i(kc_s\tau+\frac{\pi}{2}(\nu+\frac{1}{2}))}.
    \eea 
Under the similar Bunch-Davies initial conditions, the canonically conjugate momentum has the following form
\bea
{\bf \Pi_\zeta}_{{\bf SRI},{\bf BD}} = 
{\bf \zeta'}_{{\bf SRI},{\bf BD}} = \frac{2^{\nu-\frac{3}{2}} c_s H (-k c_s \tau )^{\frac{3}{2}-\nu}}{i \tau  \sqrt{2 \epsilon}(k c_s)^{\frac{3}{2}}\sqrt{2} M_p }\Bigg|\frac{\Gamma(\nu)}{\Gamma(\frac{3}{2})}\Bigg|\Bigg\{\Bigg(\frac{3}{2}-\nu \Bigg)(1+i k c_s \tau ) + k^2 c_s^2 \tau^2 \Bigg\}e^{-i(k c_s\tau+\frac{\pi}{2}(\nu+\frac{1}{2}))}.
\eea
On further implementing the limiting case of $\nu = 3/2$ in the above derived result one can get the simplified expressions in the case of exact de Sitter space-time. In this limit, the comoving curvature perturbation attains the following simplified form:
\bea
{\bf \zeta}_{{\bf SRI},{\bf dS}}=\frac{i c_s H }{\sqrt{2 \epsilon}(k c_s)^{\frac{3}{2}}\sqrt{2} M_p}(1+i k c_s\tau) e^{-ik c_s\tau}.
\eea

Similarly, the exact de Sitter solution for the canonically conjugate momentum looks as, 
\bea 
{\bf \Pi_\zeta}_{{\bf SRI},{\bf dS}} = {\bf \zeta'}_{{\bf SRI},{\bf dS}} = \frac{i c_s H }{ \tau  \sqrt{2 \epsilon}(k c_s)^{\frac{3}{2}}\sqrt{2} M_p }\; k^2 c_s^2 \tau^2 \;e^{-ik c_s\tau}.
\eea  

The figure fig. \ref{sr1dSmodes} describes evolution of scalar curvature perturbation mode and its conjugate momenta as function of the dimensionless variable $-kc_{s}\tau$. The solution used is a result of choosing the Bunch-Davies initial vacuum conditions, $(\alpha_{1}=1,\beta_{1}=0)$ and taking the limiting case of $\nu\sim 3/2$. In \ref{zetasr1mode} we notice from the blue line how the modulus, $|\zeta_{\rm SRI}|$, for a given mode behaves when inside the Horizon, after which this amplitude decays upon reaching the boundary where stochastic effects remain active. In the stochastic inflation formalism, there is an additional stochastic parameter $\sigma$ ($\sigma\ll 1$) which acts as a coarse-graining factor and creates a region in the super-Horizon through which short-wavelength modes transit and observe a quantum-to-classical transition until finally get treated as long-wavelength modes in the super-Horizon. In the super-Horizon, where classical effects dominates, the quantity $|\zeta_{\rm SRI}|$ becomes a constant. The conjugate momenta related quantity $|\Pi_{\rm SRI}|$ is shown in \ref{pisr1mode} where throughout the three regimes in the quantum-to-classical transition, the said quantity observes drastic decrease in its value as the conjugate momenta mode finally becomes super-Horizon where it is most suppressed.  
 
\subsection{Ultra Slow Roll phase (USR)}
\label{appBb}

\begin{figure*}[ht!]
    	\centering
    \subfigure[]{
      	\includegraphics[width=8.5cm,height=7.5cm]{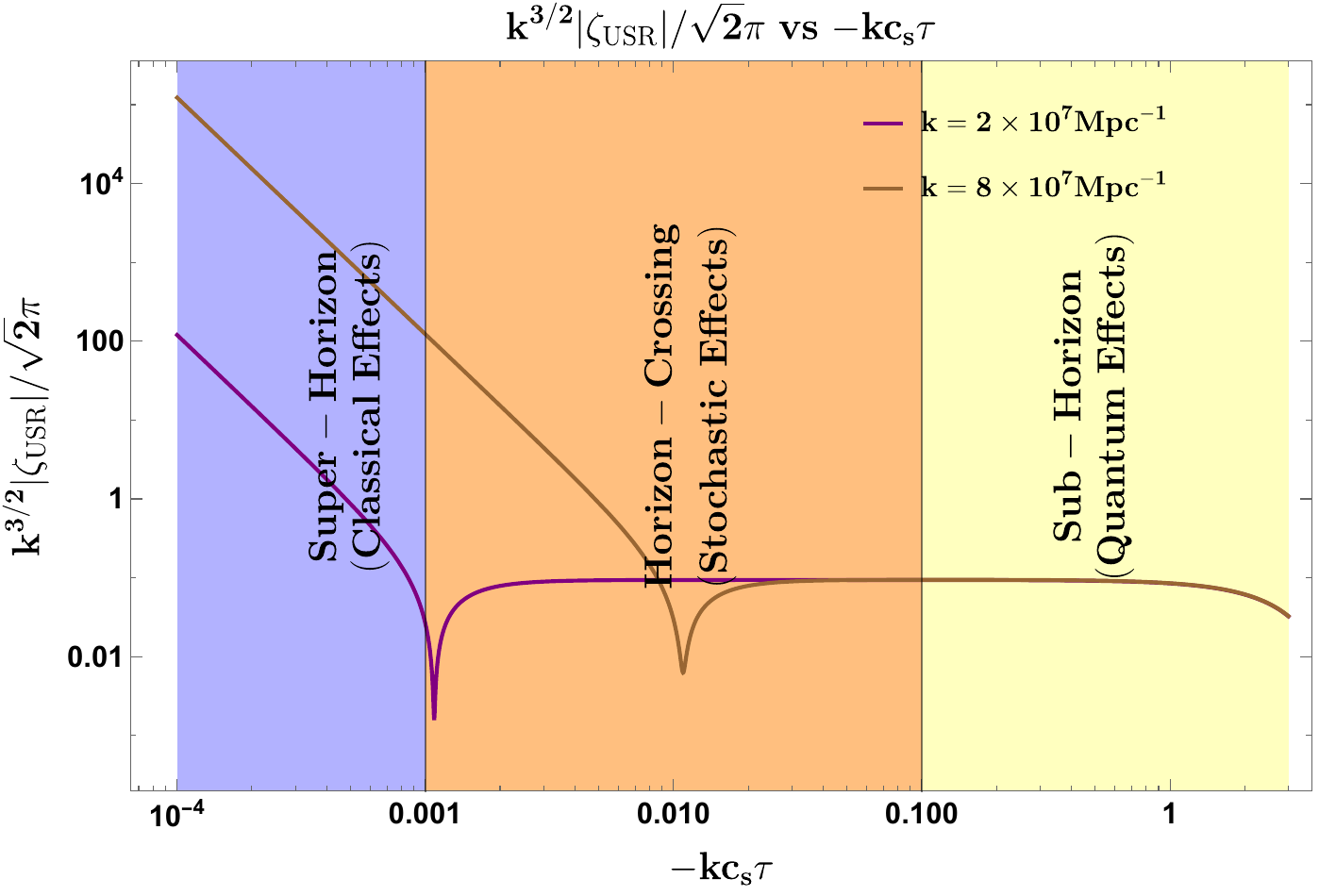}
        \label{zetausrmode}
    }
    \subfigure[]{
        \includegraphics[width=8.5cm,height=7.5cm]{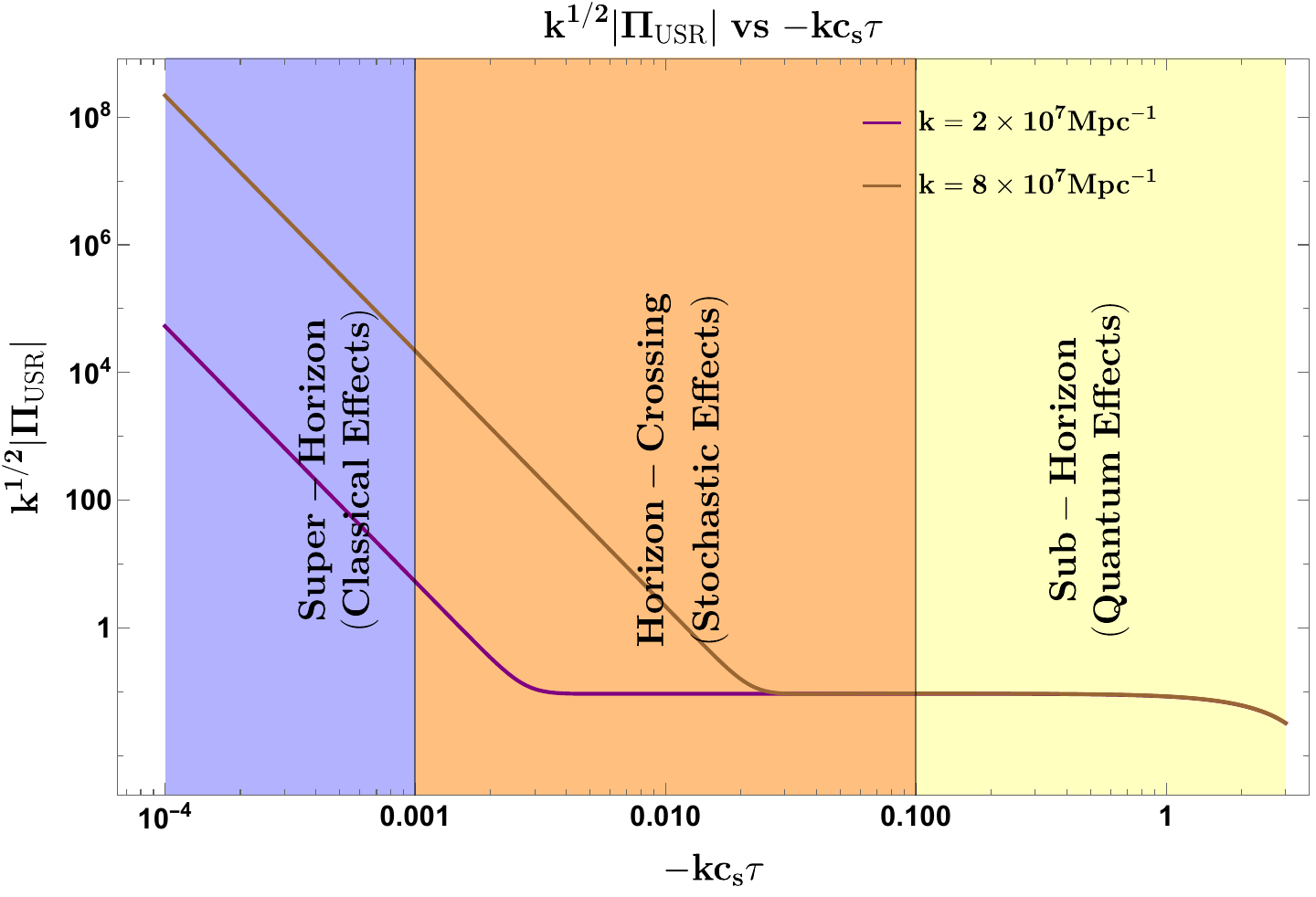}
        \label{piusrmode}
    }
    	\caption[Optional caption for list of figures]{Behaviour of the curvature perturbation and its conjugate momentum mode in the USR as a function of $-kc_{s}\tau$. The \textit{left-panel} shows evolution of $k^{3/2}|\zeta_{\rm USR}|/\sqrt{2}\pi$ after choosing the Bunch-Davies initial conditions $(\alpha_{2,{\bf dS}},\beta_{2,{\bf dS}})$ and $\nu\sim 3/2$. Similarly, the \textit{right-panel} shows evolution of $|k^{1/2}\Pi_{\rm USR}|/\sqrt{2}\pi$. The orange shaded region represents the stochastic effects where $-kc_{s}\tau = \sigma \in (0.001,0.1)$. The stochastic parameter $\sigma$ acts as a coarse-graining parameter near Horizon corssing where it the quantum-to-classical transition of the modes take place. The blue and yellow shaded regions highlights the quantum and classical effects, respectively. } 
    	\label{usrdSmodes}
    \end{figure*}

The USR phase operates during the conformal time interval $\tau_s \leq \tau \leq \tau_e$ where, $\tau_s$ is the time when SRI ends and USR begins and $\tau_e $ is the time when USR ends and the following SRII begins.  The slow-roll parameter $\epsilon $ in the USR now become extremely suppressed and time-dependent which can be described in terms of $\epsilon$ as, 
\bea \label{epsilonusr}
\epsilon(\tau) = \epsilon \Bigg(\frac{a(\tau_s)}{a(\tau )} \Bigg )^6 =  \epsilon \Bigg(\frac{\tau}{\tau_s} \Bigg)^6.
\eea 
Important is the behaviour of the second slow-roll parameter $\eta(\tau)$ that varies with time in the USR as follows,
\bea \label{etausr}
\eta(\tau)\sim \eta -6 \approx {\cal O}(-6).
\eea
It is to be noted here that we have considered a sharp transition when going from the SRI to USR. Now, the  general solution for the MS equation (\ref{MSfourier}) in the USR region with an arbitrary quantum vacuum state can be expressed using the following simplified form,
\bea \label{modezetaUSR}
{\bf \zeta}_{\bf USR}=\frac{2^{\nu-\frac{3}{2}} c_s   H (-k c_s \tau )^{\frac{3}{2}-\nu}}{i \sqrt{2 \epsilon}(k c_s)^{\frac{3}{2}}\sqrt{2} M_p}\Bigg[\frac{\tau_s}{\tau}\Bigg]^3\Bigg|\frac{\Gamma(\nu)}{\Gamma(\frac{3}{2})}\Bigg |\Bigg\{\alpha_2 (1+i k c_s\tau) e^{-i(kc_s\tau+\frac{\pi}{2}(\nu+\frac{1}{2}))}-\beta_2(1-i k c_s \tau)e^{i(k c_s\tau+\frac{\pi}{2}(\nu+\frac{1}{2}))}\Bigg\}.
\eea
Also, the corresponding canonically conjugate momentum mode function can be obtained by differentiating $\zeta_{\bf USR}$,
\bea 
\label{modepiUSR}
{\bf \Pi_\zeta}_{\bf  USR} = {\bf \zeta'}_{\bf USR} = \frac{2^{\nu-\frac{3}{2}} c_s  H (-k c_s \tau )^{\frac{3}{2}-\nu} 
 \tau_s^3 }{i  \sqrt{2 \epsilon}(k c_s)^{\frac{3}{2}}\sqrt{2} M_p  \, \tau^4} 
  \Bigg|\frac{\Gamma(\nu)}{\Gamma(\frac{3}{2})}\Bigg|\Bigg[ 
 \Bigg\{k^2 c_s^2 \tau^2 + (1+i k c_s \tau )\Bigg(\Bigg(\frac{3}{2}-\nu\Bigg)-3\Bigg)\Bigg\} \alpha_2 e^{-i(k c_s\tau+\frac{\pi}{2}(\nu+\frac{1}{2}))}\nonumber \\ 
 - \Bigg\{k^2 c_s^2 \tau^2 + (1- i k c_s \tau )\Bigg(\Bigg(\frac{3}{2}-\nu\Bigg)-3\Bigg)\Bigg\}\beta_2 e^{i(k c_s\tau+\frac{\pi}{2}(\nu + \frac{1}{2}))}\Bigg],\quad\quad
\eea

where, $\alpha_2 $ and $\beta_2$ are the Bogoliubov coefficients in the USR region that  can be found in terms of $\alpha_1$ and $\beta_1$ using the two  boundary conditions interpreted as  Israel junction conditions at time scale $\tau = \tau_s $ . 

The first condition implies that the scalar modes obtained are continuous at the sharp transition point, $\tau=\tau_s$  between SRI and USR
\bea
[\zeta(\tau)]_{\bf SRI,\tau=\tau_s} = [\zeta(\tau)]_{\bf USR,\tau=\tau_s}.
\eea 

The second condition implies that momentum modes are continuous at the sharp transition point $\tau=\tau_s$ between SRI and USR

\bea
[\zeta'(\tau)]_{{\bf SRI},\, {\bf \tau=\tau_s}} = [\zeta'(\tau)]_{\bf USR, \, \tau=\tau_s}.
\eea 

After applying the above two junction conditions we obtain two constraint equations for Bogoliubov coefficients in the USR region 
\bea \label{alpha2qds}
&& \alpha_2 = \frac{1}{2 k^3 \tau_s^3 c_s^3} \Bigg\{ \Bigg(3 i + 3 i  k^2 c_s^2 \tau_s ^2  + 2  k^3 c_s^3  \tau_s ^3  \Bigg  )\alpha_1 - \Bigg(3 i +6  k c_s \tau_s  -3 i k^2  c_s^2 \tau_s ^2    \Bigg) \beta_1 e^{i \left(2  k \tau_s  c_s +  \pi  \left(\nu +\frac{1}{2}\right)\right)}\Bigg\},
 \\
&&  \label{beta2qds} \beta_2 = \frac{1}{2 k ^3 c_s ^3 \tau_s ^3} \Bigg\{ \Bigg( 3i -6 k c_s \tau_s -3i k^2 c_s ^2 \tau_s^2 \Bigg) \alpha_1 e^{-i(\pi(\nu+\frac{1}{2})+ 2k c_s \tau_s)}- \Bigg(3i +3 i k^2 c_s ^2 \tau_s ^ 2 - 2 k^3c_s^3 \tau_s ^ 3  \Bigg )\beta_1
 \Bigg\}.
\eea 
Implementing the Bunch Davies initial vacuum state $(\alpha_1=1 , \, \beta_1=0)$ into the USR Bogoliubov coefficients we get the simplified form of $\alpha_2 $ and $\beta_2$
\bea \label{alpha2bd}
&& \alpha_{{2},{\bf BD}} =  \frac{1}{2 k^3 \tau_s^3 c_s^3}  \Bigg(3 i + 3 i  k^2 c_s^2 \tau_s ^2  + 2  k^3 c_s^3  \tau_s ^3  \Bigg  ), \\
&& \label{beta2bd}  \beta_{{2},{\bf BD}} = \frac{1}{2 k ^3 c_s ^3 \tau_s ^3}  \Bigg( 3i -6 k c_s \tau_s -3i k^2 c_s ^2 \tau_s^2 \Bigg) e^{-i(\pi(\nu+\frac{1}{2})+ 2k c_s \tau_s)}.
\eea

Further implementing the limiting case $\nu=\frac{3}{2}$ in Bunch Davies vacuum derived result one can get the simplified
expression for USR Bogoliubov coefficients in the case of de Sitter space
\bea  \label{alpha2ds}
&& \alpha_{{2},{\bf dS}} =  \frac{1}{2 k^3 \tau_s^3 c_s^3}  \Bigg(3 i + 3 i  k^2 c_s^2 \tau_s ^2  + 2  k^3 c_s^3  \tau_s ^3  \Bigg  ), \\
&& \label{beta2ds} \beta_{{2},{\bf dS}} = \frac{1}{2 k ^3 c_s ^3 \tau_s ^3}  \Bigg( 3i -6 k c_s \tau_s -3i k^2 c_s ^2 \tau_s^2 \Bigg) e^{- 2 i k c_s \tau_s}.
\eea 
The mode  function in the limiting case $\nu=\frac{3}{2}$ in Bunch Davies vacuum gives rise to the  de Sitter result which can be further written as:
\bea
{\bf \zeta}_{{\bf USR},{\bf dS} }=\frac{i c_s   H }{ \sqrt{2 \epsilon}(k c_s)^{\frac{3}{2}}\sqrt{2} M_p}\Bigg[\frac{\tau_s}{\tau}\Bigg]^3\Bigg\{\alpha_{{2},{\bf dS}} (1+i k c_s\tau) e^{-ikc_s\tau}-\beta_{{2},{\bf dS}}(1-i k c_s \tau)e^{ik c_s\tau}\Bigg\}.
\eea
The momentum mode function for the above case is:
\bea
{\bf \Pi_\zeta}_{{\bf USR},{\bf dS} } = {\bf \zeta'}_{{\bf USR},{\bf dS} } = \frac{ i c_s  H  
 \tau_s^3 }{ \sqrt{2 \epsilon}(k c_s)^{\frac{3}{2}}\sqrt{2} M_p  \, \tau^4} 
  \Bigg[ 
 \Bigg\{k^2 c_s^2 \tau^2 -3(1+i k c_s \tau )\Bigg\} \alpha_{{2},{\bf dS}} e^{-ik c_s\tau}\nonumber \\
 - \Bigg\{k^2 c_s^2 \tau^2 -3 (1- i k c_s \tau )\Bigg\}\beta_{{2},{\bf dS}} e^{i k c_s\tau}\Bigg].
\eea 

The figure fig. \ref{usrdSmodes} describes evolution in the USR of the scalar curvature perturbation mode and its conjugate momenta as function of the dimensionless variable $-kc_{s}\tau$. Similar to SRI, we continue with our choice of Bunch-Davies initial condition for the Bogoliubov coefficients in this phase and taking the limit $\nu\sim 3/2$, which leads to the coefficients $(\alpha_{2,{\bf dS}},\beta_{2,{\bf dS}})$, see eqn. (\ref{alpha2ds},\ref{beta2ds}). Here we plot the behaviour for different wavenumbers where the additional momentum dependence comes from using the quantity $k_{s}c_{s}\tau_{s}$ in the mode solutions. The plot is shown after choosing $k_{s}c_{s}\tau_{s}\sim {\cal O}(-0.01)$. From \ref{zetausrmode}, we notice that the magnitude $|\zeta_{\rm USR}|$ stays constant when in the sub-Horizon. Near the Horizon crossing, various short-scale modes encounter stochastic effects and they undergo a quantum-to-classical transition until finally in the super-Horizon regime. In this regime, the quantity $|\zeta_{\rm USR}|$ tends to increase greatly, however, when concerned with the behaviour of the modes near the Horizon crossing, including stochastic effects the strength of the modulus is sufficient enough to give $k^{3}|\zeta_{\rm USR}|^{2}\sim {\cal O}(10^{-3}-10^{-2})$. In fig. \ref{piusrmode}, the evolution of the conjugate momenta mode in the USR is shown. This time the conjugate momenta has a significant role to play, as seen from the magnitude of $|\Pi_{\rm USR}|$. During Horizon crossing, in presence of the stochastic effects, the mode starts to increase in strength and keeps going monotonically as they become of long-wavelength in the super-Horizon scales. 

\subsection{Second Slow Roll phase (SRII)}
\label{appBc}

\begin{figure*}[ht!]
    	\centering
    \subfigure[]{
      	\includegraphics[width=8.5cm,height=7.5cm]{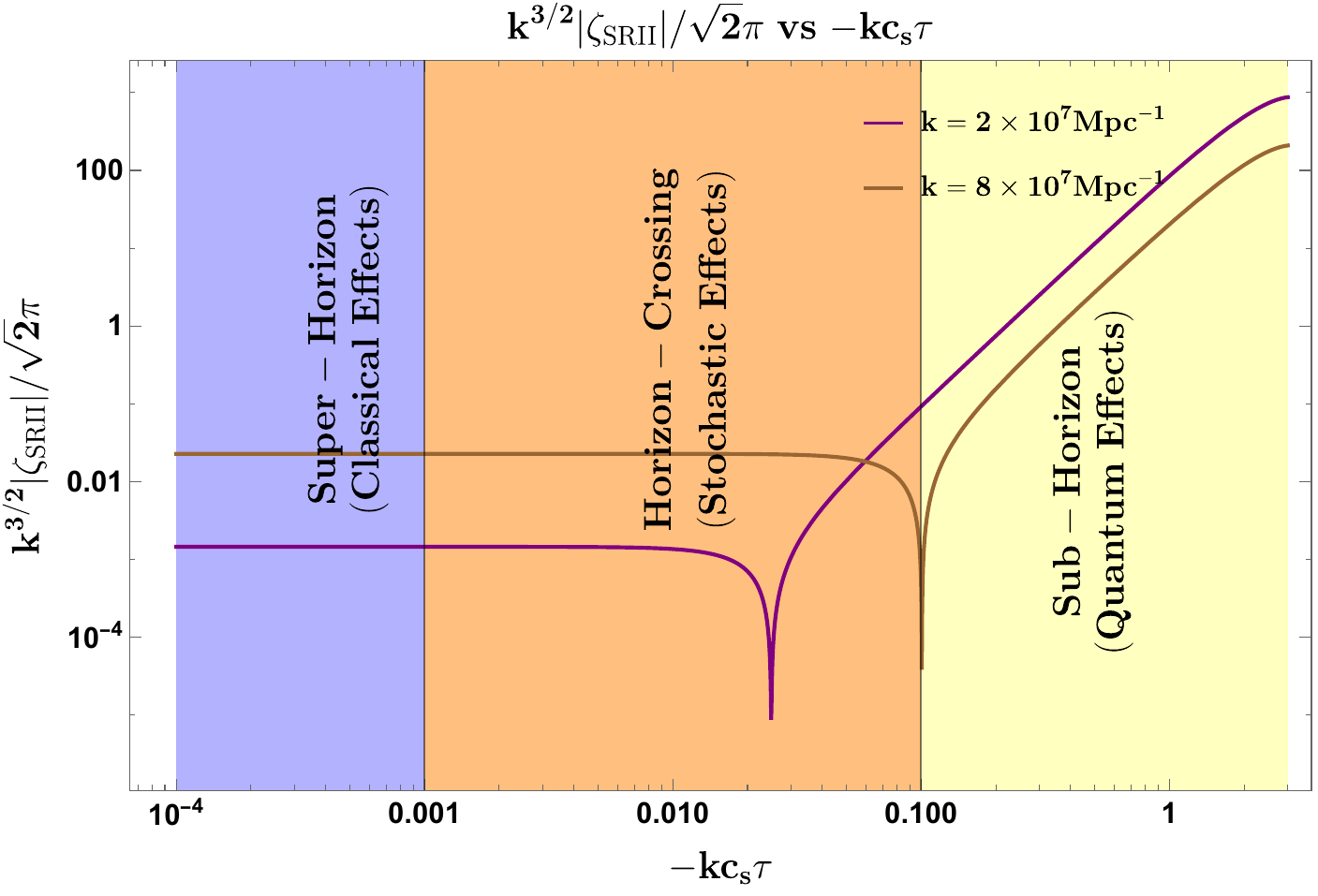}
        \label{zetasr2mode}
    }
    \subfigure[]{
        \includegraphics[width=8.5cm,height=7.5cm]{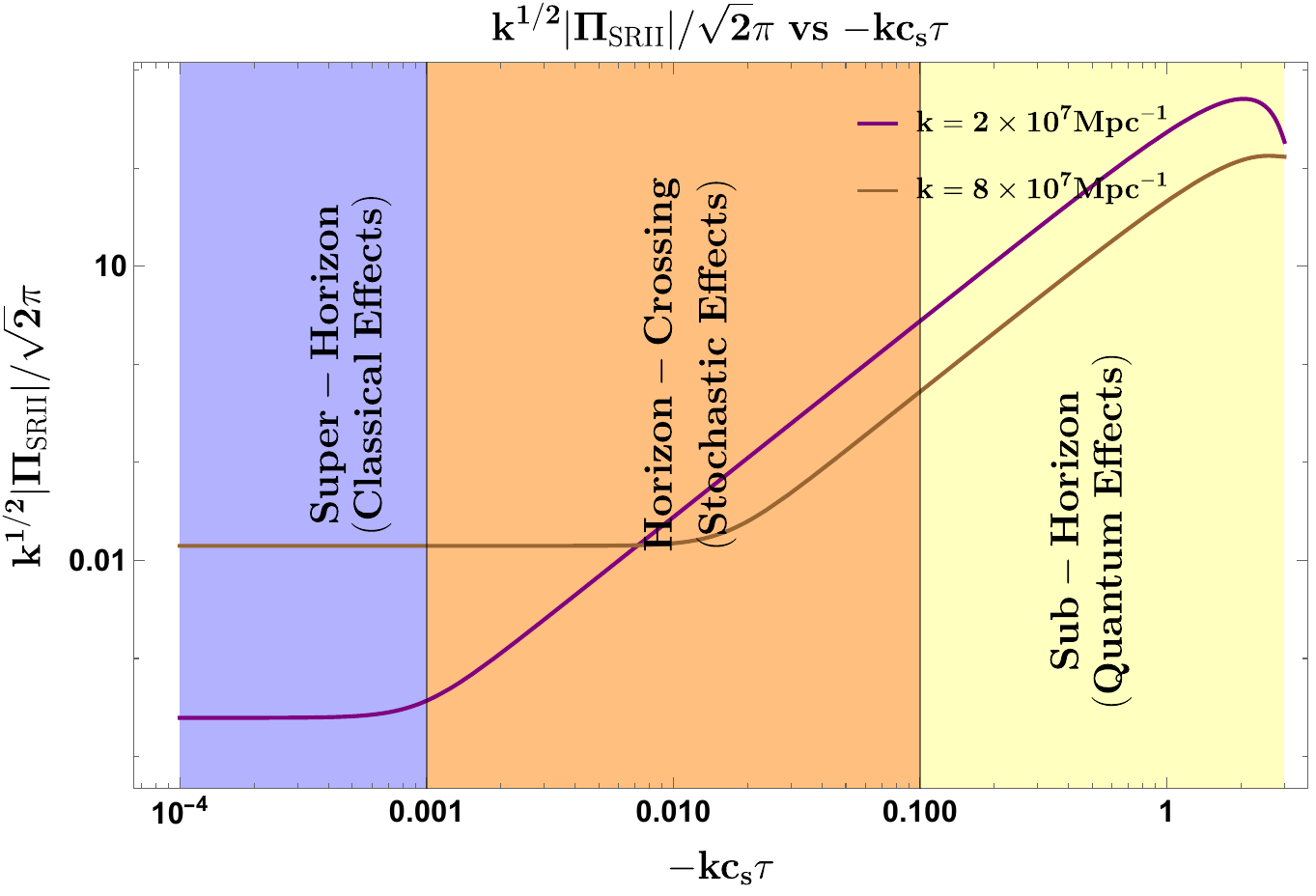}
        \label{pisr2mode}
    }
    	\caption[Optional caption for list of figures]{Behaviour of the curvature perturbation and its conjugate momentum mode in the SRII as a function of $-kc_{s}\tau$. The \textit{left-panel} shows evolution of $k^{3/2}|\zeta_{\rm SRII}|/\sqrt{2}\pi$ after choosing the Bunch-Davies initial conditions $(\alpha_{3,{\bf dS}},\beta_{3,{\bf dS}})$ and $\nu\sim 3/2$. Similarly, the \textit{right-panel} shows evolution of $|k^{1/2}\Pi_{\rm SRII}|/\sqrt{2}\pi$. The orange shaded region represents the stochastic effects where $-kc_{s}\tau = \sigma \in (0.001,0.1)$. The stochastic parameter $\sigma$ acts as a coarse-graining parameter near Horizon corssing where it the quantum-to-classical transition of the modes take place. The blue and yellow shaded regions highlights the quantum and classical effects, respectively. } 
    	\label{sr2dSmodes}
    \end{figure*}

The second slow roll phase persists for the conformal time scale $\tau_e\leq\tau \leq \tau_{\rm end} $ where, $\tau_e$ marks the end of USR phase and beginning of SRII phase whereas $\tau_{\rm end}$ marks the end of inflationary paradigm. The dependence of first and second slow roll parameters can be described as :
\bea 
&& \epsilon(\tau)= \epsilon \Bigg (\frac{a(\tau_s)}{a(\tau_e)}\Bigg)^6 = \epsilon \Bigg(\frac{\tau_e}{\tau_s}\Bigg)^6\longrightarrow \epsilon(\tau=\tau_{end})=1,  \\
&& \eta(\tau) =\eta \longrightarrow \eta(\tau=\tau_{end})\sim\mathcal{O}(-1).
\eea 
where, $\epsilon$ is slow roll parameter in SRI region .  In the preceding section we have considered sharp transition at the boundary of transition between SRI to USR here also we considered the  sharp transition from USR to SRII region. Imposing the  two boundary conditions termed as Israel junction condition the constraint relation for Bogoliubov coefficients in the SRII region can be deduced. The  mode solution for SRII region is:
\bea \label{modezetaSR2}
{\bf \zeta}_{\bf SRII}=\frac{2^{\nu-\frac{3}{2}} c_s   H (-k c_s \tau )^{\frac{3}{2}-\nu}}{i \sqrt{2 \epsilon}(k c_s)^{\frac{3}{2}}\sqrt{2} M_p}\Bigg[\frac{\tau_s}{\tau_e}\Bigg]^3\Bigg|\frac{\Gamma(\nu)}{\Gamma(\frac{3}{2})}\Bigg |\Bigg\{\alpha_3 (1+i k c_s\tau) e^{-i\left (kc_s\tau+\frac{\pi}{2}(\nu+\frac{1}{2})\right )}-\beta_3(1-i k c_s \tau)e^{i\left(k c_s\tau+\frac{\pi}{2}(\nu+\frac{1}{2})\right)}\Bigg\}.
\eea
The canonically conjugate momentum mode function can be obtained by differentiating $\zeta_{\bf SRII}$ and is expressed as:
\bea \label{modepiSR2}
{\bf \Pi_\zeta}_{\bf  SRII} = {\bf \zeta'}_{\bf SRII} =  \frac{2^{\nu-\frac{3}{2}} c_s  H (-k c_s \tau )^{\frac{3}{2}-\nu}}{i \sqrt{2 \epsilon}(k c_s)^{\frac{3}{2}}\sqrt{2} M_p \, \tau } 
 \Bigg[\frac{\tau_s}{\tau_e}\Bigg]^3 \Bigg|\frac{\Gamma(\nu)}{\Gamma(\frac{3}{2})}\Bigg|\Bigg \{\alpha_3 \Bigg\{k^2 c_s^2 \tau^2 + (1+i k c_s \tau )\Bigg(\frac{3}{2}-\nu\Bigg)\Bigg\} e^{-i\left(k c_s\tau+\frac{\pi}{2}(\nu+\frac{1}{2})\right)}\nonumber \\
 -\beta_3 \Bigg\{k^2 c_s^2 \tau^2 + (1- i k c_s \tau )\Bigg(\frac{3}{2}-\nu\Bigg)\Bigg\} e^{i \left (k c_s\tau+\frac{\pi}{2}(\nu+\frac{1}{2})\right )}\Bigg\},
\eea
where, $\alpha_3$ and $\beta_3$ are the Bogoliubov coefficients. The constraint relation for these coefficients can be found in terms of $\alpha_2$ and $\beta_2$  using the boundary conditions termed as Israel junction conditions. The underlying vacuum state in SRII changes compared to USR. The transition takes place at $\tau=\tau_e$ .

The first boundary condition implies that the scalar modes obtained from USR and SRII are continuous at the  sharp transition point, $\tau = \tau_e$
\bea
[\zeta(\tau)]_{{\bf USR},{\tau=\tau_e}}= [\zeta(\tau)]_{{\bf SRII},{\tau=\tau_e}}. 
\eea 
The second boundary condition implies that the momentum modes for the scalar perturbation obtained from USR and SRII become continuous at transition point $\tau=\tau_e$ 
\bea
[\zeta'(\tau)]_{{\bf USR},{\bf \tau = \tau_e}}= [\zeta'(\tau)]_{{\bf SRII},{\bf \tau = \tau_e}}.
\eea 
After imposing these   conditions we get two constraint equations for Bogoliubov coefficients in the SRII region 
\bea \label{alpha3qds}
 && \alpha _3 =  \frac{1}{2 k^3 \tau_e^3 c_s^3}\Bigg\{\left(-3 i -3 i  k^2 \tau_e^2 c_s^2 +2  k^3 \tau_e^3 c_s^3 \right)\alpha _2 - \left(-3 i -6  k \tau_e c_s   +3 i  k^2 \tau_e^2 c_s^2 \right)\beta _2 e^{\left(2 i k \tau_e c_s+i \pi  \left(\nu +\frac{1}{2}\right)\right)}\Bigg\},\\
&&  \label{beta3qds}  \beta _3 =  \frac{1}{2 k^3 \tau_e^3 c_s^3}\Bigg\{ \left(-3 i  +6  k \tau_e c_s +3 i k^2 \tau_e^2 c_s^2 \right)\alpha _2  e^{-\left(2 i k \tau_e c_s + i \pi  \left(\nu +\frac{1}{2}\right)\right)}+ \left(2  k^3 \tau_e^3 c_s^3 +3 i  k^2 \tau_e^2 c_s^2 +3 i  \right)\beta _2\Bigg\}, 
\eea 

where, $\alpha_2$ and $\beta_2$ are the Bogoliubov coefficients in the USR region  mentioned in eq.(\ref{alpha2qds}) and eq.(\ref{beta2qds}) respectively. Substituting these values in eq.(\ref{alpha3qds}) and eq.(\ref{beta3qds})
\bea
  \alpha _3 &=&  \frac{1}{(2 k^3 \tau_e^3 c_s^3)(2 k^3 \tau_s^3 c_s^3)}\Bigg\{\left(-3 i -3 i  k^2 \tau_e^2 c_s^2 +2  k^3 \tau_e^3 c_s^3 \right) \Bigg\{ \left(3 i + 3 i  k^2 c_s^2 \tau_s ^2  + 2  k^3 c_s^3  \tau_s ^3   \right )\alpha_1   - \left(3 i +6  k c_s \tau_s  -3 i k^2  c_s^2 \tau_s ^2   \right)
 \nonumber\\ 
&& \quad\quad\quad\quad\quad\quad\quad\quad\quad\quad\quad \times \beta_1  e^{i \left(2  k \tau_s  c_s +  \pi  \left(\nu +\frac{1}{2}\right)\right)}\Bigg\}
 \quad  - \left(-3 i -6  k \tau_e c_s   +3 i  k^2 \tau_e^2 c_s^2 \right) \quad \Bigg\{ \left( 3i -6 k c_s \tau_s -3i k^2 c_s ^2 \tau_s^2 \right)  
 \nonumber \\
  &&\quad\quad\quad \quad\quad\quad \quad\quad\quad\quad\quad \times \alpha_1 e^{-i(\pi(\nu+\frac{1}{2})+ 2k c_s \tau_s)} - \left (3i +3 i k^2 c_s ^2 \tau_s ^ 2 - 2 k^3c_s^3 \tau_s ^ 3  \right )\beta_1 \quad 
 \Bigg\} e^{\left(2 i k \tau_e c_s+i \pi  \left(\nu +\frac{1}{2}\right)\right)}\Bigg\},
 \\
 \beta_3 &=& \frac{1}{(2 k^3 \tau_e^3 c_s^3)(2 k^3 \tau_s^3 c_s^3)}\Bigg\{ \left(-3 i  +6  k \tau_e c_s +3 i k^2 \tau_e^2 c_s^2\right) \Bigg\{ \left(3 i + 3 i  k^2 c_s^2 \tau_s ^2  + 2  k^3 c_s^3  \tau_s ^3  \right)\alpha_1 - \left(3 i +6  k c_s \tau_s  -3 i k^2  c_s^2 \tau_s ^2    \right) \nonumber \\
 && \quad \quad\quad\quad\quad\quad\quad \quad\quad\times \beta_1 e^{i \left(2  k \tau_s  c_s +  \pi  \left(\nu +\frac{1}{2}\right)\right)}\Bigg\}  e^{-\left(2 i k \tau_e c_s + i \pi  \left(\nu +\frac{1}{2}\right)\right)}+ \left(2  k^3 \tau_e^3 c_s^3 +3 i  k^2 \tau_e^2 c_s^2 +3 i  \right )  \nonumber \\
 &&  \quad \quad\quad\quad\quad\quad\quad\quad\quad\times\Bigg\{ \left( 3i -6 k c_s \tau_s -3i k^2 c_s ^2 \tau_s^2 \right) \alpha_1 e^{-i\left(\pi(\nu+\frac{1}{2})+ 2k c_s \tau_s\right)}
 - \left(3i +3 i k^2 c_s ^2 \tau_s ^ 2 - 2 k^3c_s^3 \tau_s ^ 3  \right )\beta_1
 \Bigg\}\Bigg\}.
\quad\quad\quad \eea 
Further implementing the Bunch Davies initial vacuum state $(\alpha_1 = 1, \beta_1 = 0)$ into the SRII  Bogoliubov coefficients we can  get
the simplified form of $\alpha_3$ and $\beta_3$ but one has to substitute $\alpha_2$ and $ \beta_2$ from eq.(\ref{alpha2bd}) and eq.(\ref{beta2bd}) respectively which are already solved in the Bunch Davies initial vacuum state.
\bea
 \alpha _{3, {\bf BD }} &=& \frac{1}{(2 k^3 \tau_e^3 c_s^3)(2 k^3 \tau_s^3 c_s^3)}\Bigg\{\left(-3 i -3 i  k^2 \tau_e^2 c_s^2 +2  k^3 \tau_e^3 c_s^3 \right) \left(3 i + 3 i  k^2 c_s^2 \tau_s ^2  + 2  k^3 c_s^3  \tau_s ^3  \right  )  \nonumber\\ &&
\quad\quad\quad\quad\quad\quad\quad\quad\quad\quad - \left(-3 i -6  k \tau_e c_s   +3 i  k^2 \tau_e^2 c_s^2 \right)
 \left( 3i -6 k c_s \tau_s -3i k^2 c_s ^2 \tau_s^2 \right) e^{2 i k c_s( \tau_e -\tau_s) }\Bigg\},
 \\
 \beta _{3,{\bf BD}} &=&  \frac{1}{(2 k^3 \tau_e^3 c_s^3)(2 k ^3 c_s ^3 \tau_s ^3)}\Bigg\{ \left(-3 i  +6  k \tau_e c_s +3 i k^2 \tau_e^2 c_s^2\right) \left(3 i + 3 i  k^2 c_s^2 \tau_s ^2  + 2  k^3 c_s^3  \tau_s ^3  \right) e^{-\left(2 i k \tau_e c_s + i \pi  \left(\nu +\frac{1}{2}\right)\right)} \nonumber \\ 
&& +\left(2  k^3 \tau_e^3 c_s^3 + 3 i  k^2 \tau_e^2 c_s^2 +3 i \right )  \left( 3i -6 k c_s \tau_s -3i k^2 c_s ^2 \tau_s^2 \right) e^{-i\left(\pi(\nu+\frac{1}{2})+ 2k c_s \tau_s \right)}\Bigg\}.
\eea 

Further implementing the limiting case $\nu = \frac{3}{2}$ in Bunch Davies vacuum derived result one can get the simplified
expression for SRII Bogoliubov coefficients in the case of de Sitter space.
\bea \label{alpha3ds}
 \alpha _{{3},{\bf dS}} &=&  \frac{1}{(2 k^3 \tau_e^3 c_s^3)(2 k^3 \tau_s^3 c_s^3)}\Bigg\{\left(-3 i -3 i  k^2 \tau_e^2 c_s^2 +2  k^3 \tau_e^3 c_s^3 \right)   \left(3 i + 3 i  k^2 c_s^2 \tau_s ^2  + 2  k^3 c_s^3  \tau_s ^3  \right ) \nonumber \\ &&
\quad\quad\quad\quad\quad\quad\quad\quad\quad\quad - \left(-3 i -6  k \tau_e c_s   +3 i  k^2 \tau_e^2 c_s^2 \right) 
  \left( 3i -6 k c_s \tau_s -3i k^2 c_s ^2 \tau_s^2 \right ) e^{ i 2  k c_s( \tau_e-\tau_s)} \Bigg\},
  \\ \label{beta3ds}
\beta _{{3},{\bf dS}} &=&  \frac{1}{(2 k^3 \tau_e^3 c_s^3)(2 k^3 \tau_s^3 c_s^3 )}\Bigg\{ \left(-3 i  +6  k \tau_e c_s +3 i k^2 \tau_e^2 c_s^2\right) \left(3 i + 3 i  k^2 c_s^2 \tau_s ^2  + 2  k^3 c_s^3  \tau_s ^3  \right  )  e^{-2 i k \tau_e c_s }\nonumber\\ &&   
 \quad \quad\quad\quad\quad\quad\quad\quad\quad\quad 
 + \left(2  k^3 \tau_e^3 c_s^3 +3 i  k^2 \tau_e^2 c_s^2 +3 i  \right) \left( 3 i - 6 k c_s \tau_s -3i k^2 c_s ^2 \tau_s^2 \right) e^{- 2 i k c_s \tau_s}\Bigg\}.
\eea 

The mode  function in the limiting case $\nu=\frac{3}{2}$ in Bunch Davies vacuum gives rise to the  de Sitter result which can be further written as: 
\bea
{\bf \zeta}_{{\bf SRII},{\bf dS}}=\frac{ i c_s   H }{ \sqrt{2 \epsilon}(k c_s)^{\frac{3}{2}}\sqrt{2} M_p}\Bigg[\frac{\tau_s}{\tau_e}\Bigg]^3\Bigg\{\alpha_{{3},{\bf dS}} (1+i k c_s\tau ) e^{-i kc_s\tau}-\beta_{{3},{\bf dS}}(1-i k c_s \tau )e^{i k c_s\tau}\Bigg\}.
\eea
The canonically conjugate momentum mode function 
in the limiting case $\nu=\frac{3}{2}$
 in Bunch Davies vacuum gives rise to the de Sitter result which can
be further written as:

\bea
{\bf \Pi_\zeta}_{{\bf  SRII},{\bf dS}} = {\bf \zeta'}_{{\bf  SRII},{\bf dS}} =  \frac{i c_s  H }{ \sqrt{2 \epsilon}(k c_s)^{\frac{3}{2}}\sqrt{2} M_p \, \tau } 
 \Bigg[\frac{\tau_s}{\tau_e}\Bigg]^3 (k^2 c_s^2 \tau^2 )\Bigg \{\alpha_{{3}, {\bf dS}}e^{-ik c_s \tau}
 -\beta_{{3},{\bf dS}}  e^{i k c_s \tau }\Bigg\}.
\eea

The figure fig. \ref{sr2dSmodes} describes evolution in the SRII of the scalar curvature perturbation mode and its conjugate momenta as function of the dimensionless variable $-kc_{s}\tau.$ Here also we employ the similar Bunch-Davies initial conditions and the limiting case of de Sitter $\nu\sim 3/2$, giving us the Bogoliubov Bogoliubov coefficients $(\alpha_{3,{\bf dS}},\beta_{3,{\bf dS}})$, see eqn. (\ref{alpha3ds},\ref{beta3ds}). The additional wavenumber effect comes from including quantities of the form $k_{s}c_{s}\tau_{s}$ and $k_{e}c_{s}\tau_{e}$ in the mode solutions, here we have chosen them of order ${\cal O}(-0.1)$. From \ref{zetasr2mode}, we notice that the quantity $|\zeta_{\rm SRII}|$ has larger magnitude when in the sub-Horizon regime. The strength steadily decreases as it approaches Horizon-crossing where the stochastic effects start become important and soon the magnitude becomes constant. After going through the quantum-to-classical transition, the mode enters into the super-Horizon regime where it stays constant throughout much like as in the case of SRI but with now having an increased magnitude of $|\zeta_{\rm SRII}|$. The fig. \ref{pisr2mode} shows evolution of the corresponding conjugate momenta. When inside the Horizon the magnitude of quantity $|\Pi_{\rm SRII}|$ is large enough, but still small relative to $|\zeta_{\rm SRII}|$, and it continues to decrease even when the stochastic effects are encountered. Different modes experience different amount of decay in strength when going from the quantum-to-classical transition. As the modes come into the super-Horizon regime, the strength of $|\Pi_{\rm SRII}|$ takes on constant values and continues as such.

\section{ Tree level Power Spectrum and Noise Matrix Elements }
\label{app:C}

The comoving curvature perturbation that takes place at late time scale  where,  $-kc_{s}\tau \rightarrow \sigma$ the relevant tree-level contribution to the two-point cosmological correlation function  for comoving scalar  curvature perturbation can be expressed as : 
\bea
\langle\hat{\zeta_{\bf k}} \hat{\zeta_{\bf k'}}\rangle_{\bf Tree} &=& (2 \pi)^3 \, \del^3({\bf k}+{\bf k'})P_{\zeta\zeta} ^{\bf Tree}(k), \\
\langle\hat{\zeta_{\bf k}}\hat{\Pi_{{\zeta}{\bf k'}}}\rangle_{\bf Tree}&=& (2 \pi)^3 \, \del^3({\bf k}+{\bf k'})P_{\zeta \Pi_{\zeta}} ^ {\bf Tree}(k), \\
\langle\hat{\Pi_{{\zeta}{\bf k'}}}\hat{\zeta_{\bf k}}\rangle_{\bf Tree}&=& (2 \pi)^3 \, \del^3({\bf k}+{\bf k'})P_{\Pi_{\zeta} \zeta } ^ {\bf Tree}(k), \\
\langle\hat{\Pi_{{\zeta}{\bf k}}}\hat{\Pi_{{\zeta}{\bf k'}}}\rangle_{\bf Tree}&=& (2 \pi)^3 \, \del^3({\bf k}+{\bf k'})P_{\Pi_{\zeta}} ^ {\bf Tree}(k).
\eea 
Here $P_{\zeta \zeta}^{\bf Tree}(k),\,  P_{\zeta \Pi_\zeta}^{\bf Tree}(k), \, P_{\Pi_\zeta \zeta}^{\bf Tree}(k), \,   P_{\Pi_\zeta \Pi_\zeta}^{\bf Tree}(k) $ represent dimensionful power spectrum in Fourier space that can be evaluated by : 
\bea
 P_{\zeta \zeta}^{\bf Tree}(k) &=& 
\langle\hat{\zeta_{\bf k}} \hat{\zeta_{\bf -k}}\rangle_{(0,0)} = [\zeta_{\bf k}(\tau) \zeta_{\bf -k}(\tau) ]_{-kc_{s}\tau \rightarrow \sigma}= |\zeta_{\bf k}(\tau)|_{-kc_{s}\tau \rightarrow \sigma }^2 \\
 P_{\zeta \Pi_\zeta}^{\bf Tree}(k)&=& \langle\hat{\zeta_{\bf k}}\hat{\Pi_{\zeta_{\bf -k}}}\rangle_{(0,0)}=[\zeta_{\bf k}(\tau ) \Pi_{\zeta_{\bf -k}}(\tau)]_{-kc_{s}\tau \rightarrow \sigma}=(\zeta_{\bf k}(\tau) \Pi_{\zeta_{\bf k}}^*(\tau) )_{-kc_{s}\tau \rightarrow \sigma}\\
  P_{\Pi_\zeta \zeta }^{\bf Tree}(k)&=& \langle\hat{\Pi_{\zeta_{\bf k}}}\hat{\zeta_{\bf -k}}\rangle_{(0,0)}=[\Pi_{\zeta_{\bf k}}(\tau) \zeta_{\bf -k}(\tau ) ]_{-kc_{s}\tau \rightarrow \sigma}=(\Pi_{\zeta_{\bf k}}(\tau) \zeta_{\bf k}^*(\tau)  )_{-kc_{s}\tau \rightarrow \sigma }\\
 P_{{\Pi_\zeta}{ \Pi_\zeta }}^{\bf Tree}(k)&=& \langle\hat{\Pi_{\zeta_{\bf k}}}\hat{\Pi_{\zeta_{\bf -k}}}\rangle_{(0,0)} = [\Pi_{\zeta_{\bf k}}(\tau) \Pi_{\zeta_{\bf -k}}(\tau ) ]_{-kc_{s}\tau \rightarrow \sigma}=|\Pi_{\zeta_{\bf k}}(\tau)  |^2_{-kc_{s}\tau \rightarrow \sigma }
\eea 
It is convenient  to deal with the dimensionless power spectrum  in Fourier space for practical purposes and to connect cosmological observations. The dimensionless form of power spectrum can be expressed as :
\bea
\Delta^2_{\zeta\zeta,{\bf Tree} }(k)&=&\frac{k^3}{2 \pi^2}P_{\zeta\zeta}^{\bf Tree}(k) = \frac{k^3}{2 \pi^2}|\zeta|^2_{-kc_{s}\tau \rightarrow \sigma }\\
\Delta^2_{\zeta \Pi_\zeta,{\bf Tree} }(k)&=&\frac{k^3}{2 \pi^2} \tau P_{\zeta \Pi_\zeta}^{\bf Tree}(k) = \frac{k^3}{2 \pi^2} \tau (\zeta \Pi_{\zeta}^*)_{-kc_{s}\tau \rightarrow \sigma}\\
\Delta^2_{\Pi_\zeta \zeta ,{\bf Tree} }(k)&=&\frac{k^3}{2 \pi^2} \tau P_{\Pi_\zeta \zeta }^{\bf Tree}(k) = \frac{k^3}{2 \pi^2} \tau (\Pi_{\zeta} \zeta^*)_{-kc_{s}\tau \rightarrow \sigma}\\
\Delta^2_{\Pi_\zeta \Pi_\zeta ,{\bf Tree} }(k)&=&\frac{k^3}{2 \pi^2} \tau^2  P_{\Pi_\zeta \Pi_\zeta }^{\bf Tree}(k) = \frac{k^3}{2 \pi^2} \tau^2 |\Pi_{\zeta}|^2 _{-kc_{s}\tau \rightarrow \sigma }
\eea 
The power spectrum elements result from using the Fourier modes of the comoving curvature perturbation for the three separate regions. In stochastic inflation, the moment of horizon crossing is when the wavenumbers satisfy $-k c_s \tau = \sigma$. The term $\sigma $ is the stochastic tuning parameter that introduces stochasticity quantitatively into the power spectrum and the noise matrix elements. 

One can get a relation between the dimensionless scalar power spectrum and the noise matrix elements by utilizing the statistical properties of the quantum noise. 
We follow the assumption for the quantum initial conditions of the fields $\zeta_{k}$ and ${\Pi_\zeta}_{k}$ where they start from the vacuum state. Now, working at the leading order in perturbation theory implies Gaussian statistics for the modes and therefore all the statistical properties of the noise can be found by analysing their two-point correlation matrix given as follows: 
\bea  \label{noisecorrlmatrix}
\Sigma({\bf x_1},\tau_1;{\bf x_2},\tau_2)=\begin{pmatrix}
\langle 0 | \Hat{\xi}_\zeta({\bf x_1},\tau_1)\Hat{\xi}_\zeta({\bf x_2},\tau_2)|0\rangle & \langle 0 | \Hat{\xi}_\zeta({\bf x_1},\tau_1)\Hat{\xi}_{\Pi_{\zeta}}({\bf x_2},\tau_2)|0\rangle \\
\langle 0 |\Hat{\xi}_{\Pi_{\zeta}}({\bf x_1},\tau_1)\Hat{\xi}_\zeta({\bf x_2},\tau_2)|0\rangle & \langle 0 |\Hat{\xi}_{\Pi_{\zeta}}({\bf x_1},\tau_1)\Hat{\xi}_{\Pi_{\zeta}}({\bf x_2},\tau_2)|0\rangle
\end{pmatrix},
\eea 
where boldface ${\bf x}$ denotes vector quantities.
We show using the stochastic canonical quantization at the correlation level the relation between the noise matrix elements and the power spectrum can be found which will justify the correctness of our approach in the classical and the quantum regime. 
After acting of the annihilation and creation operators on the vacuum state $\ket{0}$, any element of the above correlation matrix can be described as:
\bea 
\Sigma_{{f_1},{g_1}}= \int_{\mathbb{R}^3}\frac{d^3 k }{(2\pi)^3}\frac{d}{d\tau}W\Bigg[\frac{k}{k_\sigma(\tau_1)}\Bigg]\frac{d}{d\tau}W\Bigg[ \frac{k}{k_\sigma(\tau_2)}\Bigg]f_k(\tau_1)g_k^*(\tau_2) e^{i{\bf k}.({\bf x_2}-{\bf x_1})},
\eea 
where \bea \Sigma_{{f_1},{g_1}}=\langle 0 | \xi_f({\bf x_1},\tau_1)\xi_g({\bf x_2},\tau_2)|0\rangle,\eea  and $\{f, g\}$ can be either $\{\zeta, \Pi_\zeta\}$. Because of having $[\xi_\zeta, \xi_{\Pi_\zeta}]\ne 0$, the order of subscripts $f$ and $g$ does matter. In the above, the dependence of mode functions $\zeta_{k}$ and ${\Pi_\zeta}_{k}$ on the norm of ${\bf k}$ makes the angular integral over ${\bf k}/k $ easier to evaluate. As a result, we obtain the following expression:
\bea
\Sigma_{f_1, g_1} = \int_{\mathbb{R}^+} \frac{k^{2} dk}{2\pi^2}\frac{d}{d\tau}W\Bigg[\frac{k}{k_\sigma(\tau_1)}\Bigg]\frac{d}{d\tau}W\Bigg[\frac{k}{k_\sigma(\tau_2)}\Bigg]f_k(\tau_1)g_k ^*(\tau_2)\frac{\sin(k|{\bf x_2}-{\bf x_1}|)}{k|{\bf x_2}-{\bf x_1}|},
\eea
here $W(k/k_{\sigma})$ acts as a window function, which for convenience is chosen to be a Heaviside Theta function, $W(k/k_\sigma)= \Theta(k/k_\sigma - 1)$. Taking its time derivative as the Dirac delta distribution and the integrand of above equation incorporates $\del(k-k_\sigma(\tau_1) )\del(k-k_\sigma(\tau_2))$ giving us $\del(\tau_1 - \tau_2)$ which indicates the presence of white noise. We are now lead to the expression:
\bea \label{noiseelement}
\Sigma_{{f_1},{ g_1}}= \frac{1}{6 \pi^2}\frac{dk_{\sigma}^3(\tau)}{d\tau }\Bigg|_{\tau_1}f_{k=k_{\sigma}(\tau)}g_{k=k_{\sigma}(\tau)}^*\frac{\sin[k_{\sigma}(\tau_1)|{\bf x_2}-{\bf x_1}|]}{k_{\sigma}(\tau_1)|{\bf x_2}-{\bf x_2}|}\delta(\tau_1 - \tau_2).
\eea 
We focus on the case where the noises are maximally correlated in space which happens at points ${\bf x_1}\rightarrow {\bf x_2}$ leading to \bea \lim_{{\bf x_1}\rightarrow {\bf x_2}} \frac{\sin[k_{\sigma}(\tau_1)|{\bf x_2}-{\bf x_1}|]}{k_{\sigma}(\tau_1)|{\bf x_2}-{\bf x_1}|}=1.\eea The correlations remain non-zero only at equal time as the noises are white. We can thus write the noise correlation matrix as:
\bea
\Sigma_{f_1,g_1}\equiv \Sigma_{f,g}(\tau_1)\del(\tau_1 - \tau_2).
\eea 
and using this crucial statement about the noises one can further use the  following definition of the power spectrum element between the quantum fluctuations as:
\bea 
P_{f,g}(k;\tau)=\frac{k^3}{2\pi^2} f_{k}(\tau)g_{k}^*(\tau),
\eea 
and this directly leads to the following expression with the time-dependent noise matrix element:
\bea \label{noisepower}
\Sigma_{f,g}(\tau) = \frac{d\ln{(k_\sigma(\tau))}}{d\tau}P_{f,g}(k_\sigma(\tau);\tau).
\eea 

The noise matrix elements are the correlators of 
of the noise correlation matrix elements can be computed using the following relation from the dimensionless power spectrum. These relations are obtained from the derivation provided in section \ref{s7} and the expression used here is given in eq.$(\ref{noisepower})$ where $f$ and $g$ correspond to $\zeta$ and $\Pi_\zeta$ respectively :
\bea
\Sigma_{\zeta\zeta}&=& (1-\epsilon)\Delta^2_{\zeta\zeta,{\bf Tree} }(k)= (1-\epsilon)\frac{k^3}{2 \pi^2}P_{\zeta\zeta}^{\bf Tree}(k) = (1-\epsilon) \frac{k^3}{2 \pi^2}|\zeta|^2_{-kc_{s}\tau \rightarrow \sigma},\\
\Sigma_{\zeta \Pi_\zeta}&=&(1-\epsilon)\Delta^2_{\zeta \Pi_\zeta,{\bf Tree} }(k)=(1-\epsilon) \frac{k^3}{2 \pi^2} \tau P_{\zeta \Pi_\zeta}^{\bf Tree}(k) =(1-\epsilon) \frac{k^3}{2 \pi^2} \tau (\zeta \Pi_{\zeta}^*)_{-kc_{s}\tau \rightarrow \sigma },\\
\Sigma_{ \Pi_\zeta \zeta }&=&(1-\epsilon) \Delta^2_{\Pi_\zeta \zeta ,{\bf Tree} }(k)=(1-\epsilon) \frac{k^3}{2 \pi^2} \tau P_{\Pi_\zeta \zeta }^{\bf Tree}(k) = (1-\epsilon) \frac{k^3}{2 \pi^2} \tau (\Pi_{\zeta} \zeta^*)_{-kc_{s}\tau \rightarrow \sigma },\\
\Sigma_{\Pi_\zeta\Pi_\zeta }&=&(1-\epsilon) \Delta^2_{\Pi_\zeta \Pi_\zeta ,{\bf Tree} }(k) = (1-\epsilon) \frac{k^3}{2 \pi^2} \tau^2  P_{\Pi_\zeta \Pi_\zeta }^{\bf Tree}(k) =  (1-\epsilon) \frac{k^3}{2 \pi^2} \tau^2 |\Pi_{\zeta}|^2 _{-kc_{s}\tau \rightarrow \sigma }.
\eea 

\subsection{Results in SRI region}
\label{appCa}

In this section we present the analytic expressions of the scalar power spectrum in the presence of stochastic effects for an arbitrary initial quantum vacuum condition and general quasi de Sitter background spacetime.

\subsubsection{Power spectrum in SRI}
\label{appCa1}

The upcoming expressions for the different elements of the scalar power spectrum are calculated using the general mode function solutions for the first slow-roll (SRI) phase as presented in eqs.(\ref{modezetaSRI}-\ref{modepiSRI}). The following power spectrum elements are evaluated at the Horizon crossing condition for an arbitrary wavenumber $k$ as $-kc_{s}\tau=\sigma$ where $\sigma$ is the stochastic, coarse-graining parameter.
\bea
 \Delta_{\zeta\zeta}^2&=&2^{2\nu-3}  (\sigma)^{3-2\nu} \frac{H^2}{8\pi^2\epsilon  c_s M_p^2}\Bigg|\frac{\Gamma{(\nu)}}{\Gamma{(\frac{3}{2})}}\Bigg|^2(1+\sigma^2)\Bigg|\alpha_1 e^{-i(\frac{\pi}{2}(\nu+\frac{1}{2})-\sigma)}-\frac{(1+i \sigma)}{(1-i \sigma)}\beta_1 e^{i(\frac{\pi}{2}(\nu+\frac{1}{2})-\sigma)}\Bigg|^2,
\\ 
\Delta_{\zeta\Pi_\zeta}^2&=&2^{2\nu-3}  (\sigma)^{3-2\nu} \frac{H^2}{8\pi^2\epsilon  c_s M_p^2}\Bigg|\frac{\Gamma{(\nu)}}{\Gamma{(\frac{3}{2})}}\Bigg|^2\Bigg[\Bigg(\frac{3}{2}-\nu \Bigg) (1+\sigma^2)+\sigma^2\Bigg]\Bigg\{\Bigg|\alpha_1 e^{-i(\frac{\pi}{2}(\nu+\frac{1}{2})-\sigma)}-\beta_1 e^{i(\frac{\pi}{2}(\nu+\frac{1}{2})-\sigma)}\Bigg|^2\nonumber
 \\&& \quad\quad\quad\quad\quad\quad\quad\quad
+\frac{2\Bigg(\frac{3}{2}-\nu\Bigg)\sigma^2}{\Bigg\{\Bigg(\frac{3}{2}-\nu \Bigg) (1+\sigma^2)+\sigma^2\Bigg\}}\Bigg\{\alpha_1^*\beta_1 e^{2i(\frac{\pi}{2}(\nu+\frac{1}{2})-\sigma)}+\beta_1^*\alpha_1 e^{-2i(\frac{\pi}{2}(\nu+\frac{1}{2})-\sigma)} \Bigg\} \Bigg\},
\\
\Delta_{\Pi_\zeta\zeta}^2&=&2^{2\nu-3}  (\sigma)^{3-2\nu} \frac{H^2}{8\pi^2\epsilon  c_s M_p^2}\Bigg|\frac{\Gamma{(\nu)}}{\Gamma{(\frac{3}{2})}}\Bigg|^2\Bigg[\Bigg(\frac{3}{2}-\nu \Bigg) (1+\sigma^2)+\sigma^2\Bigg]\Bigg\{\Bigg|\alpha_1 e^{-i(\frac{\pi}{2}(\nu+\frac{1}{2})-\sigma)}-\beta_1 e^{i(\frac{\pi}{2}(\nu+\frac{1}{2})-\sigma)}\Bigg|^2\nonumber
 \\&& \quad\quad\quad\quad\quad\quad\quad\quad
+ \frac{2\Bigg(\frac{3}{2}-\nu\Bigg)\sigma^2}{\Bigg\{\Bigg(\frac{3}{2}-\nu \Bigg) (1+\sigma^2)+\sigma^2\Bigg\}}\Bigg\{\alpha_1^*\beta_1 e^{2i(\frac{\pi}{2}(\nu+\frac{1}{2})-\sigma)}+\beta_1^*\alpha_1 e^{-2i(\frac{\pi}{2}(\nu+\frac{1}{2})-\sigma)} \Bigg\}\Bigg\},
\\
\Delta_{\Pi_\zeta\Pi_\zeta}^2&=&2^{2\nu-3}(\sigma)^{3-2\nu}\frac{H^2}{8\pi^2 \epsilon  
 c_s M_p^2}\Bigg|\frac{\Gamma{(\nu)}}{\Gamma{(\frac{3}{2})}}\Bigg|^2 \Bigg\{ \Bigg(\Bigg(\frac{3}{2}-\nu\Bigg)+\sigma^2 \Bigg)^2+\Bigg(\frac{3}{2}-\nu\Bigg)^2 \sigma^2 \Bigg \} \nonumber
 \\ && \quad\quad\quad\quad\quad\quad\quad\quad\quad
 \times\Bigg|\alpha_1 e^{-i(\frac{\pi}{2}(\nu+\frac{1}{2})-\sigma)}-\beta_1 \frac{\Bigg\{\Bigg(\frac{3}{2}-\nu\Bigg)(1+i \sigma )+\sigma^2 \Bigg \}}{\Bigg\{\Bigg(\frac{3}{2}-\nu\Bigg)(1-i \sigma)+\sigma^2 \Bigg \}} e^{i(k c_s\tau+\frac{\pi}{2}(\nu+\frac{1}{2}))}\Bigg|^2.
\eea

The above mentioned power spectrum elements can be expressed using the Bunch Davies initial vacuum state, where $\alpha_1=1$ and $\beta_1=0$ which is a more common choice of initial condition when talking about inflationary observables. The expressions for general de Sitter condition are as follows:

\bea \label{pspecsr1BD}
\Delta_{{\zeta\zeta},{\bf BD}}^2&=&2^{2\nu-3}  (\sigma)^{3-2\nu} \frac{H^2}{8\pi^2\epsilon  c_s M_p^2}\Bigg|\frac{\Gamma{(\nu)}}{\Gamma{(\frac{3}{2})}}\Bigg|^2(1+\sigma^2),
\\ 
\Delta_{{\zeta\Pi_\zeta},{\bf BD}}^2&=&2^{2\nu-3}  (\sigma)^{3-2\nu} \frac{H^2}{8\pi^2\epsilon  c_s M_p^2}\Bigg|\frac{\Gamma{(\nu)}}{\Gamma{(\frac{3}{2})}}\Bigg|^2\Bigg[\Bigg(\frac{3}{2}-\nu \Bigg) (1+\sigma^2)+\sigma^2\Bigg],
\\
\Delta_{{\Pi_\zeta\zeta},{\bf BD}}^2&=&2^{2\nu-3}  (\sigma)^{3-2\nu} \frac{H^2}{8\pi^2\epsilon  c_s M_p^2}\Bigg|\frac{\Gamma{(\nu)}}{\Gamma{(\frac{3}{2})}}\Bigg|^2\Bigg[\Bigg(\frac{3}{2}-\nu \Bigg) (1+\sigma^2)+\sigma^2\Bigg],
\\
\Delta_{{\Pi_\zeta\Pi_\zeta},{\bf BD}}^2&=&2^{2\nu-3}(\sigma)^{3-2\nu}\frac{H^2}{8\pi^2 \epsilon  c_s M_p^2}\Bigg|\frac{\Gamma{(\nu)}}{\Gamma{(\frac{3}{2})}}\Bigg|^2 \Bigg[\Bigg(\Bigg(\frac{3}{2}-\nu\Bigg)+\sigma^2 \Bigg)^2+\Bigg(\frac{3}{2}-\nu\Bigg)^2 \sigma^2 \Bigg]. 
\eea
Further implementing the limiting value $\nu=\frac{3}{2}$ in Bunch Davies vacuum derived result one can get the simplified
expression for Power spectrum elements  in the case of de Sitter space:

\bea \label{pspecsr1dS}
\Delta_{{\zeta\zeta},{\bf dS}}^2&=&   \frac{H^2}{8\pi^2\epsilon  c_s M_p^2}(1+\sigma^2),
\\
\Delta_{{\zeta\Pi_\zeta},{\bf dS}}^2&=&  \frac{H^2}{8\pi^2\epsilon  c_s M_p^2}\sigma^2,
\eea\bea
\Delta_{{\Pi_\zeta\zeta},{\bf dS}}^2&=&  \frac{H^2}{8\pi^2\epsilon  c_s M_p^2}\sigma^2,
\\
\Delta_{{\Pi_\zeta\Pi_\zeta},{\bf dS}}^2&=&\frac{H^2}{8\pi^2 \epsilon  c_s M_p^2} \sigma^4.
\eea

\subsubsection{Noise Matrix elements in SRI}
\label{appCa2}

The noise correlation matrix elements in the SRI for an arbitrary de Sitter background spacetime, characterized by `$\nu$' and a general initial quantum vacuum state specified with coefficients $(\alpha_{1},\beta_{1})$  can be written using the following expressions:
\bea 
\Sigma_{\zeta\zeta}&=&(1-\epsilon) 2^{2\nu-3}  (\sigma)^{3-2\nu} \frac{H^2}{8\pi^2\epsilon  c_s M_p^2}\Bigg|\frac{\Gamma{(\nu)}}{\Gamma{(\frac{3}{2})}}\Bigg|^2(1+\sigma^2)\Bigg|\alpha_1 e^{-i(\frac{\pi}{2}(\nu+\frac{1}{2})-\sigma)}-\frac{(1+i \sigma)}{(1-i \sigma)}\beta_1 e^{i(\frac{\pi}{2}(\nu+\frac{1}{2})-\sigma)}\Bigg|^2,
\\
\Sigma_{\zeta\Pi_\zeta}&=&(1-\epsilon) 2^{2\nu-3}  (\sigma)^{3-2\nu} \frac{H^2}{8\pi^2\epsilon  c_s M_p^2}\Bigg|\frac{\Gamma{(\nu)}}{\Gamma{(\frac{3}{2})}}\Bigg|^2\Bigg[\Bigg(\frac{3}{2}-\nu \Bigg) (1+\sigma^2)+\sigma^2\Bigg]\Bigg\{\Bigg|\alpha_1 e^{-i(\frac{\pi}{2}(\nu+\frac{1}{2})-\sigma)}-\beta_1 e^{i(\frac{\pi}{2}(\nu+\frac{1}{2})-\sigma)}\Bigg|^2\nonumber
 \\  && \quad \quad \quad \quad \quad \quad \quad \quad \quad 
+\frac{2\Bigg(\frac{3}{2}-\nu\Bigg)\sigma^2}{\Bigg\{\Bigg(\frac{3}{2}-\nu \Bigg) (1+\sigma^2)+\sigma^2\Bigg\}}\Bigg\{\alpha_1^*\beta_1 e^{2i(\frac{\pi}{2}(\nu+\frac{1}{2})-\sigma)}+\beta_1^*\alpha_1 e^{-2i(\frac{\pi}{2}(\nu+\frac{1}{2})-\sigma)} \Bigg\}\Bigg\},
\\
\Sigma_{\Pi_\zeta\zeta}&=&(1-\epsilon) 2^{2\nu-3}  (\sigma)^{3-2\nu} \frac{H^2}{8\pi^2\epsilon  c_s M_p^2}\Bigg|\frac{\Gamma{(\nu)}}{\Gamma{(\frac{3}{2})}}\Bigg|^2\Bigg[\Bigg(\frac{3}{2}-\nu \Bigg) (1+\sigma^2)+\sigma^2\Bigg]\Bigg\{\Bigg|\alpha_1 e^{-i(\frac{\pi}{2}(\nu+\frac{1}{2})-\sigma)}-\beta_1 e^{i(\frac{\pi}{2}(\nu+\frac{1}{2})-\sigma)}\Bigg|^2\nonumber
 \\ && \quad \quad \quad \quad \quad \quad \quad \quad \quad 
+ \frac{2\Bigg(\frac{3}{2}-\nu\Bigg)\sigma^2}{\Bigg\{\Bigg(\frac{3}{2}-\nu \Bigg) (1+\sigma^2)+\sigma^2\Bigg\}}\Bigg\{\alpha_1^*\beta_1 e^{2i(\frac{\pi}{2}(\nu+\frac{1}{2})-\sigma)}+\beta_1^*\alpha_1 e^{-2i(\frac{\pi}{2}(\nu+\frac{1}{2})-\sigma)} \Bigg\}\Bigg\},
\\
\Sigma_{\Pi_\zeta\Pi_\zeta}&=&(1-\epsilon) 2^{2\nu-3}(\sigma)^{3-2\nu}\frac{H^2}{8\pi^2 \epsilon  
 c_s M_p^2}\Bigg|\frac{\Gamma{(\nu)}}{\Gamma{(\frac{3}{2})}}\Bigg|^2 \Bigg[\Bigg\{\Bigg(\frac{3}{2}-\nu\Bigg)+\sigma^2 \Bigg\}^2+\Bigg(\frac{3}{2}-\nu\Bigg)^2 \sigma^2 \Bigg ] \nonumber
 \\ && \quad \quad \quad \quad \quad \quad  \quad \quad \quad \times
 \Bigg|\alpha_1 e^{-i(\frac{\pi}{2}(\nu+\frac{1}{2})-\sigma)}-\beta_1 \frac{\Bigg\{\Bigg(\frac{3}{2}-\nu\Bigg)(1+i \sigma )+\sigma^2 \Bigg \}}{\Bigg\{\Bigg(\frac{3}{2}-\nu\Bigg)(1-i \sigma)+\sigma^2 \Bigg \}} e^{i(\frac{\pi}{2}(\nu+\frac{1}{2})-\sigma)}\Bigg|^2.
\eea

The noise correlation matrix elements in Bunch Davies initial vacuum state where, $\alpha_1 = 1$ and $\beta_1 = 0$ can be expressed by
substituting these values in the above set of noise matrix elements.
\bea 
\Sigma_{{\zeta\zeta},{\bf BD}}&=&(1-\epsilon) 2^{2\nu-3}  (\sigma)^{3-2\nu} \frac{H^2}{8\pi^2\epsilon  c_s M_p^2}\Bigg|\frac{\Gamma{(\nu)}}{\Gamma{(\frac{3}{2})}}\Bigg|^2(1+\sigma^2),
\\
\Sigma_{{\zeta\Pi_\zeta},{\bf BD}}&=&(1-\epsilon) 2^{2\nu-3}  (\sigma)^{3-2\nu} \frac{H^2}{8\pi^2\epsilon  c_s M_p^2}\Bigg|\frac{\Gamma{(\nu)}}{\Gamma{(\frac{3}{2})}}\Bigg|^2\Bigg[\Bigg(\frac{3}{2}-\nu \Bigg) (1+\sigma^2)+\sigma^2\Bigg],
\\
\Sigma_{{\Pi_\zeta\zeta},{\bf BD}}&=&(1-\epsilon) 2^{2\nu-3}  (\sigma)^{3-2\nu} \frac{H^2}{8\pi^2\epsilon  c_s M_p^2}\Bigg|\frac{\Gamma{(\nu)}}{\Gamma{(\frac{3}{2})}}\Bigg|^2\Bigg[\Bigg(\frac{3}{2}-\nu \Bigg) (1+\sigma^2)+\sigma^2\Bigg],
\\
  \Sigma_{{\Pi_\zeta\Pi_\zeta},{\bf BD}}&=&(1-\epsilon) 2^{2\nu-3}(\sigma)^{3-2\nu}\frac{H^2}{8\pi^2 \epsilon  
 c_s M_p^2}\Bigg|\frac{\Gamma{(\nu)}}{\Gamma{(\frac{3}{2})}}\Bigg|^2 \Bigg[ \Bigg\{\Bigg(\frac{3}{2}-\nu\Bigg)+\sigma^2 \Bigg\}^2+\Bigg(\frac{3}{2}-\nu\Bigg)^2 \sigma^2 \Bigg].
\eea
Further implementing the limiting value $\nu = \frac{3}{2}$ 
in Bunch Davies vacuum derived result one can get the simplified
expression for noise correlation matrix elements in the case of de Sitter space.
\bea 
\Sigma_{{\zeta\zeta},{\bf dS}}&=&(1-\epsilon)   \frac{H^2}{8\pi^2\epsilon  c_s M_p^2}(1+\sigma^2),
\\
\Sigma_{{\zeta\Pi_\zeta},{\bf dS}}&=&(1-\epsilon)  \frac{H^2}{8\pi^2\epsilon  c_s M_p^2}\sigma^2,
\\
\Sigma_{{\Pi_\zeta\zeta},{\bf dS}}&=&(1-\epsilon)\frac{H^2}{8\pi^2\epsilon  c_s M_p^2} \sigma^2,
\\
  \Sigma_{{\Pi_\zeta\Pi_\zeta},{\bf dS}}&=&(1-\epsilon) \frac{H^2}{8\pi^2 \epsilon  
 c_s M_p^2}\sigma^4.
\eea

\subsection{Results in USR region}
\label{appCb}

\subsubsection{Power spectrum in USR}
\label{appCb1}

The upcoming expressions for the different elements of the scalar power spectrum are calculated using the general mode function solutions for the ultra slow-roll (USR) phase as presented in eqs.(\ref{modezetaUSR}-\ref{modepiUSR}). The following power spectrum elements are evaluated at the Horizon crossing condition for an arbitrary wavenumber $k$ as $-kc_{s}\tau=\sigma$ where $\sigma$ is the stochastic, coarse-graining parameter.
\bea
\Delta_{\zeta\zeta}^2&=&2^{2\nu-3}  (\sigma)^{3-2\nu} \frac{H^2}{8\pi^2\epsilon  c_s M_p^2} 
 \Bigg(\frac{k_e}{k_s}\Bigg)^6 \Bigg|\frac{\Gamma{(\nu)}}{\Gamma{(\frac{3}{2})}}\Bigg|^2(1+\sigma^2)\Bigg|\alpha_2 e^{-i(\frac{\pi}{2}(\nu+\frac{1}{2})-\sigma)}-\frac{(1+i \sigma)}{(1-i \sigma)}\beta_2 e^{i(\frac{\pi}{2}(\nu+\frac{1}{2})-\sigma)}\Bigg|^2,
 \eea\bea \Delta_{\zeta\Pi_\zeta}^2&=&2^{2\nu-3}  (\sigma)^{3-2\nu} \frac{H^2}{8\pi^2\epsilon  c_s M_p^2}\Bigg(\frac{k_e}{k_s}\Bigg)^6 
 \Bigg|\frac{\Gamma{(\nu)}}{\Gamma{(\frac{3}{2})}}\Bigg|^2\Bigg[\Bigg(\Bigg(\frac{3}{2}-\nu \Bigg)-3\Bigg) (1+\sigma^2)+\sigma^2\Bigg]\Bigg\{\Bigg|\alpha_2 e^{-i(\frac{\pi}{2}(\nu+\frac{1}{2})-\sigma)}-\beta_2 e^{i(\frac{\pi}{2}(\nu+\frac{1}{2})-\sigma)}\Bigg|^2\nonumber
 \\ && \quad \quad \quad \quad \quad \quad \quad\quad  \quad \quad \quad 
+\frac{2\Bigg(\frac{3}{2}+\nu\Bigg)\sigma^2}{\Bigg\{\Bigg(\frac{3}{2}+\nu \Bigg) (1+\sigma^2)-\sigma^2\Bigg\}}\Bigg\{\alpha_2^*\beta_2 e^{2i(\frac{\pi}{2}(\nu+\frac{1}{2})-\sigma)}+\beta_2^*\alpha_2 e^{-2i(\frac{\pi}{2}(\nu+\frac{1}{2})-\sigma)} \Bigg\}\Bigg\},
\\
\Delta_{\Pi_\zeta\zeta}^2&=&2^{2\nu-3}  (\sigma)^{3-2\nu} \frac{H^2}{8\pi^2\epsilon  c_s M_p^2}\Bigg(\frac{k_e}{k_s}\Bigg)^6 
\Bigg|\frac{\Gamma{(\nu)}}{\Gamma{(\frac{3}{2})}}\Bigg|^2\Bigg[\Bigg(\Bigg(\frac{3}{2}-\nu \Bigg) -3\Bigg)(1+\sigma^2)+\sigma^2\Bigg]\Bigg\{\Bigg|\alpha_2 e^{-i(\frac{\pi}{2}(\nu+\frac{1}{2})-\sigma)}-\beta_2 e^{i(\frac{\pi}{2}(\nu+\frac{1}{2})-\sigma)}\Bigg|^2\nonumber
 \\&&\quad \quad \quad \quad \quad \quad \quad \quad \quad \quad \quad 
+ \frac{2\Bigg(\frac{3}{2}+\nu\Bigg)\sigma^2}{\Bigg\{\Bigg(\frac{3}{2}+\nu \Bigg) (1+\sigma^2)-\sigma^2\Bigg\}}\Bigg\{\alpha_2^*\beta_2 e^{2i(\frac{\pi}{2}(\nu+\frac{1}{2})-\sigma)}+\beta_2^*\alpha_2 e^{-2i(\frac{\pi}{2}(\nu+\frac{1}{2})-\sigma)} \Bigg\}\Bigg\},
\\
\Delta_{\Pi_\zeta\Pi_\zeta}^2&=&2^{2\nu-3}(\sigma)^{3-2\nu}\frac{H^2}{8\pi^2 \epsilon  
 c_s M_p^2}\Bigg(\frac{k_e}{k_s}\Bigg)^6 
 \Bigg|\frac{\Gamma{(\nu)}}{\Gamma{(\frac{3}{2})}}\Bigg|^2 \Bigg[\Bigg\{\Bigg(\frac{3}{2}+ \nu\Bigg)- \sigma^2 \Bigg\}^2+\Bigg(\frac{3}{2}+\nu\Bigg)^2 \sigma^2 \Bigg ] \nonumber
 \\&& \quad \quad\quad\quad\quad\quad\quad\quad\quad\quad\quad\quad \times
 \Bigg|\alpha_2 e^{-i(\frac{\pi}{2}(\nu+\frac{1}{2})-\sigma)}-\beta_2 \frac{\Bigg\{\Bigg(\frac{3}{2}+ \nu\Bigg)(1+i \sigma )- \sigma^2 \Bigg \}}{\Bigg\{\Bigg(\frac{3}{2}+ \nu\Bigg)(1-i \sigma)- \sigma^2 \Bigg \}} e^{i(\frac{\pi}{2}(\nu+\frac{1}{2})-\sigma)}\Bigg|^2.
\eea
The results with an arbitrary `$\nu$' under the choice of initial Bunch Davies vacuum condition have their Bogoliubov coefficients replaced from $(\alpha_{2},\beta_{2})$ to $(\alpha_{2,{\bf BD}},\beta_{2,{\bf BD}})$.
Further implementing the limiting value $\nu=3/2 $ in the derived result of $(\alpha_{2,{\bf BD}},\beta_{2,{\bf BD}})$ for the USR, one can get the simplified expression for power spectrum elements in the case of exact de Sitter space with the new Bogoliubov coefficients $(\alpha_{2, {\bf dS}},\beta_{2, {\bf dS}})$:
\bea \label{pspecusrdS}
\Delta_{{\zeta\zeta},{\bf dS}}^2&=&  \frac{H^2}{8\pi^2\epsilon  c_s M_p^2} 
 \Bigg(\frac{k_e}{k_s}\Bigg)^6 (1+\sigma^2)\Bigg|\alpha_{2,{\bf dS}} e^{i\sigma}-\frac{(1+i \sigma)}{(1-i \sigma)}\beta_{2,{\bf dS}} e^{-i\sigma}\Bigg|^2,
 \\
 \Delta_{{\zeta\Pi_\zeta},{\bf dS}}^2&=&\frac{H^2}{8\pi^2\epsilon  c_s M_p^2}\Bigg(\frac{k_e}{k_s}\Bigg)^6 
 \Bigg[-3 (1+\sigma^2)+\sigma^2\Bigg]\Bigg\{\Bigg|\alpha_{2,{\bf dS}} e^{i\sigma}-\beta_{2,{\bf dS}} e^{-i\sigma}\Bigg|^2\nonumber
 \\&& \quad \quad \quad \quad \quad\quad  \quad \quad \quad \quad \quad 
+\frac{6\sigma^2}{\Bigg\{3 (1+\sigma^2)-\sigma^2\Bigg\}}\Bigg\{\alpha_{2,{\bf dS}}^*\beta_{2,{\bf dS}} e^{-2i\sigma}+\beta_{2,{\bf dS}}^*\alpha_{2,{\bf dS}} e^{2i\sigma} \Bigg\}\Bigg\},
\\
\Delta_{{\Pi_\zeta\zeta},{\bf dS}}^2&=& \frac{H^2}{8\pi^2\epsilon  c_s M_p^2}\Bigg(\frac{k_e}{k_s}\Bigg)^6 
\Bigg[-3(1+\sigma^2)+\sigma^2\Bigg]\Bigg\{\Bigg|\alpha_{2,{\bf dS}} e^{i\sigma}-\beta_{2,{\bf dS}} e^{-i\sigma}\Bigg|^2\nonumber
\\&&\quad \quad \quad \quad \quad \quad \quad \quad \quad \quad \quad 
+ \frac{6\sigma^2}{\Bigg\{3(1+\sigma^2)-\sigma^2\Bigg\}}\Bigg\{\alpha_{2,{\bf dS}}^*\beta_{2,{\bf dS}} e^{-2i\sigma}+\beta_{2,{\bf dS}}^* \alpha_{2,{\bf dS}} e^{2i\sigma} \Bigg\}\Bigg\},
\\
\Delta_{{\Pi_\zeta\Pi_\zeta},{\bf dS}}^2&=&\frac{H^2}{8\pi^2 \epsilon  
 c_s M_p^2}\Bigg(\frac{k_e}{k_s}\Bigg)^6 
  \Bigg[\Bigg\{3- \sigma^2 \Bigg\}^2+9\sigma^2 \Bigg ]
 \Bigg|\alpha_{2,{\bf dS}} e^{i\sigma}-\beta_{2,{\bf dS}} \frac{\Bigg\{3(1+i \sigma )- \sigma^2 \Bigg \}}{\Bigg\{3(1-i \sigma)- \sigma^2 \Bigg \}} e^{-i\sigma}\Bigg|^2.
\eea
\subsubsection{Noise Matrix elements in USR}
\label{appCb2}

The noise correlation matrix elements in the USR for an arbitrary de Sitter background spacetime characterized by `$\nu$' and general initial quantum vacuum state specified with coefficients $(\alpha_{2},\beta_{2})$ is written as follows:
\bea
\Sigma_{\zeta\zeta}&=&(1-\epsilon)2^{2\nu-3}  (\sigma)^{3-2\nu} \frac{H^2}{8\pi^2\epsilon  c_s M_p^2} 
 \Bigg(\frac{k_e}{k_s}\Bigg)^6 \Bigg|\frac{\Gamma{(\nu)}}{\Gamma{(\frac{3}{2})}}\Bigg|^2(1+\sigma^2)\Bigg|\alpha_2 e^{-i(\frac{\pi}{2}(\nu+\frac{1}{2})-\sigma)}-\frac{(1+i \sigma)}{(1-i \sigma)}\beta_2 e^{i(\frac{\pi}{2}(\nu+\frac{1}{2})-\sigma)}\Bigg|^2,
\\
\Sigma_{\zeta\Pi_\zeta}&=&(1-\epsilon) 2^{2\nu-3}  (\sigma)^{3-2\nu} \frac{H^2}{8\pi^2\epsilon  c_s M_p^2}\Bigg(\frac{k_e}{k_s}\Bigg)^6 
 \Bigg|\frac{\Gamma{(\nu)}}{\Gamma{(\frac{3}{2})}}\Bigg|^2\Bigg\{\Bigg(\Bigg(\frac{3}{2}-\nu \Bigg)-3\Bigg) (1+\sigma^2)+\sigma^2\Bigg\}\Bigg\{\Bigg|\alpha_2 e^{-i(\frac{\pi}{2}(\nu+\frac{1}{2})-\sigma)}\nonumber
 \\&&
 -\beta_2 e^{i(\frac{\pi}{2}(\nu+\frac{1}{2})-\sigma)}\Bigg|^2
+\frac{2\Bigg(\frac{3}{2}+\nu\Bigg)\sigma^2}{\Bigg\{\Bigg(\frac{3}{2}+\nu \Bigg) (1+\sigma^2)-\sigma^2\Bigg\}}\Bigg\{\alpha_2^*\beta_2 e^{2i(\frac{\pi}{2}(\nu+\frac{1}{2})-\sigma)}
 +\beta_2^*\alpha_2 e^{-2i(\frac{\pi}{2}(\nu+\frac{1}{2})-\sigma)} \Bigg\}\Bigg\},
\\ 
\Sigma_{\Pi_\zeta\zeta}&=&(1-\epsilon) 2^{2\nu-3}  (\sigma)^{3-2\nu} \frac{H^2}{8\pi^2\epsilon  c_s M_p^2}\Bigg(\frac{k_e}{k_s}\Bigg)^6 
 \Bigg|\frac{\Gamma{(\nu)}}{\Gamma{(\frac{3}{2})}}\Bigg|^2\Bigg\{\Bigg(\Bigg(\frac{3}{2}-\nu \Bigg) -3\Bigg)(1+\sigma^2)+\sigma^2\Bigg\}\Bigg\{\Bigg|\alpha_2 e^{-i(\frac{\pi}{2}(\nu+\frac{1}{2})-\sigma)}\nonumber
 \\&&
 -\beta_2 e^{i(\frac{\pi}{2}(\nu+\frac{1}{2})-\sigma)}\Bigg|^2
+ \frac{2\Bigg(\frac{3}{2}+\nu\Bigg)\sigma^2}{\Bigg\{\Bigg(\frac{3}{2}+\nu \Bigg) (1+\sigma^2)-\sigma^2\Bigg\}}\Bigg\{\alpha_2^*\beta_2 e^{2i(\frac{\pi}{2}(\nu+\frac{1}{2})-\sigma)}+\beta_2^*\alpha_2 e^{-2i(\frac{\pi}{2}(\nu+\frac{1}{2})-\sigma)} \Bigg\}\Bigg\},
\eea\bea
\Sigma_{\Pi_\zeta\Pi_\zeta}&=&(1-\epsilon) 2^{2\nu-3}(\sigma)^{3-2\nu}\frac{H^2}{8\pi^2 \epsilon  
 c_s M_p^2}\Bigg(\frac{k_e}{k_s}\Bigg)^6 
 \Bigg|\frac{\Gamma{(\nu)}}{\Gamma{(\frac{3}{2})}}\Bigg|^2 \Bigg\{\Bigg(\Bigg(\frac{3}{2}+ \nu\Bigg)- \sigma^2 \Bigg)^2+\Bigg(\frac{3}{2}+\nu\Bigg)^2 \sigma^2 \Bigg \} \nonumber
\\ && \quad \quad\quad\quad\quad
 \Bigg|\alpha_2 e^{-i(\frac{\pi}{2}(\nu+\frac{1}{2})-\sigma)}-\beta_2 \frac{\Bigg\{\Bigg(\frac{3}{2}+ \nu\Bigg)(1+i \sigma )- \sigma^2 \Bigg \}}{\Bigg\{\Bigg(\frac{3}{2}+ \nu\Bigg)(1-i \sigma)- \sigma^2 \Bigg \}} e^{i(\frac{\pi}{2}(\nu+\frac{1}{2})-\sigma)}\Bigg|^2.
\eea
Similar to the previous case in SRI, we further implement the Bunch Davies vacuum conditions followed by the limiting value of $\nu=3/2 $ in the Bunch Davies vacuum-derived result to get the new set $(\alpha_{2, {\bf dS}},\beta_{2, {\bf dS}})$ and the simplified expression for noise matrix elements in the exact de Sitter space becomes:
\bea
\Sigma_{{\zeta\zeta},{\bf dS}}&=&(1-\epsilon) \frac{H^2}{8\pi^2\epsilon  c_s M_p^2} 
 \Bigg(\frac{k_e}{k_s}\Bigg)^6 (1+\sigma^2)\Bigg|\alpha_{2,{\bf dS}} e^{i\sigma}-\frac{(1+i \sigma)}{(1-i \sigma)}\beta_{2,{\bf dS}} e^{-i\sigma}\Bigg|^2,
\\
\Sigma_{{\zeta\Pi_\zeta},{\bf dS}}&=&(1-\epsilon)  \frac{H^2}{8\pi^2\epsilon  c_s M_p^2}\Bigg(\frac{k_e}{k_s}\Bigg)^6 
\Bigg[-3 (1+\sigma^2)+\sigma^2\Bigg]\Bigg\{\Bigg|\alpha_{2,{\bf dS}} e^{i\sigma}
 -\beta_{2,{\bf dS}} e^{-i\sigma}\Bigg|^2   \nonumber
 \\&& \quad\quad\quad\quad\quad\quad\quad\quad\quad \quad \quad \quad 
+\frac{6\sigma^2}{\Bigg\{3 (1+\sigma^2)-\sigma^2\Bigg\}}\Bigg\{\alpha_{2,{\bf dS}}^*\beta_{2,{\bf dS}}e^{-2i\sigma}
 +\beta_{2,{\bf dS}}^*\alpha_{2,{\bf dS}} e^{2i\sigma} \Bigg\}\Bigg\},
\\
\Sigma_{{\Pi_\zeta\zeta},{\bf dS}}&=&(1-\epsilon) \frac{H^2}{8\pi^2\epsilon  c_s M_p^2}\Bigg(\frac{k_e}{k_s}\Bigg)^6 
 \Bigg[-3(1+\sigma^2)+\sigma^2\Bigg]\Bigg\{\Bigg|\alpha_{2,{\bf dS}} e^{i\sigma}
 -\beta_{2,{\bf dS}} e^{-i\sigma}\Bigg|^2 \nonumber
 \\&& \quad  \quad  \quad  \quad \quad\quad\quad\quad\quad\quad\quad\quad
+ \frac{6\sigma^2}{\Bigg\{3(1+\sigma^2)-\sigma^2\Bigg\}}\Bigg\{\alpha_{2,{\bf dS}}^*\beta_{2,{\bf dS}} e^{-2i\sigma}+\beta_{2,{\bf dS}}^*\alpha_{2,{\bf dS}} e^{2i\sigma} \Bigg\}\Bigg\},
\\
\Sigma_{{\Pi_\zeta\Pi_\zeta},{\bf dS}}&=&(1-\epsilon) \frac{H^2}{8\pi^2 \epsilon  
 c_s M_p^2}\Bigg(\frac{k_e}{k_s}\Bigg)^6 
  \Bigg[\Bigg(3- \sigma^2 \Bigg)^2+ 9 \sigma^2 \Bigg ]
 \Bigg|\alpha_{2,{\bf dS}} e^{i\sigma}-\beta_{2,{\bf dS}} \frac{\Bigg\{3(1+i \sigma )- \sigma^2 \Bigg \}}{\Bigg\{3(1-i \sigma)- \sigma^2 \Bigg \}} e^{-i\sigma}\Bigg|^2.
\eea

\subsection{Results in SRII region}
\label{appCc}

\subsubsection{Power spectrum in SRII}
\label{appCc1}

The upcoming expressions for the different elements of the scalar power spectrum are calculated using the general mode function solutions for the second slow-roll (SRII) phase as presented in eqs.(\ref{modezetaSR2}-\ref{modepiSR2}). The following power spectrum elements are evaluated at the Horizon crossing condition for an arbitrary wavenumber $k$ as $-kc_{s}\tau=\sigma$ where $\sigma$ is the stochastic, coarse-graining parameter.
\bea
\Delta_{\zeta\zeta}^2&=&2^{2\nu-3}  (\sigma)^{3-2\nu} \frac{H^2}{8\pi^2\epsilon  c_s M_p^2} 
 \Bigg(\frac{k_e}{k_s}\Bigg)^6 \Bigg|\frac{\Gamma{(\nu)}}{\Gamma{(\frac{3}{2})}}\Bigg|^2(1+\sigma^2)\Bigg|\alpha_3 e^{-i(\frac{\pi}{2}(\nu+\frac{1}{2})-\sigma)}-\frac{(1+i \sigma)}{(1-i \sigma)}\beta_3 e^{i(\frac{\pi}{2}(\nu+\frac{1}{2})-\sigma)}\Bigg|^2,
\\
\Delta_{\zeta\Pi_\zeta}^2&=&2^{2\nu-3}  (\sigma)^{3-2\nu} \frac{H^2}{8\pi^2\epsilon  c_s M_p^2}\Bigg(\frac{k_e}{k_s}\Bigg)^6 
 \Bigg|\frac{\Gamma{(\nu)}}{\Gamma{(\frac{3}{2})}}\Bigg|^2\Bigg\{\Bigg(\frac{3}{2}-\nu \Bigg)(1+\sigma^2)+\sigma^2\Bigg\}\Bigg\{\Bigg|\alpha_3 e^{-i(\frac{\pi}{2}(\nu+\frac{1}{2})-\sigma)}-\beta_3 e^{i(\frac{\pi}{2}(\nu+\frac{1}{2})-\sigma)}\Bigg|^2\nonumber
 \\&& \quad \quad \quad \quad \quad \quad \quad \quad \quad 
+\frac{2\Bigg(\frac{3}{2}-\nu\Bigg)\sigma^2}{\Bigg\{\Bigg(\frac{3}{2}-\nu \Bigg) (1+\sigma^2)+\sigma^2\Bigg\}}\Bigg\{\alpha_3^*\beta_3 e^{2i(\frac{\pi}{2}(\nu+\frac{1}{2})-\sigma)}+\beta_3^*\alpha_3 e^{-2i(\frac{\pi}{2}(\nu+\frac{1}{2})-\sigma)} \Bigg\}\Bigg\},
\eea\bea
\Delta_{\Pi_\zeta\zeta}^2&=&2^{2\nu-3}  (\sigma)^{3-2\nu} \frac{H^2}{8\pi^2\epsilon  c_s M_p^2}\Bigg(\frac{k_e}{k_s}\Bigg)^6 
 \Bigg|\frac{\Gamma{(\nu)}}{\Gamma{(\frac{3}{2})}}\Bigg|^2\Bigg\{\Bigg(\frac{3}{2}-\nu \Bigg)(1+\sigma^2)+\sigma^2\Bigg\}\Bigg\{\Bigg|\alpha_3 e^{-i(\frac{\pi}{2}(\nu+\frac{1}{2})-\sigma)}-\beta_3 e^{i(\frac{\pi}{2}(\nu+\frac{1}{2})-\sigma)}\Bigg|^2\nonumber
 \\ && \quad \quad \quad \quad \quad \quad \quad \quad \quad 
+ \frac{2\Bigg(\frac{3}{2}-\nu\Bigg)\sigma^2}{\Bigg\{\Bigg(\frac{3}{2}-\nu \Bigg) (1+\sigma^2)+\sigma^2\Bigg\}}\Bigg\{\alpha_3^*\beta_3 e^{2i(\frac{\pi}{2}(\nu+\frac{1}{2})-\sigma)}+\beta_3^*\alpha_3 e^{-2i(\frac{\pi}{2}(\nu+\frac{1}{2})-\sigma)} \Bigg\}\Bigg\},
\\
\Delta_{\Pi_\zeta\Pi_\zeta}^2&=&2^{2\nu-3}(\sigma)^{3-2\nu}\frac{H^2}{8\pi^2 \epsilon  
 c_s M_p^2}\Bigg(\frac{k_e} {k_s}\Bigg)^6 
 \Bigg|\frac{\Gamma{(\nu)}}{\Gamma{(\frac{3}{2})}}\Bigg|^2 \Bigg[\Bigg(\Bigg(\frac{3}{2}- \nu\Bigg)+ \sigma^2 \Bigg)^2+\Bigg(\frac{3}{2}-\nu\Bigg)^2 \sigma^2 \Bigg ] \nonumber
 \\ && \quad \quad \quad \quad \quad \quad \quad \quad \quad \quad \quad 
 \Bigg|\alpha_3 e^{-i(\frac{\pi}{2}(\nu+\frac{1}{2})-\sigma)}-\beta_3 \frac{\Bigg\{\Bigg(\frac{3}{2}- \nu\Bigg)(1+i \sigma )+ \sigma^2 \Bigg \}}{\Bigg\{\Bigg(\frac{3}{2}- \nu\Bigg)(1-i \sigma)+ \sigma^2 \Bigg \}} e^{i(\frac{\pi}{2}(\nu+\frac{1}{2})-\sigma)}\Bigg|^2.
\eea

The results with an arbitrary `$\nu$' under the choice of initial Bunch Davies vacuum condition have their Bogoliubov coefficients replaced from $(\alpha_{3},\beta_{3})$ to $(\alpha_{3,{\bf BD}},\beta_{3,{\bf BD}})$.
Further implementing the limiting value $\nu=3/2 $ in the derived result of $(\alpha_{3,{\bf BD}},\beta_{3,{\bf BD}})$ for the SRII, one can get the simplified expression for power spectrum elements in the case of exact de Sitter space with the new Bogoliubov coefficients $(\alpha_{3,{\bf dS}},\beta_{3,{\bf dS}})$:
\bea \label{pspecsr2dS}
\Delta_{{\zeta\zeta},{\bf dS}}^2&=&\frac{H^2}{8\pi^2\epsilon  c_s M_p^2} 
\Bigg(\frac{k_e}{k_s}\Bigg)^6 (1+\sigma^2)\Bigg|\alpha_{3,{\bf dS}} e^{i\sigma}-\frac{(1+i \sigma)}{(1-i \sigma)}\beta_{3,{\bf dS}} e^{-i\sigma}\Bigg|^2,
\\
\Delta_{{\zeta\Pi_\zeta},{\bf dS}}^2&=& \frac{H^2}{8\pi^2\epsilon  c_s M_p^2}\Bigg(\frac{k_e}{k_s}\Bigg)^6 
\sigma^2 \Bigg|\alpha_{3,{\bf dS}} e^{i\sigma}-\beta_{3,{\bf dS}} e^{-i\sigma}\Bigg|^2,
\\
\Delta_{{\Pi_\zeta\zeta},{\bf dS}}^2&=& \frac{H^2}{8\pi^2\epsilon  c_s M_p^2}\Bigg(\frac{k_e}{k_s}\Bigg)^6 
\sigma^2\Bigg|\alpha_{3,{\bf dS}} e^{i\sigma}-\beta_{3,{\bf dS}} e^{-i\sigma}\Bigg|^2,
\\
\Delta_{{\Pi_\zeta\Pi_\zeta},{\bf dS}}^2&=&\frac{H^2}{8\pi^2 \epsilon  
 c_s M_p^2}\Bigg(\frac{k_e} {k_s}\Bigg)^6 
 \sigma^4 
 \Bigg|\alpha_{3,{\bf dS}} e^{i\sigma}-\beta_{3,{\bf dS}}  e^{-i\sigma}\Bigg|^2.
\eea

\subsubsection{Noise Matrix elements in SRII}
\label{appCc2}

The noise correlation matrix elements in the SRII for an arbitrary de Sitter background spacetime, characterized by `$\nu$' and a general initial quantum vacuum state specified with coefficients $(\alpha_{3},\beta_{3})$  can be written using the following expressions:
\bea
\Sigma_{\zeta\zeta}&=&(1-\epsilon) 2^{2\nu-3}  (\sigma)^{3-2\nu} \frac{H^2}{8\pi^2\epsilon  c_s M_p^2} 
 \Bigg(\frac{k_e}{k_s}\Bigg)^6 \Bigg|\frac{\Gamma{(\nu)}}{\Gamma{(\frac{3}{2})}}\Bigg|^2(1+\sigma^2)\Bigg|\alpha_3 e^{-i(\frac{\pi}{2}(\nu+\frac{1}{2})-\sigma)}-\frac{(1+i \sigma)}{(1-i \sigma)}\beta_3 e^{i(\frac{\pi}{2}(\nu+\frac{1}{2})-\sigma)}\Bigg|^2,
\\
\Sigma_{\zeta\Pi_\zeta}&=&(1-\epsilon) 2^{2\nu-3}  (\sigma)^{3-2\nu} \frac{H^2}{8\pi^2\epsilon  c_s M_p^2}\Bigg(\frac{k_e}{k_s}\Bigg)^6 
 \Bigg|\frac{\Gamma{(\nu)}}{\Gamma{(\frac{3}{2})}}\Bigg|^2\Bigg\{\Bigg(\frac{3}{2}-\nu \Bigg)(1+\sigma^2)+\sigma^2\Bigg\}\Bigg\{\Bigg|\alpha_3 e^{-i(\frac{\pi}{2}(\nu+\frac{1}{2})-\sigma)}-\beta_3 e^{i(\frac{\pi}{2}(\nu+\frac{1}{2})-\sigma)}\Bigg|^2, \nonumber
 \\  && \quad \quad \quad \quad \quad \quad \quad \quad \quad \quad \quad 
+\frac{2\Bigg(\frac{3}{2}-\nu\Bigg)\sigma^2}{\Bigg\{\Bigg(\frac{3}{2}-\nu \Bigg) (1+\sigma^2)+\sigma^2\Bigg\}}\Bigg\{\alpha_3^*\beta_3 e^{2i(\frac{\pi}{2}(\nu+\frac{1}{2})-\sigma)}+\beta_3^*\alpha_3 e^{-2i(\frac{\pi}{2}(\nu+\frac{1}{2})-\sigma)} \Bigg\}\Bigg\},
\eea\bea
\Sigma_{\Pi_\zeta\zeta}&=&(1-\epsilon) 2^{2\nu-3}  (\sigma)^{3-2\nu} \frac{H^2}{8\pi^2\epsilon  c_s M_p^2}\Bigg(\frac{k_e}{k_s}\Bigg)^6 
 \Bigg|\frac{\Gamma{(\nu)}}{\Gamma{(\frac{3}{2})}}\Bigg|^2\Bigg\{\Bigg(\frac{3}{2}-\nu \Bigg)(1+\sigma^2)+\sigma^2\Bigg\}\Bigg\{\Bigg|\alpha_3 e^{-i(\frac{\pi}{2}(\nu+\frac{1}{2})-\sigma)}-\beta_3 e^{i(\frac{\pi}{2}(\nu+\frac{1}{2})-\sigma)}\Bigg|^{2}\nonumber
 \\ && \quad \quad \quad \quad \quad \quad \quad \quad \quad \quad \quad 
+ \frac{2\Bigg(\frac{3}{2}-\nu\Bigg)\sigma^2}{\Bigg\{\Bigg(\frac{3}{2}-\nu \Bigg) (1+\sigma^2)+\sigma^2\Bigg\}}\Bigg\{\alpha_3^*\beta_3 e^{2i(\frac{\pi}{2}(\nu+\frac{1}{2})-\sigma)}+\beta_3^*\alpha_3 e^{-2i(\frac{\pi}{2}(\nu+\frac{1}{2})-\sigma)} \Bigg\}\Bigg\},\quad\quad
\\
\Sigma_{\Pi_\zeta\Pi_\zeta}&=&(1-\epsilon) 2^{2\nu-3}(\sigma)^{3-2\nu}\frac{H^2}{8\pi^2 \epsilon  
 c_s M_p^2}\Bigg(\frac{k_e}{k_s}\Bigg)^6 
 \Bigg|\frac{\Gamma{(\nu)}}{\Gamma{(\frac{3}{2})}}\Bigg|^2 \Bigg[\Bigg(\Bigg(\frac{3}{2}- \nu\Bigg)+ \sigma^2 \Bigg)^2+\Bigg(\frac{3}{2}-\nu\Bigg)^2 \sigma^2 \Bigg ] \nonumber
 \\ &&  \quad \quad \quad \quad \quad \quad \quad \quad \quad \quad \quad \quad \quad \quad 
 \Bigg|\alpha_3 e^{-i(\frac{\pi}{2}(\nu+\frac{1}{2})-\sigma)}-\beta_3 \frac{\Bigg\{\Bigg(\frac{3}{2}- \nu\Bigg)(1+i \sigma )+ \sigma^2 \Bigg \}}{\Bigg\{\Bigg(\frac{3}{2}- \nu\Bigg)(1-i \sigma)+ \sigma^2 \Bigg \}} e^{i(\frac{\pi}{2}(\nu+\frac{1}{2})-\sigma)}\Bigg|^{2}.
\eea
We further implement the Bunch Davies vacuum conditions followed by the limiting value of $\nu=3/2$ in the Bunch Davies vacuum derived result to get the new set $(\alpha_{3, {\bf dS}},\beta_{3, {\bf dS}})$ and the simplified expression for noise matrix elements in exact de Sitter space becomes:
\bea
\Sigma_{{\zeta\zeta},{\bf dS}}&=&(1-\epsilon) \frac{H^2}{8\pi^2\epsilon  c_s M_p^2} 
 \Bigg(\frac{k_e}{k_s}\Bigg)^6 (1+\sigma^2)\Bigg|\alpha_{3,{\bf dS}} e^{i\sigma}-\frac{(1+i \sigma)}{(1-i \sigma)}\beta_{3,{\bf dS}} e^{-i\sigma}\Bigg|^2,
\\
\Sigma_{{\zeta\Pi_\zeta},{\bf dS}}&=&(1-\epsilon) \frac{H^2}{8\pi^2\epsilon  c_s M_p^2}\Bigg(\frac{k_e}{k_s}\Bigg)^6 
 \sigma^2 \Bigg|\alpha_{3,{\bf dS}} e^{i\sigma}-\beta_{3,{\bf dS}} e^{-i\sigma}\Bigg|^2, 
\\
\Sigma_{{\Pi_\zeta\zeta},{\bf dS}}&=&(1-\epsilon)  \frac{H^2}{8\pi^2\epsilon  c_s M_p^2}\Bigg(\frac{k_e}{k_s}\Bigg)^6 
\sigma^2 \Bigg|\alpha_{3,{\bf dS}} e^{i\sigma}-\beta_{3,{\bf dS}} e^{-i\sigma}\Bigg|^2,
\\
\Sigma_{{\Pi_\zeta\Pi_\zeta},{\bf dS}}&=&(1-\epsilon) \frac{H^2}{8\pi^2 \epsilon  
 c_s M_p^2}\Bigg(\frac{k_e}{k_s}\Bigg)^6 
  \sigma^4  
 \Bigg|\alpha_{3,{\bf dS}} e^{i\sigma}-\beta_{3,{\bf dS}} e^{-i\sigma}\Bigg|^2.
\eea

\newpage

\bibliography{Refs}
\bibliographystyle{utphys}

\end{document}